\documentclass[rivista]{cimento_nostro}

\usepackage{graphicx}  
\usepackage{ifpdf}
\usepackage{amsmath}
\usepackage{bm}
\usepackage{color}
\usepackage[english]{babel}
\usepackage{amsfonts}
\usepackage{amssymb}
\usepackage{enumerate}
\usepackage{color}
\usepackage{slashed}
\usepackage[numbers, sort]{natbib}
\usepackage{multirow}
\usepackage{float}
\usepackage{lineno}
\usepackage{rotating}

\newcommand{\meg}{\ifmmode\mu^+ \to e^+ \gamma\else$\mu^+ \to e^+ \gamma$\fi}

\newcommand{\muenng}{$\mu^{+}\rightarrow
e^{+}\nu_e\overline{\nu}_{\mu}\gamma$~}

\newcommand{\muenn}{$\mu^{+}\rightarrow e^{+}\nu_e\overline{\nu}_{\mu}$~}

\newcommand{\ord}[1]{\mathcal{O}\left( #1 \right)}

\newcommand{\eps}{\epsilon}
\newcommand{\vp}{\Phi}
\newcommand{\vpj }{\mbox{${\vp^\dag i~\raisebox{1.5mm}{${}^\leftrightarrow$}\hspace{-3mm} D_\mu\,\vp}$}}
\newcommand{\vpjt}{\mbox{${\vp^\dag i~\raisebox{1.5mm}{${}^\leftrightarrow$}\hspace{-3mm} D_\mu^{\,I}\,\vp}$}}

\title{Charged Lepton Flavour Violation: \\An Experimental and Theoretical Introduction}
\shorttitle{An Experimental and Theoretical Introduction to CLFV}
\author{Lorenzo Calibbi\from{ins:x}\thanks{{calibbi@itp.ac.cn}}
\atque
Giovanni Signorelli\from{ins:y}\thanks{giovanni.signorelli@pi.infn.it}}

\instlist{\inst{ins:x} CAS Key Laboratory of Theoretical Physics, Institute of Theoretical Physics, \\Chinese Academy of Sciences - Beijing, China

\inst{ins:y} Istituto Nazionale di Fisica Nucleare, Sezione di Pisa - Largo Bruno Pontecorvo 3 - 56127 Pisa, Italy}

\begin{document}

\maketitle

\begin{abstract}
Charged lepton flavour violating transitions would be a clear signal of new physics beyond the Standard Model. Their search has been carried out in a variety of channels, the most sensitive being those involving a muon: however no positive evidence has been found so far. The MEG experiment has recently set the best limit on such processes by investigating the existence of the $\mu \to e\gamma$ decay. 
In the next decade several experiments are planned to pursue the search for  $\mu \to e\gamma$, $\mu \to eee$, $\mu \to e$ conversion in nuclei, as well as on processes involving the $\tau$, to an unprecedented level of precision. In this review we want to give a pedagogical introduction on the theoretical motivations for such searches as well as on the experimental aspects upon which they are based.
\end{abstract}

\begin{center}
{\it Invited review for La Rivista del Nuovo Cimento}
\end{center}

\newpage
\tableofcontents

\section{Introduction}

\subsection{A brief history of flavour}
The concept of flavour, {\em i.e.}~the existence of three replicas of each family of elementary fermions, is a cornerstone in the physics of elementary particles, and it is realized in the Standard Model introducing three copies of the same gauge representations of the fermion fields.

We can track the origin of the flavour concept to exactly 70 years ago. At the beginning of 1947, in fact, the experiment of Conversi, Pancini and Piccioni (CPP)~\cite{CPP} showed that negative muons (then called negative ``mesotrons''), discovered by Anderson and Neddermeyer in cosmic radiation~\cite{Neddermeyer}, did not undergo nuclear capture in light materials, but decayed in a fashion similar to those of positive charge, yielding electrons. Therefore, contrarily to what it was thought for over a decade, they could not be the quanta mediating the strong force as predicted by Yukawa~\cite{Yukawa}. 

In May~1947 Powell and his group reported the evidence~\cite{Powell} of the two-step decay which we now write as $\pi \to \mu \to e$, telling the $\pi$-meson (now called pion) from the $\mu$-meson (muon), which reconciled the Yukawa meson theory with the CPP measurement and with the microsecond lifetime of the particle generating the electrons measured at sea level.

A couple of weeks earlier Bruno Pontecorvo, in a letter to Gian Carlo Wick dated 8 May 1947~\cite{wick}, had made the hypothesis that the mesotron (meaning the muon) be a ``sort of isomer'' of the electron, and discussed the opportunity to search for its decay in electron plus photon or electron plus a spin-zero particle.
The idea that there could be a ``second generation'' of elementary fermions gained further momentum that year, after the presentation of Powell's results in Copenhagen, when the properties of the muon started to be seen as suggestive that there could be a lepton spectrum with the electron being a sort of ground state~\cite{Pais}.

In December 1947 Rochester and Butler reported~\cite{rochester} the discovery of unusual events in their cloud chamber cosmic ray pictures that they called the V-particles because they appeared as two opposite-charge tracks originating from nothing (implying V-particles were neutral). Despite it took a few years more for physicists to realize it, we now know they discovered the neutral kaon and with it the second generation of quarks: year 1947 ended with the hint of a second generation of elementary fermions.

At the end of 1950 it was already clear~\cite{powell-rpp} that muon and pion were two distinct particles with different masses, lifetimes and decays ($m$ expressed in electron masses, $\tau$ in seconds; modern values in parentheses):
\begin{align}
\mu^+ : \qquad  m = 212~(207), \qquad \tau = 2.1 (2.2) \times 10^{-6}, \qquad \mu^+ \to e^+ + 2\nu\\
\pi^+ : \qquad  m = 276~(277), \qquad \tau = 1.6 (2.6) \times 10^{-8}, \qquad \pi^+ \to \mu^+ + \nu\
\end{align}
where ``$\nu$ represents any particle of small rest-mass not a photon\footnote{Funnily enough in the original caption to Table 4 in~\cite{powell-rpp} Powell wrote ``$\nu$ represents any particle of small rest-mass  not a {\em proton}''!}''. The decay $\mu \to e \gamma$ was sought for but not found at the percent level (more on this later) and moreover it was not then clear whether the two neutral decay products of the muon where two identical particles or a particle and an antiparticle. This would have given two different energy spectra for the decay positron~\cite{Michel1}.
As clarified in the next years, the decay was $\mu^\pm \to e^\pm \nu \bar \nu$ giving raise to the principle of conservation of leptons, a concept originated with a paper by Konopinski and Mahmoud~\cite{konopinski} in 1953. 

While on the hadron side we had to wait until the mid-70s to build up the complete second generation of quarks (and the quark concept itself as elementary fermion!) passing first through the GIM mechanism \cite{Glashow:1970gm} and the final discovery of the charm quark, on the leptonic side things were proceeding quicker.

In 1955 the upper limit on the branching ratio of  $\mu^+ \to e^+ \gamma$ was set to below $2 \times 10^{-5}$ at Columbia University Nevis cyclotron~\cite{loko55}. In the meanwhile, with the discovery of parity violation, it was suggested that the weak interaction takes place through the exchange of charged intermediate bosons, and Feinberg~\cite{fein58} in 1958 pointed out that, if those bosons existed, would cause $\mu^+ \to e^+ \gamma$ at a level of $10^{-4}$ through a loop involving the neutrino and the (then hypothetical) heavy boson. 
This directly lead to the two-neutrino hypothesis~\cite{numunue}, 
in which the neutrino coupled to the muon is different from that coupled to the electron, and the \meg\ process  forbidden. The two-neutrino hypothesis
was verified experimentally at the Brookhaven National Laboratory (BNL) by
confirming muon production, and no electron production, from the
scattering of neutrinos coming from pion decays~\cite{danb62}. 
The fact that the existence of two neutrinos implied two conserved quantum numbers, muon number ($L_{\mu} = +1$ for $\mu^-$ and $\nu_\mu$)  and electron number ($L_{e} = +1$ for $e^-$ and $\nu_e$) seems obvious now, but had to wait a few years to be clarified~\cite{pontecorvo-lf}.

In contrast, in the hadronic sector, transitions between different generations \cite{Cabibbo:1963yz} were observed albeit restricted to the weak interaction. The unexpected observation of CP violation in the decay of neutral kaons~\cite{ccft} brought forward the hypothesis of third generation of quarks~\cite{kobayashi-maskawa}, which gained strength after the discovery of the $\tau$ and of third family of leptons, and the realization that only an equal number of quark and
lepton generations make the theory safe from quantum anomalies \cite{Bouchiat:1972iq}.
It took a few more years to get the picture complete (the $b$ quark made its appearance in 1977~\cite{lederman}, the $t$ quark in 1995~\cite{top, top2} and finally the $\nu_\tau$ in 2000~\cite{nutau}).
\begin{table}[bt]
\begin{center}
\begin{tabular}{lcc}
\hline
up-type quarks ``$u$''& $3_{+2/3}$ & $u$, $c$, $t$\\
down-type quarks ``$d$'' & $3_{-1/3}$ & $d$, $s$, $b$ \\
charged leptons ``$\ell$'' or ``$e$'' & $1_{-1}$ & $e$, $\mu$, $\tau$ \\
neutrinos ``$\nu$''& $1_{0} $ & $\nu_e$, $\nu_\mu$, $\nu_\tau$\\
\hline
\end{tabular}
\caption{\label{tab:flavours}Elementary fermions of the Standard Model, with corresponding representation with respect $SU(3)_{c} \times U(1)_{em}$. In the following we will refer to the up-type quarks and down-type quarks collectively by $u$ and $d$ respectively.}
\end{center}
\end{table}

Now we include the existence of the three generations in the Standard Model (see Table~\ref{tab:flavours}) where the only terms that distinguish between different generations are the Yukawa interactions. As we will discuss in some detail in the next section, the conservation of lepton flavour is accidental, not related to the gauge structure of the theory but arising from its particle content, in particular from the absence of right-handed neutrinos. The discovery of neutrino oscillations~\cite{Fukuda:1998mi,Ahmad:2001an,Ahmad:2002jz} has however demonstrated that the lepton family numbers are not conserved even though, as we shall see later, this has no practical effect in inducing transitions among charged leptons in the framework of the Standard Model. 

\begin{table}[phbt!]
\renewcommand{\arraystretch}{1.1}
\begin{center}
\begin{tabular}{llcccc}
\hline
Reaction & Present limit & C.L. & Experiment & Year & Reference \cr
\hline
$\mu^+ \to e^+ \gamma$ & $< 4.2 \times 10^{-13}$ & 90\% & MEG at PSI & 2016 & \cite{MEGLimit} \cr
$\mu^+ \to e^+ e^- e^+$  & $< 1.0 \times 10^{-12}$ & 90\%  &  SINDRUM & 1988 & \cite{sindrum} \cr
$\mu^{-}{\rm Ti}\rightarrow e^{-}{\rm Ti}~{}^\dag$ &  $< 6.1 \times 10^{-13}$ &  90\% & SINDRUM II& 1998 & \cite{wint98} \cr
$\mu^{-}{\rm Pb}\rightarrow e^{-}{\rm Pb}~{}^\dag$ &  $< 4.6 \times 10^{-11}$ & 90\% & SINDRUM~II & 1996  &  \cite{honecker} \cr
$\mu^{-}{\rm Au}\rightarrow e^{-}{\rm Au}~{}^\dag$  & $< 7.0 \times 10^{-13}$ & 90\% & SINDRUM~II & 2006 & \cite{Bertl} \cr
$\mu^{-}{\rm Ti}\rightarrow e^{+}{\rm Ca}^{\ast}~{}^\dag$ & $< 3.6 \times 10^{-11}$ & 90\%  & SINDRUM~II & 1998 & \cite{mucalcio} \cr
$\mu^{+}e^{-} \rightarrow \mu^{-}e^{+}$ & $< 8.3 \times 10^{-11}$ &  90\% & SINDRUM & 1999 &  \cite{will99} \cr
$\tau\rightarrow e\gamma$   & $< 3.3 \times 10^{-8}$ & 90\% & BaBar & 2010 & \cite{babar-tau} \cr
$\tau\rightarrow \mu\gamma$ & $< 4.4 \times 10^{-8}$ &  90\% & BaBar & 2010 & \cite{babar-tau} \cr
$\tau\rightarrow eee$ & $< 2.7 \times 10^{-8}$ &90\% &  Belle & 2010&  \cite{belle-3mu} \cr 
$\tau\rightarrow \mu\mu\mu$ & $< 2.1 \times 10^{-8}$ &  90\% & Belle & 2010 & \cite{belle-3mu} \cr
$\tau\rightarrow  \pi^0 e$ & $< 8.0 \times 10^{-8}$ &90\% &  Belle & 2007&  \cite{tau-pi-e} \cr 
$\tau\rightarrow \pi^0 \mu $ & $< 1.1 \times 10^{-7}$ &90\% &  BaBar & 2007&  \cite{tau-pi-mu} \cr 
$\tau\rightarrow \rho^0 e$ & $< 1.8 \times 10^{-8}$ &90\% &  Belle & 2011&  \cite{tau-rho} \cr 
$\tau\rightarrow  \rho^0 \mu$ & $< 1.2 \times 10^{-8}$ &90\% &  Belle & 2011&  \cite{tau-rho} \cr 
\hline
$\pi^{0} \rightarrow \mu e$ & $< 3.6 \times 10^{-10}$ & 90\% & KTeV & 2008 & \cite{pi0-mue} \cr
$K^{0}_{L}\rightarrow \mu e$ & $< 4.7 \times 10^{-12}$ & 90\%  & BNL E871 & 1998 & \cite{ambr98} \cr
$K^{0}_{L}\rightarrow\pi^{0}\mu^{+}e^{-}$ & $< 7.6 \times 10^{-11}$ & 90\% & KTeV & 2008 & \cite{pi0-mue} \cr
$K^{+}\rightarrow\pi^{+}\mu^{+}e^{-}$ & $< 1.3 \times 10^{-11}$ & 90\% & BNL E865 & 2005 & \cite{kplus} \cr
$J/\psi\to \mu e$    & $< 1.5 \times 10^{-7}$ & 90\% & BESIII & 2013 &\cite{Jpsi-emu} \cr
$J/\psi\to \tau e$    & $< 8.3 \times 10^{-6}$ & 90\% & BESII & 2004 &\cite{Jpsi-tau} \cr
$J/\psi\to \tau \mu$    & $< 2.0 \times 10^{-6}$ & 90\% & BESII & 2004 &\cite{Jpsi-tau} \cr
$B^0\rightarrow \mu e$    & $< 2.8 \times 10^{-9}$ & 90\% & LHCb & 2013 &\cite{LHCb-emu} \cr
$B^0\rightarrow \tau e$   & $< 2.8 \times 10^{-5}$ & 90\% &BaBar & 2008 &\cite{Btaue} \cr
$B^0\rightarrow \tau\mu$  & $< 2.2 \times 10^{-5}$ & 90\% &BaBar & 2008 &\cite{Btaue} \cr
$B \rightarrow K \mu e ~{}^\ddag$    & $< 3.8 \times 10^{-8}$ & 90\% & BaBar & 2006 &\cite{BtoK} \cr
$B \rightarrow K^{*} \mu e~{}^\ddag$    & $< 5.1 \times 10^{-7}$ & 90\% & BaBar & 2006 &\cite{BtoK} \cr
$B^+ \rightarrow K^+ \tau\mu $    & $< 4.8 \times 10^{-5}$ & 90\% & BaBar & 2012 &\cite{BtoKtau} \cr
$B^+ \rightarrow K^+ \tau e $    & $< 3.0 \times 10^{-5}$ & 90\% & BaBar & 2012 &\cite{BtoKtau} \cr
$B_s^0\rightarrow \mu e$    & $< 1.1 \times 10^{-8}$ & 90\% & LHCb & 2013 &\cite{LHCb-emu} \cr
$\Upsilon(1s)\rightarrow \tau\mu $    & $< 6.0 \times 10^{-6}$ & 95\% & CLEO & 2008 &\cite{Y-taumu} \cr
\hline
$Z\rightarrow \mu e$    & $< 7.5 \times 10^{-7}$ & 95\% & LHC ATLAS & 2014 & \cite{atlasZ0} \cr
$Z\rightarrow \tau e$   & $< 9.8 \times 10^{-6}$ & 95\%  & LEP OPAL & 1995 & \cite{opalZ0} \cr
$Z\rightarrow \tau\mu$  & $< 1.2 \times 10^{-5}$ & 95\%  &LEP DELPHI & 1997 &\cite{delphiZ0} \cr
$h\rightarrow e\mu$  & $< 3.5 \times 10^{-4}$ &  95\%  & LHC CMS & 2016 & \cite{higgsmue} \cr
$h\rightarrow \tau\mu$  & $< 2.5 \times 10^{-3}$ &95\% & LHC CMS & 2017 & \cite{higgsmutau} \cr
$h\rightarrow \tau e$  & $< 6.1 \times 10^{-3}$ & 95\% & LHC CMS &  2017 &\cite{higgsmutau} \cr
\hline
\end{tabular}
\end{center}
\caption{Limits for the branching ratio of charged lepton flavour violating 
processes of leptons, mesons, and heavy bosons. More extensive lists of  $B$-meson and $\tau$ CLFV decays (including all hadronic modes) can be found in \cite{PDG,HFLAV}.
${}^\dag$Rate normalised to the muon capture rate by the nucleus, see Eq.~(\ref{eq:CRdef}). ${}^\ddag$$B$-charge averaged modes.}
\label{tab:limits}
\end{table}
\subsection{Enter Charged Lepton Flavour Violation}
There is  still no evidence of Lepton Flavour Violation (LFV)\footnote{Unlike in the quark sector, where the concept of flavour (associated to quantum numbers preserved by strong and electromagnetic interactions, but not by the weak ones: {\em strangeness}, {\em charm} etc.)~and that of family or generation are somewhat different, in the lepton sector the terms flavour, family, and generation are analogous, so we will use them interchangeably. In fact, in the early literature it is not uncommon to read about `Lepton Family Violation'.}  in processes involving charged leptons, despite the search for such a violation has been pursued to date in a host of channels both at dedicated and general purpose experiments, some of which are reported in Table~\ref{tab:limits}.

Why it is so and which are the present and future characteristics 
of experiments searching for Charged Lepton Flavour Violation (CLFV) is the aim of the present review. We first give a theoretical introduction to set the stage and to see in a more formal and detailed way what we mentioned above, as well as to discuss how and why Lepton Flavour can be violated in extensions of the Standard Model: what, in other words, makes CLFV processes so sensitive to new physics. 
\begin{figure} [t]
\begin{center}
\includegraphics[width=0.99\columnwidth]{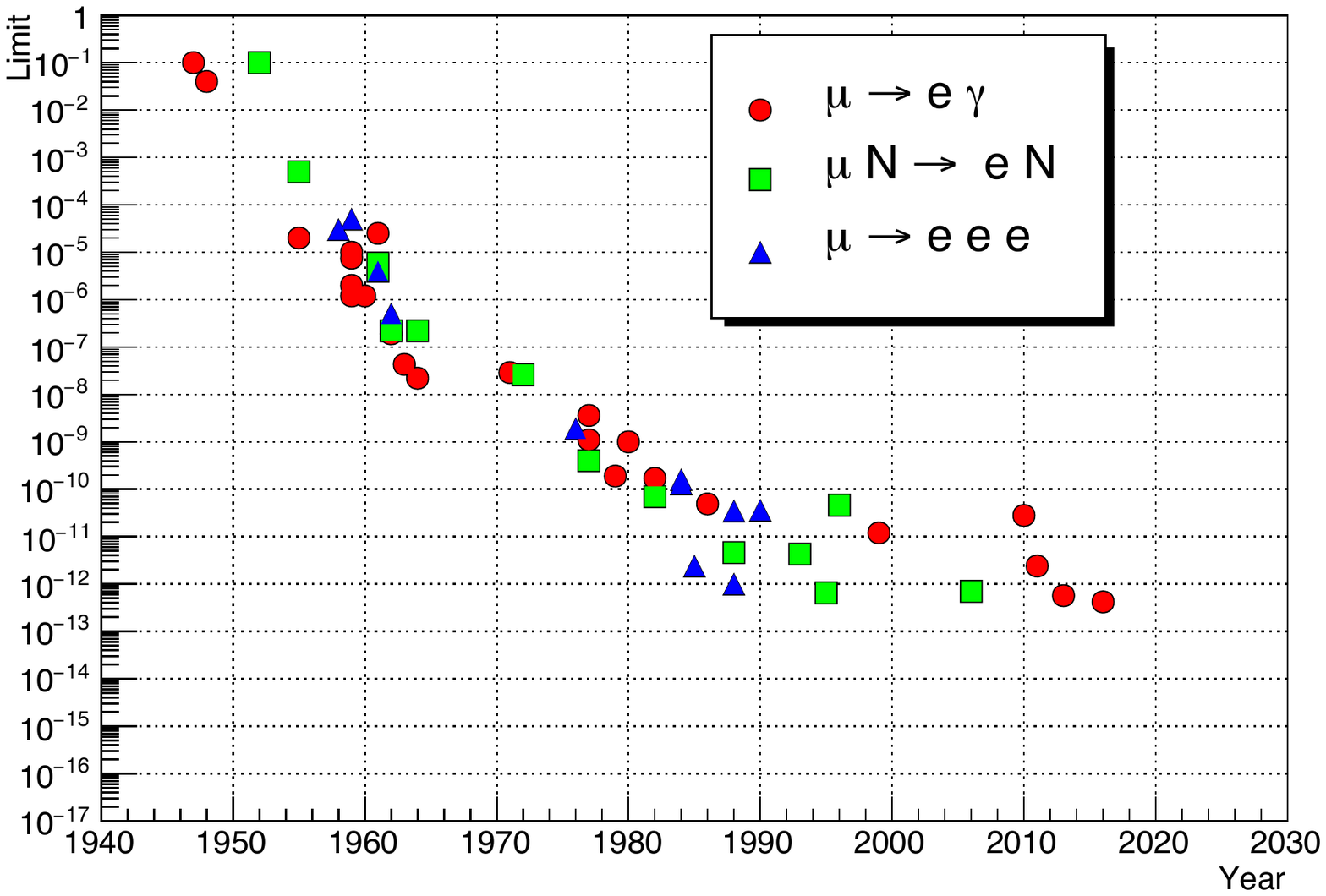}
\caption{\label{fig:calando}Limit on the branching ratio of flavour violating muon decays as a function of the year. The three main clusters correspond to the usage of cosmic ray muons (until the 1950s), stopped pion beams (until the 1970s) and stopped muon beams. Presently the best limit is that on the \meg\ decay set by the MEG experiment~\cite{MEGLimit}.
}
\end{center}
\end{figure}

We will then review the general aspects of the experimental searches and discuss some of the present and planned experiments with particular emphasis on the transition between the first and the second family of leptons.
To this class, in fact, belong the three most searched modes -- $\mu^+ \to e^+ \gamma$ (``mu-to-e-gamma''), $\mu^- N \to e^- N$ (``mu-e-conversion''), and $\mu^+ \to e^+ e^- e^+$ (``mu-to-three-e'') -- due to the copious availability of the parent particle in the cosmic radiation first and at dedicated accelerators afterwards. The history of the limit on the probability of these processes is shown in Figure~\ref{fig:calando}, which starts with the first experiment performed by Hinks and Pontecorvo in 1947~\cite{pontecorvo}. They stopped cosmic ray muons in a lead absorber and measured the coincidence between signals from two Geiger-M\"uller counters: having seen no such coincidence they gave as a limit essentially the inverse of the number of observed muons. The limits on the three processes improved as artificial muons were produced, stopping pion beams first (until the 1970s) and starting directly with muon beams afterwards.

These experiments give the best constraints to date to possible extensions of the Standard Model inducing CLFV, therefore they play a prominent role in this review. 
There exist in fact a number of reviews on this subject, which concentrate on theoretical~\cite{Lindner:2016bgg,deGouveaVogel} or experimental aspects~\cite{bernstein,fabrizio-donato,MarcianoMori,MoriOotani,pireddone, Ootani} or both~\cite{okadakuno, Raidal:2008jk}, some including CLFV in the broader subjects of muon physics~\cite{hertzog} or theories of massive neutrinos \cite{Bilenky:1987ty}.
We tried to be complementary to these excellent reviews, presenting things in a somewhat pedagogical spirit with the hope that newcomers to the field of muon CLFV might find it useful for a first tour. 

In the theoretical part we try to emphasize the conditions, under which a new physics model leads to sizeable rates for CLFV processes. In order to do so, in section \ref{sec:SM}, we first briefly review the lepton sector of the Standard Model and show why lepton family numbers are conserved in absence of right-handed neutrinos. We then show how minimal departures from the original formulation of the Standard Model (such as introducing a second Higgs doublet) can drastically change the picture and induce large CLFV; the interesting and (experimentally) new subject of LFV decays of the Higgs boson is also briefly discussed.
In section \ref{sec:neutrinos} we introduce neutrino masses to the game and review some details of the standard calculation of $\mu\to e\gamma$ from oscillating neutrinos, showing how this leads to negligible rates unless the new dynamics at the origin of neutrino masses lies at rather low energy scales. Before moving to discuss predictions of some specific models, we present a brief discussion of model-independent approaches to CLFV based on effective operators (section \ref{sec:EFT}). We then discuss, in section \ref{sec:susy}, CLFV in supersymmetric models as a case study of model-dependent predictions, on which we give more general remarks in the subsequent section \ref{sec:models}. We conclude the theory part reviewing in section \ref{sec:LFU}
the possible link between CLFV and lepton flavour non-universality in semi-leptonic $B$-meson decays that several experiments have recently hinted at. 

In the experimental part we highlight the general aspects of the design of experiments aimed at the most sensitive searches for CLFV, in particular of those involving muons (mu-e-gamma, mu-to-three-e, and mu-to-e-conversion) albeit in a unified picture we try to involve also the heaviest lepton, the $\tau$.  Section~\ref{sec:window} starts by introducing general concepts regarding the search for rare decays, such as ``single event sensitivity'' and ``signal region''. In section~\ref{sec:muons} we discuss the generation and decay of positive and negative muons, free and in orbit, and in section~\ref{sec:positrons} and~\ref{sec:photons} we analyse the general aspects, and the difficulties, of detecting low energy electrons, positrons and photons. Before starting the review of the experiments, in section~\ref{sec:calibrations} we make some comments on the need of calibrations and monitoring of the experiments, and the care that should be used in their simulation through Monte Carlo codes. In section~\ref{sec:exps} we review present and programmed muon experiments. In this description we did not want to be comprehensive: those eager of full details are referred to the technical articles describing the experiments. But each experiment is built around some clever ideas, and those we have tried to pass to our readers. 
We will have a look at tau channels in section~\ref{sec:tau} before looking at future directions in section~\ref{sec:future}. 
Our conclusions are drawn in section~\ref{sec:conclusions} in the hope to have made
the theory part accessible to experimentalists as well as the experimental part readable by theorists!  

\section{The lepton sector of the Standard Model and its simple extensions}\label{sec:SM}
The flavour sector of the Standard Model (SM) -- {\em i.e.}~the fermion masses and the mixing among different generations -- arises from 
the Yukawa couplings of the fermion fields with the Higgs field $\Phi$:
\begin{equation}
\label{eq:SM}
- \mathcal{L}_{Y} = (Y_u)_{ij} ~\overline{Q}_{L\,i}\, u_{R\,j} \,\widetilde{\Phi} 
+(Y_d)_{ij} ~\overline{Q}_{L\,i} \,d_{R\,j}\, \Phi +(Y_e)_{ij} ~\overline{L}_{L\,i} \,e_{R\,j} \,\Phi + { h.c.}
\end{equation}
where $SU(2)_L$ indices were omitted and $i$ and $j$ run over the three families,
such that $Y_f$ ($f= u,d,e$) are in general complex $3\times 3$ matrices\footnote{As already noted in the caption of Table~\ref{tab:flavours} we collectively indicate the up-type quarks, the down-type quarks and the charged leptons with $u$, $d$ and $e$ respectively.}. 
The fields are defined as follows:
$Q_L$ are the left-handed (LH) quark doublets, $Q^T_L= (u_{L}~d_{L})$, $u_R$ and $d_R$ the right-handed (RH) up and down quarks respectively,
$L_L$ the LH letpon doublets, $L^T_L= (\nu_{L}~e_{L})$, and $e_R$ the RH leptons.
The conjugate Higgs field is as usually defined as $\widetilde{\Phi} \equiv i \tau_2 \Phi^*$, where $\tau_2$ is the second Pauli matrix. Fermion mass terms of the kind $m_f \overline{f}_L f_R$ arise upon the breaking of the electro-weak (EW) symmetry $SU(2)_L\times U(1)_Y$ by the vacuum expectation value (vev) of the Higgs field, $\langle \Phi \rangle^T = (0~v)/\sqrt{2}$ ($v\simeq 246$ GeV), such that:
\begin{equation}
(m_f)_{ij} = \frac{v}{\sqrt{2}} (Y_f)_{ij},\quad ~\quad f= u,d,e
\end{equation}
In the original formulation of Standard Model, the Lagrangian in Eq.~(\ref{eq:SM}) does not give rise to mass terms for the neutrinos, which are thus exactly massless.
The Yukawa matrices and thus the fermion mass matrices can be diagonalised by unitary rotations of the fields, as follows:
\begin{equation}
\label{eq:mass-basis}
Y_f = V_f \hat{Y}_f W^\dag_f,~\quad f= u,d,e
\end{equation}
where $\hat{Y}_f$ denotes diagonal Yukawa matrices. 
Given the unitarity of the matrices $V_f$ and $W_f$, applying these transformations does not modify the kinetic terms and the neutral-current interactions, such as the fermion couplings to the photon and the $Z$ boson, which then result flavour conserving.
Similarly the fermion couplings to the physical Higgs $h$ are proportional to the mass matrix, thus they can be diagonalised in the same basis and no flavour violation is induced in the interactions with the Higgs either:
\begin{equation}
\label{eq:SMhff}
- \mathcal{L}_{h\bar{f}f} =\frac{m_f}{v}\, \bar{f}_L f_R\,h  + { h.c.}
\end{equation}

On the other hand, the two rotations in Eq.~(\ref{eq:mass-basis}) do induce flavour violation in the charged-current interactions with the $W$ bosons:
\begin{equation}
\label{eq:cc}
\mathcal{L}_{cc} = \frac{g}{\sqrt{2}}\left( \overline{u}_L  \gamma^\mu (V_u^\dag V_d) d_L  + \overline{\nu}_L \gamma^\mu 
(V_\nu^\dag V_e)  e_L \right) W_\mu^+ + { h.c.}
\end{equation}
As we can see, flavour violation in quark sector arises from the fact that, in general, diagonalising $Y_u$ and $Y_d$ requires $V_u \neq V_d$. Such a misalignment gives rise to flavour-changing transitions controlled by the matrix 
$V_{\rm CKM}\equiv V_u^\dag V_d$, which is nothing but the Cabibbo-Kobayashi-Maskawa (CKM) matrix \cite{Cabibbo:1963yz,kobayashi-maskawa}.
On the other hand, in the lepton sector of the original Standard Model with massless neutrinos, one can choose $V_\nu = V_e$, because no other term in the Lagrangian involves the lepton doublets, and the leptonic flavour is exactly conserved. 
Clearly, this feature does not hold any longer in extensions of the Standard Model addressing the generation of mass terms for the neutrinos, as we will discuss in the next section.
In other words, the Lagrangian in Eq.~(\ref{eq:SM}) is invariant under
three independent global $U(1)$ rotations associated to each lepton family, which implies three conserved charges: the lepton family numbers $L_e$, $L_\mu$, and $L_\tau$\footnote{This is in contrast to the hadronic sector that is only invariant under a common global phase rotation of the quark fields, which implies the conservation of the total baryon number alone.}.

To summarize, in the Standard Model the lepton family numbers are individually conserved because of the minimality of the construction, which also implies that neutrinos are massless. 
In fact, the matrix of the lepton Yukawa couplings $Y_e$ defines a single direction in the space of leptonic flavour. Hence, as we have seen, one can use the freedom of rotating LH and RH lepton fields to make the matrix diagonal without inducing flavour-changing effects in other sectors of the theory. This is in contrast to the quark sector where there are two different Yukawa matrices, $Y_u$ and $Y_d$, both involving $Q_L$, such that they can not be simultaneously diagonalised in the same basis. 

From the above discussion, we can immediately see under which condition an extension of the Standard Model features flavour violation in the leptonic sector: the presence in the Lagrangian of at least another term involving the lepton fields, {\em i.e.}~of another non-trivial direction in the flavour space. This is for instance the case of a neutrino mass term (Dirac or Majorana), as we will see in the next section. Here we mention another minimal extension of the Standard Model leading to lepton flavour violation, namely the introduction of a second Higgs doublet \cite{Branco:2011iw}. In fact, in presence of two scalar doublets $\Phi_1$ and $\Phi_2$, the gauge symmetries allow couplings of the fermion fields to both: 
\begin{align}
\label{eq:2HDM}
- \mathcal{L}_{\rm 2HDM} =\sum_{a=1,2}\Big[ & (Y^{(a)}_u)_{ij} ~\overline{Q}_{L\,i}\, u_{R\,j} \,\widetilde{\Phi_a} 
+(Y^{(a)}_d)_{ij} ~\overline{Q}_{L\,i} \,d_{R\,j}\, \Phi_a  + \nonumber \\
& (Y^{(a)}_e)_{ij} ~\overline{L}_{L\,i} \,e_{R\,j} \,\Phi_a + { h.c.}
\Big].
\end{align}
In particular, we see that in the lepton sector there are two Yukawa matrices $Y^{(1)}_e$ and $Y^{(2)}_e$, hence from the above discussion we expect that lepton family numbers are violated in this framework. Indeed, while the fermion masses are now generated by the vevs of both Higgs fields, $v_1$ and $v_2$,
\begin{equation}
m_f = \frac{1}{\sqrt{2}}\left( Y^{(1)}_f v_1 + Y^{(2)}_f v_2 \right),\quad ~\quad f= u,d,e
\end{equation}
the couplings of the physical Higgs particles -- which are now five: two neutral
CP-even states, one neutral CP-odd state, two charged states -- 
are no longer aligned to the fermion masses as in Eq.~(\ref{eq:SMhff}). Let us consider as an example the case of the two CP-even states, $h$ and $H$, which are mixtures of the real parts of the neutral components of the two doublets:
\begin{align}
h =& \sqrt{2} ({\rm Re}(\Phi^0_1) \sin \alpha + {\rm Re}(\Phi^0_2) \cos \alpha), \\
H = &\sqrt{2} ({\rm Re}(\Phi^0_1) \cos \alpha - {\rm Re}(\Phi^0_2) \sin \alpha),
\end{align}
where the mixing angle $\alpha$ depends on the couplings of the scalar potential that we have assumed to be CP conserving.
 In order to highlight the above-mentioned misalignment, it is convenient to rotate the scalar doublets such that only $\Phi_1$ has a vev, i.e.~$v_1= v$, $v_2= 0$. The couplings of these two physical CP-even Higgses to the fermions are then:
\begin{align}
\label{eq:2Hhff}
- \mathcal{L}_{h\bar{f}f} =&\left(\frac{m_f}{v}\sin\alpha + Y^{(2)}_f \cos\alpha \right)\, \bar{f}_L f_R\,h  + { h.c.}\\
- \mathcal{L}_{H\bar{f}f} =&\left(\frac{m_f}{v}\cos\alpha - Y^{(2)}_f \sin\alpha \right)\, \bar{f}_L f_R\,H + { h.c.}
\end{align}
While only the matrices $Y^{(1)}_f$ contribute to the fermion mass terms, 
the above couplings depend on the $Y^{(2)}_f$ too. Hence,
as we can see, these couplings are in general not diagonal in the mass basis where $m_f$ are. 
As a consequence, flavour-changing neutral currents (FCNC) are generated 
already at the tree level by diagrams exchanging $h$ and $H$. 
In the lepton sector in particular, CLFV decays of the type  $\ell_i \to \ell_j \ell_k \ell_k$, {\em e.g.}~$\mu \to eee$, receive tree-level contributions of such kind. The same occurs for $\mu \to e$ conversion in nuclei.
This is in contrast to the Standard Model, where FCNC processes arise at loop level only (in the hadronic sector) or are forbidden (in the leptonic one). This is the reason why a generic flavour structure of the matrices $Y^{(2)}_f$ easily gives unacceptably large FCNC effects and thus it is experimentally excluded -- for a recent overview see \cite{Crivellin:2013wna}.
The two Higgs doublet models most commonly studied in the literature feature additional symmetries
that control the structure of the Yukawa couplings in Eq.~(\ref{eq:2HDM}), in order to suppress the FCNCs. 
One possibility is introducing a $\mathbb{Z}_2$ symmetry and assign parities to the fields
such that the fermions of each sector can couple to one of the two Higgs doublets only, for instance up quarks to $\Phi_1$, down quarks and leptons to $\Phi_2$ (the so-called ``type II'' model). In this framework -- labelled as ``natural flavour conservation'' \cite{Glashow:1976nt,Paschos:1976ay} -- 
an alignment of the kind of Eq.~(\ref{eq:SMhff}) is restored and no new effects are generated at the tree level. 
This is the situation of the minimal supersymmetric models,
whose Higgs sector is in fact a two-Higgs-doublet model of
type II\footnote{Higgs-mediated CLFV effects that arise by introducing small couplings of the leptons to the `wrong' Higgs doublet -- which are indeed radiatively generated in supersymmetric models -- have been studied in \cite{H-CLFV}.}.
Less radical approaches -- but effective in suppressing FCNC processes at acceptable levels -- 
consist in imposing the ansatz of minimal flavour violation \cite{DAmbrosio:2002vsn} to the Yukawa couplings \cite{Buras:2010mh} (which implies that the new physics flavour-violating effects are controlled by the CKM matrix, as the SM contributions), or a flavour symmetry addressing the hierarchy of the Yukawas dynamically \cite{Dery:2016fyj}. 

\begin{figure}
\begin{center}
\includegraphics[width=.9\columnwidth]{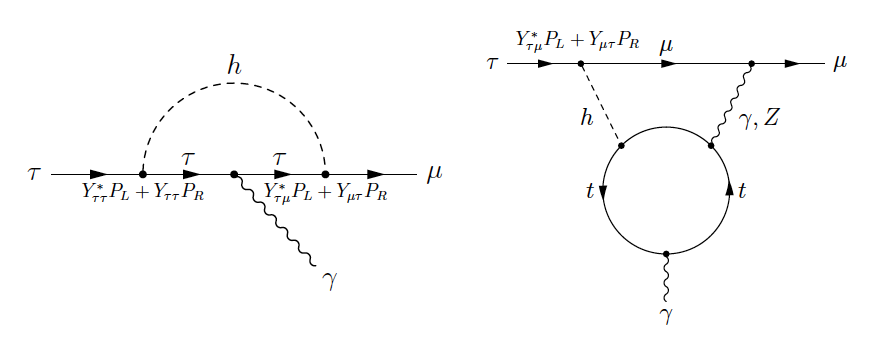}
\caption{Examples of one and two loop diagrams giving comparable contributions to $\tau\to \mu\gamma$ in presence of flavour-violating Higgs couplings $Y^h_{\tau\mu}$ and $Y^h_{\mu\tau}$. From \cite{Harnik:2012pb}.  \label{fig:htaumu}}
\end{center}
\end{figure}

\begin{table}[t]
\renewcommand{\arraystretch}{1.5}
  \centering
  \begin{tabular}{@{\qquad}lcc@{\qquad}}
  \hline
   Process   &    Coupling                               & Bound        \\ \hline
    $h\to \mu e $   & $\sqrt{|Y^h_{\mu e}|^2 + |Y^h_{e \mu}|^2}$ & $< 5.4 \times 10^{-4} $ \\ 
    $\mu \to e\gamma$         & $\sqrt{|Y^h_{\mu e}|^2 + |Y^h_{e \mu}|^2}$ & $< 2.1 \times 10^{-6} $ \\
    $\mu \to eee$              & $\sqrt{|Y^h_{\mu e}|^2 + |Y^h_{e \mu}|^2}$ & $\lesssim 3.1 \times 10^{-5}$ \\
    $\mu\, {\rm Ti} \to e\, {\rm Ti}$    & $\sqrt{|Y^h_{\mu e}|^2 + |Y^h_{e \mu}|^2}$ & $< 1.2 \times 10^{-5} $ \\
    \hline
    $h\to \tau e $   & $\sqrt{|Y^h_{\tau e}|^2 + |Y^h_{e \tau}|^2}$ & $< 2.3\times10^{-3} $ \\
    $\tau \to e\gamma$        & $\sqrt{|Y^h_{\tau e}|^2 + |Y^h_{e \tau}|^2}$ & $< 0.014$ \\
    $\tau \to eee$            & $\sqrt{|Y^h_{\tau e}|^2 + |Y^h_{e \tau}|^2}$ & $\lesssim 0.12$ \\
    \hline
    $h\to \tau \mu$   & $\sqrt{|Y^h_{\tau \mu}|^2 + |Y^h_{\mu \tau}|^2}$ & $< 1.4 \times10^{-3}$ \\
    $\tau\to \mu\gamma$       & $\sqrt{|Y^h_{\tau\mu}|^2 + |Y^h_{\mu\tau}|^2}$ & $<0.016$ \\
    $\tau\to \mu\mu\mu$            & $\sqrt{|Y^h_{\tau\mu}|^2 + |Y^h_{\mu\tau}|^2} $ & $\lesssim 0.25$  \\
    \hline
%
  \end{tabular}
  \caption{Bounds on flavour violating Higgs couplings to leptons assuming that the flavour diagonal Yukawa couplings are as predicted by the SM, $Y^h_{ii} = m_i/v$ ($i=e,\mu,\tau$); adapted from \cite{Harnik:2012pb}. 
   The constraints from $h\to \ell_i\ell_j$ are reported in \cite{higgsmue,higgsmutau}.}
   \label{tab:Yukawa-bounds}
\end{table}
%
A rather model-independent parametrisation of Higgs-induced CLFV effects is the following \cite{Blankenburg:2012ex,Harnik:2012pb,Dery:2013rta,Dorsner:2015mja}:
\begin{equation}
\label{eq:LFVyuk}
- \mathcal{L} \supset (m_e)_{i} \bar{e}_{L\,i}\, e_{R\,i}  + (Y^h_{e})_{ij}\,\bar{e}_{L\,i}\, e_{R\,j} \,h  + { h.c.}
\end{equation}
where, as we can see, the matrix of the lepton couplings to the physical Higgs $Y^h_{e}$ is in general not aligned to the lepton mass matrix.
As we have seen, a Lagrangian of this kind arises from a two Higgs doublet model, of which
it is an approximation, the better the heavier the masses of the extra Higgses are -- as only one SM-like physical Higgs is considered here. 
Alternatively, it can be induced by higher-dimensional operators featuring the SM leptons and the Higgs field of the type $\overline{L}_L e_R \Phi (\Phi^\dag \Phi)$, which result from integrating out heavier degrees of freedom.
The off-diagonal Higgs couplings $(Y^h_{e})_{i\neq j}$ give contributions to CLFV processes, such as 
$h\to \ell^+_i\ell^-_j$ (directly), $\ell_i \to \ell_j \gamma$ (via a $h$-$\ell_i$ loop and two-loop diagrams of the Barr-Zee type \cite{Barr:1990vd}, {\em i.e.}~featuring one loop of heavy particles,
see Fig.~\ref{fig:htaumu}), 
$\ell_i \to \ell_j \ell_k \ell_k$ (at tree level, as we have seen above, although the dominant contributions turn out to be the same as for the radiative decay \cite{Harnik:2012pb}), as well as $\mu N\to e N$ (with again the loop contributions dominant). 
The bounds that experimental searches for these processes, cf.~Tab~\ref{tab:limits}, set on the flavour-violating Higgs couplings $Y^h_{ij}\equiv(Y^h_{e})_{i\neq j}$ are shown in Table~\ref{tab:Yukawa-bounds}.
As we can see present limits already set stringent constraints to such couplings, in particular in the $\mu-e$ sector: as a reference, the SM values of the Higgs-lepton couplings are $Y^h_{ee} \simeq 1.6\times 10^{-6}$, $Y^h_{\mu\mu} \simeq 4.3\times 10^{-4}$, $Y^h_{\tau\tau} \simeq 7.2\times 10^{-3}$.
The bound on $Y^h_{\mu e}$ and $Y^h_{e\mu}$ is dominated by $\mu\to e\gamma$, but given that the rates of all these processes
are $\propto |Y^h_{\mu e}|^2 + |Y^h_{e \mu}|^2$ a future increase by two orders of magnitude of the sensitivity to
 $\mu \to eee$  and  $\mu \to e$ conversion in nuclei would translate to a comparable constraint. 
As a consequence of these low-energy constraints, $h\to e\mu$ is way too suppressed to be observable at the LHC, namely ${\rm BR}(h\to e\mu) \lesssim 10^{-8}$. On the other hand, the Table shows that in the $\tau-e$ and  $\tau-\mu$  sectors
the leading constraints come from the LHC searches for  $h\to e\tau $ and $h\to \mu\tau $ themselves, so that the discovery
of CLFV at the LHC, or future leptonic colliders, is still an open possibility.

\section{CLFV in the Standard Model with massive neutrinos}
\label{sec:neutrinos}
The observation of neutrino oscillations on the one hand provides evidence that the lepton family numbers are violated. 
 On the other hand, it calls for an extension of the SM to include neutrino mass terms. As we have discussed in the previous section, these two points are related, for the conservation of lepton flavours is a mere consequence of the minimality of Eq.~(\ref{eq:SM}), where $Y_e$ is the only term involving leptonic fields. A departure from minimality is required to generate neutrino masses and it can 
occur in two ways: either by introducing RH neutrinos $\nu_R$, hence extending the field content, or by considering
non-renormalisable operators (for reviews, see {\em e.g.}~\cite{Bilenky:1987ty,Bilenky:1998dt}).
In the first case, Dirac mass terms arise upon electroweak symmetry breaking from neutrino Yukawa couplings, in perfect analogy to what occurs to the other fermions: 
\begin{equation}
\label{eq:dirac}
\mathcal{L}_D = -(Y_\nu)_{ij} ~\overline{\nu}_{R\,i} \,\widetilde{\Phi}^\dag\, L_{L\,j} + {\ h.c.} \quad\Longrightarrow \quad
({m}^D_\nu)_{ij} = \frac{v}{\sqrt{2}}  (Y_\nu)_{ij}\,.
\end{equation}
In such a case, the smallness of neutrino masses can be accounted for only by assuming tiny values for the 
entries of $Y_\nu$, the largest of which should be $\lesssim 10^{-12}$\footnote{Moreover, since $\nu_{R}$ are complete singlets under the SM gauge symmetries, nothing forbids Majorana mass terms of the kind $\overline{\nu^c_{R}} \nu_{R}$, which would change the picture as we will see below.}.
As a consequence, the second option is perhaps more appealing, namely introducing non-renomalisabe operators giving rise to Majorana masses for the LH neutrinos alone:
\begin{equation}
\mathcal{L}_M = -\frac{1}{2} \,m^M_\nu\,  \overline{\nu^c_{L}}\nu_{L} + {\ h.c.}
\end{equation}
Terms of this kind violate the total lepton number $L$ and can be already generated by the only dimension-5 operators 
compatible with the SM symmetries that one can write \cite{Weinberg:1979sa}:
\begin{equation}
\label{eq:weinb}
\mathcal{L} \supset  \frac{C_{ij}}{\Lambda} \left(\overline{L^c_{L\,i}}\, \tau_2 \Phi \right) \left( \Phi^T \tau_2 L_{L\,j} \right) +{ h.c.}
\quad \Longrightarrow \quad ({m}^M_\nu)_{ij} = \frac{C_{ij} v^2}{\Lambda},
\end{equation}
where $\Lambda$ corresponds to the mass scale of extra degrees of freedom -- associated to the breaking of $L$ --  that have been integrated out. The fact that $({m}^M_\nu)_{ij} \ll v$ is then naturally explained if $\Lambda \gg v$.

In either way neutrino masses are accounted for, lepton flavour violation in the charged current interactions of Eq.~(\ref{eq:cc})
becomes physical and controlled by the matrix $U_{\rm PMNS} \equiv (V_e^\dag V_\nu)$, which is usually called Pontecorvo-Maki-Nakagawa-Sakata (PMNS) matrix \cite{Pontecorvo:1957cp,Maki:1962mu}. Also notice that $U_{\rm PMNS}$ is the matrix that diagonalises the neutrino mass matrix in the basis of diagonal charged lepton masses, thus connecting neutrino flavour and mass eigenstates:
\begin{equation}
\nu_\alpha = \sum_{k=1,3} U_{\alpha k} \,\nu_k,\quad \alpha = e,\mu,\tau.
\end{equation}

\begin{figure}
\begin{center}
\includegraphics[width=.9\columnwidth]{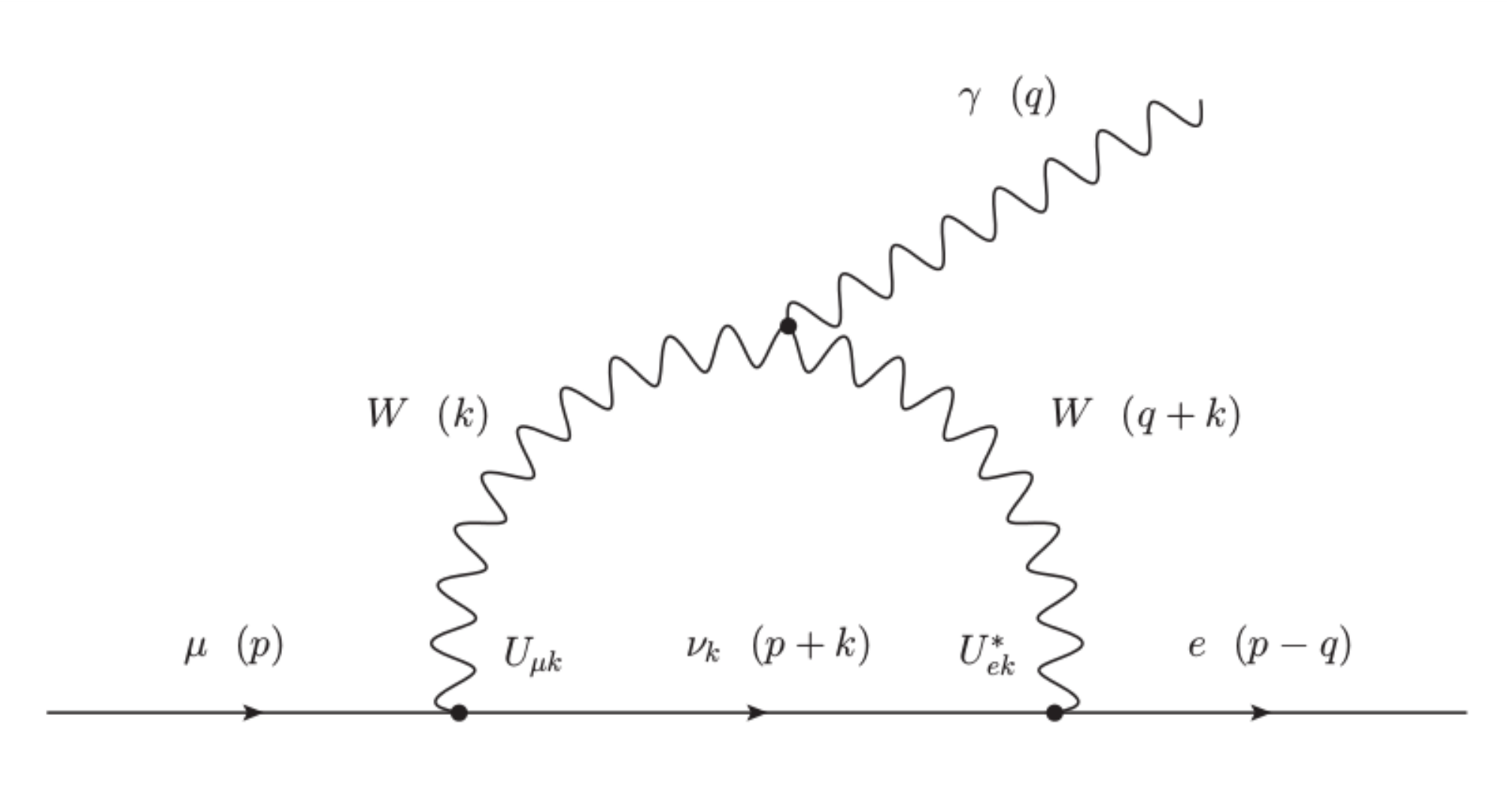}
\caption{Diagram contributing to $\mu\to e\gamma$ in the SM with massive neutrinos. The four-momenta of the particles are indicated in parenthesis. \label{fig:diag}}
\end{center}
\end{figure}
As is well known, besides inducing neutrino oscillations, $U_{\rm PMNS}$ can in principle give rise to CLFV as well.
This can only occur through loop diagrams involving neutrinos and $W$ bosons. 
For example, a diagram contributing to the $\mu \to e \gamma$  decay is shown in Figure \ref{fig:diag}.
A detailed description of the calculation of the resulting $\ell_i \to \ell_j \gamma$ rate -- originally published in \cite{clfv-sm} -- 
can be be found in \cite{Cheng:1985bj}. 
In the following we review some important features of it, specialising in $\mu \to e\gamma$, 
being $ \tau\to e\gamma$ and $ \tau\to \mu\gamma$ completely analogous. 

In general, the transition is described by an effective muon-electron-photon interaction that we denote $V_\alpha$, where  $\alpha$ is a Lorentz index.
The decay amplitude is then given by:
\begin{equation}
\mathcal{M}(\mu\to e \gamma) = i\, \bar{u}_e (p-q) V_\alpha u_\mu(p)  \epsilon^{*\alpha}(q)\,
\label{eq:meg-amp}
\end{equation}
where $p$ and $q$ are the four-momenta of the muon and the photon respectively, $u_e$ and $u_\mu$ are the Dirac spinors for the electron and the muon, and $\epsilon_\lambda$ the polarisation vector of the photon. 
Written in terms of Lorentz-invariant amplitudes $V_\alpha$ has the following general structure:
\begin{equation}
V_\alpha = \sigma_{\alpha\beta} q^\beta (F_1 + F_2 \gamma_5) + \gamma_\alpha (F_3 + F_4 \gamma_5)
+  q_\alpha (F_5 + F_6 \gamma_5),
\label{eq:V}
\end{equation}
where $\sigma_{\alpha\beta} = \frac{i}{2}[\gamma_\alpha,\gamma_\beta]$. The invariant amplitudes $F_1,\dots,F_6$
are not independent quantities. In fact, imposing the conservation of the electric charge, 
$\frac{\partial}{\partial q_\alpha} V_\alpha = 0$,  Eqs.~(\ref{eq:meg-amp}) and~(\ref{eq:V}) give:
\begin{equation}
-m_e (F_3 + F_4 \gamma_5) + m_\mu (F_3 - F_4 \gamma_5) +q^2  (F_5 + F_6 \gamma_5) = 0\,,
\label{eq:charge}
\end{equation}
where $\{\gamma_\alpha,\gamma_5\}=0$ has been used, as well as the Dirac equation for the spinors, 
$(\slashed{p} - m) u(p) = \bar{u} (p)  (\slashed{p} - m) =0$.
From this expression, we see that for an on-shell photon, $q^2 =0$, $F_3 = F_4 = 0$, and $F_5$ and $F_6$ 
do not contribute\footnote{Instead, after imposing Eq.~(\ref{eq:charge}), two independent combinations of these form factors 
contribute to processes mediated by an off-shell photon, such as $\mu\to e e e$.}.
The resulting amplitude is thus:
\begin{align}
\mathcal{M}(\mu\to e \gamma) &= i\, \bar{u}_e (p-q) \sigma_{\alpha \beta} q^\beta  (F_1 + F_2 \gamma_5) u_\mu(p)  \epsilon^{*\alpha}(q)\, \nonumber \\
&= i\, \bar{u}_e (p-q) \sigma_{\alpha \beta} q^\beta  (A_R P_R + A_L P_L) u_\mu(p)  \epsilon^{*\alpha}(q)\,,
\label{eq:meg-amp2}
\end{align}
where in the second line we defined left-handed and right-handed amplitudes for later convenience, $A_{R,L} = F_1\pm F_2$,
and the chirality projectors are $P_{R,L} = (1\pm \gamma_5)/2$. 

Computing the unpolarised ({\em i.e.}~averaged over initial polarisations, summed over the final ones)
squared amplitude $|\mathcal{M}|^2$ one gets:
\begin{equation}
|\mathcal{M}|^2 = m^4_\mu \left(|A_R|^2 + |A_L|^2\right)\,
\end{equation}
hence the resulting decay rate is (taking the limit $m_e\ll m_\mu$ ):
\begin{equation}
\Gamma(\mu\to e\gamma) = \frac{|\mathcal{M}|^2}{16\pi \,m_\mu} = \frac{m_\mu^3}{16 \pi}  \left(|A_R|^2 + |A_L|^2\right),
\label{eq:meg-rate}
\end{equation}
where we can see that $A_{L,R}$ have mass dimension $[M]^{-1}$. 

The amplitudes $A_{L,R}$ are obtained by computing diagrams such as the one in Figure \ref{fig:diag}.
Some comments are in order. 
\begin{itemize}
\item Possible diagrams with the photon vertex on the external lepton lines correspond to a Lorentz structure of the
kind $\bar{u}_e \gamma_\alpha u_\mu \epsilon^{* \alpha}$, which does not match the one we inferred
above, cf.~Eq.~(\ref{eq:meg-amp2}).
Therefore they can not contribute to our decay and, in fact, they will be cancelled by some terms of the diagrams
with the internal photon vertex.
\item Since $P_{R,L}$ and $\sigma_{\alpha \beta}$ commute, 
we can see from Eqs.~(\ref{eq:meg-amp2}, \ref{eq:meg-rate}) that the electron and muon spinors must have opposite chiralities.
Since only LH fields interact with $W$, in the diagram of Figure \ref{fig:diag} such chirality flip can only be due to the mass terms
of the external leptons. As a consequence, we have $A_R \propto m_\mu$ and $A_L \propto m_e$, which means
that $|A_L|^2 \ll |A_R|^2$, thus negligible. 
As a consequence, it is sufficient to compute $A_R$, which corresponds to set $m_e=0$ in the calculation.
\item It is convenient to perform the calculation within a class of gauges called $R_\xi$ gauges, instead of choosing
the unitary gauge (in the unitary gauge, the Goldstone bosons associated to the electro-weak symmetry breaking are integrated out from the physical spectrum, and appear as longitudinal degrees of freedom of the gauge bosons). 
 The reason is that the loop diagrams shown in Figure \ref{fig:diag} are divergent in the unitary gauge, 
 because the propagator of the $W$ is of the form $-i(g_{\mu\nu}-k_\mu k_\nu/M_W^2)/(k^2 -M_W^2)$, 
 which does not vanish for $k\to \infty$.
On the contrary, the diagrams are manifestly finite in $R_\xi$ gauges, 
as the $W$ propagator behaves like $1/k^2$ in the limit  $k\to \infty$.
The price to pay is to include additional diagrams, besides that of Figure \ref{fig:diag}, involving 
the unphysical Goldstone bosons. 
\end{itemize}
Let us write the amplitude $\mathcal{M}^{(1)}$ corresponding to the diagram of Figure \ref{fig:diag}
summed over the three neutrino eigenstates:
\begin{align}
\mathcal{M}^{(1)} = i \sum_{k=1,3} \int& \frac{d^4 k}{(2\pi)^4} 
\bar{u}_e(p-q)\left(-i \frac{ g\, U^*_{ek}} {\sqrt{2}} \gamma_\alpha P_L\right)
\frac{\slashed{p}+\slashed{k}}{(p+k)^2-m^2_{\nu_k}}  
\left(i\frac{g\, U_{\mu k} } {\sqrt{2}}  \gamma_\beta P_L\right) u_{\mu} (p) 
\nonumber \\ 
 & \times \left(i \Delta^{\alpha \mu}(k+q)\right) \left(-ie\Gamma_{\lambda\mu\nu}(-q,k+q,-k)\right)  
(i \Delta^{\nu\beta}(k))\, \epsilon^{*\lambda}(q)\,,
\label{eq:integral}
\end{align} 
where the $R_\xi$-gauge form of the $W$ propagator is:
\begin{equation}
\Delta_{\alpha\beta}(k) = -\frac{g_{\alpha\beta} - (1-\xi) \frac{k_\alpha k_\beta}{k^2 -\xi M_W^2}}{k^2-M_W^2},
\end{equation}
and the photon-$W$ vertex is given by:
\begin{equation}
\Gamma_{\lambda\mu\nu}(-q,k+q,-k) = 
\left( (q-k)_\mu g_{\lambda\nu} + (2k+q)_\lambda g_{\mu\nu} - (2q+k)_\nu g_{\lambda\mu}\right).
\end{equation}

The sum over the neutrino eigenstates gives:
\begin{equation}
\sum_{k=1,3} \frac{U_{\mu k}U^*_{ek}}{(p+k)^2-m^2_{\nu_k}} = 
\sum_{k=1,3} \left[\frac{U_{\mu k}U^*_{ek}}{(p+k)^2} + \frac{U_{\mu k}U^*_{ek}m^2_{\nu_k}}{(p+k)^4}+\dots  \right],
\label{eq:gim}
\end{equation}
where the expansion is justified because $m_{\nu_k} \ll m_\mu$ (and $p =(m_\mu,0,0,0)$ in the muon rest frame).
The first term in the expansion vanishes because of the unitarity of the PMNS matrix, $U U^\dagger = \bf 1$,
thus the leading term is the second one proportional to differences of the neutrino squared masses, 
$\Delta m^2_{ij} \equiv m^2_{\nu_i} - m^2_{\nu_j}$: 
\begin{equation}
\sum_{k=1,3}  U_{\mu k}U^*_{ek}m^2_{\nu_k} = 
U_{\mu 3}U^*_{e3} \Delta m^2_{31} + U_{\mu 2}U^*_{e2} \Delta m^2_{21}.
\end{equation}
Such GIM-like cancellation \cite{Glashow:1970gm} is eventually responsible for the negligible CLFV rates induced by diagrams of this kind.
As we will see below, this conclusion can change in presence of low-energy RH neutrinos. 

Finally, in order to compute the contribution of our diagram to $A_R$, 
one has to perform the integration over the internal momentum $k$ and pick up
the terms with the Lorentz structure as in Eq.~(\ref{eq:meg-amp2}).
The result is \cite{Cheng:1985bj}
\begin{equation}
A_R^{(1)}= \frac{g^2 e} {128\pi^2} \frac{m_\mu}{M_W^4}\sum_{k=1,3}  U_{\mu k}U^*_{ek}m^2_{\nu_k} 
\left[1 - \frac{1}{3} \frac{\ln \xi}{\xi -1} + \frac{1}{\xi-1}\left( \frac{\xi \ln\xi}{\xi-1} -1\right) \right],
\end{equation}
which is proportional to the muon mass, reflecting the necessary chirality flip, as we discussed above.
Computing the missing diagrams that feature the Goldstone bosons, one can check that they exactly 
cancel the $\xi$-dependent terms in the above expression, so that the total amplitude $A_R$
is gauge invariant. 
This is consistent with the fact that taking the limit $\xi\to \infty$, corresponding to the unitary gauge, 
the Goldstone bosons decouple, hence the additional contributions as well as the $\xi$-dependent 
terms in $A_R^{(1)}$ vanish. In other words $A_R^{(1)} \to A_R$ for $\xi\to \infty$.

Substituting the obtained $A_R$ into Eq.~(\ref{eq:meg-rate}), and considering that the rate of the 
ordinary muon decay is $\Gamma(\mu \to e \nu \bar{\nu})= m_\mu^5 G_F^2/192\pi^3$, where
the Fermi constant is $G_F/\sqrt{2}= g^2/8M_W^2$, we get to the final result for the branching ratio of $\mu \to e \gamma$ \cite{clfv-sm}:
\begin{equation}
\label{eq:meg-mnu}
{\rm BR}(\mu\to e \gamma) \simeq \frac{\Gamma(\mu\to e \gamma)}{\Gamma(\mu \to e \nu \bar{\nu})}
= \frac{3\alpha}{32\pi}\left| \sum_{k=1,3} \frac{ U_{\mu k}U^*_{ek}m^2_{\nu_k} }{ M_W^2 } \right|^2.
\end{equation}

Considering the uncertainties in the neutrino parameters \cite{Esteban:2016qun}, the resulting numerical value is ${\rm BR}(\mu\to e \gamma)= 10^{-55}\div 10^{-54}$, about forty orders of magnitudes smaller than the sensitivity of present-day experiments!

To summarise, CLFV is indeed induced by adding neutrino mass terms to the SM Lagrangian -- as follows from
the general considerations of the previous section -- but at a level of no phenomenological interest.
This occurs although lepton family numbers are badly broken, as the $\mathcal{O}(1)$ mixing angles among
neutrino generations show. It is rather a consequence of the above-mentioned GIM cancellation and the tiny values of the neutrino masses compared to $M_W$, hence to some extent `accidental' rather than `fundamental'.
On a more positive note, the above result means that an experimental observation of CLFV would be
an unambiguous sign of new dynamics related to a non-trivial extension of the lepton sector of the SM.
We conclude this section briefly discussing an example of such a non-trivial extension.

Let us consider the case of Majorana neutrinos. The non-renormalisable operator shown in  Eq.~(\ref{eq:weinb})
must arise from integrating out the degrees of freedom of some high-energy renormalisable theory.
The simplest example of such a theory is provided by the so-called seesaw mechanism \cite{seesaw}, which consists in extending
the SM Lagrangian by adding RH neutrinos with Majorana mass terms $M_R$ that break the total lepton number $L$\footnote{Notice that such terms are compatible with the SM gauge symmetries, being the RH neutrinos completely neutral under them. Therefore, the following Lagrangian is more general than the one in Eq.~(\ref{eq:dirac}) that leads to Dirac neutrinos. In fact, the latter one requires additional symmetries have to be imposed to forbid the RH neutrino masses.}:
\begin{align}
\mathcal{L} =\mathcal{L}_{\rm SM} + i \overline{\nu}_R\slashed{\partial}\nu_R -\left(Y_\nu \overline{\nu}_R \widetilde{\Phi}^\dag L_L +
\frac{1}{2}M_R \overline{\nu}_R \nu_R^c + {\rm h.c.}\right).
\label{eq:Lseesaw}
\end{align}
The spontaneous breaking of the EW symmetry induces Dirac mass terms through $Y_\nu$ as in Eq.~(\ref{eq:dirac}), as well as mixing among LH and RH neutrinos. However, due to $M_R$, once the total (LH+RH) neutrino mass matrix is diagonalised, the 
resulting eigenstates are Majorana fields.
If the scales in $M_R$ are much heavier than the EW scale, 
the $\nu_R$ decouple from the Lagrangian and the light neutrinos result to be mostly LH with Majorana mass terms given by the celebrated seesaw formula:
\begin{equation}
\label{eq:seesaw}
m_\nu = -\frac{v^2}{2} Y_\nu^T M_R^{-1} Y_\nu\,,
\end{equation}
which is of the same form as in Eq.~(\ref{eq:weinb})\footnote{The seesaw mechanism is the simplest way of generating the neutrino mass operator of Eq.~(\ref{eq:weinb}). As we will mention in the next sections, there are other possibilities also based on the tree-level exchange of heavy fields (but non-singlet under $SU(2)_L$). Moreover, in certain models neutrino masses only arise at the loop level: for a recent review of such radiative neutrino mass models, including their
CLFV phenomenology, see \cite{radiative}.}.

Another effect of the mixing between LH and RH neutrinos is that the unitary matrix that 
diagonalises $m_\nu$ -- the PMNS matrix $U$ -- does not coincide anymore with the matrix that appears in the
charged current, Eq.~(\ref{eq:cc}), being this latter one -- which we call  $\mathcal{U}$ -- 
a submatrix of a higher dimensional matrix for both LH and RH neutrinos. 
Most importantly, $\mathcal{U}$ is not unitary and can be written in terms of the matrices $Y_\nu$ and $M_R$ as \cite{Antusch:2006vwa,Abada:2007ux}: 
\begin{equation}
\label{eq:nonuni}
\mathcal{U} = \left( 1- \frac{v^2}{2} Y_\nu^\dag M_R^{-2} Y_\nu \right) U\,.
\end{equation}

We can now understand why this setup possibly has a large impact on CLFV: entries of  $\mathcal{U}$ -- instead of $U$ --
now appear at the vertices of Figure \ref{fig:diag}, and the GIM cancellation of Eq.~(\ref{eq:gim}) does not occur anymore.
Taking this into account, Eq.~(\ref{eq:meg-mnu}) becomes
\begin{equation}
\label{eq:meg-seesaw}
{\rm BR}(\mu\to e \gamma) =
\frac{3\alpha}{32\pi} \frac{\left| \sum_k \mathcal{U}_{\mu k} \mathcal{U}^*_{e k} F(x_k) \right|^2}{(\mathcal{U}\mathcal{U}^\dag)_{\mu\mu}(\mathcal{U}\mathcal{U}^\dag)_{ee}},
\end{equation}
where $x_k = m^2_{\nu_k}/M_W^2$ and the loop function is $F(x_k)  = \frac{10}{3} - x_k + \ord{x_k^2}$. As we can see, in the limit
$\mathcal{U} \to U$, the expression in Eq.~(\ref{eq:meg-mnu}) is recovered, but for a sizeable deviations of
$\mathcal{U}$ from unitarity the first term of $F(x_k)$ -- which is not suppressed by the small neutrino masses -- 
dominate.
As a consequence, CLFV rates can be raised at observable levels for low-scale RH neutrinos \cite{Antusch:2006vwa,Abada:2007ux,lowEseesaw,1209.2679}.  
However, from Eq.~(\ref{eq:nonuni}), we can see that sizeable CLFV requires both $M_{R}$ not too far above the EW scale and 
large neutrino Yukawa couplings. These two conditions are not compatible with the smallness of 
the neutrino masses in generic realisations of the seesaw mechanism, as Eq.~(\ref{eq:seesaw}) naively gives
$M_{R} \sim 10^{14}\div 10^{15}$ GeV for $Y_\nu \sim \ord{1}$.
It is clear then that some large cancellations are required in Eq.~(\ref{eq:seesaw})  in order to have
a theory of phenomenological relevance for low-energy observables including CLFV processes. However, this
does not necessarily require fine tunings, but it can be the natural consequence of some symmetry giving a 
special texture to $Y_\nu$ and $M_R$, for instance the approximately conserved lepton number itself. 
Indeed, as vanishing neutrino masses lead to augmented symmetry, small $L$-breaking terms 
(responsible of the smallness of $m_\nu$) are technically natural. An instance of this idea is the so-called inverse seesaw \cite{GonzalezGarcia:1988rw,Mohapatra:1986bd,Mohapatra:1986aw}. 

\section{Model-independent approach to CLFV: non-renormalisable operators}
\label{sec:EFT}
We now turn to discuss a largely model-independent approach to describe possible deviations from the SM giving rise to CLFV, 
which consists in extending the SM Lagrangian with higher-dimensional gauge-invariant operators that only involve the SM fields:
\begin{align}
\mathcal{L}=\mathcal{L}_{\rm SM} + \frac{1}{\Lambda}\sum_a C^{(5)}_a Q_a^{(5)} 
+ \frac{1}{\Lambda^2}\sum_a C^{(6)}_a Q_a^{(6)}  + \ldots
\end{align}
We denoted the non-renormalisable operators $Q_a^{(D)}$ according to their dimensionality $D$.
The coefficients $C^{(D)}_a$ are dimensionless, and $\Lambda \gg v$ is the cut-off of this effective field theory, 
{\em i.e.}~the energy scale of new degrees of freedom that give rise to the operators once integrated out.

\begin{figure}[t]
\begin{center}
\includegraphics[width=0.8\columnwidth]{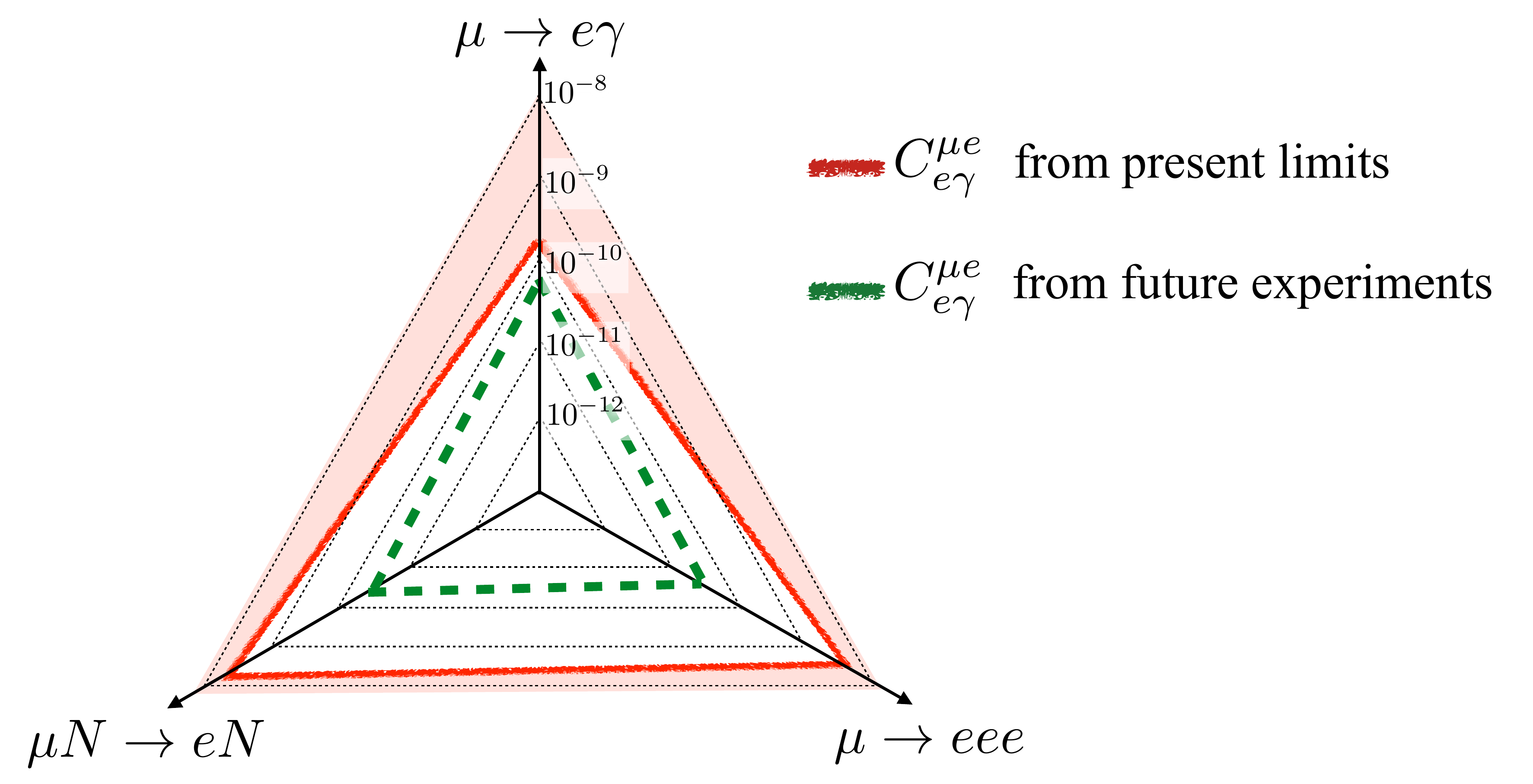}
\caption{Graphical representation of the bounds on the coefficient of the dipole operator $C^{\mu e}_{e \gamma}$  for $\Lambda =1$ TeV coming from the present and future limits (allowed region inside the triangle, data from Table~\ref{tab:bounds}). It is clear that presently $\mu\to e \gamma$ is the most sensitive process  on this parameter but future experiments will have a comparable sensitivity. \label{fig:spiderweb}}
\end{center}
\end{figure}

The only operators with $D=5$ one can write
are those given in Eq.~(\ref{eq:weinb}) that generate neutrino Majorana mass terms.
As we have seen in the previous section, they induce CLFV effects only at loop level and with negligibly small rates.
The first relevant CLFV effects thus arise at the dimension-6 level from the operators shown in Table \ref{tab:ops} \cite{Crivellin:2013hpa}.
These $D=6$ operators -- which represent a subset of the full basis given for instance in \cite{Grzadkowski:2010es} -- can induce 
CLFV processes either directly or at loop level\footnote{Recent discussions of the phenomenology of the CLFV operators and the experimental bounds on their coefficients have been presented in \cite{Raidal:2008jk,Cirigliano:2009bz,Carpentier:2010ue,Petrov:2013vka,Crivellin:2013hpa,Pruna:2014asa,Beneke:2015lba,Feruglio:2015gka,Davidson:2016edt,Hazard:2016fnc,Crivellin:2016ebg,Crivellin:2017rmk,Dinh:2017smk}.}. 
\begin{table}[t] 
\centering
\renewcommand{\arraystretch}{1.5}
\begin{tabular}{cccc} 
 \hline
\multicolumn{2}{c}{4-leptons operators} &
\multicolumn{2}{c}{Dipole operators} \\
\hline
$Q_{\ell\ell }$ & $(\bar L_L \gamma_\mu L_L)(\bar L_L\gamma^\mu L_L)$ & 
$Q_{eW}$ & $(\bar L_L \sigma^{\mu\nu}e_R) \tau_I \vp W_{\mu\nu}^I$  \\
$Q_{ee}$ & $(\bar e_R \gamma_\mu e_R)(\bar e_R \gamma^\mu e_R)$ &
$Q_{eB}$ & $(\bar L_L \sigma^{\mu\nu} e_R) \vp B_{\mu\nu}$ \\
$Q_{\ell e}$ & $(\bar L_L \gamma_\mu L_L)(\bar e_R \gamma^\mu e_R)$ & & \\
\hline
\multicolumn{4}{c}{2-lepton 2-quark operators} \\
\hline 
$Q_{\ell q}^{(1)}$ & $(\bar L_L \gamma_\mu L_L)(\bar Q_L \gamma^\mu Q_L)$ & 
  $Q_{\ell u}$ & $(\bar L_L \gamma_\mu L_L)(\bar u_R \gamma^\mu u_R)$ \\
$Q_{\ell q}^{(3)}$ & $(\bar L_L \gamma_\mu \tau_I L_L)(\bar Q_L \gamma^\mu \tau_I Q_L)$ & 
$Q_{eu}$ & $(\bar e_R \gamma_\mu e_R)(\bar u_R \gamma^\mu u_R)$ \\
$Q_{eq}$ & $(\bar e_R \gamma^\mu e_R)(\bar Q_L \gamma_\mu Q_L)$ &
$Q_{\ell edq}$ & $(\bar L_L^a e_R)(\bar d_R Q_L^a)$ \\
$Q_{\ell d}$ & $(\bar L_L \gamma_\mu L_L)(\bar d_R \gamma^\mu d_R)$ & 
$Q_{\ell equ}^{(1)}$ & $(\bar L_L^a e_R) \eps_{ab} (\bar Q_L^b u_R)$ \\
$Q_{ed}$ & $(\bar e_R \gamma_\mu e_R)(\bar d_R\gamma^\mu d_R)$ & 
$Q_{\ell equ}^{(3)}$ & $(\bar L _i^a \sigma_{\mu\nu} e_R) \eps_{ab} (\bar Q_L^b \sigma^{\mu\nu} u_R)$ \\
\hline
\multicolumn{4}{c}{Lepton-Higgs operators}\\ 
\hline
$Q_{\vp \ell }^{(1)}$ & $(\vpj)(\bar L_L\gamma^\mu L_L)$ &
$Q_{\vp \ell}^{(3)}$ & $(\vpjt)(\bar L_L \tau_I \gamma^\mu L_L)$\\
$Q_{\vp e}$ & $(\vpj)(\bar e_R \gamma^\mu e_R)$ &
$Q_{e\vp 3}$ & $(\bar L_L e_R\vp)(\vp^\dag\vp)$\\
\hline
\end{tabular}
\caption{Complete list of the CLFV dimension-6 operators from \cite{Crivellin:2013hpa}.
The SM fields are denoted as in Eq.~(\ref{eq:SM}), and $B_{\mu\nu}$ and $W^I_{\mu\nu}$ ($I=1,2,3$) are 
the $U(1)_Y$ and $SU(2)_L$ field strengths. Family indices are not shown, while $a,b=1,2$ are $SU(2)_L$ indices,
and $\tau_I$ are the Pauli matrices.
Flavour indices of the fermions are not indicated.
\label{tab:ops}}
\end{table}
Expressions of the rates of CLFV processes in terms of the coefficients of these operators can be found in \cite{Crivellin:2013hpa,Pruna:2014asa,Beneke:2015lba,Feruglio:2015gka}.
Here, we report for illustration the decay rate of $\mu \to e \gamma$ as induced by the dipole operators of Table~\ref{tab:ops} only.
Upon EW-symmetry breaking $Q_{eW}$ and $Q_{eB}$ mix and give rise to photon-dipole operators $Q_{e\gamma}$.\footnote{The flavour-conserving dipole operators contribute to leptonic anomalous magnetic moments and electric dipole moments, hence these observables are typically related to CLFV processes. For a review on the interplay between the muon $g-2$ and CLFV see \cite{Lindner:2016bgg}.} 
Those that are relevant to $\mu \to e \gamma$ read:
\begin{equation}
\label{eq:Ldipole}
{\mathcal L} \supset \frac{C^{e\mu}_{e\gamma}}{\Lambda^2} \frac{v}{\sqrt{2}}\,  \bar{e}\, \sigma_{\mu \nu} P_R \,\mu\, F^{\mu \nu} + 
\frac{C^{\mu e}_{e\gamma}}{\Lambda^2} \frac{v}{\sqrt{2}}\,  \bar{\mu}\, \sigma_{\mu \nu} P_R \,e \,F^{\mu \nu} + {h.c.},
\end{equation}
with $C^{ij}_{e\gamma} = \cos\theta_W\,C^{ij}_{eB} - \sin\theta_W\,C^{ij}_{eW}$ ($\sin \theta_W\simeq 0.23$ being the weak
mixing). Matching the above Lagrangian to the decay amplitude written in Eq.~(\ref{eq:meg-amp2}), we find:
\begin{equation}
A_R = \frac{\sqrt{2}\, v}{\Lambda^2} C^{e\mu}_{e\gamma},\quad A_L = \frac{\sqrt{2}\, v}{\Lambda^2} C^{\mu e\,*}_{e\gamma}.
\end{equation}
Thus employing these amplitudes in the expression for the decay rate in Eq.~(\ref{eq:meg-rate}), we get:
\begin{align}
\label{eq:dipole-rate}
\Gamma(\mu \to e \gamma) = \frac{m^3_{\mu} v^2}{8\pi \Lambda^4} \left( |C_{e\gamma}^{e\mu}|^2 +  |C_{e\gamma}^{\mu e}|^2 \right).
\end{align}
\begin{figure}[t]
\begin{center}
\includegraphics[width=.6\columnwidth]{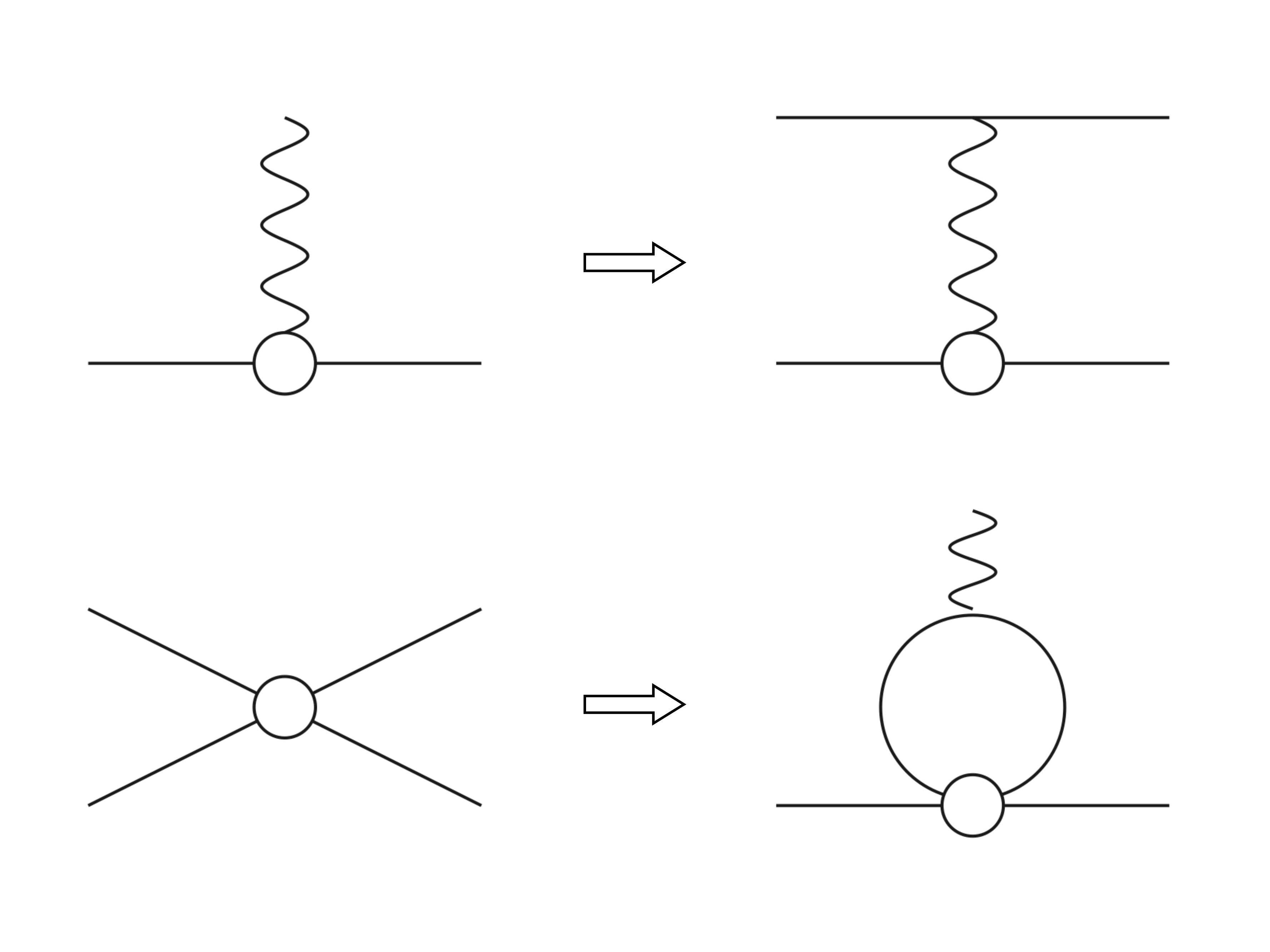}
\caption{Schematic representation of the contribution to processes such as $\ell_i\to\ell_j \ell_k\ell_k$ and $\mu\to e$ conversion arising from a flavour-violating dipole operator and, conversely,  to $\ell_i\to\ell_j \gamma$ from 4-fermion operators. \label{fig:ops-blobs}}
\end{center}
\end{figure}
We can now make use of this last expression -- and the analogous formulae for $\mu \to eee$, $\mu \to e$ in nuclei, and $\tau$ decays~\cite{okadakuno,Crivellin:2013hpa,Petrov:2013vka,Celis:2014asa,Pruna:2014asa,Beneke:2015lba,Feruglio:2015gka} -- to translate the experimental limits on CLFV processes reported in Table \ref{tab:limits} into bounds on the coefficients $|C_a| /\Lambda^2$ of the relevant operators. 
The diagrams of Figure \ref{fig:ops-blobs} schematically show how the dipole operator give also rise to contributions to processes
such as $\ell_i\to\ell_j \ell_k\ell_k$ and $\mu\to e$ conversion,
and vice versa how a 4-fermion operator can contribute to $\ell_i\to\ell_j \gamma$.
In Table \ref{tab:bounds}, we show the bounds for a subset of the operators considered in \cite{Crivellin:2013hpa}, obtained by switching on only one operator at a time.
\begin{table}[t]
\renewcommand{\arraystretch}{1.4}
\centering
\begin{tabular}{cccc}
\hline
 & $|C_a|$ [$\Lambda=1$ TeV] & $\Lambda$ (TeV) [$|C_a| =1$] & CLFV Process \\
 \hline 
 $C_{e\gamma}^{\mu e}$& $2.1\times 10^{-10}$ & $6.8\times 10^{4}$ & $\mu\to e \gamma$\\
   $C_{\ell e}^{\mu \mu\mu e, e \mu \mu\mu}$& $1.8\times 10^{-4}$ & $75$ & $\mu\to e\gamma$ {[\tt 1-loop}] \\
  $C_{\ell e}^{\mu \tau\tau e, e \tau\tau\mu}$& $1.0\times 10^{-5}$ & $312$ & $\mu\to e\gamma$ {[\tt 1-loop}]\\
  \hline 
 $C_{e\gamma}^{\mu e}$& $4.0\times 10^{-9}$ & $1.6\times 10^{4}$ & $\mu\to eee$\\
$C_{\ell\ell,ee}^{\mu eee}$& $2.3\times 10^{-5}$ & $207$ & $\mu\to eee$\\
$C_{\ell e}^{\mu eee, e e\mu e}$& $3.3\times 10^{-5}$ & $174$ & $\mu\to eee$\\
 \hline 
$C_{e\gamma}^{\mu e}$& $5.2\times 10^{-9}$ & $1.4\times 10^{4}$ & $\mu^- {\rm Au} \to e^- {\rm Au}$ \\
${C}_{\ell q,\ell d,e d}^{e\mu}$& $1.8 \times 10^{-6}$ & $745$ & $\mu^- {\rm Au} \to e^- {\rm Au}$ \\
 $C_{eq}^{e\mu}$& $9.2 \times 10^{-7}$ & $1.0 \times 10^3$ & $\mu^- {\rm Au} \to e^- {\rm Au}$ \\
$C_{\ell u,eu}^{e\mu}$& $2.0\times 10^{-6}$ & $707$ & $\mu^- {\rm Au} \to e^- {\rm Au}$ \\
\hline 
 $C_{e\gamma}^{\tau\mu}$& $2.7\times 10^{-6}$ & $610$ & $\tau\to \mu \gamma$\\
$C_{e\gamma}^{\tau e}$& $2.4\times 10^{-6}$ & $650$ & $\tau\to e \gamma$\\
$C_{\ell\ell,ee}^{\mu\tau \mu\mu}$& $7.8\times 10^{-3}$ & $11.3$ & $\tau \to \mu\mu\mu$\\
$C_{\ell e}^{\mu\tau\mu\mu,\mu\mu\mu\tau}$& $1.1\times 10^{-2}$ & $9.5$ & $\tau \to \mu\mu\mu$\\
$C_{\ell\ell,ee}^{e\tau ee}$& $9.2\times 10^{-3}$ & $10.4$ & $\tau\to eee$\\
$C_{\ell e}^{e\tau ee,e e e\tau}$& $1.3\times 10^{-2}$ & $8.8$ & $\tau\to eee$\\
  \hline
\end{tabular}
\caption{
\label{tab:bounds}
Bounds on the coefficients of some of the flavour-violating operators of  e \ref{tab:ops} for $\Lambda=1~{\rm TeV}$ ,
and corresponding bounds on $\Lambda$ (in $\rm TeV$) for $|C_a|=1$. 
Superscripts refer to the flavour indices
of the leptons appearing in the operators.
Adapted from \cite{Feruglio:2015gka,Crivellin:2013hpa,Pruna:2014asa}. }
\end{table}
The second column reports bounds on $|C_a|$ for $\Lambda =1$ TeV, the third column limits on $\Lambda$ for $|C_a|=1$. 
The well-known conclusion of an analysis of this kind is that any new dynamics generating CLFV operators must either (i) feature a strong suppression mechanism (due to loops, small mixing angles, etc.)~yielding $|C_a|\ll 1$, especially if the new particles have TeV-scale masses, or (ii) lie at very high energy scales, in particular if some of the operators in Table \ref{tab:ops} are induced at tree level by unsuppressed couplings between new physics and SM fields, {\em i.e.}~$|C_a|=\ord{1}$. 
Moreover, we can appreciate the complementarity among the different observables, which are more or less sensitive to different operators, hence they are able to test different new physics sources of CLFV. 
In other words, observing or not CLFV in different channels
would provide precious information on the underlying new physics. 
Given the experimental situation, this is especially true for the observables involving $\mu\to e$ transitions. 
Let us for instance consider the case of the dipole operator $Q_{e\gamma}$ as the main source of CLFV. In such a case, Tables \ref{tab:bounds} and \ref{tab:limits} show us that the experimental sensitivity to $\mu\to eee$ and $\mu\to e$ conversion in nuclei has to be improved by more than two orders of magnitudes, in order to provide a more stringent constraint than the one currently given by $\mu \to e \gamma$. 
This is due to the fact that, if the dipole operator dominates, the rates of
 $\mu\to eee$ and $\mu~{\rm N}\to e~{\rm N}$ are suppressed by a factor of order $\alpha$ with respect to $\mu \to e \gamma$ \cite{Hisano:1995cp}, as it can be intuitively understood from Figure \ref{fig:ops-blobs}\footnote{For full calculations of the $\mu\to e$ conversion rates in different nuclei, see \cite{Kitano:2002mt,Cirigliano:2009bz,Crivellin:2014cta}.}:
\begin{align}
\label{eq:m3e}
{\rm BR}(\mu\to eee)
&\simeq
\frac{\alpha}{3\pi}
\bigg(\log\frac{m^2_{\mu}}{m^2_{e}}-3\bigg)\times
{\rm BR}(\mu\to e\gamma)\,, \\
{\rm CR}(\mu~{\rm N}\to e~{\rm N})
&\simeq\alpha \times {\rm BR}(\mu\rightarrow e\gamma)\,.
\label{eq:m2e}
\end{align}
Therefore the MEG bound on BR($\mu \to e \gamma$) translates -- within this scenario -- 
to a limit to the above observables at the $10^{-15}$ level. 
Conversely, a measurement of the rates of $\mu\to eee$ and $\mu~{\rm N}\to e~{\rm N}$
much above that value would clearly signal that the source of CLFV is not the dipole operator $Q_{e\gamma}$, rather some of the 4-fermion operators listed in Table \ref{tab:ops}\footnote{As a matter of fact, there are several new physics models where such operators arise at the tree level, thus with much larger coefficients than the dipoles that can only be loop induced. Some examples will be mentioned in section \ref{sec:models}.}. This would rule out large classes of models, such as the typical supersymmetric frameworks that we will discuss in section \ref{sec:susy}.	A graphical representation
of present and forecast limits on the coefficient of the dipole operators from $\mu\to e$ observables is shown in Figure \ref{fig:spiderweb}.

The above considerations are based on the rather unrealistic hypothesis that new physics effects are encoded in a single operator. Although this can be approximately true in certain
scenarios, yet the coefficients of the operators in Table~\ref{tab:ops} are in general not independent due to radiative effects. Such effects -- summarised by the renormalisation group (RG) equations -- can mix the operators, for instance generating at low energies some
that vanish at the scale $\Lambda$. The effects of the RG running above and below the EW scale -- where a basis of operators invariant under $SU(3)_c\times U(1)_{Q}$ only has to
be employed -- and the matching have been discussed in detail in \cite{Davidson:2016edt,Crivellin:2016ebg,Crivellin:2017rmk}, where
several examples of the resulting correlations among operators are provided. 

Whereas the effective field theory approach briefly introduced in this section 
is certainly a useful tool to describe CLFV effects in a generic model-independent way and study the impact of experimental searches, it is also affected by a limited predictive power. In fact, within the effective field theory, the coefficients of different operators at high-energy scales
are unrelated, while in a specific model they can be instead
correlated, since several operators are typically generated by integrating out heavy degrees of freedom (let's think for instance at the muon decay and the $\beta$-decay 4-fermion operators both generated by integrating out the $W$ boson). It is therefore fruitful to consider in addition some specific 
high-energy theories. The next two sections are devoted to such a discussion.
Finally, let us recall that the effective field theory
is a valid approximation of the full underlying theory only if 
there is a substantial separation between the energy scale of the new degrees of freedom and that associated with CLFV processes. This is not the case if the flavour-violating interactions are mediated by a light new field, {\it e.g.}~the gauge boson of a new symmetry. For recent related studies, see \cite{Farzan:2015hkd,Heeck:2016xkh}.

\section{Supersymmetric CLFV}
\label{sec:susy}
Supersymmetry (SUSY) has represented one of the main theoretical frameworks to motivate experimental efforts 
in search for CLFV, since thirty years at least \cite{Borzumati:1986qx}\footnote{For a general review on SUSY see \cite{Martin:1997ns}, while for a discussion of its phenomenological implications for flavour observables we refer to \cite{Altmannshofer:2009ne}.}. 
The reason is probably twofold. On the one side, models with SUSY broken at energies close to the electro-weak scale have given the prototypical solution to the gauge hierarchy problem ({\em i.e.}~how to keep the Higgs mass 17 orders of magnitude smaller than the Planck scale, avoiding extremely fine-tuned cancellations), which oriented the theoretical speculations for more than thirty years. On the other hand, we have seen that the extreme suppression of CLFV, as shown in Eq.~(\ref{eq:meg-mnu}), is due to the wide separation of the electro-weak and the neutrino mass scale, which clearly does not take place once
the SUSY partners of neutrinos and $W$ (all with masses approximately given by a SUSY breaking scale $m_{\rm SUSY}$) are replaced in the loop of Figure \ref{fig:diag}. Indeed, the contributions of SUSY particles with generic flavour mixing are such that we should have already observed CLFV processes decades ago, unless $m_{\rm SUSY}\gg v$, as we are going to see. 
\begin{table}[t] 
\centering
\renewcommand{\arraystretch}{1.5}
\begin{tabular}{cccc} 
 \hline
{Gauge Eigenstates} &
{Mass Eigenstates}  & Spin & SM partners \\
\hline
 $\widetilde B$, ${\widetilde W}^{0}$, ${\widetilde H}_u^{0}$, ${\widetilde H}_d^{0}$ & 
  ${\widetilde \chi}^0_1$, ${\widetilde \chi}^0_2$,  ${\widetilde \chi}^0_3$, ${\widetilde \chi}^0_4$   &  \multirow{2}{*}{1/2} &  \multirow{2}{*}{$\gamma$, $Z$, $h$}\\
Bino, neut.~Wino and Higgsinos & neutralinos  & & \\ \hline
${\widetilde W}^{\pm}$, ${\widetilde H}_u^{+}$, ${\widetilde H}_d^{-}$ & 
  ${\widetilde \chi}^\pm_1$, ${\widetilde \chi}^\pm_2$ & \multirow{2}{*}{1/2} &  \multirow{2}{*}{$W^\pm$} \\
charg.~Winos and Higgsinos & charginos & & \\ \hline
${\widetilde e}^\pm_R$, ${\widetilde e}^\pm_L$, ${\widetilde \nu}_e$ &
\multirow{3}{1.2cm}{${\widetilde \ell}^\pm_1,\ldots,~{\widetilde \ell}^\pm_6$  \\ $~{\widetilde \nu}_1, {\widetilde \nu}_2, {\widetilde \nu}_3$}
&  \multirow{3}{*}{1}   &  
\multirow{3}{1.2cm}{$e^\pm,\mu^\pm,\tau^\pm$\\  $~\nu_e,\nu_\mu,\nu_\tau$}\\
${\widetilde \mu}^\pm_R$, ${\widetilde  \mu}^\pm_L$, ${\widetilde \nu}_\mu$ &  
 & & \\
${\widetilde \tau}^\pm_R$, ${\widetilde \tau}^\pm_L$, ${\widetilde \nu}_\tau$ &   & & \\
RH/LH selectron, smuon, stau & sleptons, sneutrinos & &  \\    \hline
\end{tabular}
\caption{ Supersymmetric particles relevant for CLFV processes. The mass eigenstates are customarily ordered from the lightest to the heaviest.
\label{tab:sparticles}}
\end{table}

Before going on, we should ask ourselves what is the status of SUSY after LHC has been running 
for several years without finding any sign of it. 
The bounds placed by the LHC experiments on the masses of strongly-interacting SUSY particles are typically well above 1 TeV. For instance, the limit on the mass of the gluino (the fermion partner of the gluon) is approaching 2 TeV in a wide range of searches with different signatures, unless the lightest SUSY particle (LSP) -- and hence the entire SUSY spectrum -- is heavier than about 1 TeV \cite{gluino1,gluino2}. Searches for direct production of stops 
(the scalar partners of the top)
give a bound on the stop mass exceeding 1 TeV for an LSP lighter than 500 GeV, and even in the case the stop-LSP mass splitting is small ($\lesssim$ 100 GeV), stops are bounded to be heavier than about 500 GeV \cite{stops1,stops2}. Furthermore, the measured Higgs mass, $m_h \simeq 125$ GeV, requires substantial radiative corrections (at least in minimal models), which translates into constraints on the masses and left-right mixing of the stops (pushing the former to several TeV if the latter is small), see {\em e.g.}~\cite{Arbey:2011ab}. 
All this surely challenges SUSY as a (fully) natural solution to the hierarchy problem, typically implying fine tuning at the per cent to per mil level, a fate that is shared by most of theories addressing the hierarchy problem. However, certain non-minimal constructions can evade the LHC bounds and be consistent with $m_h \simeq 125$ GeV at a much lower tuning price, for recent proposals see {\em e.g.}~\cite{Calibbi:2016qwt,Ding:2015epa,Katz:2016wtw}.
We also observe that, as we will see below, the CLFV processes depend on masses and couplings of the uncoloured SUSY particles: sleptons, electroweak gauginos (Bino and Wino, {\em i.e.}~the SUSY partners of $U(1)_Y$ and $SU(2)_L$ gauge bosons), and Higgsinos (fermion partners of the two Higgs doublets), see Table \ref{tab:sparticles}.
Hence the above-mentioned constraints have no direct impact on the CLFV effects, unless some specific construction (such as a grand-unified theory) relating the strong and electroweak sectors of the theory is assumed. 
For this reason and for the historical importance of SUSY models for the development of CLFV phenomenology,
we find it sound to revise the constraints from CLFV processes in SUSY, but rather than assuming a specific SUSY-breaking scenario or UV completion, we are going to follow a simplified-model approach and focus only on the particles relevant for our CLFV observables, disregarding whether or not the fine tuning problem is accounted for by another sector of the theory.

Direct searches for the electroweak production at the LHC of the particles we are interested in have set limits substantially beyond those by LEP, at least for a light enough LSP, although weaker than those on the strongly-interacting particles. Searches for events with opposite-sign same-flavour leptons $\ell^+\ell^-$ and large missing transverse momentum $\slashed{E}_T$ address the simple topology of a slepton (selectron or smuon) pair production, followed by decays of the sleptons to the corresponding leptons and the invisible neutralino LSP, $ \widetilde{\chi}^0_1$ : $pp\to \widetilde{\ell}^+ \widetilde{\ell}^- \to \ell^+ \ell^-+ \widetilde{\chi}^0_1 \widetilde{\chi}^0_1$ ($\ell=e,~\mu$). The resulting limits on the slepton masses -- based on the dataset of the $\sqrt{s}=$ 8 TeV run -- are about 250 (300) GeV for RH (LH) sleptons, if $m_{ \widetilde{\chi}^0_1} \lesssim 100$ (150) GeV \cite{Aad:2014vma}.
Much stronger is the recent 13 TeV run limit on the Wino mass, based on events with three leptons and  $\slashed{E}_T$
interpreted in terms of associate production of heavy chargino-neutralino, $pp\to\widetilde{\chi}^\pm_1\widetilde{\chi}^0_2$, decaying through intermediate on-shell sleptons: $ \widetilde{\chi}^\pm_1 \to \nu+ \widetilde{\ell}^\pm  \to\nu + \ell^\pm  \widetilde{\chi}^0_1 $,
$ \widetilde{\chi}^0_2 \to \widetilde{\ell}^\pm  \ell^\mp \to \ell^+\ell^-  +\widetilde{\chi}^0_1$. In this case, the bound on Wino-like chargino and neutralino is up to $1.1\div1.2$ TeV for $m_{\widetilde{\chi}^0_1}\lesssim 600\div700$ GeV \cite{CMS-Wino}. The limit is substantially relaxed if the decays to slepton are kinematically forbidden and $\widetilde{\chi}^\pm_1$ and $\widetilde{\chi}^0_2$ decay directly to the LSP: $\widetilde{\chi}^\pm_1\to W^\pm\widetilde{\chi}^0_1$, $\widetilde{\chi}^0_2 \to Z/h~\widetilde{\chi}^0_1$. This gives 
a constraint on the Wino mass of 450 GeV at most and only for a rather light LSP,  $m_{\widetilde{\chi}^0_1}\lesssim 150$ GeV \cite{CMS-Wino}.
If interpreted in terms of production of Higgsino-like chargino and neutralinos, this constraint is further relaxed as a consequence of the reduced production cross section.

In supersymmetric extensions of the SM, new sources of CLFV stem from the couplings among the SM leptons, their SUSY partners ({\em i.e.}~the LH and RH sleptons and the sneutrinos), and the neutralinos and charginos (which are mixture of, respectively, the neutral and charged gauginos and Higgsinos). These interactions are flavour violating, unless the lepton and slepton mass matrices are aligned and thus can be simultaneously diagonalised, which does not occur in general.
The $6\times6$ slepton mass matrix is given by:
\begin{align}
\label{eq:Mslep}
\mathcal{M}_{\widetilde \ell}= \left( 
\begin{array}{cc}
\Delta_{LL} & \Delta_{LR} \\
\Delta_{RL} & \Delta_{RR} 
\end{array}
 \right),
\end{align}
\begin{align}
\label{eq:deltaLL}
\left(\Delta_{LL}\right)_{ij} & = \left(\tilde{m}^2_L \right)_{ij} + \left(Y_e^\dagger Y_e \right)_{ij} \frac{v^2_d}{2}
+  m_Z^2  \cos2\beta \left(-\frac{1}{2}+\sin^2\theta_W \right) \delta_{ij}\,, \\
\left(\Delta_{RR}\right)_{ij} & = \left(\tilde{m}^2_{E} \right)_{ij} + \left(Y_e^\dagger Y_e \right)_{ij} \frac{v^2_d}{2}
-  m_Z^2  \cos2\beta \sin^2\theta_W \delta_{ij}\,, \\
\left(\Delta_{LR}\right)_{ij} & = \left(\Delta^{*}_{RL}\right)_{ji} =
 \frac{v_d}{\sqrt{2}} \left( (A_e)_{ij} - \mu^*(Y_e)_{ij} \tan\beta \right)\,,
\label{eq:deltaLR}
\end{align}
where $\mu$ is the Higgsino mass parameter, and
$\tan\beta \equiv v_u/v_d$ is the ratio of the vevs of the two Higgs doublets (thus $v_d=v \cos\beta$, $v_u=v \sin\beta$).
The $3\times3$ matrices $\tilde{m}^2_L$, $\tilde{m}^2_{E}$,  and $A_e$ contain respectively SUSY-breaking mass terms of LH sleptons $\widetilde L_i$, RH sleptons $\widetilde E_i$, and trilinear scalar terms $\widetilde L$-$\widetilde E$-Higgs.
If these SUSY-breaking matrices have off-diagonal entries in the basis in which $Y_e$ is diagonal,
 $\mathcal{M}_{\widetilde \ell}$ can not be simultaneously diagonalised with the lepton mass matrix. As a consequence, the physical sleptons are mixtures of different flavours.
 
\begin{figure}
\begin{center}
\includegraphics[width=.65\columnwidth]{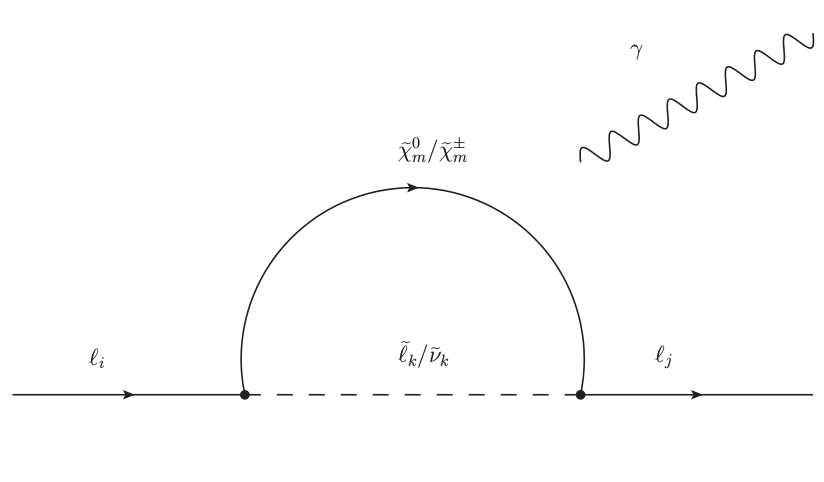}
\caption{Supersymmetric contribution to $\ell_i\to \ell_j \gamma$.
It is understood that the photon attaches to the fermion or the scalar in the loop, depending on which field is charged.
\label{fig:susy-loop}}
\end{center}
\end{figure}
In standard SUSY models with conserved R-parity, the dominant CLFV effects are captured by dipole operators like those shown in Eq.~(\ref{eq:Ldipole}) for $\mu\to e \gamma$, which arise from sneutrino-chargino and slepton-neutralino loops, as shown in
Figure \ref{fig:susy-loop}.
Defining for convenience the normalisation of the invariant amplitudes as follows:
\begin{equation}
\frac{C_{e\gamma}^{ij}}{\Lambda^2} =y_i  \frac{e}{2} \mathcal{A}_R^{ij},\quad
\frac{C_{e\gamma}^{ji*}}{\Lambda^2}= y_i \frac{e}{2}  \mathcal{A}_L^{ij},\quad
\left[y_i =(m_e,m_\mu,m_\tau)\sqrt{2}/v\right],
\end{equation}
the expression for BR($\ell_i \to \ell_{j}\gamma$) is obtained from a generalisation of Eq.~(\ref{eq:dipole-rate}):
\begin{equation}
\label{eq:BR_LFV}
\frac{{\rm BR}(\ell_i \to \ell_{j}\gamma)}{{\rm BR}(\ell_i \to \ell_{j} \nu_{i} \bar\nu_{j})} =
\frac{48\pi^3\alpha}{G^2_F}
\left( |\mathcal{A}_{L}^{ij}|^2 + |\mathcal{A}_{R}^{ij}|^2\right)\,.
\end{equation}
Furthermore, the fact that dipole diagrams as in Figure \ref{fig:susy-loop} dominate implies that, for $\mu\to eee$ and $\mu\to e$ conversion, 
the correlations of Eqs.~(\ref{eq:m3e}, \ref{eq:m2e}) hold.
The full 1-loop expressions of $\mathcal{A}_{L}$ and $\mathcal{A}_{R}$ can be found in \cite{Hisano:1995cp,Ilakovac:1994kj,Arganda:2005ji,Arganda:2007jw}\footnote{The first calculation of the neutralino-slepton and chargino-sneutrino loop contributions to $\mu\to e \gamma$ has been presented in \cite{Lee} in the context of a theory with R-parity violation where LFV arises from the mixing of sleptons (sneutrinos) with charged (neutral) Higgses.}.
For our discussion, it is sufficient to report the expressions in the so-called Mass Insertion Approximation (MIA) \cite{Hall:1985dx}.
Working in a basis where lepton-slepton couplings are flavour diagonal, the flavour violation instead appears in a non-diagonal slepton propagator that can be expressed as a series expansion in terms of the off-diagonal slepton mass terms, 
Eqs.~(\ref{eq:deltaLL}-\ref{eq:deltaLR}).
The approximation consists in retaining the lowest term of such expansion only 
and it is better the smaller the slepton mass splitting is.
The expressions of the amplitude can be further simplified if a similar approximation is applied to neutralinos and 
charginos, working in the gauge eigenbasis and treating Bino-Higgsino and 
Wino-Higgsino mixing as mass insertions of the order $v/M_{1,2}$ ($M_1$ and $M_2$ being the Bino and Wino mass terms) 
and $v/\mu$ \cite{Hisano:1998fj,Paradisi:2005fk}. 
In the Appendix, we show the MIA expressions for the different contributions to the amplitudes (from \cite{Calibbi:2015kja}) classified according to the source of the chirality flip required to match the dipole operator\footnote{As we have seen where sketching the calculation in section \ref{sec:neutrinos} and it is apparent from the form of the dipole operator, see Eq.~(\ref{eq:Ldipole}), the leptons in Figure \ref{fig:susy-loop} need to have opposite chiralites.} 
and the kind of gauge interaction -- $SU(2)_L$ or $U(1)_Y$ -- involved.

\subsection{Simplified SUSY models}
As is customary, we adopt as a convenient parameterisation of the amount of CLFV the off-diagonal entries of the matrices in  Eqs.~(\ref{eq:deltaLL}-\ref{eq:deltaLR}) ({\em i.e.}~of  $\tilde{m}^2_L$, $\tilde{m}^2_{E}$,  
and $A_e$), normalised by the diagonal entries, which are approximately equal to the slepton masses:
\begin{align}
\label{eq:deltas}
(\delta_{\rm LL})_{ij}= \frac{(\Delta_{\rm LL})_{ij}}{\sqrt{(\tilde{m}^2_{L})_{ii}(\tilde{m}^2_{L})_{jj}}},~
(\delta_{\rm RR})_{ij}= \frac{(\Delta_{\rm RR})_{ij}}{\sqrt{(\tilde{m}^2_{E})_{ii}(\tilde{m}^2_{E})_{jj}}},~
(\delta_{\rm LR})_{ij}=  \frac{(\Delta_{\rm LR})_{ij}}{\sqrt{(\tilde{m}^2_{L})_{ii}(\tilde{m}^2_{E})_{jj}}},
\end{align}
where $i\neq j$.
From the formulae in the Appendix, we see that the above quantities can be constrained by the experimental limits on CLFV
once information on the SUSY spectrum (and other parameters such as $\tan\beta$) are provided. 
Such an exercise gives, on the one hand, a useful indication of the extent to which lepton and slepton mass matrices have to be aligned to fulfil the CLFV bounds for a given SUSY scale (which sets a overall suppression of the rates), on the other hand,  it
allows a quick assessment of the sensitivity of present and future experiments once a specific model for the flavour structure
of the SUSY-breaking matrices is considered.
Early discussions of bounds on the CLFV parameters of Eq.~(\ref{eq:deltas}) can be found in \cite{Gabbiani:1988rb,Gabbiani:1996hi,Masina:2002mv,Paradisi:2005fk}. 
Here, following \cite{Calibbi:2015kja}, we are going to revise and update these bounds within the same simplified models (defined by a subset of the relevant particles and couplings) employed by the LHC collaboration to interpret the above-mentioned searches for EW production of SUSY particles. This allows a simple and direct comparison of the discovery potential of high-energy and low-energy experiments,
and at the same time, it highlights the impact of the different contributions listed in the Appendix.

\paragraph{Bino and RH or LH sleptons}
\begin{figure}[t]
\begin{center}
\includegraphics[width=.42\columnwidth]{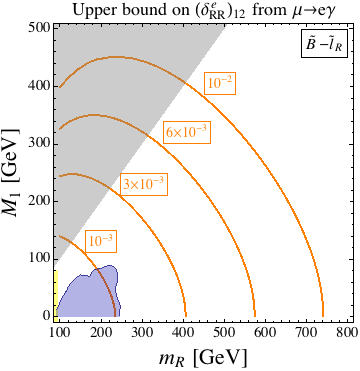}
\hspace{0.5cm}	
\includegraphics[width=.42\columnwidth]{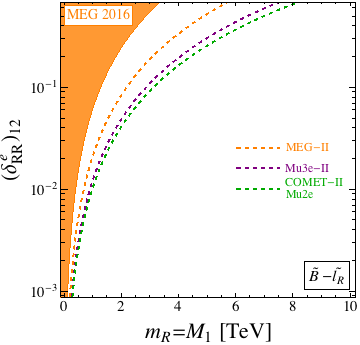}\\
\vspace{0.5cm}
\includegraphics[width=.42\columnwidth]{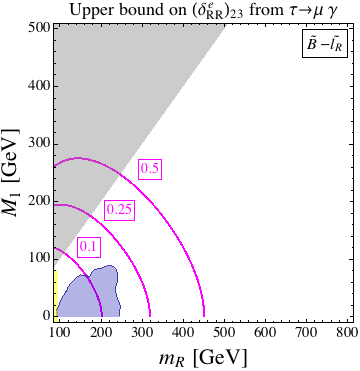}
\hspace{0.5cm}	
\includegraphics[width=.42\columnwidth]{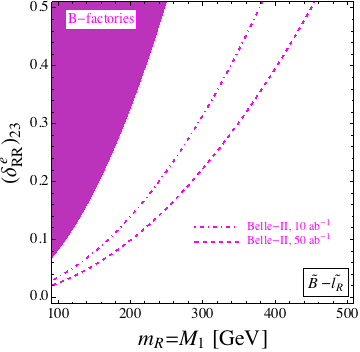}
\caption{
Bounds on the CLFV parameters $\delta_{RR}$ and prospects for $\mu\to e$ observables (first row) and 
$\tau\to\mu\gamma$ (second row), for a model with Bino and RH sleptons as the only light SUSY particles.
The blue-shaded region is excluded by direct searches for sleptons at the LHC \cite{Aad:2014vma}.
\label{fig:BE}}
\end{center}
\end{figure}
We first consider a model where the only light SUSY particles are the Bino and the RH sleptons, while the rest of the spectrum is decoupled. 
Under this hypothesis, among the possibility shown the Appendix, the only possible contribution to $\ell_i \to \ell_{j}\gamma$ requires 
flavour-mixing among RH sleptons and the chirality flip on an external lepton line, hence the amplitudes appearing in Eq.~(\ref{eq:BR_LFV}) are simply:
\begin{equation}
\label{eq:BE}
\mathcal{A}_{L} = 0,\quad \mathcal{A}_{R} = (\mathcal{A}^{n_1}_R)_{\small U(1)} = \frac{\alpha_Y}{\pi} \frac{\delta_{RR}}{m_R^2}
f_{1n} \left(\frac{|M_1|^2}{m_R^2}\right),
\end{equation}
where $M_1$ and $m_R$ are respectively the Bino and the common slepton mass, and the loop function is given in the Appendix. 
As we can see, this contribution does depend only on the masses of the particles in the loop and on the flavour mixing. We show
the resulting bounds and prospects in Figure \ref{fig:BE} for $\mu\to e$ observables (first row) and $\tau\to\mu\gamma$ (second row).
The prospected CLFV limits shown here correspond to ${\rm BR}(\mu\to e\gamma) =5\times 10^{-14}$, ${\rm BR}(\mu\to eee) = 10^{-16}$,
${\rm CR}(\mu~N\to e~N) =6\times 10^{-17}$, 
${\rm BR}(\tau\to \mu\gamma) =8(4)\times 10^{-9}$ for 10 (50) ab$^{-1}$ of integrated luminosity at Belle 2 (see sections \ref{sec:exps} and \ref{sec:tau} for details).
The present exclusion from the LHC search in \cite{Aad:2014vma} is also shown in blue for a comparison\footnote{The shown interpretation of the search in \cite{Aad:2014vma} assumes mass-degenerate selectron and smuon with no substantial flavour mixing. How the resulting mass limit can change in presence of large mass splitting and/or flavour-violating decays of the sleptons has been discussed in \cite{Calibbi:2015kja}.}. 

\noindent From the figure, we see that the present bound on ${\rm BR}(\mu\to e\gamma)$ set by MEG is able to exclude the model beyond 
the direct searches for sleptons at the LHC if $(\delta_{RR})_{12}\gtrsim 10^{-3}$.
Future experiments on $\mu\to e$ transitions are sensitive to sleptons as heavy as $\approx 8$ TeV for $\mathcal{O}(1)$
values of the CLFV parameter $(\delta_{RR})_{12}$.

\noindent In a similar model with only Bino and LH sleptons instead of RH sleptons, we have
$\mathcal{A}_{L} =  (\mathcal{A}^{n_1}_L)_{\small U(1)}$,  $\mathcal{A}_{R} =0$. As one can see from the Appendix, 
the total amplitude is then a factor 4 smaller than in the previous case, due to the different hypercharges of LH and RH leptons. 
Hence the resulting bounds on $(\delta_{LL})_{12}$ are numerically as those in Figure \ref{fig:BE} multiplied by 4.

\paragraph{Bino, LH and RH sleptons}
\begin{figure}[t]
\begin{center}
\includegraphics[width=.42\columnwidth]{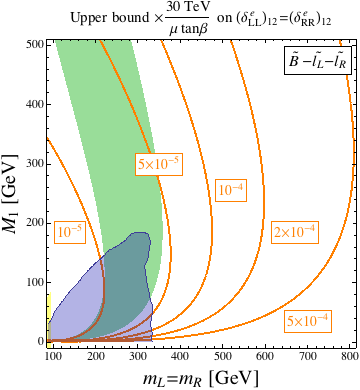}
\hspace{0.5cm}	
\includegraphics[width=.42\columnwidth]{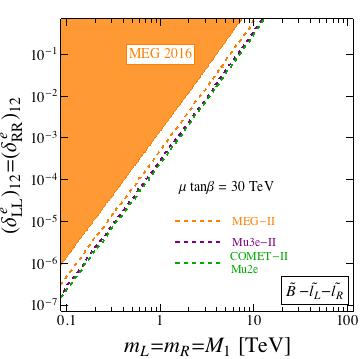}\\
\vspace{0.5cm}
\includegraphics[width=.42\columnwidth]{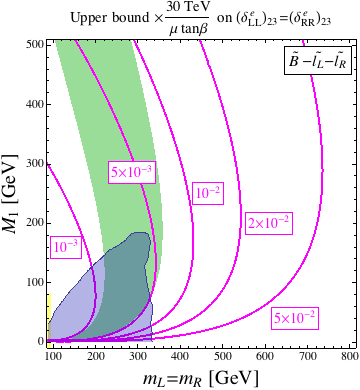}
\hspace{0.5cm}	
\includegraphics[width=.42\columnwidth]{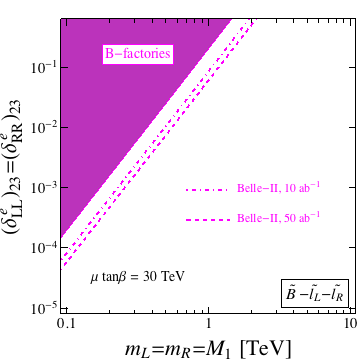}
\caption{Bounds and prospects for a model with Bino and both LH and RH sleptons.
The blue-shaded region is excluded by direct searches for sleptons at the LHC \cite{Aad:2014vma}.
The green region corresponds to a contribution to the muon $g-2$ at the level of the observed discrepancy.
\label{fig:BLE}}
\end{center}
\end{figure}
If both kinds of sleptons are light, there are additional and qualitatively different contributions to the $\ell_i\to\ell_j\gamma$ amplitudes:
\begin{equation}
\mathcal{A}_{L} =  (\mathcal{A}^{n_1}_L)_{\small U(1)}+(\mathcal{A}^{n_3}_L)_{\small U(1)},
\quad \mathcal{A}_{R} = (\mathcal{A}^{n_1}_R)_{\small U(1)} +  (\mathcal{A}^{n_3}_R)_{\small U(1)}.
\end{equation}
In particular, one can draw diagrams -- corresponding to $(\mathcal{A}^{n_3}_{L,R})_{\small U(1)}$ -- with the chirality flip inside the loop on the slepton line due to left-right slepton mixing, thus avoiding the suppression from small lepton masses occurring in cases like the previous one where the chirality flip is only possible on the external lepton lines.
Bounds and prospects for this model are shown in Figure \ref{fig:BLE}, where, for illustration, we set the $\delta_{LR}$ to zero 
and take $\delta_{LL}=\delta_{RR}$ as well as equal LH and RH slepton masses. 
As indicated in the Figure, the bounds scale linearly with the product 
$\mu\times\tan\beta$ that controls the smuon or stau left-right mixing. Moreover, if either $\delta_{LL}$ or $\delta_{RR}$
vanish, the bounds are approximately relaxed by a factor $\sqrt{2}$.
As we can see, the MEG limit requires very stringent bounds on $(\delta_{LL})_{12}$ and $(\delta_{RR})_{12}$  -- at the level $10^{-5}\div10^{-4}$ or smaller -- for SUSY particles in the ballpark of the LHC sensitivity.
Interestingly, unlike is the previous model, there is a region of the parameter space (shaded in green in the figure) where 
the muon $g-2$ anomaly can be fitted at the level of 2 $\sigma$ or better, {\em i.e.}~$\Delta a_\mu^{\rm SUSY}= (2.9\pm 1.8)\times
10^{-9}$ \cite{Jegerlehner:2009ry}\footnote{Contributions to leptonic magnetic and electric dipole moments arise from flavour-conserving diagrams similar to that of Figure \ref{fig:susy-loop}, thus for a given slepton mixing these observables are correlated to CLFV processes. For a recent discussion on the correlation between the muon $g-2$ and $\mu\to e\gamma$ in SUSY, see \cite{Kersten:2014xaa}. For the contributions to the electron EDM from flavour-mixing sleptons see \cite{Hisano:2008hn}.}.
\paragraph{Wino, Bino, and LH sleptons}
 \begin{figure}[t]
\begin{center}
\includegraphics[width=.42\columnwidth]{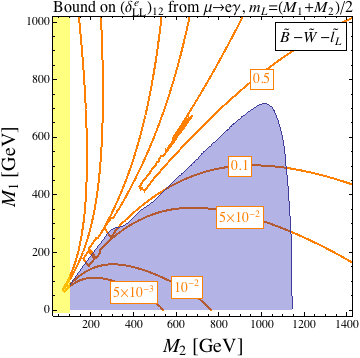}
\hspace{0.5cm}	
\includegraphics[width=.42\columnwidth]{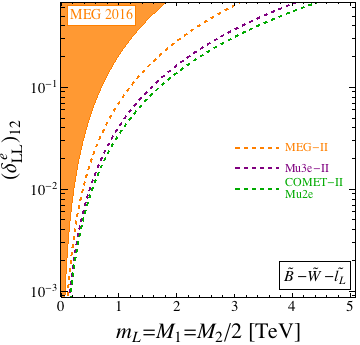}
\caption{
Bounds and prospects for a model with Bino, Wino, and LH sleptons.
The blue-shaded region is excluded by direct searches for sleptons at the LHC \cite{CMS-Wino}.
\label{fig:WBL}}
\end{center}
\end{figure}
In this model, the presence of the Wino introduces contributions controlled by $SU(2)_L$ interactions, corresponding
to loops of charged Wino and sneutrinos, and neutral Wino and charged sleptons.
\begin{equation}
\mathcal{A}_{L} =  (\mathcal{A}^{n_1}_L)_{\small U(1)}+(\mathcal{A}^{n_1}_L)_{\small SU(2)}+(\mathcal{A}^{c_1}_L)_{\small SU(2)},
\quad \mathcal{A}_{R} = 0.
\end{equation}
This model has  been employed by the LHC collaborations for the interpretation of searches based on multi-leptons plus missing energy, assuming a Bino LSP, and sleptons half-way between the
Bino and Wino. The highest sensitivity is reached in the case of heavy neutralino-chargino ({\em i.e.}~neutral and charged Wino) associated production, followed by decays to the Bino LSP through intermediate on-shell sneutrinos and charged sleptons, 
leading to events with three leptons and $\slashed{E}_T$.
Again, we compare the LHC exclusion -- given by a recent CMS analysis of 13 TeV run data \cite{CMS-Wino} -- with
the CLFV bounds. The result is shown in Figure \ref{fig:WBL} for the $\mu\to e$ processes only, as 
$\tau\to\mu\gamma$ gives irrelevant bounds at present. As we can see, the $\mu\to e \gamma$ constraint itself
is quite mild, which is due to cancellations among the neutralino and chargino contributions. As a consequence, this model
provides an example where at present direct constraints from the LHC tend to be stronger than those from CLFV experiments even for large flavour mixing. The situation may be reversed given the sensitivity of the future experiments, as shown in the second plot of Figure \ref{fig:WBL}.
  
\paragraph{Wino, Higgsinos, and LH sleptons}
\begin{figure}[t]
\begin{center}
\includegraphics[width=.42\columnwidth]{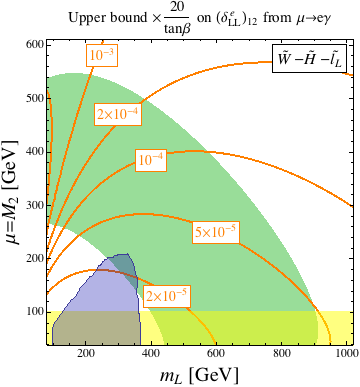}
\hspace{0.5cm}	
\includegraphics[width=.42\columnwidth]{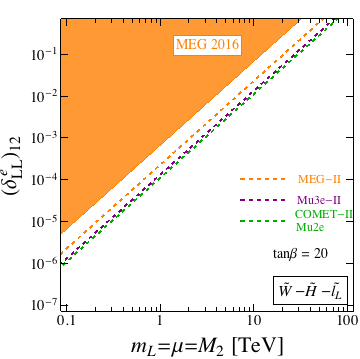}\\
\vspace{0.5cm}
\includegraphics[width=.42\columnwidth]{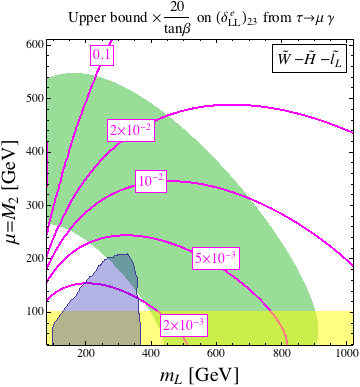}
\hspace{0.5cm}	
\includegraphics[width=.42\columnwidth]{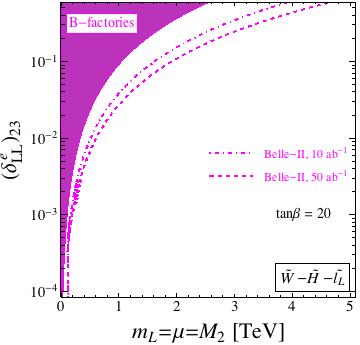}
\caption{Bounds and prospects for a model with Wino, Higgsinos, and LH sleptons.
The blue-shaded region is excluded by direct searches for sleptons at the LHC \cite{Aad:2014vma}, according to the reinterpretation in \cite{Eckel:2014dza}.\label{fig:WHL}}
\end{center}
\end{figure}
The last model we consider here features Wino, Higgsinos and LH sleptons. The presence of light Higgsinos allows diagrams where the chirality flip occurs at one diagram's vertex, since the Higgsino coupling involves LH sleptons and RH leptons (and vice versa).
These contributions are proportional to the lepton Yukawa couplings, hence they are enhanced by a factor $\sim\tan\beta$
compared to those with the external chirality flip. In total, the amplitude reads:
\begin{equation}
\mathcal{A}_{L} = (\mathcal{A}^{n_1}_L)_{\small SU(2)}+(\mathcal{A}^{c_1}_L)_{\small SU(2)} +  (\mathcal{A}^{n_2}_L)_{\small  SU(2)}+  (\mathcal{A}^{c_2}_L)_{\small  SU(2)},
\quad \mathcal{A}_{R} = 0.
\end{equation}
The resulting stringent bounds on $(\delta_{LL})_{12}$  and $(\delta_{LL})_{23}$ are displayed in Figure \ref{fig:WHL},
as well as the prospects of future experiments. These will be able to test SUSY masses up to $\mathcal{O}(100)$ TeV,
in presence of large smuon-selectron mixing. The figure also shows that a large contribution to the muon $g-2$
can be easily achieved for SUSY masses still beyond the present LHC sensitivity.

\subsection{Specific SUSY setups}
As it is clear from the above discussion, we need to know two basic ingredients if we want to assess the rates of CLFV observables in a SUSY model: (i) the masses of the sleptons and the other SUSY particles appearing in diagrams like that in Figure \ref{fig:susy-loop}; (ii) the compositions of the sleptons in terms of flavour eigenstates or --  which is equivalent in the the limit of degenerate slepton masses -- the CLFV parameters of Eq.~(\ref{eq:deltas}). In particular, if we lack the latter piece of information, we can just say that SUSY can give CLFV rates at the level of the present limits as well as vanishingly small! Similarly, we have no 
clue of the relative size of different flavour transitions, {\it e.g.}~$\mu\to e$ and $\tau\to \mu$ observables. Can we do better than this (which is not much)?  
Indeed we can if, unlike in the previous subsection, we work in a specific high-energy framework, such as a grand-unified theory, a SUSY-breaking mechanism, a model of fermion masses and mixing based on a new symmetry of flavour\footnote{There is a plethora of possible flavour models depending 
on the the employed symmetry: from a simple $U(1)$ as in the original proposal
by Froggatt and Nielsen \cite{Froggatt:1978nt,Leurer:1992wg,Leurer:1993gy}, to non-abelian groups, continuous (such as $U(2)$ \cite{Pomarol:1995xc,Barbieri:1995uv}, or $SU(3)$ \cite{King:2001uz}) or discreet ({\em e.g.}~$A_4$ \cite{Altarelli:2005yx}). Besides shaping the Yukawa matrices thus accounting for their hierarchical structure, in supersymmetric extensions of the SM, these symmetries also dictate the flavour structure of the squark and slepton mass matrices.}, and so on. 
In setups of this kind, we have better control on the flavour structure of SUSY-breaking terms appearing in the slepton mass 
matrix, cf.~Eqs.~(\ref{eq:deltaLL}-\ref{eq:deltaLR}), and thus on the prediction of the CLFV observables. 

On one extreme, there are models that predict an anarchical flavour structure of the slepton mass terms, {\em i.e.}~$\mathcal{O}(1)$
values for the CLFV parameters in Eq.~(\ref{eq:deltas}), such as generic gravity-mediated SUSY breaking and certain 
flavour models based on a $U(1)$ flavour symmetry, see {\em e.g.}~\cite{Dudas:2010yh}. In such a case, as we have seen in the previous subsection, 
the present bounds set a lower bound on the SUSY masses from several TeV up to several dozens of TeV, certainly beyond the reach of the LHC (see also \cite{Altmannshofer:2013lfa}).

A more interesting case is represented by models, in which the flavour structure of the matrices in Eqs.~(\ref{eq:deltaLL}-\ref{eq:deltaLR}) is controlled -- and to some extent protected -- by the same dynamics that generates the
hierarchical structure of the SM Yukawa matrices, {\em e.g.}~a flavour symmetry such as $SU(3)$, $U(2)$, $A_4$  (for studies focusing on CLFV within these classes of models, see \cite{Calibbi:2008qt,Calibbi:2009ja,Blankenburg:2012nx,Feruglio:2009hu,Altarelli:2012bn}),
or the partial compositeness of the fermions in strongly-interacting theories with a composite Higgs/extra-dimensional Randall-Sundrum models \cite{Nomura:2007ap,Nomura:2008pt,Dudas:2010yh,KerenZur:2012fr}.

The opposite extreme is given by scenarios that predict a trivial flavour structure of the SUSY-breaking terms, {\em i.e.}~aligned lepton and slepton mass matrices and no CLFV effects at all. This situation can be realised by a flavour-blind SUSY-breaking mechanism
such as in Gauge Mediation models \cite{Giudice:1998bp}: supersymmetry is broken in a hidden sector and the breaking is transmitted to the visible sector through loop diagrams only involving gauge interactions that do not distinguish among flavours, 
so that the matrices $\tilde{m}^2_L$ and $\tilde{m}^2_E$ result proportional to the identity matrix $\mathbf 1$. 
This situation -- rather frustrating from the point of view of CLFV experiments -- can however change in at least two cases.
Gauge Mediation models can be extended by allowing couplings among the SM fields and the messenger fields that
communicate the breaking to the visible sector. Such couplings -- that are in fact allowed by the gauge symmetries so that are present unless {\it ad hoc} symmetries are introduced --
help Gauge Mediation models to achieve the observed Higgs mass without superheavy stops \cite{Chacko:2001km,Evans:2011bea,Evans:2012hg,Evans:2013kxa}, and induce new contributions
to the sfermion mass matrices spoiling the above-mentioned flavour universality although in a controlled way \cite{Shadmi:2011hs,Abdullah:2012tq,Calibbi:2013mka}. As a consequence, CLFV processes are potentially observables, as has been discussed in \cite{Calibbi:2014yha}.
A second way to induce interesting slepton flavour mixing even in presence of a flavour-blind SUSY-breaking mechanism such as Gauge Mediation is through radiative corrections. Indeed, this unavoidably occurs if there is some intermediate scale featuring fields couplings to the SM leptons below the scale that SUSY-breaking is mediated at. 
Examples of such radiative breaking of flavour universality of the slepton masses -- that is summarised, as we will see, by the renormalisation group evolution of the SUSY-breaking terms -- are given by flavour models where the breaking of the flavour symmetry occurs at a scale lower than the SUSY-breaking mediation scale \cite{Calibbi:2012yj}, or by a seesaw model as the one of Eqs.~(\ref{eq:Lseesaw}, \ref{eq:seesaw}) again at a scale lower than the mediation scale. Starting from \cite{Borzumati:1986qx}, this latter case has been object of a large number of studies, see {\em e.g.}~\cite{Ilakovac:1994kj,Hisano:1995cp,Hisano:1998fj,Casas:2001sr,Masiero:2002jn,Arganda:2005ji,susy-seesaw}. We will give some details about it in the next subsection. 
 
\subsection{SUSY seesaw}
The seesaw mechanism provides a natural explanation of the smallness of the neutrino masses, as well as a convincing 
framework -- called leptogenesis -- to address the asymmetry between baryons and anti-baryons observed in the universe \cite{Buchmuller:2004nz,Fong:2013wr}.
In its simplest version just requires the introduction of RH neutrino fields, but marrying it with low-energy supersymmetry
gives rich implications in terms of CLFV phenomenology. 
As we have mentioned at the end of the previous subsection, in supersymmetric seesaw models, mixing among different-flavour sleptons is radiatively induced even in presence of a flavour-blind SUSY-breaking mechanism. 
This effect can be computed by solving the renormalisation group equation (RGE) for the LH slepton mass terms \cite{Hisano:1995cp}:
\begin{align}
\frac{d}{dt} \tilde{m}^2_L = \left(\frac{d}{dt} \tilde{m}^2_L\right)_{\rm \small MSSM} +&
\frac{1}{16\pi^2} \left( \tilde{m}^2_L  Y_\nu^\dag Y_\nu + Y_\nu^\dag Y_\nu  \tilde{m}^2_L\right)+ \nonumber\\
& \frac{1}{8\pi^2}\left(  Y_\nu^\dag \tilde{m}^2_\nu Y_\nu+
 \tilde{m}^2_{H_d} Y_\nu^\dag Y_\nu + A_\nu^\dag A_\nu \right), 
 \label{eq:rge}
\end{align}
where $t=\ln \mu/\overline{\mu}$ with $\mu$ being the renormalisation scale and $\overline{\mu}$ any reference scale, {\em e.g.}~$m_{\rm SUSY}$. The first term on the right-hand side of Eq.~(\ref{eq:rge}) collects ordinary flavour-diagonal terms,
$\tilde{m}^2_\nu$ and $\tilde{m}^2_{H_d}$ are SUSY-breaking masses of the RH sneutrinos and the down-type Higgs respectively, and $A_\nu$ is the matrix of the trilinear scalar couplings Higgs-slepton-sneutrino. 
The above RGE shows that the size of the entries $(\tilde{m}^2_L )_{i\neq j}$,  and thus the slepton mixing effects, does depend on the flavour structure of the neutrino Yukawa matrix $Y_\nu$ defined in Eq.~(\ref{eq:Lseesaw}).
Assuming a common SUSY-breaking mass for the scalars at the mediation scale $M_X$, namely $\tilde{m}^2_L = \tilde{m}^2_\nu = \tilde{m}^2 \mathbf{1}$ and $\tilde{m}^2_{H_d} = \tilde m^2$, and approximately integrating the Eq.~(\ref{eq:rge}), we can estimate the low-energy value of the CLFV parameters:
\begin{align}
(\delta_{\rm LL})_{i\neq j} \approx  - \frac{3 +a_\nu^2}{8\pi^2} (Y^*_\nu)_{ki}   (Y_\nu)_{kj} \ln\frac{M_X}{M_{R_k}},
\label{eq:delta-seesaw}
\end{align}
where we neglected the running of the diagonal slepton masses, {\em i.e.}~we took $(\tilde{m}^2_L)_{ii}\approx \tilde m^2$ 
at low energy, and we defined $A_\nu \equiv a_\nu \tilde{m} Y_\nu$. 
$M_{R_k}$ are the masses of the RH neutrinos.
Similarly off-diagonal entries of the lepton A-term $A_e$, hence $(\delta_{\rm LR})_{i\neq j}$, are generated. In contrast, no flavour violation is induced in the RH sector, 
{\em i.e.}~$(\delta_{\rm RR})_{i\neq j}=0$, as the RGE for $\tilde{m}^2_E$ only depends on gauge couplings and lepton Yukawa couplings $Y_e$, hence $\tilde{m}^2_E$ remains diagonal in the basis in which $Y_e$ is.

As mentioned above, the size of our CLFV effects depend on the flavour structure of $Y_\nu$, which unfortunately can not be inferred by the seesaw formula, Eq.~(\ref{eq:seesaw}), even if we had a perfect knowledge of neutrino masses and mixing (including the phases and the absolute mass scale). The reason is that the formula involves more parameters than those that can be measured at low energies from neutrino experiments. This can be seen by solving Eq.~(\ref{eq:seesaw}) for $Y_\nu$ and parameterzing such a mismatch by means of a orthogonal complex matrix $R$ (hence containing 6 real parameters) \cite{Casas:2001sr}:
\begin{equation}
Y_\nu = \frac{1}{v_u}  \sqrt{ \hat{M}_R} \,R\, \sqrt{\hat{m}_\nu}\, U^\dagger,
\label{eq:CI}
\end{equation}
where $\hat{m}_\nu$ and $\hat{M}_R$ are diagonal mass matrices for the light neutrinos and heavy RH neutrinos respectively,
and $U$ is the PMNS matrix.
It is then clear that a definite prediction of the CLFV rates requires further assumptions. 
\begin{figure}[t]
\begin{center}
\includegraphics[width=.42\columnwidth]{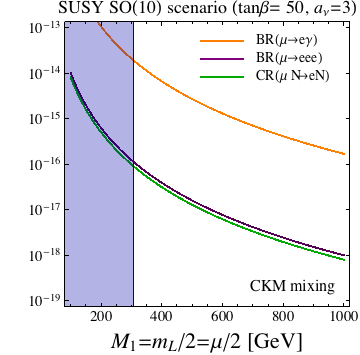}
\hspace{0.5cm}	
\includegraphics[width=.42\columnwidth]{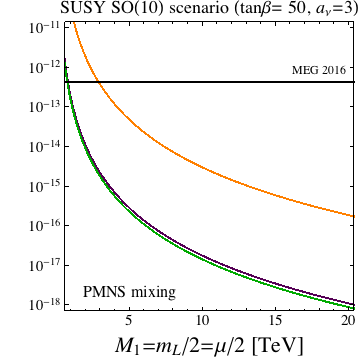}
\caption{Left: Rates of $\mu\to e$ processes as predicted in a SUSY seesaw with the `minimal' $SO(10)$ relation in Eq.~(\ref{eq:so10}). $SO(10)$ mass relations among SUSY particles are taken. The blue-shaded area is excluded by the latest searches for gluinos at the LHC \cite{gluino1,gluino2}. The chosen parameters aim at maximising the CLFV effects, {\em e.g.}~${\rm BR}(\mu\to e\gamma)\propto (3+a_\nu)^2\tan^2\beta$. Right: the same for the case of $SO(10)$ with large slepton mixing 
as given by Eq.~(\ref{eq:so10pmns}).
\label{fig:SUSY-SO10}}
\end{center}
\end{figure} 

The first example we want to show here is inspired by $SO(10)$ GUT. In $SO(10)$, each generation of SM fermions  plus one RH neutrino are unified in a 16-dimensional representation, which makes it a very natural high-energy completion
of the seesaw mechanism. Another consequence is that quark and lepton masses are correlated. 
The minimal setup simply gives at the $SO(10)$ scale:
\begin{equation}
Y_\nu = Y_u\quad[{\rm ``minimal"}~SO(10)],
\label{eq:so10}
\end{equation}
which means that our matrix features the strong hierarchy of up quarks -- hence only the third generation Yukawa matters
for the sake of the CLFV effects -- and the same mixing structure controlled by the CKM matrix.
Therefore Eq.~(\ref{eq:delta-seesaw}) becomes:
\begin{align}
(\delta_{\rm LL})_{i\neq j} \approx  - \frac{3 +a_\nu^2}{8\pi^2}   y_t^2 V^*_{ti} V_{tj} \ln\frac{M_U}{M_{R_3}},
\label{eq:deltaSO10}
\end{align}
where $y_t$ is the top Yukawa coupling ($\approx 0.5$ at high energies), $V_{ij}$ are entries of the CKM, $M_U$ is the $SO(10)$-breaking scale ({\em i.e.}~the gauge-coupling unification scale, $\approx 10^{16}$ GeV) , and $M_{R_3}$ is the mass of the heaviest RH neutrino. From Eq.~(\ref{eq:seesaw}), this latter scale results $M_{R_3} \approx 10^{15}$ GeV, if we take
hierarchical light neutrinos: $\hat{m}_\nu \simeq (10^{-3},~9\times10^{-3},~5\times 10^{-2})$ eV. 
Despite the large coupling, the effect is suppressed by the small CKM angles and 
it is additionally challenged by the strong constraints
on the mass of coloured SUSY particles that we mentioned at the beginning of this section. The reason is that sleptons and Bino/Wino masses are related to squarks and gluino masses by the $SO(10)$ unification. For instance, gaugino mass unification implies
the following low-energy relation between Bino mass $M_1$ and gluino mass $M_3$ (obtained by solving the corresponding RGEs): 
$M_3 \approx 6 M_1$. These limitations are clearly depicted in the left panel of Figure \ref{fig:SUSY-SO10}, where the predicted rates of the $\mu\to e$ transitions are shown and compared with the LHC bound: we take for simplicity as limit on the gluino mass $M_3 > 2$ TeV.
As we can see, as a consequence of this constraint together with the small mixing, this scenario does predict rates below the present and future sensitivities of MEG, even with a choice of the parameters aimed at maximise the CLFV effects ($\tan\beta=50$, $a_\nu=3$): $ {\rm BR}(\mu\to e\gamma) \lesssim 2\times 10^{-14}$. For the other channels we have
$ {\rm BR}(\mu\to eee) \lesssim 10^{-16}$, and $ {\rm CR}(\mu N\to e N) \lesssim 10^{-16}$, so that these processes could
be observable only if the SUSY spectrum lies close to the present bounds set by the LHC experiments.
The situation does not improve in the $\tau-\mu$ and $\tau-e$ sectors that are tightly correlated to the $\mu-e$ sector by Eq.~(\ref{eq:deltaSO10}). For instance, we have for $\tau\to\mu\gamma$:
\begin{equation}
\frac{{\rm BR}(\tau\to\mu\gamma)}{ {\rm BR}(\mu\to e\gamma)} = \frac{|V^*_{tb} V_{ts} |^2}{|V^*_{ts} V_{td} |^2}\times
{\rm BR}(\tau\to\mu\bar{\nu}\nu) \approx 5\times 10^3,
\end{equation}
so that the LHC bound in Figure \ref{fig:SUSY-SO10} implies ${\rm BR}(\tau\to\mu\gamma)\lesssim 10^{-10}$. 

Fortunately $SO(10)$ unification does not necessarily imply that the CLFV effects are suppressed by the small CKM angles. In fact, one can modify the relation in Eq.~(\ref{eq:so10}) by introducing multiple Higgs representations and/or non-renormalisable operators, which on the other hand is required to correct the relations among charged lepton and quark masses that work poorly, 
especially for the first two generations. In our context, this has been discussed in \cite{Chang:2002mq,Masiero:2002jn}, with a particular focus on the possibility of obtaining large PMNS-like mixing:
\begin{equation}
Y_\nu = U^\dagger \hat{Y}_u\quad[{\rm ``large-mixing"}~SO(10)],
\label{eq:so10pmns}
\end{equation}
where $\hat{Y}_u$ is the diagonal up-quark Yukawa matrix and $U$ is the PMNS.
In this case, the induced slepton mixing is controlled by the large angles of the PMNS:
\begin{align}
(\delta_{\rm LL})_{i\neq j} \approx  - \frac{3 +a_\nu^2}{8\pi^2}   y_t^2 U_{i3} U^*_{j3} \ln\frac{M_U}{M_{R_3}}.
\label{eq:deltaSO10pmns}
\end{align}
The result of this is shown in the right panel of Figure \ref{fig:SUSY-SO10}, where we can see that the present MEG limit already excludes the model way beyond the LHC. 

\begin{figure}[t]
\begin{center}
\includegraphics[width=.42\columnwidth]{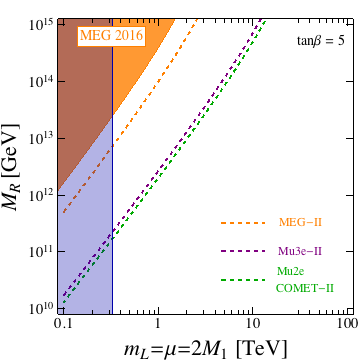}
\hspace{0.5cm}	
\includegraphics[width=.42\columnwidth]{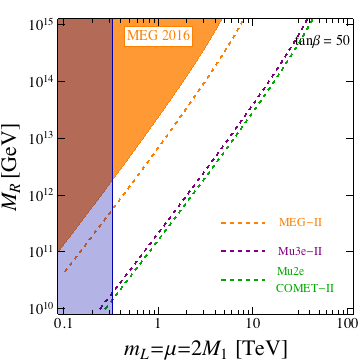}\\
\caption{Bounds and prospects for a SUSY seesaw model with degenerate RH neutrinos,
as in Eq.~(\ref{eq:deg}), for $\tan\beta=5$ (left) and 50 (right). 
The blue region is excluded by LHC searches for sleptons \cite{Aad:2014vma}. 
\label{fig:SUSYseesaw}}
\end{center}
\end{figure}

By inspecting Eq.~(\ref{eq:CI}), we can easily spot other scenarios, for which we can expect large CLFV effects. 
For instance, if the RH neutrinos are degenerate with mass $M_R$, {\em i.e.}~$\hat{M}_R = M_R \mathbf{1}$, they can not contribute to the PMNS mixing, the matrix $R$ becomes trivial, and again the mixing structure of $Y_\nu$ is given by the PMNS:
\begin{equation}
Y_\nu = \frac{\sqrt{M_R}}{v_u}\sqrt{\hat{m}_\nu}\, U^\dagger\quad[{\rm degenerate}~\nu_{R\,i}].
\label{eq:deg}
\end{equation}
Similarly, even if $\hat{M}_R$ is hierarchical but with a trivial flavour structure, namely $R=1$ (which can be enforced by a flavour symmetry), one gets $Y_\nu \sim  U^\dagger$. Both cases give quantitatively similar predictions in terms of CLFV processes,
hence we will focus on the one of Eq.~(\ref{eq:deg}) in the following. 
In this case, the RGE-induced CLFV parameters shown in Eq.~(\ref{eq:delta-seesaw}) result:
\begin{align}
(\delta_{\rm LL})_{i\neq j} \approx  - \frac{3 +a_\nu^2}{8\pi^2} \sum_k  y_{\nu_k}^2 U_{ik} U^*_{jk} \ln\frac{M_X}{M_{R}}
\quad \left[y_{\nu_k} = \frac{\sqrt{M_R}}{v_u}\left(\sqrt{\hat{m}_\nu}\right)_{kk}\right].
\label{eq:delta-deg}
\end{align}
Notice that, because of the unitarity of the PMNS, $(\delta_{\rm LL})_{i\neq j}$ is strongly suppressed if the light neutrinos are almost degenerate, {\em i.e.}~the $y_{\nu_k}$ are almost equal. In the following we consider hierarchical light neutrinos, choosing for illustration $m_{\nu_1} =10^{-3}$ eV, as above. 
As we can see, $(\delta_{\rm LL})_{i\neq j}$ grows linearly with $M_R$, as heavier RH neutrinos mean larger Yukawa couplings. 
The resulting CLFV constraints and future prospects are shown in Figure \ref{fig:SUSYseesaw}
on a plane displaying the LH slepton mass and the degenerate RH neutrino mass. Here we dropped any GUT relation among
EW-interacting and coloured SUSY particles and considered only the limit from direct searches for sleptons \cite{Aad:2014vma}, shown as a blue-shaded region. 
\begin{figure}[t]
\begin{center}
\includegraphics[width=.42\columnwidth]{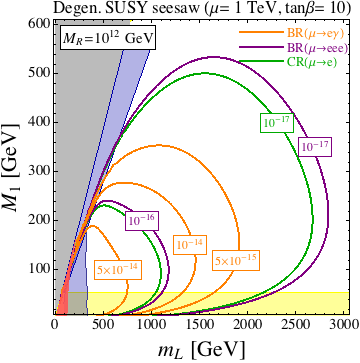}
\caption{Prediction for $\mu\to e$ processes for degenerate RH neutrinos with $M_R=10^{12}$ GeV 
in the LH slepton mass/Bino mass plane. 
For illustration, we took $\mu =1$ TeV, $\tan\beta=10$, and $M_2 =M_1$ and $m_E =m_L$ at high energies. 
LHC bounds for slepton \cite{Aad:2014vma} and chargino/neutralino \cite{CMS-Wino} searches are shown in blue. The yellow area is excluded by LEP chargino searches. \label{fig:SUSYseesaw-deg}}
\end{center}
\end{figure} 
As we can see, the MEG limit already set bounds on the degenerate RH neutrino mass at the level of 
$M_R\lesssim 10^{12\div13}$ GeV for SUSY masses close to the LHC sensitivity, 
which future experiments will further improve down to $M_R\lesssim 10^{10\div11}$ GeV. This range is of particular interest as successful leptogenesis typically requires for the lightest RH neutrino $M_{R_1}>10^8\div10^9$ GeV \cite{Buchmuller:2004nz,Fong:2013wr}\footnote{Moreover, scenarios with strongly-hierarchical RH neutrinos give results that are very close to those of Figs.~\ref{fig:SUSYseesaw} and \ref{fig:SUSYseesaw-deg} with the substitution $M_R\to M_{R_3}$.}. 
The predicted CLFV rates for a case with a rather low value of $M_R$ ($10^{12}$ GeV) are shown in 
Figure \ref{fig:SUSYseesaw-deg}.
On the other hand, we see from Figure \ref{fig:SUSYseesaw} that scenarios with high seesaw scales are already excluded unless $m_L >\mathcal{O}(1)$ TeV and will be tested up to $m_L =\mathcal{O}(10)$ TeV.
Given that, in contrast to the CKM, there are no large hierarchies among PMNS mixing angles, $\tau\to\mu/e$ observables play no significant role in this scenario, as we can see from:
\begin{equation}
\frac{{\rm BR}(\tau\to\mu\gamma)}{ {\rm BR}(\mu\to e\gamma)} = \frac{\left|\sum_k y^2_{\nu_k}U_{\tau k} U^*_{\mu k} \right|^2}{\left|\sum_k y^2_{\nu_k}U_{\mu k} U^*_{e k} \right|^2}\times {\rm BR}(\tau\to\mu\bar{\nu}\nu) \lesssim 10,
\end{equation}
which shows that the MEG bound implies that ${\rm BR}(\tau\to\mu\gamma)$ can not be above the $10^{-12}$ level. 

For simplicity, here we have only considered the first and simplest kind of seesaw mechanism based on extending the SM Lagrangian with RH neutrino fields. 
There are other two ways to generate the neutrino mass operator of Eq.~(\ref{eq:weinb}) at the tree level: 
by introducing one scalar $SU(2)_L$ triplet with hypercharge $Y=1$, or two or more fermion $SU(2)_L$ triplets with $Y=0$. 
These further options have been respectively labelled as type II \cite{typeII} and type III \cite{typeIII} seesaw.
In both cases, supersymmetric CLFV effects are radiatively generated in a similar way as in Eq.~(\ref{eq:delta-seesaw}), as was shown
in \cite{Rossi:2002zb,Joaquim:2006mn} for type II and in \cite{Biggio:2010me,Esteves:2010ff} for type III. Particularly interesting is the case of type II seesaw, where the mixing structure
of the seesaw Yukawa couplings is directly controlled by the PMNS matrix, so that, without the need of further assumptions, 
the predictions are similar to those displayed above in Figs.~\ref{fig:SUSYseesaw} and \ref{fig:SUSYseesaw-deg} 
for the case of degenerate RH neutrinos. 
Finally, let us remark that, in extended versions of the seesaw models, seesaw scales as low as the TeV scale
does not necessarily imply so small Yukawa couplings that the induced CLFV parameters of Eq.~(\ref{eq:delta-seesaw}) vanishes, as it occurs
in the ordinary seesaw, cf.~Eq.~(\ref{eq:CI}).
In fact, as we mentioned in section \ref{sec:neutrinos}, there are setups, such as the so-called inverse seesaw, that allow
large couplings and low seesaw scales at the same time. CLFV processes in these scenarios have been discussed {\em e.g.}~in \cite{Deppisch:2004fa,Abada:2014kba}. For a recent review of CLFV effects in SUSY seeesaw models, including non-minimal extensions of this kind, see \cite{Vicente:2015cka}.

\section{Some remarks on model-dependent predictions}
\label{sec:models}
In the previous section we discussed low-energy SUSY as a case study to show advantages and limitations of considering a specific framework compared to more model-independent approaches as in section \ref{sec:EFT}.
A general conclusion we can draw is that a given framework is not going to provide sharp predictions for our CLFV processes unless 
both the mass scale of the new physics and its flavour mixing structure are somehow specified. That's a trivial statement but serves to remind us that no `guaranteed minimum' rates are to be expected -- barring  the irrelevant SM contributions as in
Eq.~(\ref{eq:meg-mnu}) -- even within a given class of models: as we have seen, SUSY might give vanishing effects (if lepton and sleptons are aligned as in pure Gauge Mediation models), as well as saturate the present bounds. For this reason,
scenarios that predict maximal CLFV rates below the present limits\footnote{An example of this class of models is given by the minimal-$SO(10)$ SUSY seesaw discussed in the previous section, another one is provided by SUSY $SU(5)$ scenarios with SUSY breaking above the unification scale \cite{Barbieri:1994pv,Barbieri:1995tw}, where, as an effect of the quark-lepton unification, the CLFV effects are likewise suppressed by CKM mixing angles.} are of some interests, as they at least naturally account for the fact that CLFV has not been observed yet and require a definite improvement of the experimental sensitivity. 

Of course, the more a scenario is specific the more unambiguously can be tested. A neat example is given by 
composite Higgs models where the hierarchical structure of the SM Yukawas is solely accounted for by the SM fermions 
being a mixture of elementary fermions and operators of the new strongly-coupled sector 
(that is what is called `partial compositeness' with flavour-anarchical couplings in the fundamental theory), see \cite{Panico:2015jxa} for a review. 
In this case, the new dynamics is required to be not much heavier than few TeV to provide a natural solution to the hierarchy
problem, hence the anarchical flavour structure of the couplings give rise to CLFV effects that the elementary-composite mixing
can suppress only to some extent, so that they result several orders of magnitude above the present bounds \cite{Redi:2013pga,Feruglio:2015gka,Panico:2015jxa}. While these minimal models are excluded, one can conceive setups where additional protection against flavour-violating effects is provided by a non-trivial structure of the couplings in the strong sector, as given for instance by some flavour symmetry \cite{Hagedorn:2011un,Hagedorn:2011pw,Feruglio:2015gka}.

\begin{figure}[b]
\includegraphics[width=0.4\textwidth]{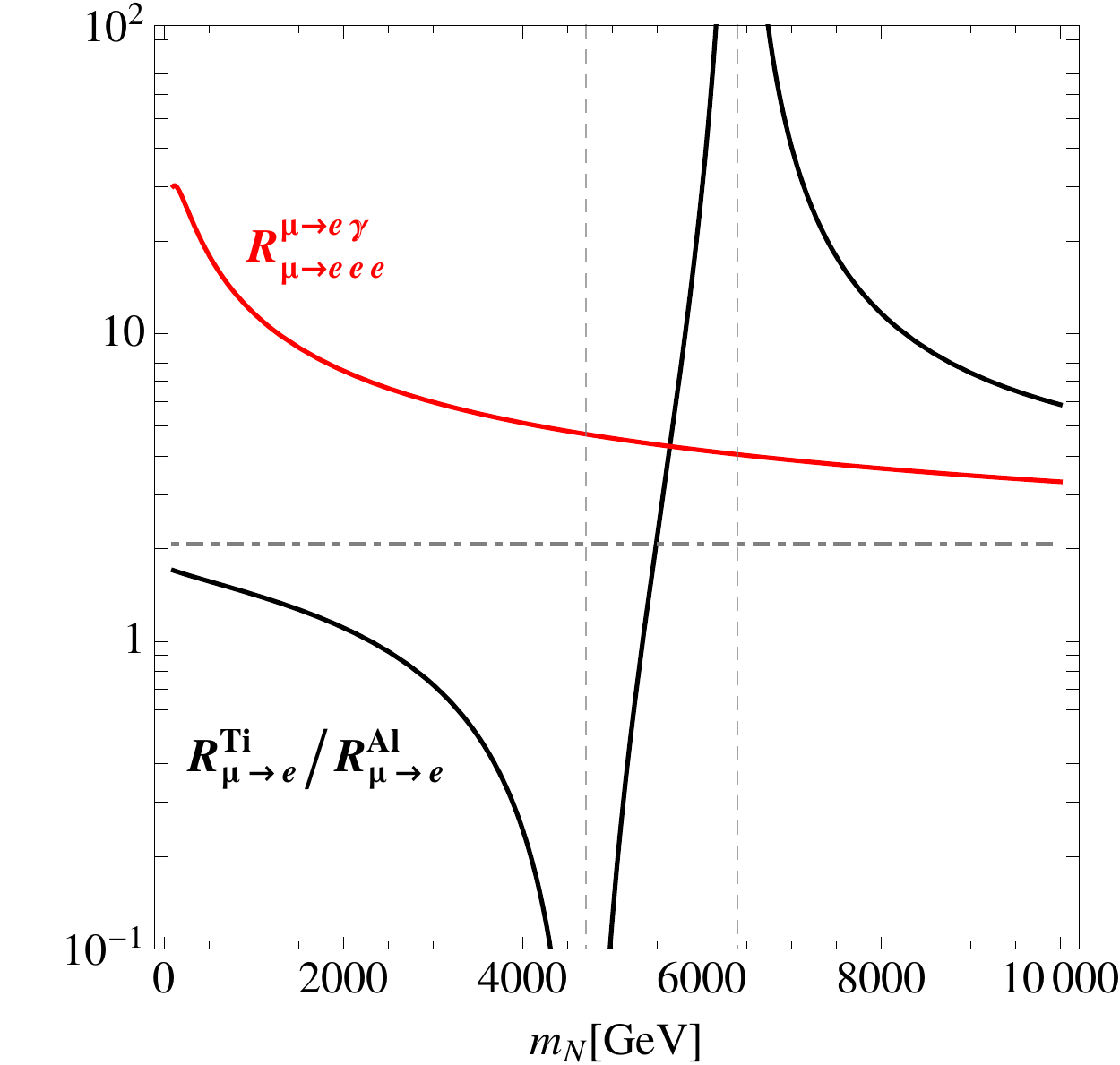}
\hspace{1cm}	
\includegraphics[width=0.42\textwidth]{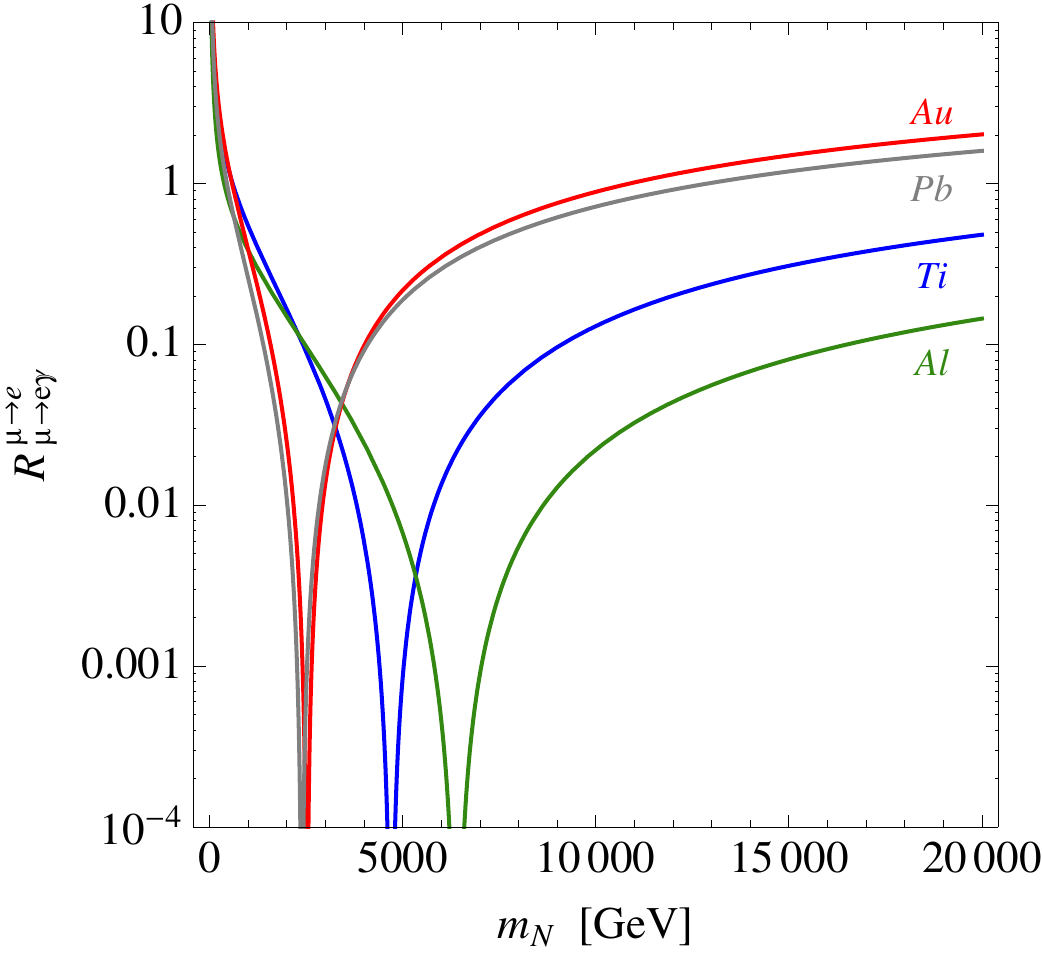}
\caption{Correlations among different $\mu\to e$ observables [$R^A_B \equiv {\rm BR}(A)/{\rm BR}(B)$] for a low-energy seesaw of type I as a function of the RH neutrino mass. From \cite{1209.2679}.
\label{fig:lowseesaw} }
\end{figure}

There are several classes of models that, although not providing sharp predictions for the absolute rates of the CLFV processes, have the appealing feature of predicting definite patterns of the relative rates in different channels, providing
in principle an experimental ``smoking gun''. As we have seen, standard SUSY models feature the correlations of Eqs.~(\ref{eq:m3e}, \ref{eq:m2e}), as any other scenario where the dominant effect is given by the dipole operator, hence 
this pattern of effects -- although rather robust -- does not provide a definite handle for model discrimination. 
Models where other operators in Table \ref{tab:ops} arise at the tree level such that the dipole is subdominant and 
Eqs.~(\ref{eq:m3e}, \ref{eq:m2e}) do not hold are therefore very interesting. This is for instance the case of SUSY models with R-parity violation, see \cite{deGouvea:2000cf,Dreiner:2006gu}. Low-energy seesaw models of the type II and III provide other remarkable examples of these kinds of scenarios, see {\em e.g.}~\cite{Abada:2007ux}. In the type II seesaw, the scalar triplet mediates $\ell_i \to \ell_j \ell_k \ell_k$, which thus set the strongest constraints (furthermore the relative rates of different flavour transitions are directly related to PMNS matrix elements, as in the SUSY-seesaw example of the previous section). In the type III seesaw, the charged component of the fermion triplets mix with the leptons generating flavour-violating couplings to $Z$. As a consequence  $\ell_i \to \ell_j \ell_k \ell_k$ and $\mu\to e$ conversion in nuclei arise at the tree level from a $Z$ exchange, resulting in very definite predictions for the ratios of the different modes \cite{Abada:2008ea}.
We illustrate the pattern of the relative predictions for the $\mu\to e$ modes within a selection of models in Table \ref{tab:models}.
The chosen examples are the minimal (in terms of low-energy particle content) SUSY model discussed in the previous section (MSSM), the three types of low-energy seesaw models, the scenario with LFV-couplings of the Higgs (LFV Higgs) that we discussed at the end of section \ref{sec:SM} (cf.~Eq.~(\ref{eq:LFVyuk}) and Table \ref{tab:Yukawa-bounds}), and the so-called `two-site' composite Higgs model studied in \cite{Feruglio:2015gka}. The displayed results highlight the model-discriminating power of the three modes and show how all the three searches are necessary. In Figure \ref{fig:lowseesaw} these correlations are nicely displayed for the case of the low-energy type I seesaw (from \cite{1209.2679}). 
\begin{table}[t]
\renewcommand{\arraystretch}{1.3}
\begin{center}
\begin{tabular}{lllcc}
\hline
Model & $\mu\to eee$ & $\mu N\to e N$ & $\frac{{\rm BR}(\mu\to eee)}{ {\rm BR}(\mu\to e\gamma)} $ & $\frac{{\rm CR}(\mu N\to eN)}{ {\rm BR}(\mu\to e\gamma)} $   \cr
\hline
 MSSM & Loop & Loop  & $\approx 6\times10^{-3}$ & $10^{-3}-10^{-2}$  \cr
Type-I seesaw  &  Loop$^*$ & Loop$^*$ & $3\times 10^{-3}-0.3$ & 0.1$-$10  \cr
Type-II seesaw  &  Tree & Loop & $(0.1-3)\times 10^3 $ & $\mathcal{O}(10^{-2})$   \cr
Type-III seesaw  &  Tree & Tree & $\approx 10^3$ & $\mathcal{O}(10^3)$ \cr
LFV Higgs  & Loop$^{\dag}$ & Loop$^{*\,\dag}$ &$\approx 10^{-2}$ & $\mathcal{O}(0.1)$     \cr
Composite Higgs  & Loop$^*$  & Loop$^*$  & $0.05-0.5$&  $2-20$  \cr
\hline
\end{tabular}
\end{center}
\caption{Pattern of the relative predictions for the $\mu\to e$ processes as predicted in several models (see the text for details). It is indicated whether the dominant contributions to $\mu\to eee$ and $\mu\to e$ conversion are at the tree or at the loop level; Loop$^*$ indicates that there are contributions that dominate over the dipole one, typically giving an enhancement compared to Eq.~(\ref{eq:m3e},~\ref{eq:m2e}). $^\dag$ A tree-level contribution to this process exists but it is subdominant.
\label{tab:models}}
\end{table}

Before moving to discuss the experimental aspects of searches for CLFV, let us comment about the possible connection between recent hints of breaking of lepton flavour universality and CLFV observables.
  
\section{Lepton Flavour (Non-)Universality and Lepton Flavour Violation}\label{sec:LFU}
In recent years, $B$-physics experiments have reported a number of interesting anomalous results that might hint to the violation of lepton flavour universality (LFU) in semi-leptonic $B$ decays. If confirmed, this would be a signal of new physics, because LFU is predicted by the SM as a consequence of the flavour-independent couplings of leptons to electroweak gauge bosons. 
These results concern two classes of observables: (i) neutral-current transitions of the 
kind $b\to s \ell^+\ell^-$, and (ii) charged-current $b\to c \ell \nu$ processes.

The following ratios of branching ratios -- belonging to the first class -- have been measured by LHCb~\cite{Aaij:2014ora,Aaij:2017vbb}: 
\begin{align}
R_K &\equiv  \frac{ {\rm BR}(B^+ \to K^+ \mu^+ \mu^-)}{ {\rm BR}(B^+ \to K^+ e^+e^-)}   =  0. 745^{+0.090}_{-0.074} \pm 0.036 \quad
{\rm for}~{q^2\in[1,6]{\rm GeV}^2},\\
R_{K^*} &\equiv  \frac{ {\rm BR}(B^0 \to K^{*\,0} \mu^+ \mu^-) }{ {\rm BR}(B^0 \to K^{*\,0} e^+ e^- )}=  0. 685^{+0.113}_{-0.069} \pm 0.047\quad {\rm for}~{q^2\in[1.1,6]{\rm GeV}^2},
\end{align}
where $q^2$ is the dilepton invariant mass squared, the first quoted errors are statistical and the second ones are systematic. 
In the SM, the above observables, which are theoretically very clean (hadronic uncertainties cancel in the ratios), are predicted to deviate from unity only at the percent level due to small radiative corrections: $(R_{K^{(*)}})^{\rm SM}  =1.00 \pm 0.01$ \cite{Bordone:2016gaq}.
Besides $R_{K^{(*)}}$, other tensions with the SM predictions (at the 3$\sigma$ level) 
are reported in $b\to s \mu^+\mu^-$ transitions, in particular, in the rate of the decays $B \to K^*\mu^+\mu^-$ \cite{Aaij:2014pli} and $B_s^0 \to \phi\mu^+\mu^-$ \cite{Aaij:2015esa}, and in the angular distributions of $B \to K^*\mu^+\mu^-$ \cite{Aaij:2014pli,Aaij:2013qta,Aaij:2015oid}. 

Coming to the second class of observables, data from Babar \cite{Lees:2013uzd}, Belle \cite{Hirose:2016wfn}, and LHCb  \cite{Aaij:2015yra} (averaged in~\cite{HFLAV}) show a combined $\approx 4\sigma$ deviation from the  theory predictions~\cite{Fajfer:2012vx,Aoki:2016frl}:
\begin{align}
R_{D } &\equiv \frac{ {\rm BR}(B \to D  \tau \overline{\nu})_{\rm exp}/{\rm BR}(B \to D  \tau \overline{\nu})_{\rm SM} }{ {\rm BR}(B \to D  \ell \overline{\nu} )_{\rm exp}/ {\rm BR}(B \to D  \ell \overline{\nu} )_{\rm SM} } =   1.34 \pm 0.17~, \\
R_{D^*} &\equiv \frac{ {\rm BR}(B \to D^* \tau \overline{\nu})_{\rm exp}/{\rm BR}(B \to D^* \tau \overline{\nu})_{\rm SM} }{ {\rm BR}(B \to D^* \ell \overline{\nu} )_{\rm exp}/ {\rm BR}(B \to D^* \ell \overline{\nu} )_{\rm SM} } = 
 1.23 \pm 0.07~,
\end{align}
where $\ell=e, \mu$ and $B$ and $D^{(*)}$ charges were averaged. 

Both classes of anomalies -- if confirmed -- would require a sizeable new-physics contribution, at the level of $\mathcal{O}(10\%)$ of the SM one. This is particularly interesting in the case of the charged-current (class-II) observables that come from tree-level processes in the SM: in fact, new physics should contribute to $B \to D^{(*)}  \tau \overline{\nu}$ at the tree-level as well and lie at a scale not far from the EW scale in order to give such large effect. In the case of the class-I observables, the new physics sector can be rather heavier, yet below $\mathcal{O}(100)$ TeV.
For a detailed recent discussion on the new-physics scales hinted at by these anomalies see \cite{DiLuzio:2017chi}.

It is interesting to note that the $b\to s \ell^+\ell^-$ ({\em i.e.}~class-I) data are in better agreement with the theory predictions if we introduce new physics as a single two-quarks-two-leptons effective operator involving only muons -- $(\bar{s}\gamma^\mu P_L b)(\bar{\mu}\gamma^\mu \mu)$ or $(\bar{s}\gamma^\mu P_L b)(\bar{\mu}\gamma^\mu P_L \mu)$ (in the latter case, only LH fields appear) --  which can lead to a destructive interference with the SM contributions and hence to a deficit of muon events \cite{Banomalies}. 
Recent global fits to the data show that this scenario is preferred to the SM at the 4-5$\sigma$ level, see \cite{Bfits} (and references therein for early theoretical interpretations). Of course, setups involving more operators -- possibly also with electrons -- can also give a good fit. 
In terms of the $SU(2)_L\times U(1)_Y$-invariant operators listed in Table \ref{tab:ops}, the above-mentioned low-energy operator
$(\bar{s}\gamma^\mu P_L b)(\bar{\mu}\gamma^\mu P_L \mu)$ arise from both $Q^{(1)}_{\ell q}$ and $Q^{(3)}_{\ell q}$ (which differ by the $SU(2)_L$ contractions) with suitable choices of flavour indices: 
\begin{align}
\label{eq:Q1}
(Q^{(1)}_{\ell q})_{\mu\mu bs} &= (\bar{L}^a_{L\,2}\,\gamma^\mu\,L^a_{L\,2})(\bar{Q}^b_{L\,2}\,\gamma_\mu\,Q^b_{L\,3}) \\
(Q^{(3)}_{\ell q})_{\mu\mu bs}&=\sum_{I=1,3} (\bar{L}^a_{L\,2}\,\gamma^\mu (\tau_I)_{ab}\,L^b_{L\,2})(\bar{Q}^c_{L\,2}\,\gamma_\mu  (\tau_I)_{cd}\,Q^d_{L\,3}), 
\label{eq:Q3}
\end{align}
where $a,b,c,d$ are summed-up $SU(2)_L$ indices.
Interestingly, the second operator also gives rise to charged-current interactions, hence it can in principle contribute to
the class-II observables, $R_{D^{(*)}}$, as well  \cite{Bhattacharya:2014wla}. Notice that these operators also induce processes with final-state neutrinos such as $B\to K^{(*)} \nu \bar{\nu}$, which give a stringent constraint that can be however relaxed if the fundamental theory generates both operators with equal coefficients \cite{Calibbi:2015kma}.

What are the consequences in the context of CLFV of possible new physics in the above form? Would a new physics contribution to these operators imply the arising of their LFV counterparts ({\em e.g.}~$(Q^{(1,3)}_{\ell q})_{\tau\mu bs}$) too?
As it should be clear from the previous sections, (lepton) flavour-conserving and flavour-violating operators are not related in general 
terms. In other words, we can not draw a definite conclusion unless the underlying new physics theory that give rise to our operators
 is fully specified. However, it is also fair to say that if such large breaking of LFU is verified -- implying new physics coupling with very different strengths to the different lepton generations --  CLFV effects are to be expected to some extent \cite{Glashow:2014iga}. Indeed, it is likely that a theory inducing operators like $(Q^{(1,3)}_{\ell q})_{\mu\mu bs}$ in the interaction basis does induce the CLFV counterparts as well, either directly or at least as a consequence of the rotation to the lepton mass basis -- 
cf.~Eq.~(\ref{eq:mass-basis}) -- unless a very specific structure of the lepton Yukawa matrix is assumed (possibly a consequence of a symmetry of flavour).
As an illustration of the above discussion, let us consider -- as in \cite{Glashow:2014iga,Bhattacharya:2014wla,Alonso:2015sja,Greljo:2015mma,Calibbi:2015kma,Feruglio:2016gvd,Feruglio:2017rjo} -- a scenario where the underlying new physics mainly generates operators involving third generations, i.e.~$b$ quarks and taus: 
$(Q^{(1,3)}_{\ell q})_{\tau\tau bb}$. 
This choice -- resembling the hierarchical structure of the SM Yukawas -- is a natural consequence of a flavour structure controlled either by the minimal flavour violation ansatz \cite{Alonso:2015sja} or by a flavour symmetry $U(2)_q\times U(2)_\ell$ acting on first and second generation quarks and leptons only \cite{Greljo:2015mma}. Our operators in Eq.~(\ref{eq:Q1}, \ref{eq:Q3}) -- and thus the wanted LFU effects -- are then generated by rotating the LH quarks and the LH leptons to the respective mass basis. As a consequence,
the coefficients of $(Q^{(1,3)}_{\ell q})_{\mu\mu bs}$ result $\propto (V_d)_{32}\times |(V_e)_{32}|^2$, where we used the notation of Eq.~(\ref{eq:mass-basis}). As we can see, it is thus unavoidable to generate at the same time the LFV operators
 $(Q^{(1,3)}_{\ell q})_{\tau\mu bs}$ with coefficients $\propto (V_d)_{32} \times(V_e)_{32}$.
These operators induce the CLFV processes $B_s\to \tau\mu$, $B\to K^{(*)} \tau\mu$, 
although with rates below the present limits (if any at all, cf.~Table~\ref{tab:limits}), 
at least for the choices of the parameters fitting the LFU anomalies and fulfilling bounds from $B\to K^{(*)} \nu \bar{\nu}$ \cite{Calibbi:2015kma}:
\begin{align}
{\rm BR}(B\to K^{*} \tau\mu)\approx 2\times{\rm BR}(B\to K \tau\mu) \approx  2\times{\rm BR}(B_s\to \tau\mu) 
\lesssim 10^{-6}.
\end{align}
More interestingly, radiative effects -- i.e.~the RG running of the operators from the new-physics scale down to low energies -- 
induce other classes of operators that contribute to, among other effects, $Z$ couplings to leptons and $\tau$ decays \cite{Feruglio:2016gvd,Feruglio:2017rjo}.
As a result, strong constraints arise from measurements of LFU observables in $Z$ and $\tau$ decays, 
as well as from the CLFV decays $\tau\to \mu \mu \mu$, $\tau\to \pi^0 \mu$, $\tau\to \rho^0 \mu$, typically predicted at the level 
of the present bounds or above \cite{Feruglio:2016gvd,Feruglio:2017rjo}. This challenges the simultaneous explanation of class-I and class-II observables, unless some cancellations among the parameters of the fundamental theory are at work \footnote{For a recent discussion about models that address both classes of anomalies see \cite{Buttazzo:2017ixm}.}. On the other hand, this means that, if confirmed, the LFU anomalies would raise the expectation of rates of the CLFV $\tau$ decay in the reach of Belle 2.

A very large number of models have been proposed as a possible origin of the effective interactions able to fit the data, like those of the kind of Eqs.~(\ref{eq:Q1},~\ref{eq:Q3}). 
These high-energy completions typically feature a $Z^\prime$, namely the gauge boson associated to a new local $U(1)$, 
or leptoquarks (LQs), {\em i.e.}~new scalars or vectors that couple to one quark and one lepton field. 
A neat example of the first kind is given by a $Z^\prime$ resulting from gauging the (anomaly-free) combination of leptonic family numbers  $L_\mu-L_\tau$ \cite{Altmannshofer:2014cfa,Crivellin:2015mga}, while a general discussion of phenomenology and constraints of a LFV $Z^\prime$ can be found in \cite{Crivellin:2015era} (see also \cite{Becirevic:2016zri}). In another interesting proposal, the new gauge boson corresponds to a gauged abelian subgroup of a $U(2)$ symmetry of flavour accounting for the observed fermion masses and mixing \cite{Falkowski:2015zwa}; as a consequence the flavour structure of the $Z^\prime$ couplings is fully specified and the tightest low-energy constraint turns out to be given by $\mu\to e$ conversion.
Detailed discussions of the LQ models with a particular focus on their implications for CLFV observables are presented in \cite{Varzielas:2015iva,Becirevic:2016oho,Becirevic:2017jtw,Crivellin:2017dsk} (for a general review on LQs see \cite{Dorsner:2016wpm}), in particular \cite{Crivellin:2017dsk} considers the consequences for $\mu\to e$ observables of introducing LQs with sizeable couplings to both muons and electrons.


\section{General considerations regarding  experimental CLFV searches}
\label{sec:window}
In this section we discuss some topics that are in common to CLFV searches with $\mu\to e \gamma$, $\mu\to eee$ and $\mu\to e$ conversion in nuclei, but are somehow general and applicable to other processes.

Searching for a LFV muon decay or conversion process implies the precise
measurement of the four-momenta of the daughter particles in an 
environment which is virtually free from Standard Model background.

In the muon rest frame the signature of all three processes is the presence
of monochromatic and/or time coincident particles, supplemented by further
constraints such as back-to-back topology (as in the case of $\mu^+ \to 
e^+ \gamma$) or all the three daughter particles lying on the same plane
(as in the case of the $\mu \to eee$ process).

Due to its peculiarity it is convenient to perform the experiment
in the muon rest frame, therefore stopping the largest number of muons in the thinnest possible target so as to define this rest frame in the best way.

In real life the ``monochromatic'' decay is an idealization. On the one
hand there are physics processes that interfere with daughter particle
trajectories during their  measurement, on the other, ordinary muon decays or captures produce particles that can mimic the decay we are searching for, because we are unable to measure their properties with sufficient precision.

In a simplified framework, the rare process search analysis consists in 
defining a multi dimensional box in the variable space ({\em e.g.}\ the 
energies of the daughter particles, relative time and directions) whose 
size is proportional to the experimental resolutions, and ``count'' the number of events falling in that box. It is understood that this number depends on the number of observed muons, on the rate of the processes we are searching for as well as the probability that normal events spill in. 
The size of the box is adjusted to minimize the background maintaining a high efficiency on the signal. 

It is usually defined ``single-event sensitivity'' (SES) the probability (branching ratio or conversion probability) of the CLFV process under test for which just one signal event is expected (on average) in the box with negligible background:
\begin{align}
\label{eq:SES1}
& N_{\rm events} = {\rm BR}_{\rm process} \times N_\mu \times {\rm window} \times {\rm efficiencies},\\
& \Rightarrow \quad SES = \left( N_\mu \times {\rm window} \times {\rm efficiencies} \right)^{-1},
\label{eq:SES2}
\end{align}
from which it follows that for a fixed probability of the process under
investigation, the better the $SES$, the larger the number of expected signal events:
\begin{equation}
N_{\rm events} = \frac{{\rm BR}_{\rm process}}{SES}.
\label{eq:SES3}
\end{equation}

The $SES$ is not to be confused with the experimental (90\% confidence level) sensitivity which is dictated by the statistical analysis of simulated data samples and in this simplified approach, {\em i.e.}\ a signal box with negligible background, is usually quoted as $2.30 \times SES$.
This descends from the fact that in absence of background the probability of getting zero events when the expectation is $\xi = {\rm BR}_{\rm process} / SES$ (see Eqs~(\ref{eq:SES1}),~(\ref{eq:SES2}) and~(\ref{eq:SES3})) is given by the Poisson distribution
\begin{equation}
{\rm Pois}(0 | \xi) = e^{-\xi},
\end{equation}
and in a Bayesian framework\footnote{The same (numerical) result holds also in a frequentist framework, but in the computation  one has to revert to the concept of $p-$value. Note that in general Bayesian and frequentist limits do not coincide.} the upper limit on $\xi$ (now seen as an independent variable given no observed event) at 90\% confidence level (C.L.) is given by the $\bar \xi$ for which
\begin{equation}
\int_0^{\bar \xi} e^{-\xi} {\rm d}\xi = 0.9
\end{equation}
which implies $\bar \xi = \ln10 \approx 2.30$.
Therefore at 90\% confidence level\footnote{The generalization to any confidence level is trivial.}
\begin{equation}
\xi < 2.30 \quad \Rightarrow \quad {\rm BR}_{\rm process} < 2.30 \times SES.
\end{equation}

In reality the analysis is based on a likelihood fit in which the probability of a single event to be signal or background is taken into account and the number of signal and background events is simultaneously fitted. This has the advantage of using a larger dataset sample as well as taking into account possible correlations among the various experimental variables (see~\cite{xenon-analysis} for a nice introduction on the subject, applied to the search for dark matter interactions, which has many things in common with rare muon decay searches). 
 
Finally we want to stress again what we mean, here and in the following, when we refer to ``window" and ``resolution" on some experimental parameter. ``Resolution" is related to the capability of measuring a physical quantity (a particle energy, direction, time\ldots) which is characterized by some uncertainty introduced by the measurement process itself. This is usually oversimplified by assuming that a Gaussian function with zero mean and standard deviation $\sigma$ is sufficient to describe the response of any detector. Actually real-life resolution functions may be asymmetric, have some bias or have long non-Gaussian tails. Nevertheless it is convenient to think of Gaussian resolutions since the area subtended by the curve can be related easily to the $\sigma$-parameter.

When searching for a particular signal we define a window (or box) in the parameter space which is large enough to contain signal events with a chosen probability. If the resolution were Gaussian functions and the window were $\pm 1\sigma$ wide we know that we expect the window to be 68\% efficient. Background events can spill in the window due to its finite size and resolution effects on background measurement itself. In general in rare event decay searches, signal windows are optimized in order to maximize the signal-to-background ratio.

In a simplified way we can think that the resolution defines the efficiency on the signal at fixed window size, while the window accounts for the background events spilling in, and can be made smaller, the better the resolution.

\section{Muons}
\label{sec:muons}
Before we present in the following the experimental searches for rare muon processes, we summarize here the ordinary muon decay. In a system of reference in which the muon is at rest the differential probability with the emission of one electron with energy $E$ and momentum $\vec p$ in a direction $\vartheta$ with respect to the muon spin is given by
\begin{equation}
{\rm d}\Gamma(x,\vartheta) \simeq \frac{1}{\tau_\mu} 
\left\{ \left( 3 - 2x \right) \mp P \left( 2x-1 \right) \cos \vartheta 
\right\} x^2 {\rm d}x\ {\rm d}\cos \vartheta,
\end{equation}
or, for unpolarized muon decays
\begin{equation}
{\rm d}\Gamma(x) \simeq \frac{1}{\tau_\mu} 
\left( 3 - 2x \right)  x^2 {\rm d}x,
\label{eq:michel}
\end{equation}
where 
\begin{equation}
\frac{1}{\tau_\mu} \simeq \frac{G_F^2 m_\mu^5}{192 \pi^3}
\left\{ 1 - 8 \left( \frac{m_e}{m_\mu}\right)^2 + \frac{\alpha}{2\pi}
\left( \frac{25}{4} - \pi^2 \right) + 
{\cal O} \left( \left( \frac{m_e}{m_\mu}\right)^3 \right) +
{\cal O} \left( \alpha^2 \right)
\right\}
\end{equation}
$x= 2E/m_\mu$ is the reduced electron energy, $P$ is the muon polarization ant the $\mp$ sign reflects the charge of the parent particle. 

The spectrum in Eq.~(\ref{eq:michel}) from a muon decaying in vacuum extends up to the kinematic
endpoint of about half the muon mass, apart from corrections due to the tiny electron mass. 
Both the mass and lifetime of the muon are known with superb experimental sensitivity~\cite{mohr, mulan, PDG}:
\begin{equation}
m_\mu = 105.6583745(24)~{\rm MeV/}c^2,
\end{equation}
\begin{equation}
\tau_\mu = 2.1969811(22) \times 10^{-6}~{\rm s},
\end{equation}
and the latter is used to extract the Fermi constant $G_F$ that, despite the develpoment of the unified electroweak theory, has survived as a convenient low energy effective coupling for the weak sector of the Standard Model and of its possible extensions. For a comprehensive review see~\cite{hertzog} and references therein.

\begin{figure}
\begin{center}
\begin{tabular}{cc}
\includegraphics[height=0.3\columnwidth]{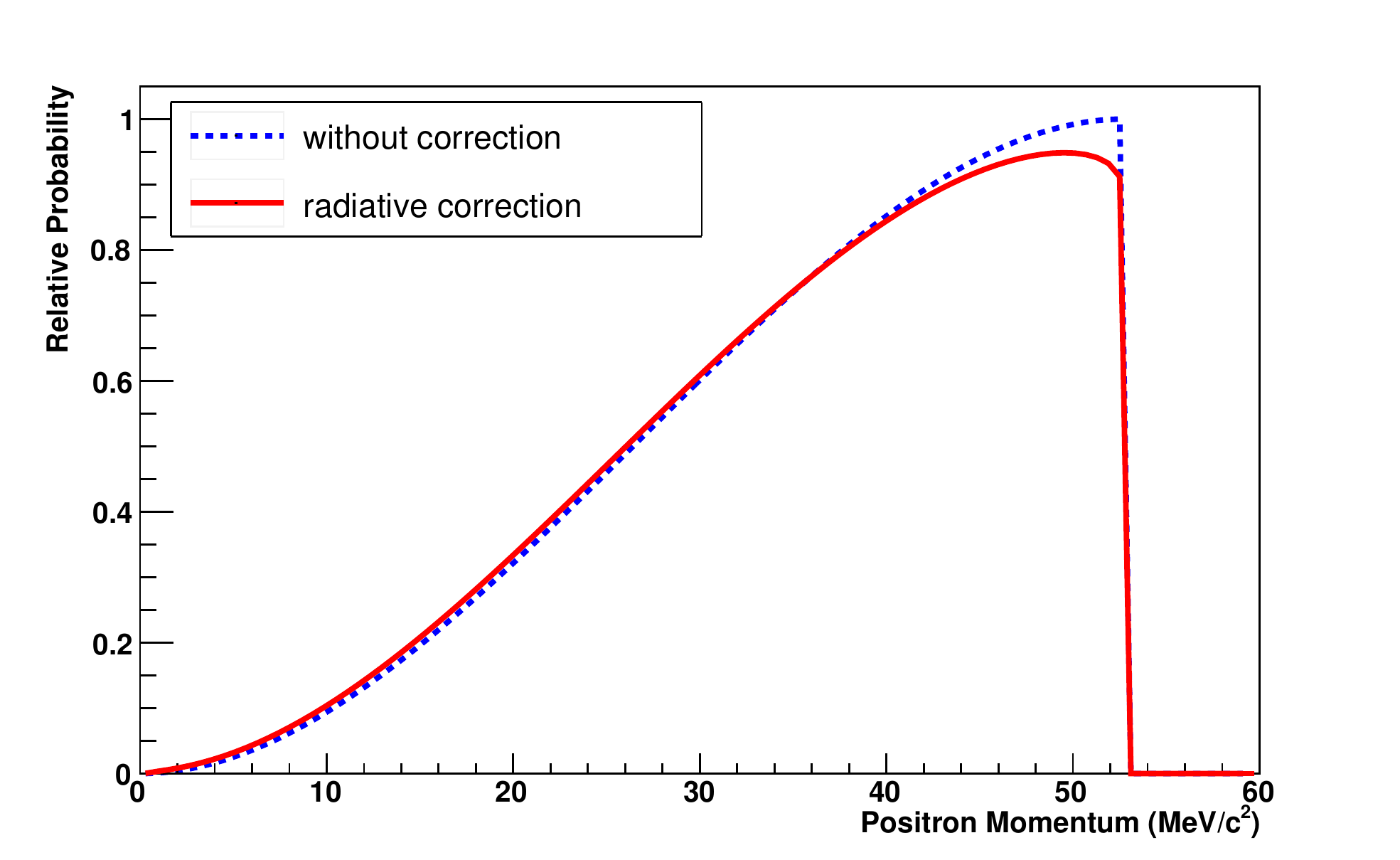} & 
\includegraphics[height=0.3\columnwidth]{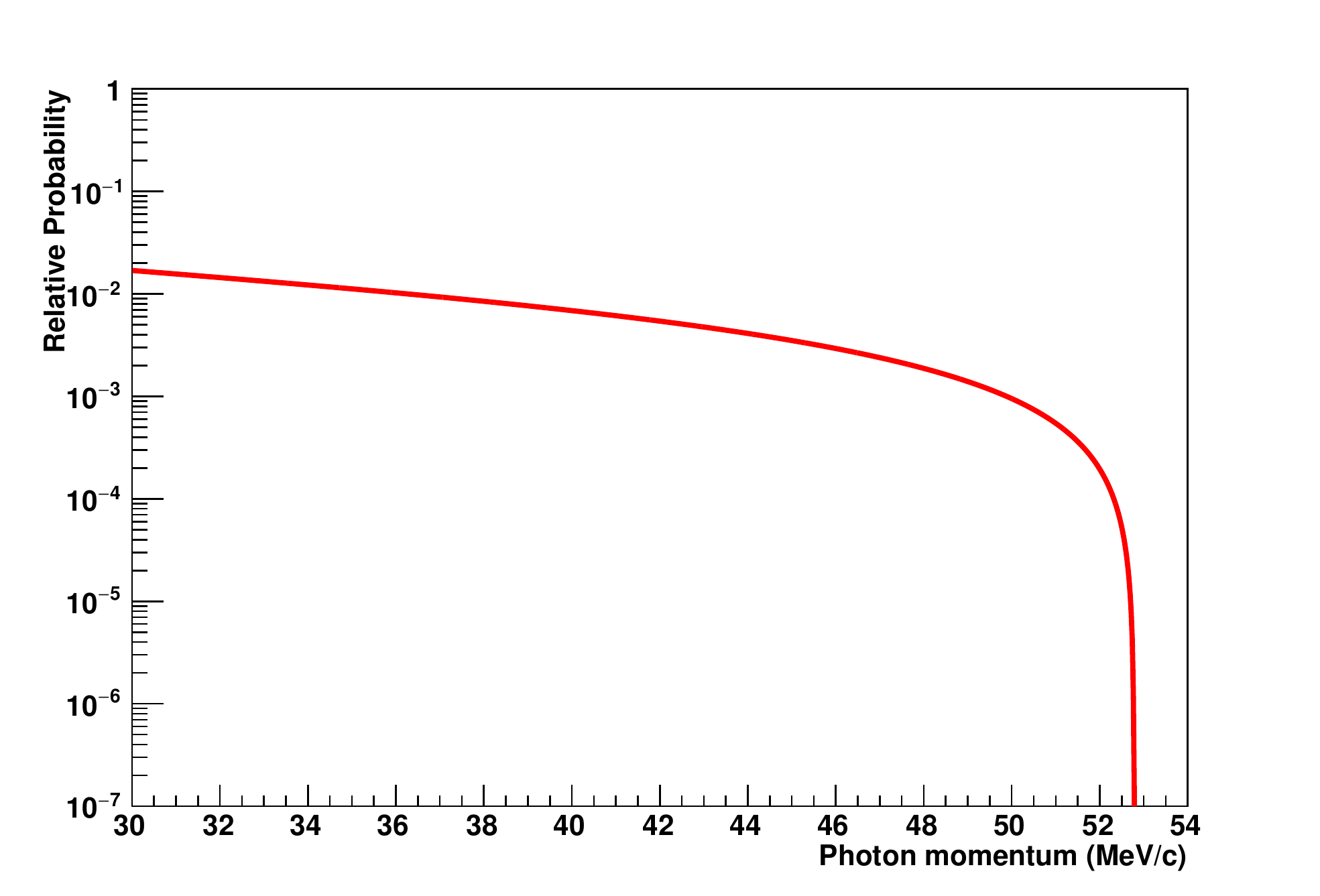}\\
(a) & (b) \\
\end{tabular}
\caption{\label{fig:michel-decay} (a) Michel spectrum, {\em i.e.} energy spectrum of the electron emitted in the normal muon decay, including radiative corrections. (b) Spectrum of the photon emitted in the Michel radiative decay.}
\end{center}
\end{figure}

In this range of energies the four-fermion contact interaction of vector and axial vector type is pratically indistinguishable from an interaction mediated by the W$^\pm$ bosons. The computation of the muon decay probability dates back to Kinoshita and Sirlin~\cite{kinoshita-sirlin} while the general analysis in terms of real decay parameters originated with L.~Michel~\cite{Michel1, Michel2} and it is for this reason that the 
$\mu \to e \nu \bar \nu$ decay is known as the ``muon Michel decay'' in jargon, and the positron energy spectrum, whose characteristic shape is depicted by the blue dashed line in Figure~\ref{fig:michel-decay}, is known as the ``Michel spectrum".
The two massless neutrinos having opposite helicity are emitted preferentially against the electron which cannot be too slow to ensure its correct handedness, hence its 
energy spectrum is peaked at $x=1$ (see~\cite{scheck} for a pedagogical computation).
Radiative corrections smooth the sharp upper edge of the spectrum: since in this configuration an extra $\gamma-$ray, carrying a unit of angular momentum, can be emitted.
The resulting spectrum is depicted by the red line in Figure~\ref{fig:michel-decay}a and its analytical form is reported in the appendix for convenience. In Figure~\ref{fig:michel-decay}b the energy spectrum of the photon emitted in the radiative decay is shown\footnote{The probability of emitting a photon with energy larger than 10~MeV in the Michel decay is $\sim 1.4\%$~\cite{crittenden}, with energy larger than 40~MeV is $\approx 6 \times 10^{-8}$~\cite{megradiative}.}.

\begin{figure}
\begin{center}
\begin{tabular}{cc}
\includegraphics[viewport= 0 250 587 603, width=0.45\columnwidth]{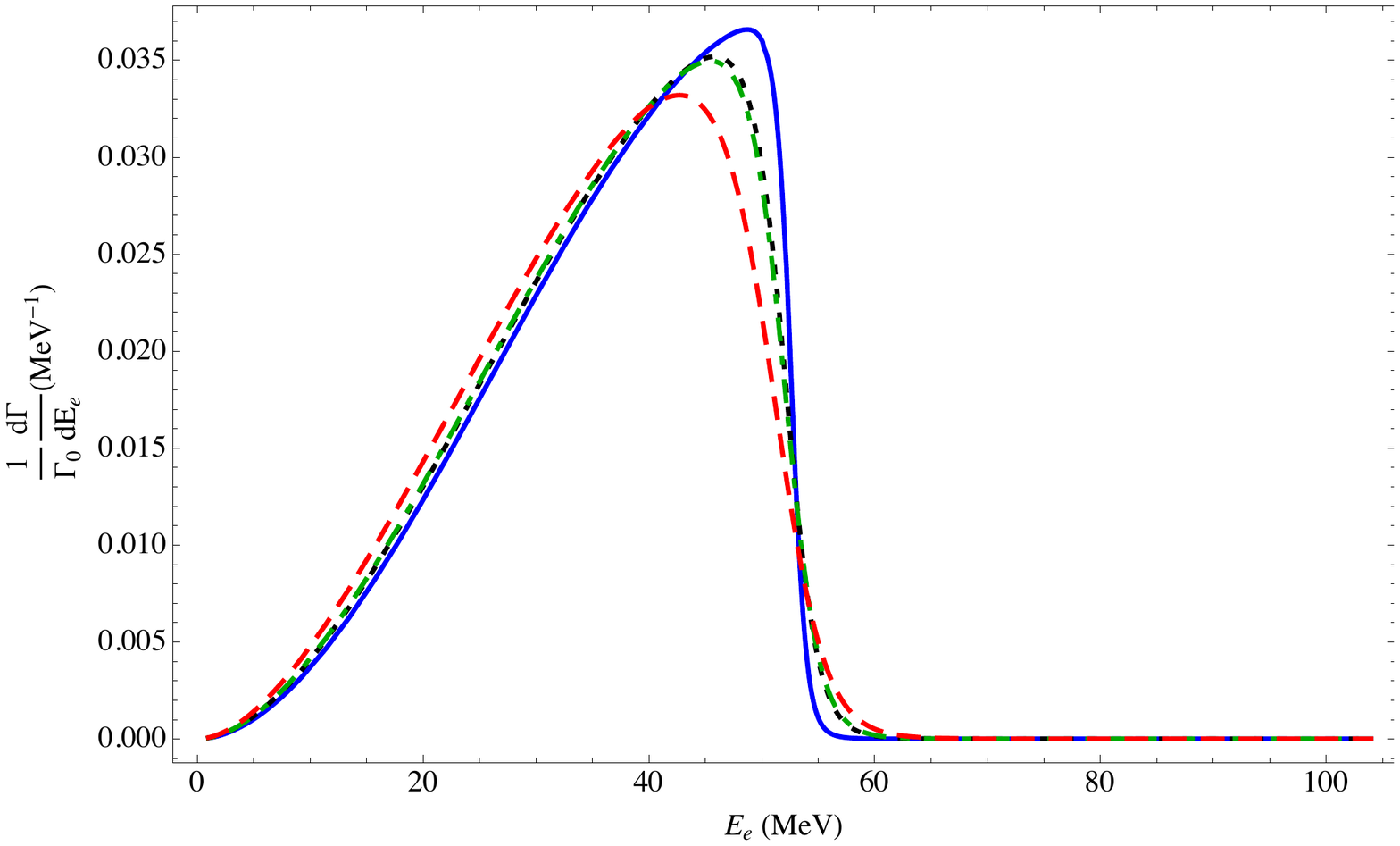} &
\includegraphics[viewport= 0 250 587 603, width=0.45\columnwidth]{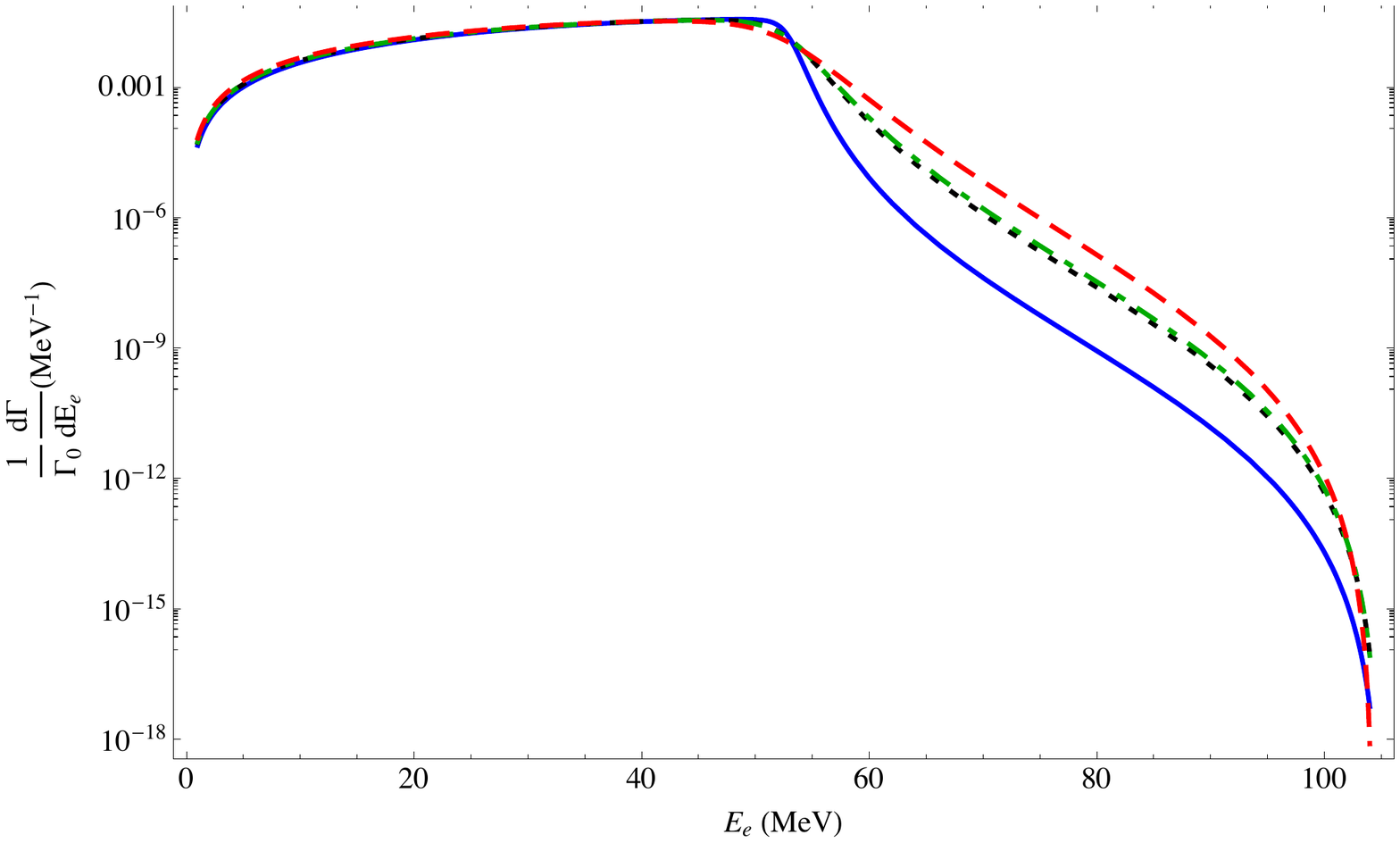} \\
$(a)$ & $(b)$ \\
\end{tabular}
\caption{\label{fig:czarnecki}Electron spectrum, normalized to the free-muon decay rate $\Gamma_0$. The solid blue line is for carbon, the
black dotted line for aluminum, the green dot-dashed line for silicon and the red dashed line for titanium. Adapted from~\cite{czarnecki2}. The computation takes into account all the effects needed in the high energy tail: relativistic effect in the $e$ and $\mu$ wave functions, Coulomb interactions between the electron and the nucleus, finite nuclear size and nuclear recoil.}
\end{center}
\end{figure}

The computation is valid for muons decaying in vacuum which is quite different from the real situation, in which the muon is brought to a stop in some material. It nevertheless holds for positive muons as we shall see in the following. 
On the other hand, electrons from the decay of bound $\mu^-$ exhibit a long tail extending up to 
\begin{equation}
\label{eq:muminus}
E_e \approx m_\mu - B_\mu - E_r
\end{equation}
where $B_\mu \approx Z^2 \alpha^2 m_\mu/2$ is the muon binding energy and $E_r \approx m^2_\mu / 2 m_{\rm N}$ the nuclear recoil energy.  In Figure~\ref{fig:czarnecki}~\cite{czarnecki, czarnecki2} the electron energy spectra from muons bound to the nucleus of an atom in common materials are presented. Note that the lighter the nucleus, the more the distribution resembles a Michel decay spectrum.
At its simplest level the electron energy spectrum is obtained by replacing the delta function of the muon at rest with the normalized momentum probability distribution of the bound $\mu^-$ moving about the point charge nucleus $Z\alpha$ in the 1S orbital~\cite{primakoff, muto, szafron} which is essentially the 3D Fourier transform of the single exponential with a Bohr radius:
\begin{equation}
\psi(r) \sim e^{-Z\alpha m_\mu r}
\qquad
\psi(p_\mu) \sim 
\frac{ 8 \sqrt{ \pi } \left( Z \alpha m_\mu \right)^{5/2} } 
{\left[ p_\mu^2 + \left( Z \alpha m_\mu \right)^2 \right]^2 }
\end{equation}
The muon momentum smears the decay electron energy distribution adding the long tail extending up to the total available energy while Coulomb attraction between nucleus and decay electron causes the decay electrons to appear with diminished momentum~\cite{uberall}.

Refined calculations~\cite{kitano} include the relativistic wave function for the electron and its phase shift due to the nuclear potential and  the muon wavefunction modified by the finite size of the nucleus. Radiative corrections modify substantially the absolute rate near the endpoint, which is the interesting region for $\mu^- N \to e^- N$ conversion searches~\cite{schanker, gilinski,czarnecki,czarnecki2}.
As it will be shown in the following this is not the only difference between positive and negative muons.

Since CLFV search experiments  deal with electron, positron and photon energies of few tens of MeVs produced by stopped muons, it is worthwhile to give a quick overview of the physics processes involved:
%
namely the production of low energy muon beams (of either charge) and the interaction of photons, electrons and positrons of energy below $\sim 100$~MeV.

\subsection{Muon production}
Muons of either charge are produced from pion and kaon decays. These are in turn produced at accelerators from a proton beam through a multitude of different channels. For instance the following processes are possible above the single pion production threshold ($T_p \sim 2 m_\pi \sim 280~$MeV, $T_p$ being the proton kinetic energy): 
\begin{equation}
{\rm p + p \to p + n + \pi^+},
\qquad
{\rm p + n \to p + p + \pi^-}, 
\qquad \ldots
\end{equation}
Above a proton energy of $\sim 600$~MeV creation of pairs of pions becomes possible and additional production channels open up. At even higher proton energies further higher multiplicity pion production channels become possible; however at traditional meson factories the energy is kept below the kaon production threshold. For instance the energy of the synchrotron at the Paul Scherrer Institut (PSI, Villigen) is $560$~MeV.

Pions  decay with a lifetime of $\sim 26$~ns to give muons via 
$\pi \to \mu \nu_\mu$ ($\sim 100\%$ probability). In this decay muons are $100\%$ polarized, in the pion rest frame, along their direction of flight.

\begin{figure}
\begin{center}
\includegraphics[width=0.7\columnwidth]{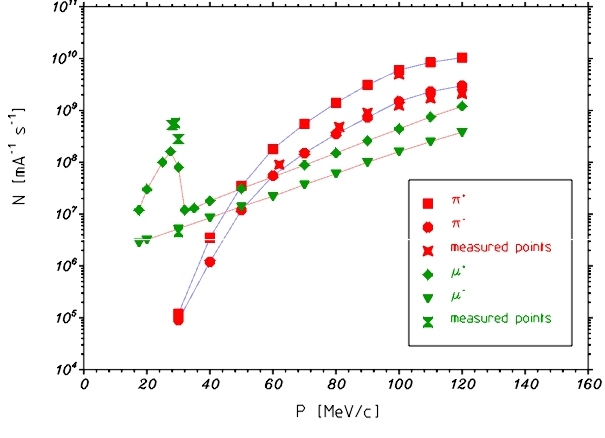}
\caption{\label{fig:pie5}Pion and muon flux at the $\pi$E5 beamline at PSI.}
\end{center}
\end{figure}

A thin (few cm) graphite target, more rarely beryllium, is used hence most protons (95\%) do not interact on the target and can be used for other purposes.
In Figure~\ref{fig:pie5} we reproduce the flux of $\pi^\pm$ and $\mu^\pm$ as a function of their momentum at at the PSI $\pi E5$ beam line that shows two general features: an increase of flux with the momentum and a significant increment in $\mu^+$ collection rate at a beam line tune of $\sim 29$~MeV/c.
This happens since at this momentum there is a preferential collection of $\mu^+$ mesons which originate from $\pi^+$ decays at rest which occur on the target skin, and are therefore named as ``surface muons"~\cite{surface}. This enhancement is only possible for positive muons since negative pions at rest in the target rapidly undergo nuclear capture.

This serendipity has a series of lucky experimental consequences. The $\mu^+$ production place on the target is well defined making it easier to obtain and transport a focussed beam ({\em i.e.} well defined in the transverse direction). Secondly it is possible to stop this beam in a very thin target ({\em i.e.} well defined in the longitudinal direction) therefore defining the position of the muon for its subsequent decay with high accuracy.
In fact, in this momentum regime, the range $R$ (the average distance travelled by a charged particle in a material) scales with momentum $P$ as $R \sim a P^{3.5}$~\cite{surface}.  At 29~MeV/c there is a 100-fold decrease of range with respect to 100~MeV/c making it possible to stop a positive muon beam in less than 1~mm of plastic with an extremely low straggling ({\em i.e.} range spread) $\Delta R = 3.5 \frac{\Delta P}{P} R$ where usually momentum 
resolution of the transport beamline is of the order of a few percents\footnote{Sometimes ``sub-surface'' muon beams may be used, if the requirement to the stopped muon beam is even more stringent. As the name suggests these are $\sim 25$~MeV/c muons which originate from pion decays deeper within the target that loose some of their energy on exiting~\cite{subsurface, MEG2, MEG2bis}.} . 
This is in extreme contrast to what happens at collider experiments where high energy muons are the most penetrating particles and escape all the way to the outer layers of detectors.

The mechanism produces muons that are intrinsecally close to be 100\% polarized. 
Polarization is decreased since multiple scattering changes the direction of the momentum but not that of the spin.
Multiple scattering depolarization  can be estimated using the characteristics of the production target and using the approximation
\begin{equation}
\langle \cos \psi \rangle \simeq 1 - \frac{1}{2} \langle \psi \rangle^2 
\end{equation}
where $\langle \psi \rangle^2$ is the average RMS multiple scattering angle which depends on target material and dimension. In the case of the PSI $\pi E5$ beamiline (that used by, among others, MEG, MEG-II, Mu3e, SINDRUM)  
for 29~MeV/c muons gives a $\langle \cos \psi \rangle \approx 0.997$ for a 4~cm thick graphite target~\cite{MEGPol}.

A more important depolarization effect is due to ``cloud muons'', {\em i.e.} muons originated from pion decays in flight inside or in the vicinity of the production target that are in  the acceptance of the transport system.
They lack, in fact, of the kinematic constraint that determines the net polarization and, if assumed to be $\sim 100\%$ unpolarized, their effect is to dilute the net polarization in proportion to their number.

Polarization can be preserved, altered or even destroyed, by the transport system and stopping target. It is important to note, however, that polarization is in principle present and usable, {\it e.g.}, for model dependent studies or possible background reduction in rare decay search experiments~\cite{kunopol}. 
\begin{figure}
\begin{center}
\includegraphics[width=0.99\columnwidth]{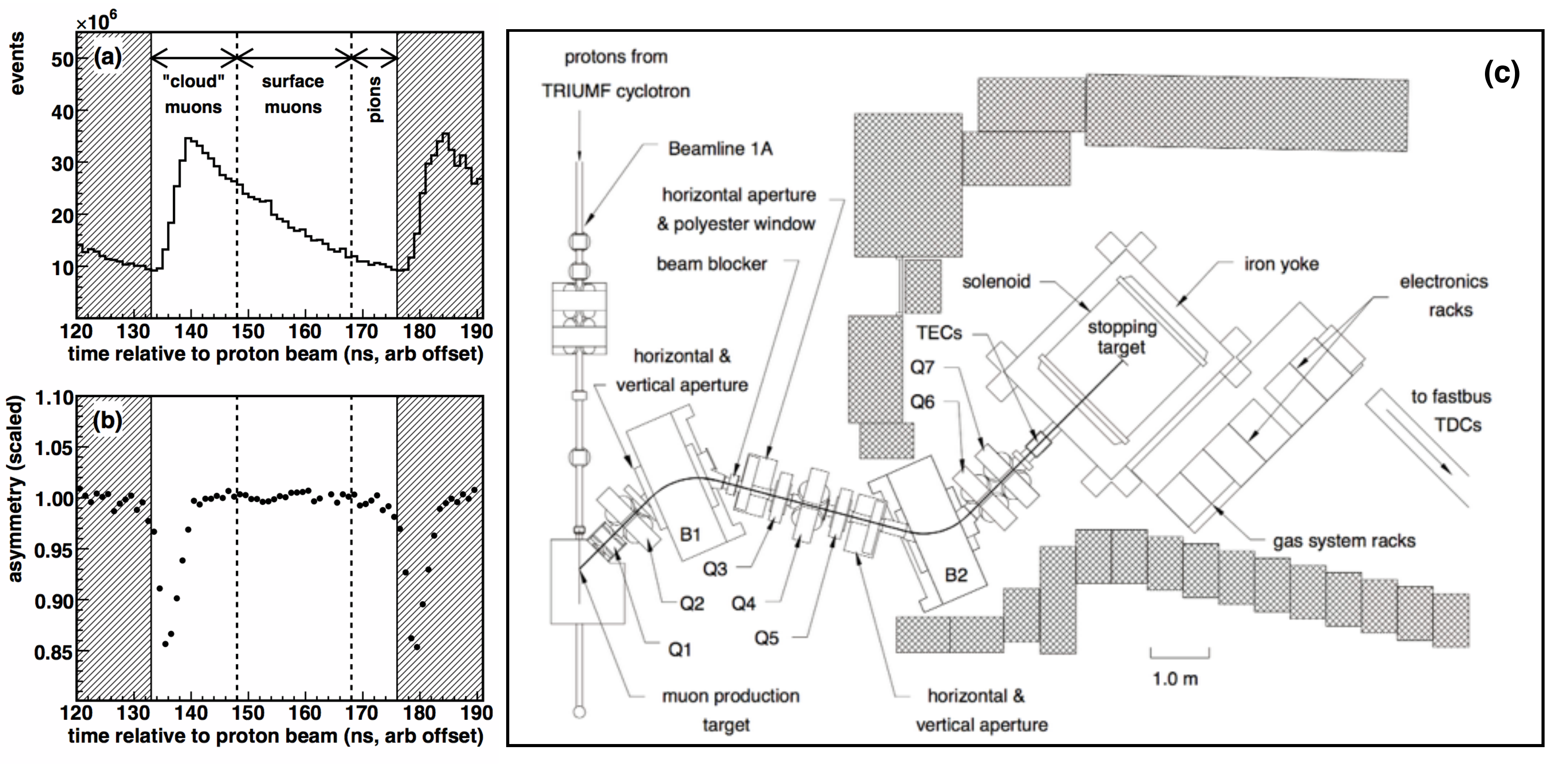}
\caption{\label{fig:twist} This picture illustrates the production of surface muons and their polarization in the TWIST experiment: (a) Number of muons as a function of the time relative to the primary proton beam, showing the pion lifetime; (b) The forward-backward asymmetry of the decay positron, corresponding to the muon polarization; (c) The M13 low energy muon beamline at LAMPF which illustrates a typical surface muon production beamline. Note that the muons are collected backward with respect to the impinging proton beam and are successively focussed and momentum selected by megnetic quadrupoles and dipoles respectively (adapted from~\cite{TWIST}.)}
\end{center}
\end{figure}

Some of the illustrated characteristics are visible in Figure~\ref{fig:twist} where we show the number of stopped $\mu^+$ on target measured by the TWIST experiment at LAMPF~\cite{TWIST}. The number of muons is modulated by the repetition rate of the primary proton beam (bunches of $\sim 3$~ns width every $43$~ns) and shaped by the $\sim 26$~ns $\pi^+$ lifetime%
\footnote{A surface muon beam from a pulsed primary proton beam is in fact at the base of the presently most accurate determination of the lifetime of the positive pion~\cite{Koptev}.}.
Their polarization (measured from the forward-backward asymmetry of decay positrons) is pratically $100\%$ for pure ``surface muons'' and it is reduced by the contamination of cloud muons during proton bunches, where decays from pions escaping the target are still in the beamline acceptance.

This shows another characteristics of this kind of beams: the fact that they provide an almost ``constant'' or ``DC'' rate of muons (as opposed to a ``pulsed'' one) where the small modulation is induced by the $26$~ns pion lifetime. This is even stronger at PSI where the primary proton repetition rate is $\sim 20$~ns, thereby reducing the time dependence of the absolute muon rate. As we shall see in the following a constant muon flux helps in search processes where accidental coincidences are dominant, while pulsed beams are preferred in searches where the beam-related background is more prominent.

Among muon sources, those at the Paul Scherrer Institute (PSI) in Switzerland and TRIUMF in Canada  provide an essentially continuous beam, modulated by the micro-time structure dictated by the cyclotron radio-frequency,  by  using high current--medium energy proton cyclotrons.  Muon beams based on lower current, higher energy, proton synchrotrons  providing a pulsed beam (10s of Hertz typical repetition rates) are present at facilities at Fermilab in the U.S. and J-PARC in Japan. Surface muon beams are being proposed to Chinese and Korean funding agiencies. Muon sources for ultra-cold muons and ultra-intense muons, based respectively on the formation and ionization of muonium atoms and on capture and transport by specially designed superconducting solenoids, are under development for future muon experiments as well.

Table~\ref{tab:stefan} shows the present and planned  intensities for some muon beams in various countries. It is clear that the present number of $\sim 10^8 \mu$/sec sets the scale for the number of available muons per year to $10^{15}$, while in the near future improvements are expected both in continuous and  in pulsed mode to bring this number to $10^{17} \div 10^{18}~\mu$/year necessary for the next generation of rare muon processes searches~\cite{SCook, BergMuBeam}.
\begin{table}[tb]
\begin{center}
\begin{tabular}{llll}
\hline
Laboratory & Beam Line & DC rate ($\mu$/sec) & Pulsed rate ($\mu$/sec) \cr
\hline \hline
PSI (CH) & $\mu E4, \pi E5$ & $2 \div 4 \times 10^8~(\mu^+)$ & \\
(590~MeV, 1.3~MW) & HiMB at EH & ${\cal O}(10^{10})~(\mu^+)$ ($>$2018) & \\
\hline

J-PARC (Japan)  & MUSE D-Line & & $3 \times 10^{7} (\mu^+)$ \\
(3~GeV, 210~kW) & MUSE U-Line & & $6.4 \times 10^{7} (\mu^+)$ \\

(8~GeV, 56~kW)  & COMET & & $1 \times 10^{11} (\mu^-) (2020)$ \\

\hline
FNAL (USA) & & & \\

(8~GeV, 25~kW) & Mu2e & & $5 \times 10^{10} (\mu^-) (2020)$ \\
\hline
TRIUMF (Canada) & & & \\

(500~MeV, 75~kW) & M13, M15, M20 & $1.8 \div 2 \times 10^{6} (\mu^+)$ & \\
\hline
RAL-ISIS (UK) & EC/RIKEN-RAL& & $7 \times 10^{4} (\mu^-)$\\

(800~MeV, 160~kW) &  & & $6 \times 10^{5} (\mu^+)$ \\

\hline
KEK (Tsukuba, Japan) &  & & \\

(500 MeV, 25~kW) & Dai Omega & & $4 \times 10^{5} (\mu^+) (2020)$ \\

\hline
RCNP (Osaka, Japan) & MuSIC & $10^{4} (\mu^-) \div 10^5 (\mu^+)$ \\

(400 MeV, 400~W) &  & $10^{7} (\mu^-) \div 10^8 (\mu^+) $($>$2018) \\

\hline
JINR (Dubna, Russia) & Phasotron & $$ &\\

(660 MeV, 1.6~kW) &  & $10^5 (\mu^+)$ &\\
\hline
RISP (Korea) & RAON & & \\

(600 MeV, 0.6~MW) &  & $2 \times 10^8 (\mu^+) $($>$2020) \\
\hline
CSNS (China) &  HEPEA& & \\

(1.6 6eV, 4~kW) &  & $1 \times 10^8 (\mu^+) $($>$2020) \\
\hline

\end{tabular}
\end{center}
\caption{Intensities of present and near future muon beams of positive and negative charge, continuous and pulsed. Note that for surface muon beams, increasing the proton beam energy does not necessarily increases the rate, since at higher proton energies  most pions escape the target rather than come to rest and decay~\cite{probeam}. Table compilation based on~\cite{mub1, mub2, mub3, mub4, SCook}
\label{tab:stefan}}
\end{table}

\subsection{Negative muons}
Reverting to negative muon beams we have to notice, first of all, that stopping negative muons in a target involves some processes which are not present when we stop positive muons: in fact negative muons rapidly displace an electron from the target nucleus (emitting characteristic X-rays that identify the process) and set into the lowest energy orbit at a radius which can be, for heavier elements, comparable to the nuclear radius (for $Z>50$ the Bohr radius for muonic atoms is within the nucleus itself!).

A calculation of the time necessary for the $\mu^-$ to approach the nucleus was performed by Fermi and Teller [55] in connection with the experiment of Conversi, Pancini and Piccioni and shows that within $t\sim 9 \times 10^{-14}$~sec on carbon ($t \sim 6 \times 10^{-14}$~sec  on iron) the muon is in the K-shell.  Such muon can subsequently
decay or be captured by the nucleus, and the total probability that it will disappear as a result of the two processes is equal to the sum of of the decay and capture probabilities, $\Lambda_d$ and $\Lambda_c$:
\begin{equation}
\Lambda = \Lambda_d + \Lambda_c.
\end{equation}
If $\tau$ is the mean negative-muon lifetime in the K shell and $\Lambda_d$ is the muon decay probability we have 
\begin{equation}
\frac{1}{\tau} = \Lambda_d + \Lambda_c.
\label{eq:muminustau}
\end{equation}
It is usually assumed that the decay probability for a muon in the K shell of the mesonic atom is the same as the decay probability for a free positive muon, {\em i.e.} $\Lambda_d = 1/\tau_0$ where $\tau_0$ is the lifetime of a free muon. In reality the decay probabilities are not equal (the two main corrections being a relativistic time dilation for the $\mu^-$ average velocity and a slight reduction in phase space~\cite{Weissemberg,gilinski}), but the difference is small.

The $\mu^-$  capture probability increases rapidly with the nuclear charge $Z$, and for
$Z\sim 11$ it is approximately equal to the muon decay probability $\approx 4.5 \times 10^5$~sec$^{-1}$. As $Z$ increases further the probability is found to be approximately proportional to Z$^4$\footnote{
The fact that the $Z^4$ rule is approximately valid at low-$Z$ is a consequence of 
the fact that the capture probability is proportional to the number of protons times the square of muon wave-function in the K-shell orbit at the origin, which is proportional to $Z^{3/2}$ for normalization reasons.
}.
The observation of decays from negative muons has therefore the following important characteristics:
\begin{enumerate}
\item the effective lifetime is reduced in proportion to the atomic number of the target nucleus and can range from 2~$\mu$s for carbon, 880~ns for aluminum, 330~ns for titanium to 73~ns in gold~\cite{muminus-capture};
\item during the disexcitation of the muonic atom to the 1S state, characteristic X-rays are emitted that allow the precise monitoring of the intensity of the muon beam;
\item  the capture of negative muons on a nucleus effectively reduces the number of emitted high energy electrons that represent a background for $\mu \to e $ conversion experiments.
\end{enumerate}

\section{Electrons and Positrons}
\label{sec:positrons}
\begin{figure}
\begin{center}
\includegraphics[width=0.75\columnwidth]{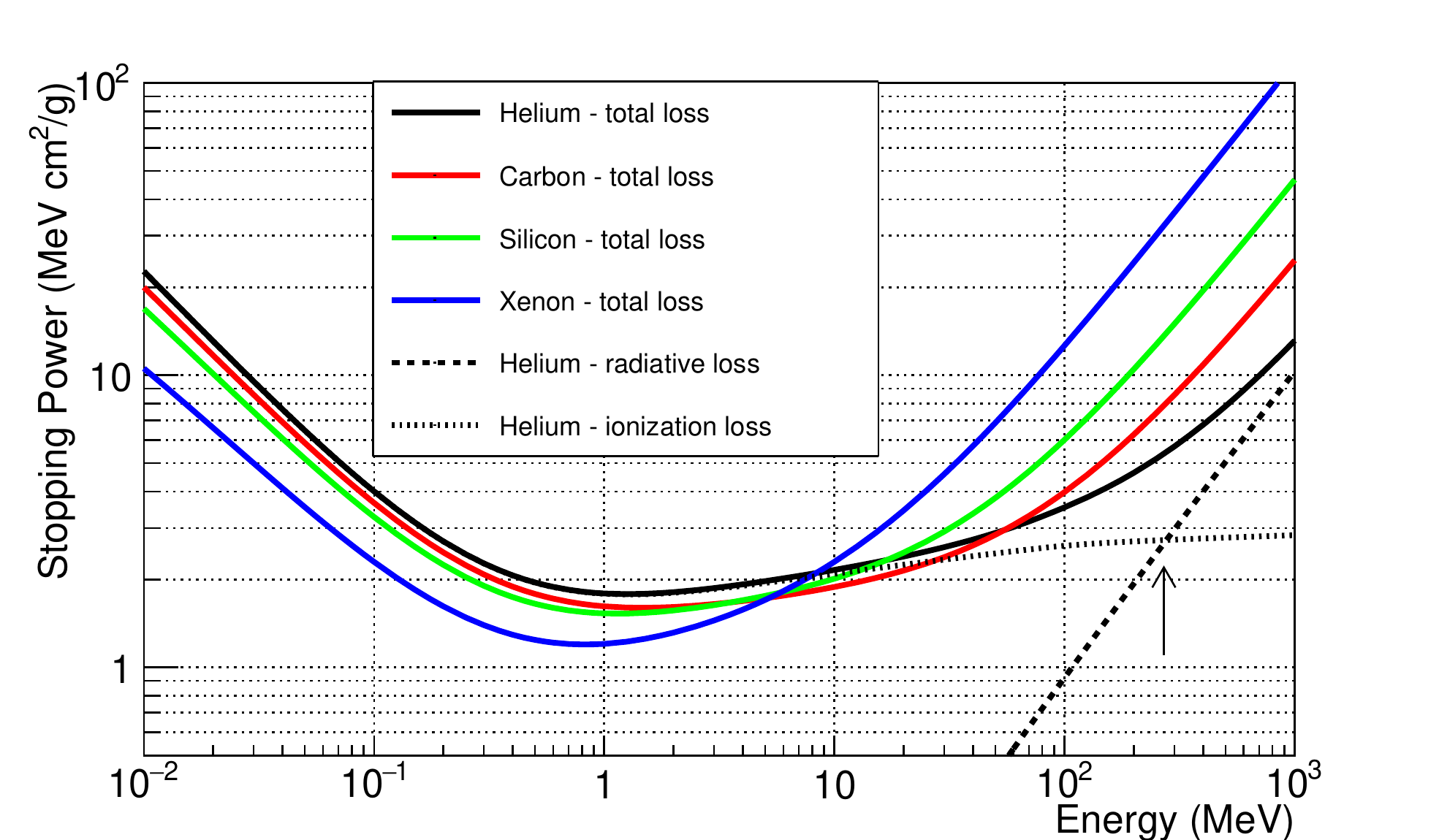}
\caption{\label{fig:eloss}Stopping power for electrons for various materials used to build electron and positron detectors. Multiplying for the material density gives the energy loss per unit length (MeV/cm). Data shown here are taken from the NIST ESTAR database~\cite{estar}.}
\end{center}
\end{figure}
Electrons and positrons loose energy by collision and radiation. Collisional loss dominates (depending on the material) up to few tens of MeVs where radiation losses start to prevail~\cite{Seltzer}. 

Two complementary techniques are usually used to measure the energy and momentum of an electron or positron: tracking and calorimetry. Tracking the particle trajectory in a magnetic field allows the measurement of its momentum at the few per mille, depending on the geometrical configuration, with a single hit resolution below 100~$\mu$m.  With as many as a few thousands channels one needs to readout signals which are comparatively slow, since ionization induced by the charged particles in the tracking medium needs to be amplified and collected, and typical drift velocities are of the order of a few mm/$\mu$s. For this reason trackers are usually coupled to faster scintillating detectors for the precise determination of timing.

Electromagnetic calorimeters measure the total energy of the charged particle by collecting the light (sometimes the charge) emitted following the energy release from the electromagnetic shower. This can be accomplished with a limited number of fast photo-sensors (typically within few hundred nanoseconds) with a resolution, dominated by photo-statistics and shower fluctuations, of the order of a few percent in energy and of a few millimeters in impinging position. 

The utilization of one or the other detector depends on the experimental requirements. There are situations in which their combined use is possible, as it is done in collider detector experiments.

In Figure~\ref{fig:eloss} we present for reference the stopping power of helium, carbon, silicon and xenon (for helium collisional and radiation losses are also presented separately, an arrow indicating were they are equal, the so-called ``critical energy"). In the energy range of interest for rare muon decay searches ($1 \div 100$~MeV) the stopping power has a plateau at the famous value of ``two" MeV/g/cm$^2$, meaning that a particle looses about 2~MeV in 1~cm of material whose density is the same as water. This value represents a huge fraction of the energy of an electron/positron coming from $\mu \to e \gamma$ or $\mu \to eee$ which implies that whatever tracker is needed for such searches must have a total thickness well below the centimeter size.

Energy loss, on the other hand, is not the only concern that must be kept in mind. By knowing the total thickness of the detector it could be possible in principle to correct for it. What cannot be corrected are the fluctuations around the average energy loss ({\em straggling}) and the change of direction of the particle induced by the material itself ({\em multiple scattering}).

\begin{figure}
\begin{center}
\begin{tabular}{ccc}
\includegraphics[width=0.3\columnwidth]{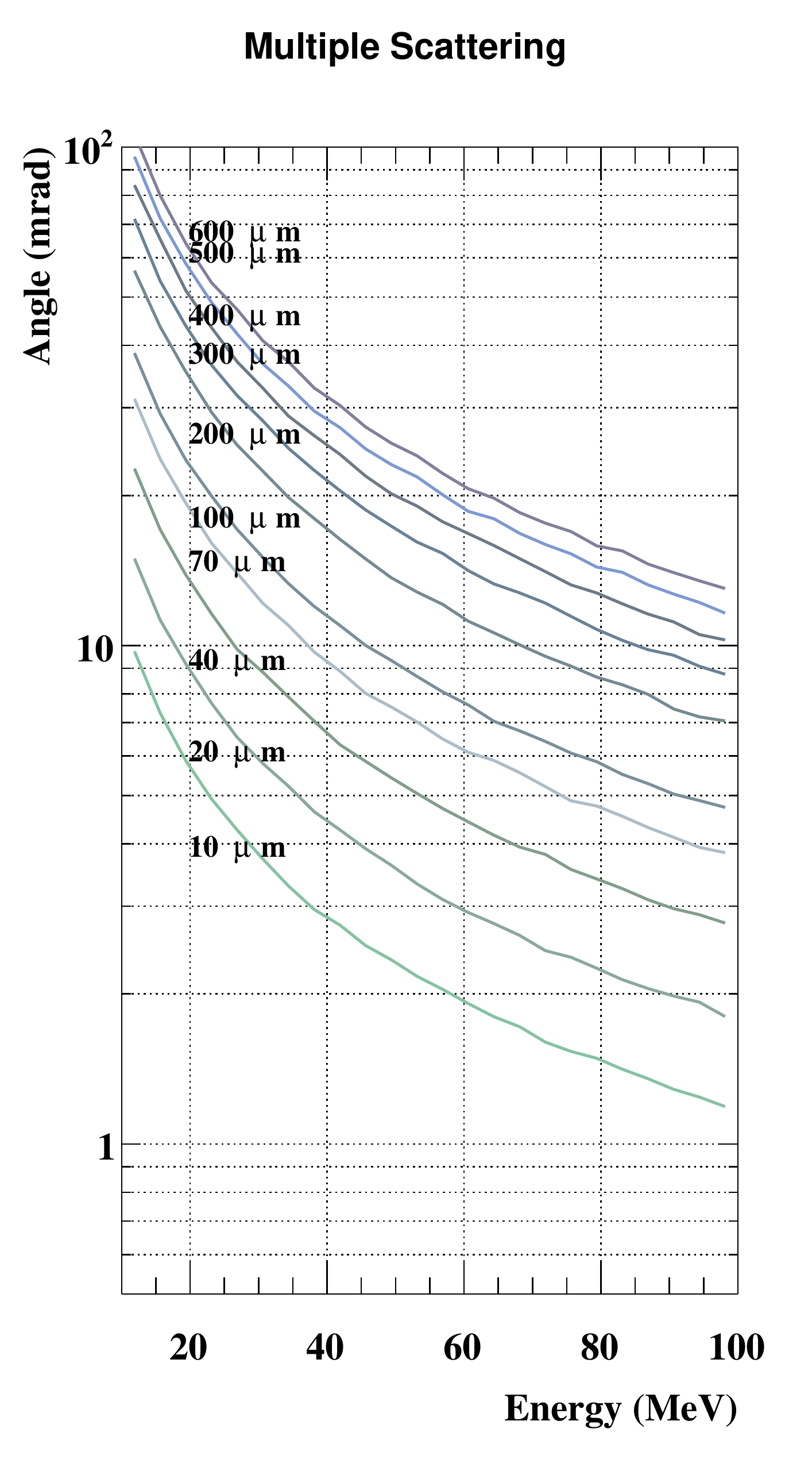}
&
\includegraphics[width=0.3\columnwidth]{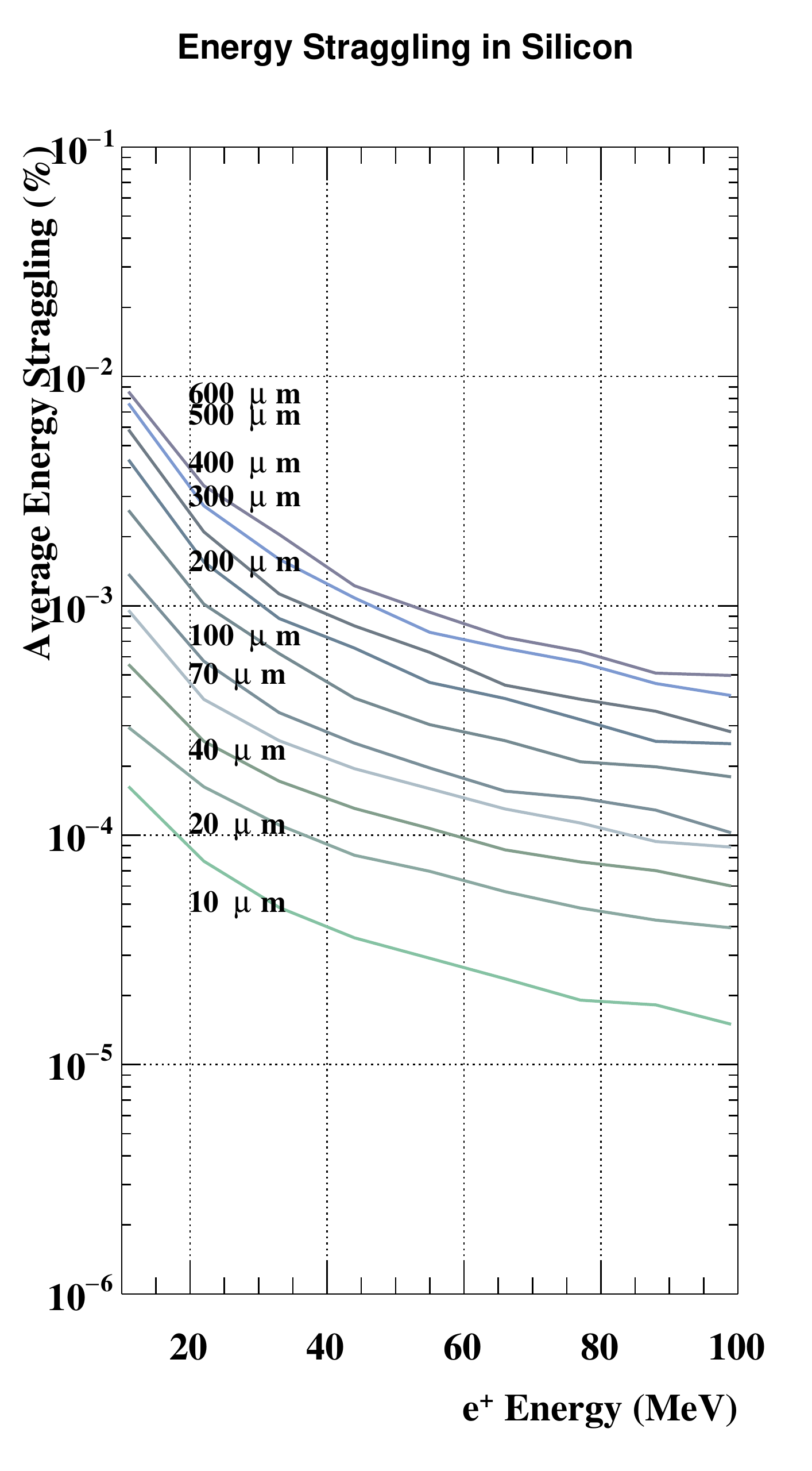}
&
\includegraphics[width=0.3\columnwidth]{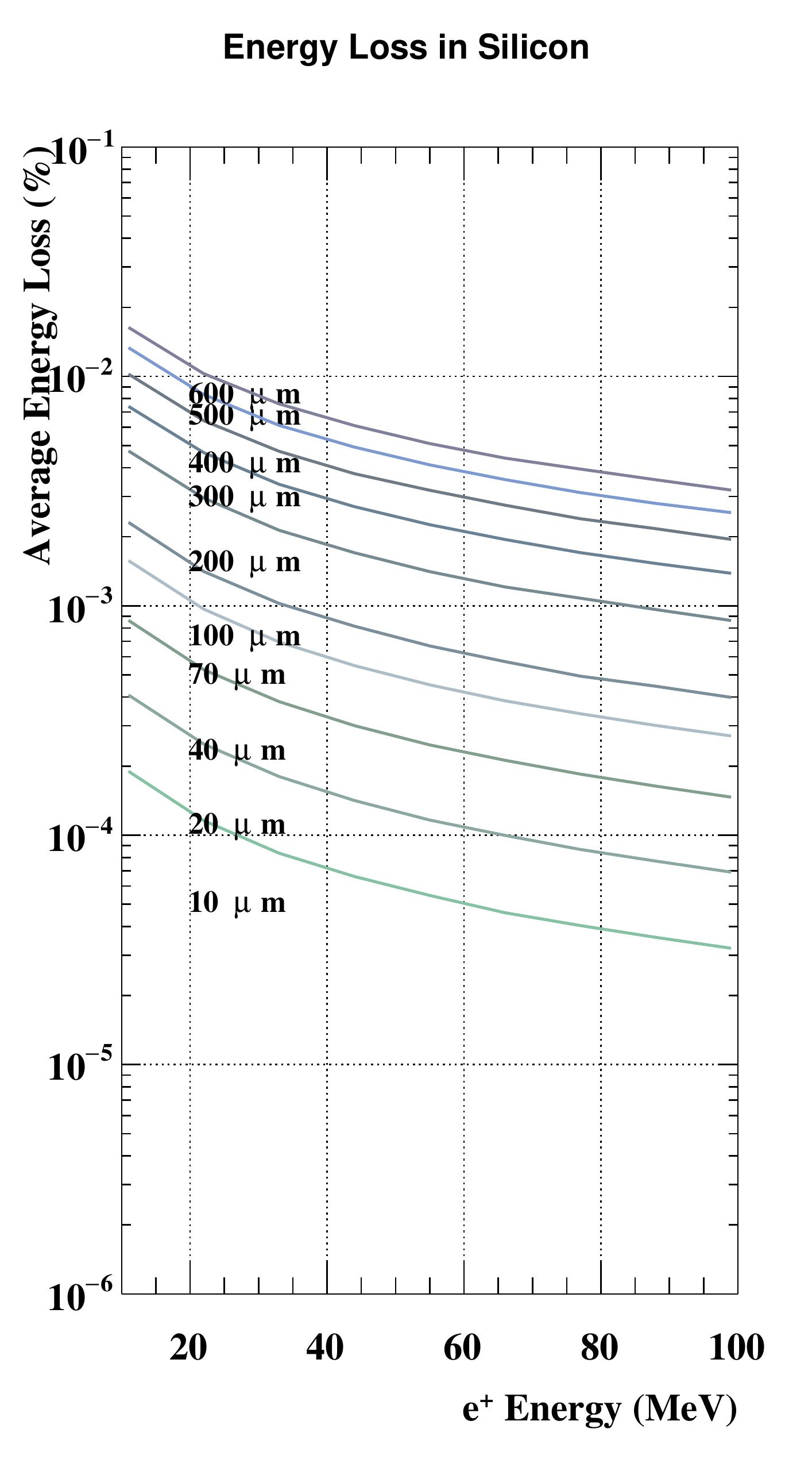}
\\
\end{tabular}
\caption{\label{fig:silicon}Impact of a silicon tracker on the trajectory of a positron/electron. The panel on the left shows the multiple scattering angle induced by different thicknesses of material for impinging particles in the energy range of interest for CLFV searches (10~MeV $\div$ 100~MeV). The central panel shows the energy straggling while the rightmost panel shows the energy loss. It appears that the most dramatic effect happens in the deflection of the electron/positron from its original direction (multiple scattering effect).}
\end{center}
\end{figure}

\begin{figure}
\begin{center}
\begin{tabular}{ccc}
\includegraphics[width=0.3\columnwidth]{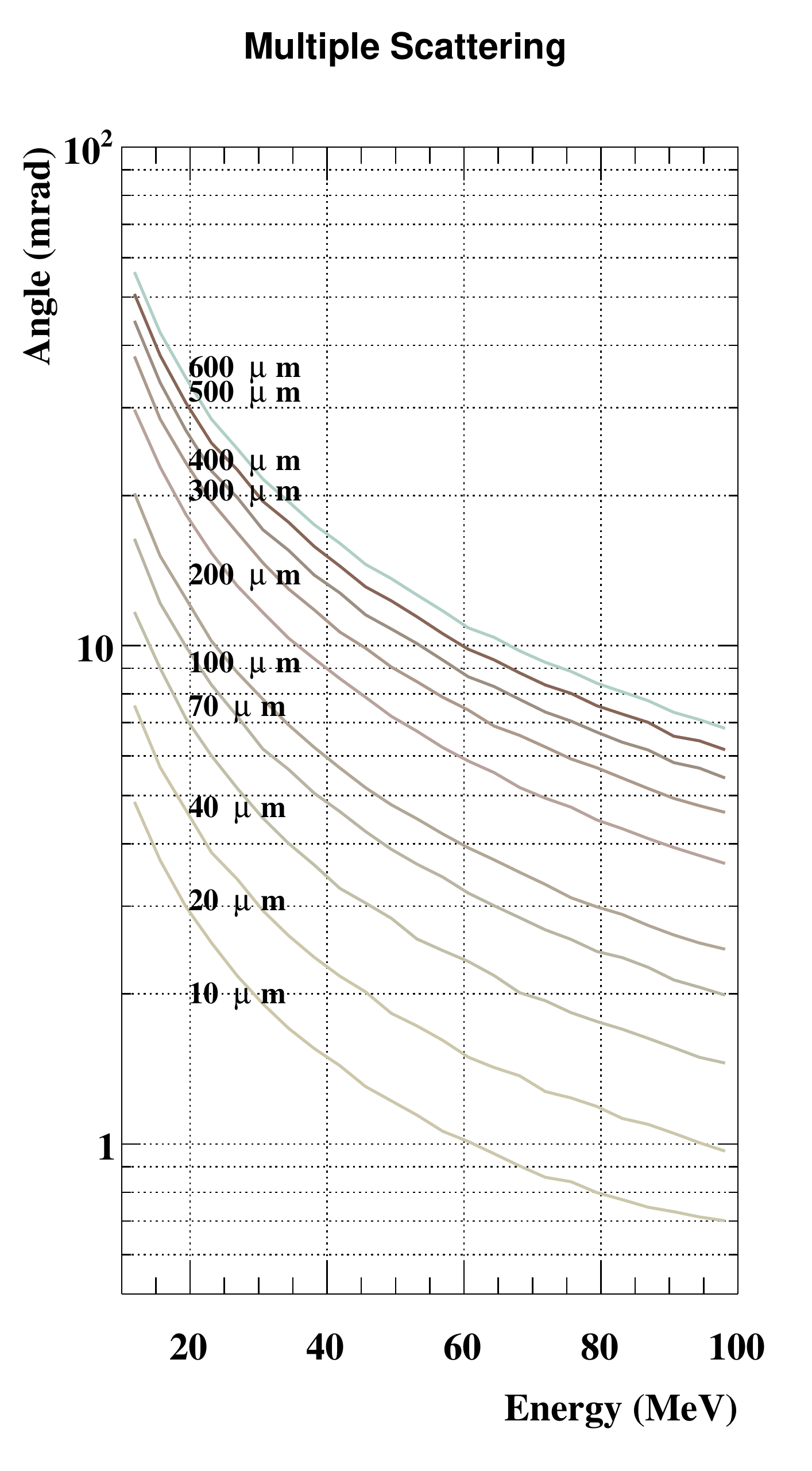}
&
\includegraphics[width=0.3\columnwidth]{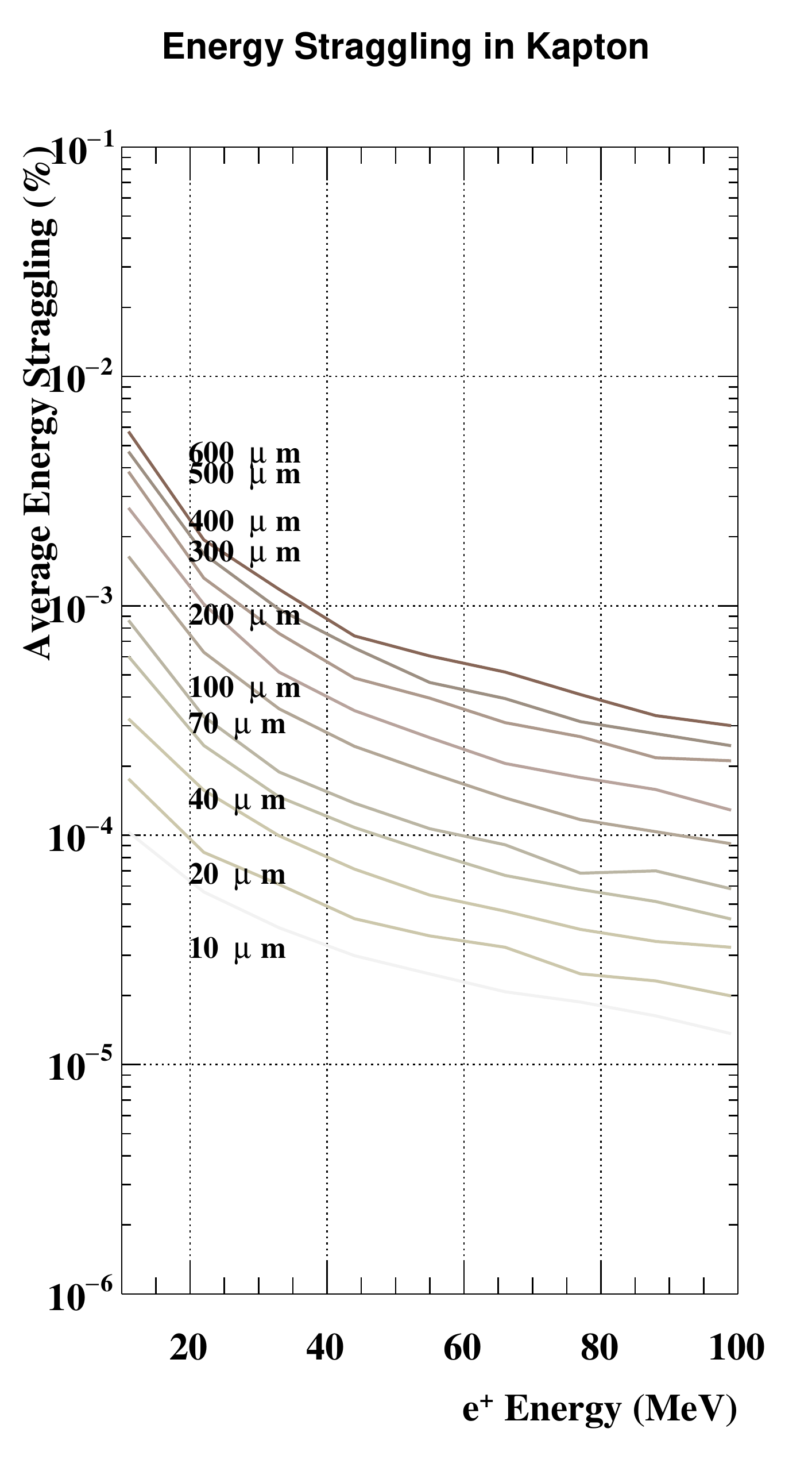}
&
\includegraphics[width=0.3\columnwidth]{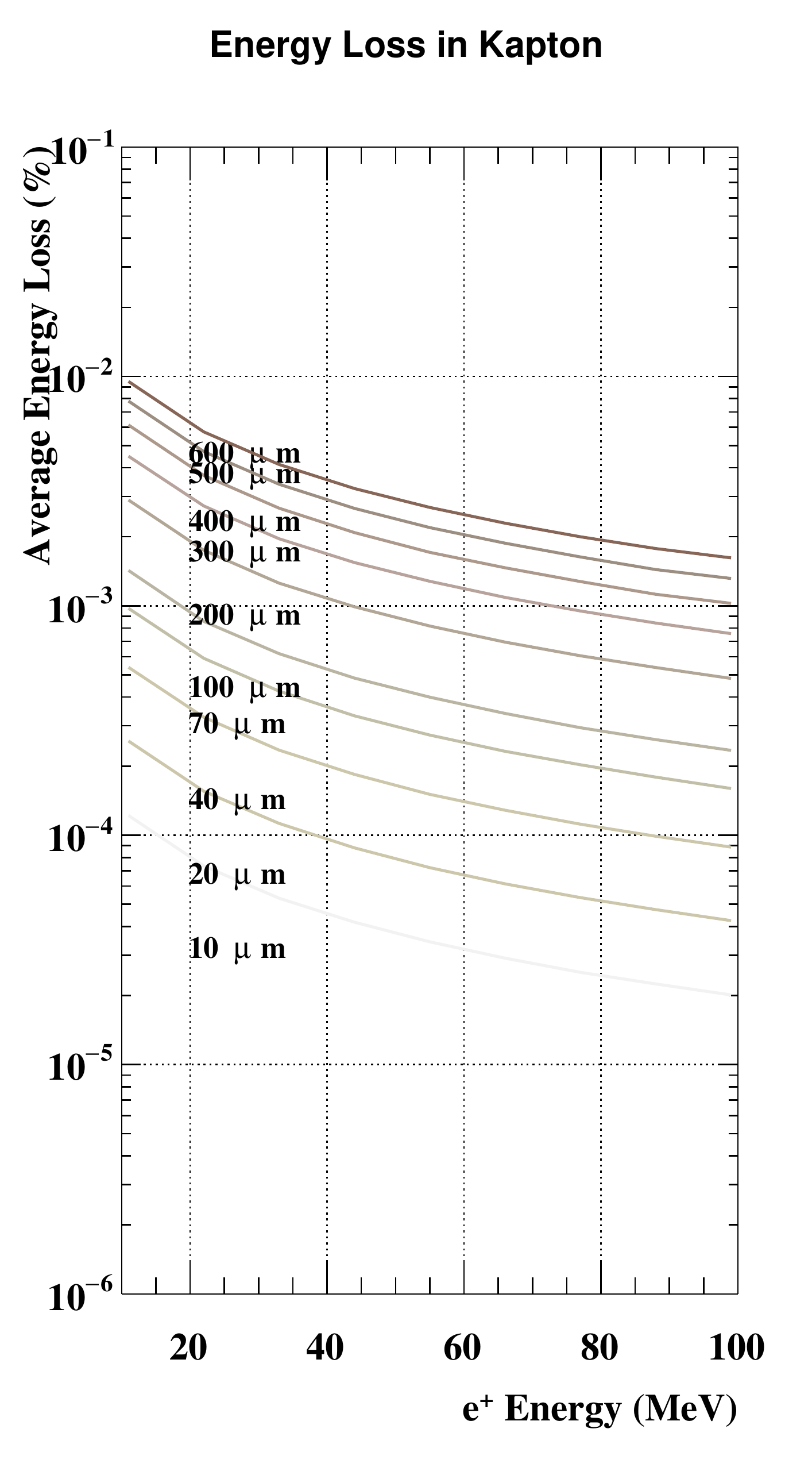}
\\
\end{tabular}
\caption{\label{fig:kapton}Impact of a polymide foil (Kapton)  tracker on the trajectory of a positron/electron. The panel on the left shows the multiple scattering angle induced by different thicknesses of material for impinging particles in the energy range of interest for CLFV searches (10~MeV $\div$ 100~MeV). The central panel shows the energy straggling while the rightmost panel shows the energy loss. }
\end{center}
\end{figure}

In Figure~\ref{fig:silicon} and~\ref{fig:kapton} we show the multiple scattering,  energy straggling and  energy loss induced on electrons or positrons in the $10 \to 100$~MeV energy range by silicon and kapton 
which are among the materials planned to be used in future experiments.
The largest effect of the material is to smear the information on the particle direction: take as an example the case of the MEG experiment~\cite{MEGdetector} where the positron tracker had an equivalent thickness of 300~$\mu$m silicon: a signal positron ($E_e \sim 52.8$~MeV) suffers an energy loss of $\Delta E/E < 0.25\%$ with a $< 0.05\%$ energy straggling, nevertheless experiences an average scattering of $15$~mrad which, as we shall see, dominates the angular resolution needed to reconstruct the original direction of the particle.

\section{Photons}
\label{sec:photons}
\begin{figure}
\begin{center}
\includegraphics[width=0.99\columnwidth]{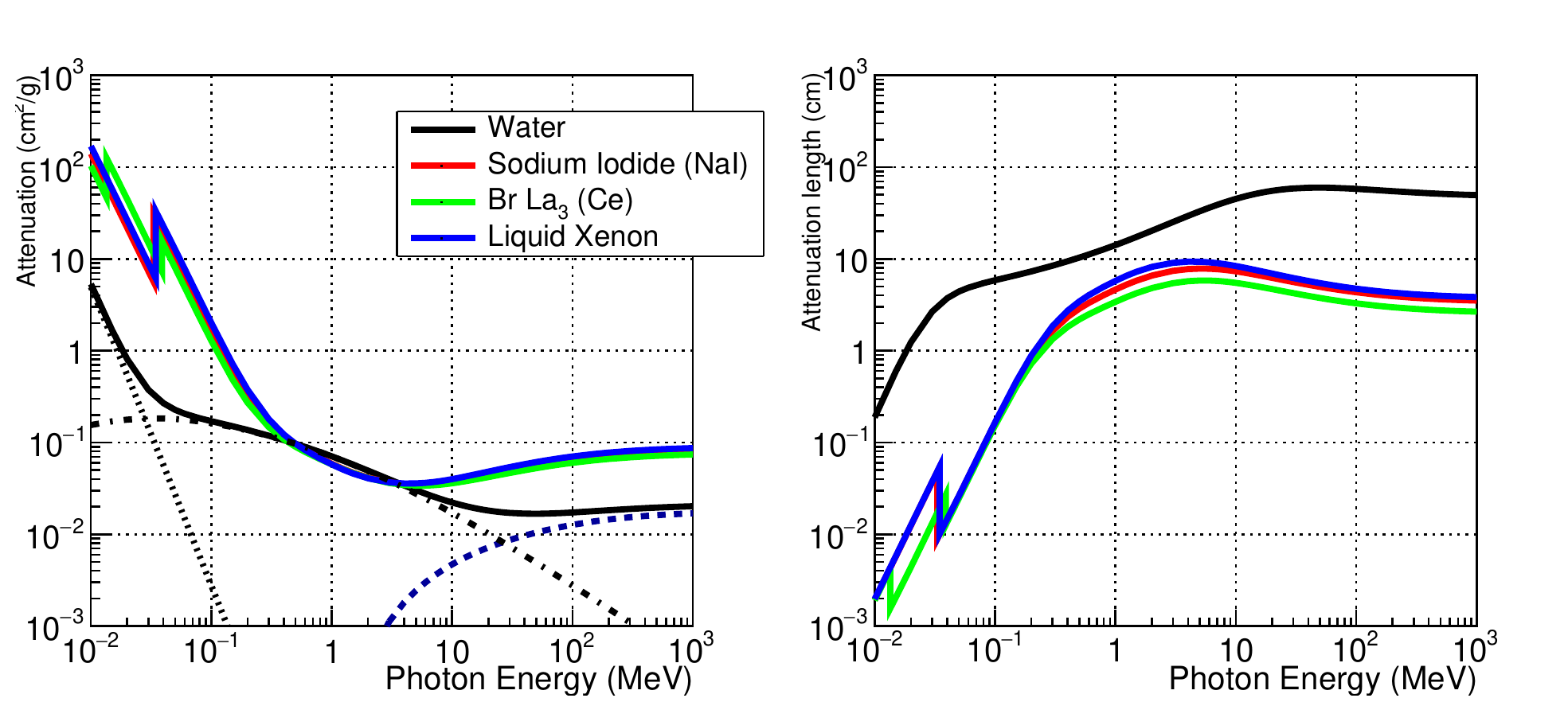}
\caption{\label{fig:gammaloss}Photon cross section and attenuation length for different materials used for detecting $\gamma$-rays in the energy range of interest for rare muon decay searches. Data shown here are taken from the NIST XCOM database~\cite{xcom}}
\end{center}
\end{figure}
The situation is symmetric for  detecting photons in the $10 \div 100$~MeV energy range~\cite{Nelson}: there is not a-priori a reason to prefer the measurement of the gamma ray by means of a calorimeter with respect to converting the photon and measure the momentum of the resulting $e^+ e^-$ pair. The trade-off is between high detection efficiency and energy and momentum resolution.

\begin{figure}
\begin{center}
\begin{tabular}{cc}
\includegraphics[width=0.51\textwidth]
{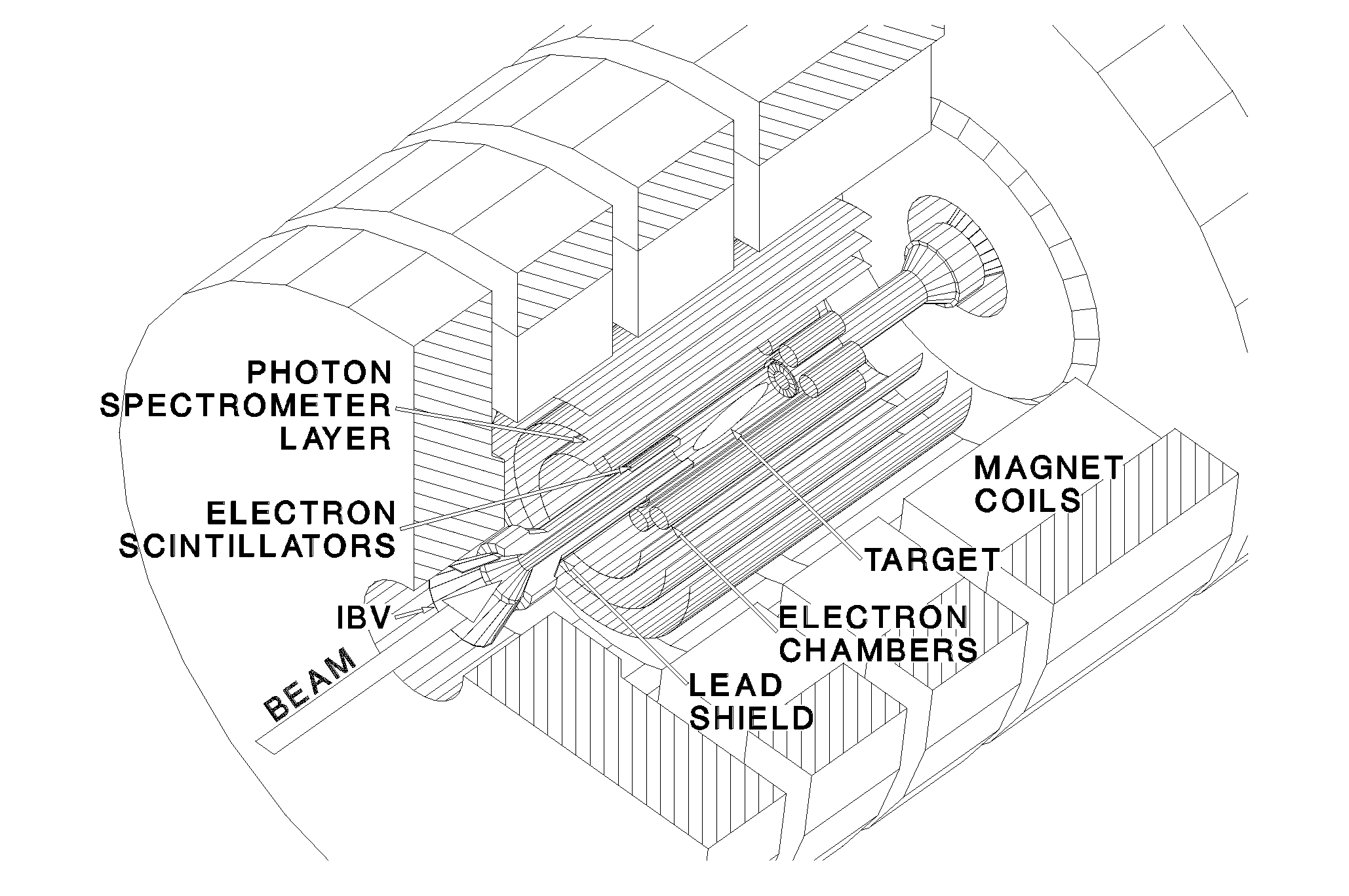} &
\includegraphics[width=0.44\textwidth]{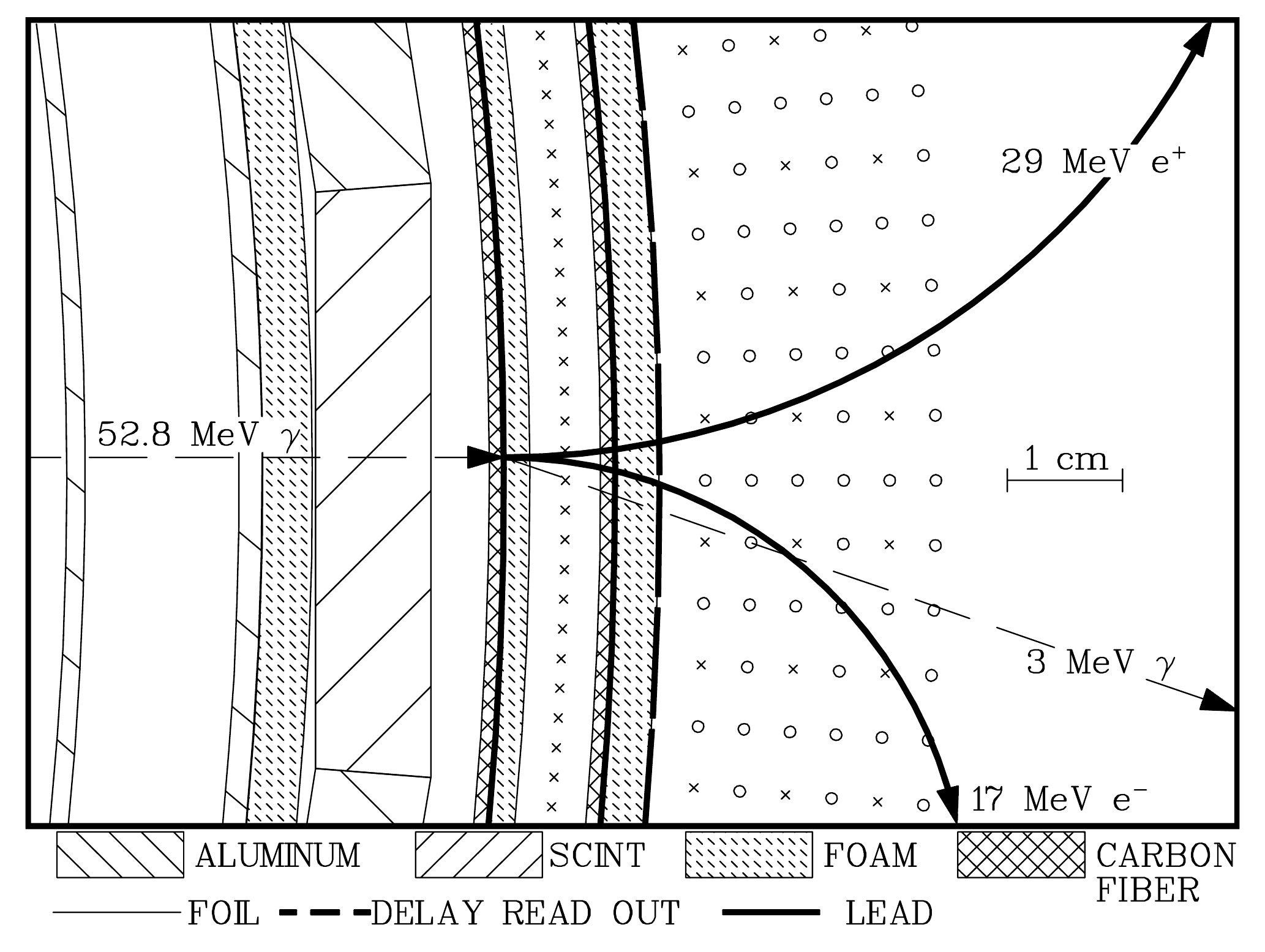} \\
$(a)$ & $(b)$
\end{tabular}
\end{center}
\caption{\label{fig:mega}The MEGA detector and an example of pair conversion of the photon.}
\end{figure}

A typical electromagnetic calorimeter can reach an energy resolution of a few percent at most with a spatial resolution determined by the granularity of its readout sensors. Conversion of the $\gamma-$ray requires the usage of thin layers of high-Z materials (usually tungsten or lead, see {\em e.g.}\ Figure~\ref{fig:mega}) which cannot be too thick in order not to interfere with the measurement of the $e^+ e^-$ pair, and is therefore limited to measure only a small (order of 1\%) fraction of the total number of produced gamma rays. Moreover, since  the partition of the  photon energy between $e^+$ and $e^-$ is flat in this energy range, one has to deal with the problems discussed in the previous section in tracking the products of the $\gamma-$ray conversion.

As for the calorimeter detection is concerned the situation is quite different since below $100$~MeV we cannot properly talk of a canonical electromagnetic shower as described in standard textbooks (see, for instance,~\cite{perkins, knoll}) in which a photon produces a huge cascade of electrons, positrons and photons. The $\gamma-$ray cross section has a minimum in the energy range $10\to 100$~MeV (see Figure~\ref{fig:gammaloss}, left panel) and consequently the attenuation length, which is inversely proportional to it, is of the order of few centimeters (Figure~\ref{fig:gammaloss}, right panel). Futhermore the comparatively high critical energy (10s of MeV) implies that the gamma energy is released, after the first pair production, in few spots where secondary photons undergo Compton scattering. This results in a highly asymmetric distribution of the energy release in the detector and in a high probability of having escaping low energy photons carrying away part of the original energy.

\begin{figure}
\begin{center}
\includegraphics[width=0.99\columnwidth]{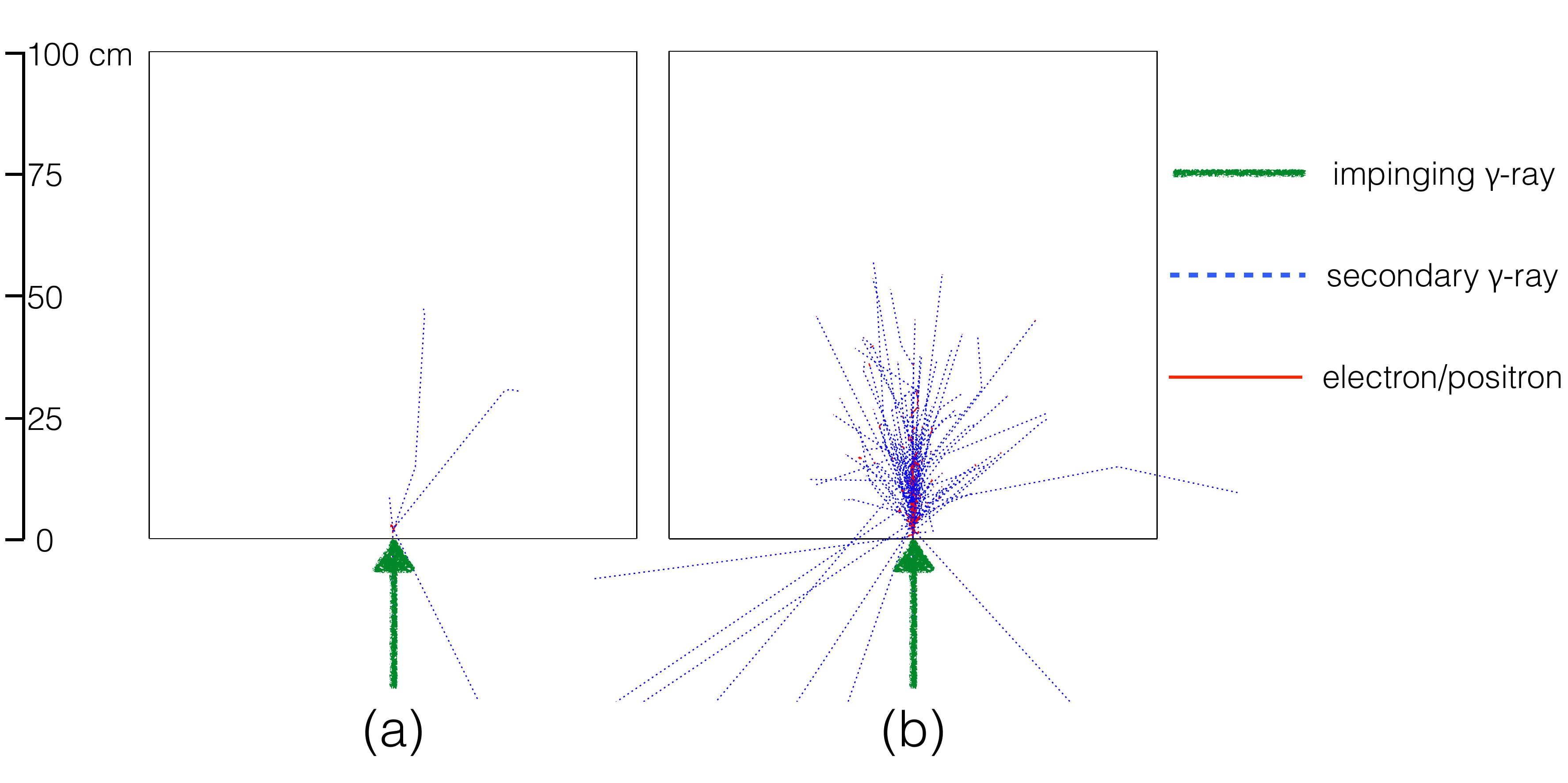}
\caption{\label{fig:singlephoton} Simulation of $(a)$ Single 52.8~MeV $\gamma$-ray and $(b)$ superposition of different 52.8 MeV photon showers. The $\gamma-$ray is entering the detector from the bottom of the page. Blue dashed lines represent photon trajectories, while red lines represent electrons and positrons. The event-by-event fluctuations in such low energy events is the most dominant term in the energy resolution as it is apparent from panel (a) where we see that the energy release is far from being symmetric, but low energy photons deliver energy even 30 cm far from the main interaction point through Compton scattering. If we superimpose several decays as in panel (b) the symmetry of the process is restored but only on a statistical basis.  Note also the unavoidable effect that some of the low energy photons are back-scattered out of the detector.}
\end{center}
\end{figure}
This is illustrated in Figure~\ref{fig:singlephoton} where we show a simulated interaction of a 52.8~MeV photon, coming from the bottom of the page, in an ideal sodium iodide detector. If on average the distribution of the energy release resembles a decent electromagnetic shower, there is a large event-by-event fluctuation on a scale of several tens of centimeters. 
Homogeneous detectors are preferred over segmented ones in order to minimize non sensitive regions even though other characteristics of the detector are important as well, such as light yield (the amount of light produced in response to an energy deposit) and time constant: a fast response is desired in order to get a better timing of the event and to be able to distinguish events with multiple - pile-up - gamma rays. Among the various materials used in the past or available for future detectors sodium iodide (NaI) plays a prominent role since it is a bright crystal available in large sizes~\cite{kinnison, ale2, ale0b,ale1} with a time constant of 250~ns. The MEG experiment has been using liquid xenon~\cite{MEGdetector} for its high light yield and fast response (different components from 4 to 45~ns) but other materials are being considered for possible future experiments, a good candidate being cerium-doped lantanum bromide -- LaBr$_3$(Ce)~(see Table~\ref{tab:scintillators} for the properties of commonly used scintillators)~\cite{mu2eTDR, Brillance1, Brillance2}.
\begin{table}[htb]
\begin{center}
\begin{tabular}{lccc}
{Scintillator} & { Density (g/cm$^3$} & { Light Yield (photons/keV) } & { Decay Time (ns)}\\
\hline  
LaBr$_3$(Ce) & 5.08 & 63 & 16 \\
LYSO & 7.1 & 27 & 41 \\
YAP & 5.35 & 22 & 26 \\
LXe & 2.89 & 40 & 45 \\
NaI(Tl) & 3.67 & 38 & 250 \\
BGO & 7.13 & 9 & 300 \\
\hline
\end{tabular}
\caption{\label{tab:scintillators} Properties of state-of-the-art scintillators being considered for CLFV searches.}
\end{center}
\end{table}

Both calorimeters and pair spectrometers have been successfully used in past and present experiments, the choice of the particular detector being dictated by the general design of the experimental search, and especially by the expected rate of background events (see~\cite{VoenaPapa} for a comparison between the two detection methods applied to the search for \meg).




\section[The role of Monte Carlo simulations in rare-decay search experiments]{The role of Monte Carlo simulations in rare-decay search experiments: the importance of calibrations}
\label{sec:calibrations}
If one wanted to stop cosmic ray muons in a hand and use them to perform a rare decay search at the level proposed by near future experiments, he or she would have to wait the entire age of the Universe to be able to observe just one event\footnote{The flux of cosmic ray muons  at sea level is $\sim 1$~muon/cm$^2$/min so it would take 10 billion years to collect $3\times10^{17}$ muons.}!
This simple comparison makes it apparent the smallness of the probability of the processes under investigation, and that it is virtually impossible to predict all sources of nuisance that may interfere with the measurement. 

Monte Carlo codes are essential to design the experimental apparatuses and to interpret the results but one must never forget their limitation when dealing with processes that happen with extremely low probability. The typical situation encountered when evaluating the sensitivity of such experiments consists in defining criteria to remove events that do not look like the signal and test them by using Monte Carlo samples that simulate the expected background. After applying all the selection cuts one is left with an empty set and it is easy to draw the conclusion that the experiment is background free. Even if it is a trivial statement, it is worthwhile to recall that Monte Carlo events are the result of the combination of processes that have been previously coded and deemed as possible. Furthermore as large as the simulated sample can be, its size is limited by storage or by computing power to a few millions of events, and even if all of them are removed from the signal region, there are still many orders of magnitude to go.

It is for this reason that the best way of knowing the background of a rare decay search experiment is to measure it with the experiment itself, and to devise, already at design stage, all possible means of measuring and calibrating the detector characteristics. 
It is of utmost importance to demonstrate, in case of positive observation of a signal, that the apparatus was in good shape before, during and after such measurement.

In the energy range of interest there exist many processes that can be used to monitor the response of the detectors to electrons, positrons and gamma rays. At low energy radioactive gamma ray sources like $^{60}$Co or $^{88}$Y (in the range 0.89 to 1.81~MeV) up to Pu-Be or Am-Be sources ($4.43$~MeV via the reaction $^{9}$Be($\alpha$,n) $^{12}$C$^* \to ^{12}$C$+ \gamma$) can be used.
\begin{table}[t]
\renewcommand{\arraystretch}{1.2}
\begin{center}
{\begin{tabular}{cccc}
\hline
\multicolumn{2}{c}{\bf Process}  & {\bf Energy(MeV)} & {\bf Frequency} \\
\hline
Charge exchange & $\pi^- p \to \pi^0 n$  & $54.9,82.9$ & yearly \\
& $ \pi^0 \to \gamma \gamma$ & & \\[1.2mm]
Charge exchange & $\pi^- p \to n \gamma$  & $129.0$ & yearly \\
Radiative $\mu^+$ decay  & \muenng & 52.83 endpoint & weekly \\[1.2mm]
Proton accelerator & $^7 {\mathrm Li} (p, \gamma_{17.6(14.8)}) ^8 {\mathrm Be}$ & 14.8, 17.6 & weekly \\[1mm]
& $^{11} {\mathrm B} (p,  \gamma_{4.4}\gamma_{11.6}) ^{12} {\mathrm C}$ & 4.4, 11.6 & weekly \\[1.2mm]
Nuclear reaction & $^{58} {\mathrm Ni}(n,\gamma_{9.0}) ^{59}{\mathrm Ni}$ & 9.0 & daily \\[1.2mm]
AmBe source & $^9\mathrm{Be}(\alpha,n)^{12}\mathrm{C}^*$ & 4.4 & daily \\
& $^{12}\mathrm{C}^*\to ^{12}\mathrm{C} \gamma_{4.4}$ & & \\[1.2mm]
\hline
\end{tabular}}
\caption{Typical calibrations performed in the MEG experiment to determine the
LXe detector performance (energy scale, linearity, etc.) with their energy
range and schedule.\label{tab:calibrations}}
\end{center}
\end{table}
Monochromatic photons at higher energy (up to $\sim 18$~MeV) can be produced by nuclear reactions of protons on fluorine, lithium or boron targets while at even higher energies neutral pions from the charge exchange reaction $\pi^- p \to n \pi^0$ produces $\gamma$ rays in the energy range $55-83$~MeV.
Not all these reactions are equally simple to be used: while placing a radioactive source in front of a detector poses no serious constraint, the usage of nuclear reactions requires a dedicated proton accelerator, while using reactions involving pions  requires a different target and a different beam tune. 

Nevertheless they were used extensively in past~\cite{carrington, kinnison}
and present~\cite{MEGdetector,CW} experiments. In particular both in the TRIUMF and in the MEG experiment a dedicated small accelerator (a Van de Graaff in~\cite{kinnison} and a Cockcroft-Walton in~\cite{CW}) was integrated in the design of the experiment and allowed a continuous monitoring of the $\gamma$ energy scale and resolution. In Table~\ref{tab:calibrations} we show some of the calibrations used in the MEG experiment. 

On the positron side the charged pion decay $\pi^\pm \to e^\pm \nu$ produces  
monochromatic electrons that were used in the past to check the energy scale and resolution at SINDRUM~II~\cite{sindrum2}. MEG, on the other hand, uses Mott scattering of monochromatic positrons from the beam on a dedicated target~\cite{mott}.

It is also important to demonstrate that a detector searching for improbable decays sees the events that it should: in this regard the muon radiative decay, with its time coincident $e-\gamma$ pair is both useful for assessing the timing resolution of an experiment and used to check the branching ratio of the process in a kinematic range where its probability is comparable to the single event sensitivity~\cite{megradiative}. Other decays used in this respect where $\pi^+ \to \pi^0 e^+ \nu$ and $\pi^+ \to e^+ \nu \gamma$ (BR$\sim 10^{-8}$) that could be observed via the $e-\gamma$ coincidence in~\cite{ale2} to confirm the experiment capability to be sensitive to $\mu \to e \gamma$ at the $10^{-8}$ level, the limit that was set at the time. 

In summary it is important to rely on Monte Carlo simulations when designing a rare-decay search experiment in order to define and optimize the detecting technique. Nevertheless the best way to study the possible backgrounds is to devise, already at the design stage, the possibility to measure it directly. If possible this has to be complemented with a reasonable number of calibration and monitoring devices, not excluding possibility to measure some existing rare process to validate the apparatus. 

\section{Review of the muon experiments}
\label{sec:exps}
\subsection{$\mu^+ \to e^+ \gamma$}
The signature of the $\mu^+ \to e^+ \gamma$ decay is a time coincident, back-to-back pair of a monoenergetic photon and a monoenergetic positron, both with an energy equal to half of the muon mass ($E_e = E_\gamma \approx 52.8$~MeV).
If one were able to measure the energies, relative time and relative angle with infinite precision this would be a bakground-free search. Finite experimental resolutions on the other hand imply that non \meg\ events can mimic its topology.

There are  two major backgrounds in this search:
one is a prompt background from radiative muon decay, \muenng, when the $e^{+}$
and the photon are emitted back-to-back with the two neutrinos carrying
away little energy. In this decay the two particles are emitted at the same time.
The other background is an accidental coincidence
of an $e^{+}$ in a normal muon decay, \muenn, accompanied by a high
energy photon. The sources of the latter might be either \muenng
decay, annihilation-in-flight or external bremsstrahlung of
$e^{+}$s from normal muon decay.

For both types of background it is possible to know the expected number of events in the signal region. 

\begin{figure}
\begin{center}
\begin{tabular}{cc}
\includegraphics[viewport= 40 210 520 625,width=0.48\columnwidth]{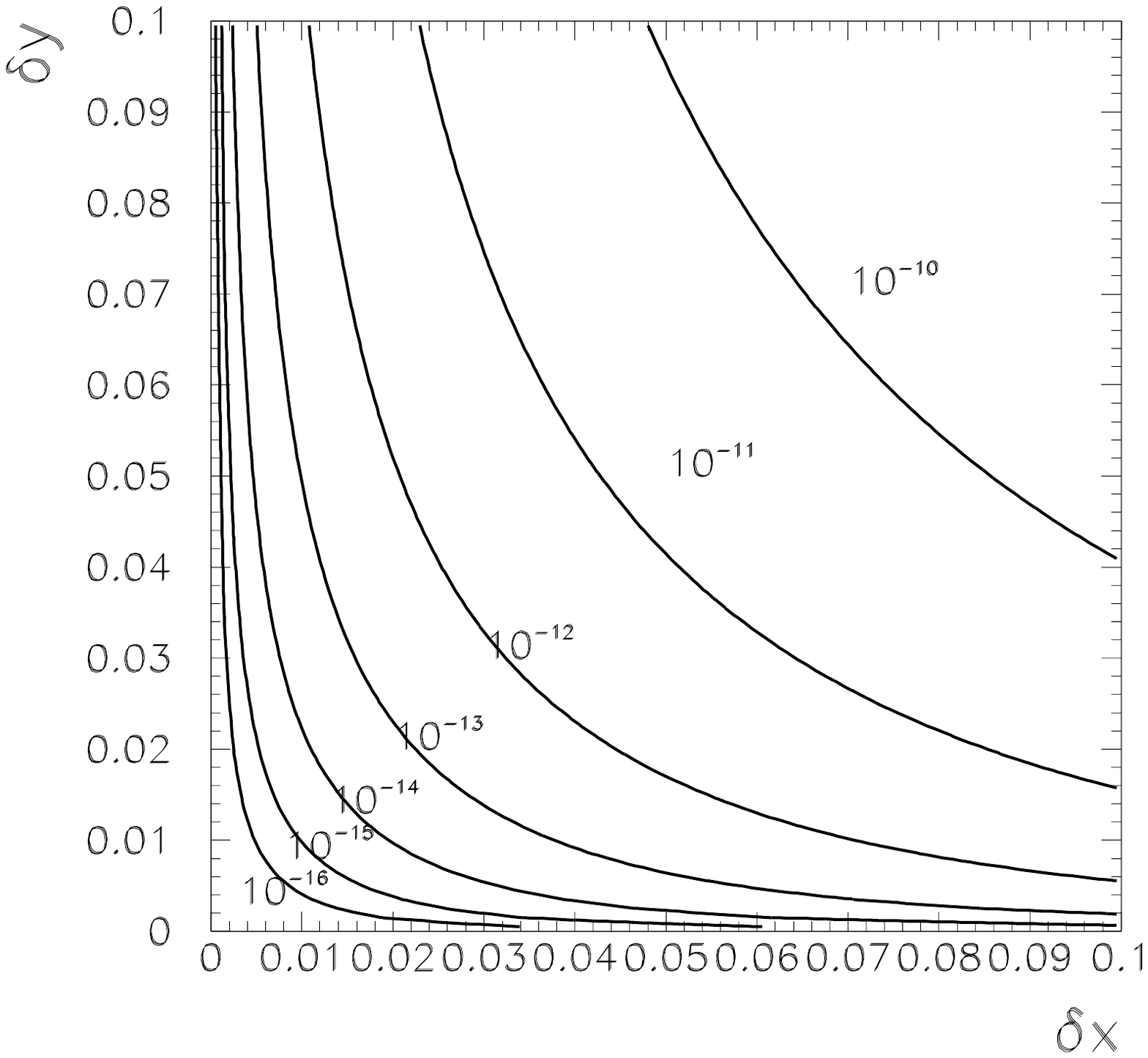}
&
\includegraphics[viewport= 40 210 520 625,width=0.48\columnwidth]{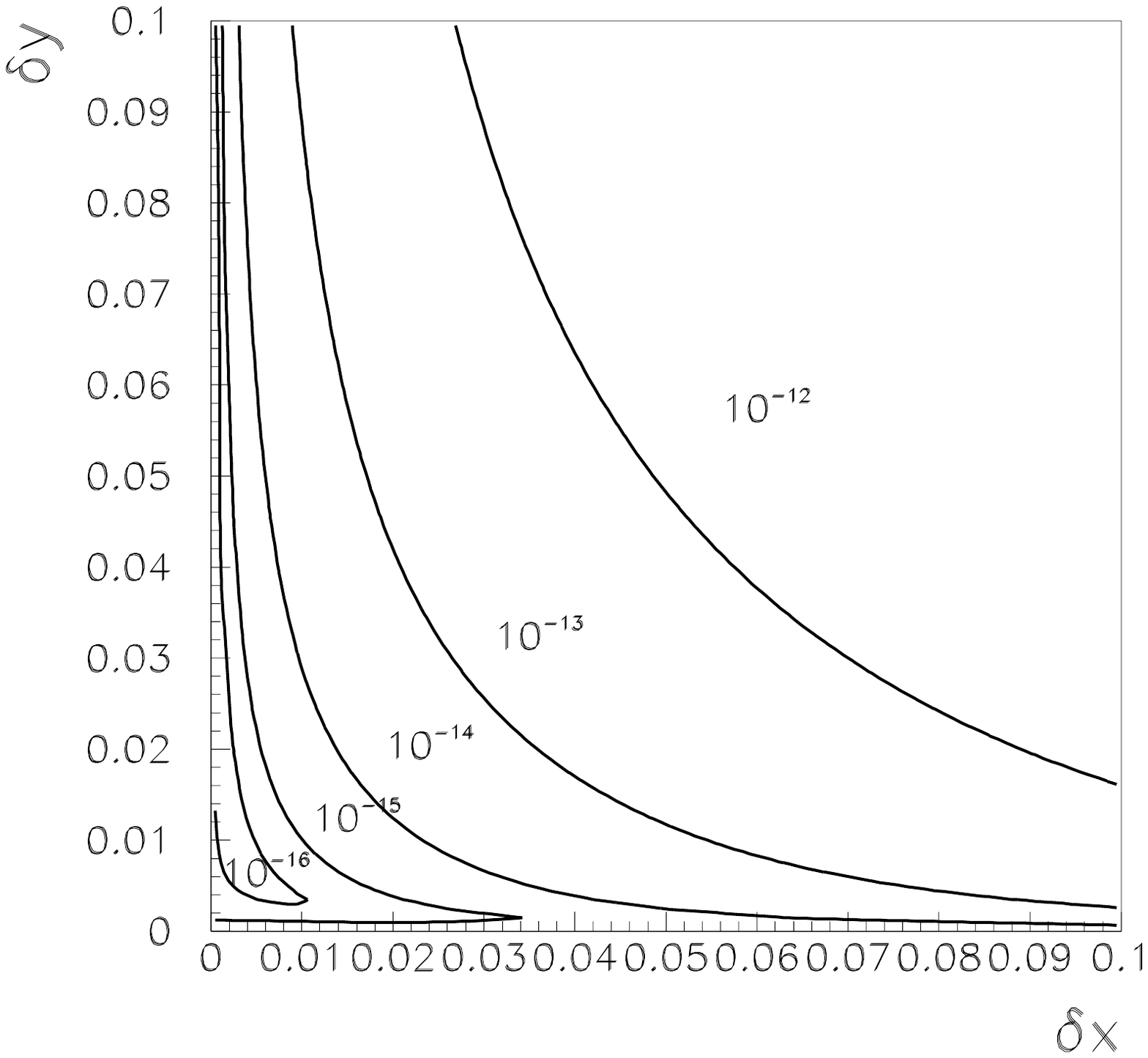}
\\
(a) & (b) \\
\end{tabular}
\caption{\label{fig:phisbg}Branching ratio of the background to the $\mu \to e \gamma$ process induced by muon radiative decays, as a function of the experimental energy resolution on positron ($\delta x$) and on photon ($\delta y$) energies. Figure $(a)$ shows the case in which the angular resolution is worse than the energetic ones, figure $(b)$ represents the reversed situation, which is the case of the MEG and MEG~II experiments. An angular resolution of 15~mrad is assumed in $(b)$.}
\end{center}
\end{figure}

The differential branching ratio of the radiative muon decay has been computed~\cite{kunopol} as a function of the energies and relative angle of the two daughter particles\footnote{Very recently a full computation appeared of the differential branching ratio of the muon radiative decay at next-to-leading order from which it emerges that corrections up to 20\% on the branching ratio are expected near the endpoint~\cite{mu3e-signer}.}. 
Given the detector resolutions, the sensitivity limitation from this
physics background can be estimated by integrating the differential
decay width over the kinematic signal box. The computation is given in~\cite{okadakuno}, and the exact analytical expression depends on which resolution is better between the angular or the energy one. The plot of the branching ratio is shown in figure~\ref{fig:phisbg}$(a)$ and $(b)$ which differ by the assumption on the relative magnitude of the energy and angular resolutions (see~\cite{kunopol, tesimia} for details). In either assumption an energy resolution for both positron and photon at the percent level reduces this background to below the $10^{-15}$ level. On the other hand the same plots show that the radiative decay represents a serious limit in searches below $10^{-15}$ since it is difficult to imagine apparatuses with an energy resolution much better than $1\%$.

Also the number of accidental background events is related to the experimental resolutions; furthermore, with a very high rate of incident muons, the accidental background
becomes more important than the radiative since the former, depending on the combined probability of having a high energy $\gamma$ and a high energy positron, grows
with the square of the muon rate. This is usually
the case for the present and future experiments: the rate of the
accidental background normalized to the total decay rate ($B_{\rm
acc}$) can be estimated by
\begin{equation}
B_{\rm acc} = R_{\mu} 
\cdot f^{0}_{e} \cdot f^{0}_{\gamma} 
\cdot (\Delta t_{e\gamma}) \cdot \left(\frac{\Delta\Omega_{e\gamma}}{4\pi}\right),
\end{equation}
where $R_{\mu}$ is the instantaneous muon intensity,
$f^{0}_{e}$ and $f^{0}_{\gamma}$ are the integrated fractions of the
spectrum of $e^{+}$ in the normal muon decay and that of photon (such
as from \muenng~decay) within the signal region, respectively\footnote{The linear dependence of the effective branching ratio on the muon rate implies a quadratic dependence on the {\em number} of random coincidences.}. 
$\Delta t_{e\gamma}$ and
$\Delta\Omega_{e\gamma}$ are, respectively, the full widths of the
signal regions for the time coincidence and angular constraint on the
back-to-back kinematics\footnote{
We denote here and in what follows $\Delta {\cal O}$ to be the width of the window centered on the expected value of the observable ${\cal O}$ with some large efficiency on the signal (say 90\%) as described in section~\ref{sec:window}; $\delta {\cal O}$, representing its half-width, is intended to play the role of the resolution, which is not assumed to be Gaussian in the discussion.}.
The accidental coincidences branching ratio $B_{acc}$ can be evaluated given the sizes of the signal region.
Let us call $\delta x$, $\delta y$, $\delta\theta_{e\gamma}$, and $\delta
t_{e\gamma}$  the half widths of the signal region for $e^{+}$ and $\gamma$
energies, relative angle $\theta_{e\gamma}$, and relative timing. $f_{e}^{0}$ can be estimated by
integrating the Michel spectrum of the normal muon decay over
$1-\delta x \leq x \leq 1$, yielding $f_{e}^{0}~\approx 2(\delta
x)$, since the Michel spectrum is constant at $x \approx 1$.
Given the angular resolution, $\delta\theta_{e\gamma}$, the
back-to-back resolution ($\Delta\Omega_{e\gamma}/4\pi$) is given
by ($\Delta\Omega_{e\gamma}/4\pi) = {(\delta\theta_{e\gamma})^2/4}$.
This descends from the fact that the positron line of flight defines a two dimensional acceptance region on the photon detector whose aperture is proportional to $\delta \theta$.

As for $f_{\gamma}^{0}$, if the radiative muon decay \muenng is
considered as a source of the 52.8 MeV photon, it can be given by
integrating 
the photon bremsstrahlung spectrum
within the width of the signal region
($1-\delta y\leq y \leq 1$). For unpolarized muons, it is given by~%
\cite{okadakuno}
\begin{equation}
f_{\gamma}^{0} = \int^{1}_{1-\delta y}{\rm d}y \int 
{\rm d}(cos\theta_{\gamma}) \frac{
{\rm d}B(\mu^{+}\rightarrow e^{+}\nu\overline{\nu}\gamma)}
{ {\rm d}y {\rm d}(\cos\theta_{\gamma})}
\approx \left(\frac{\alpha}{2\pi}\right) (\delta y)^2 
\Bigl[\ln(\delta y) + 7.33 \Bigr].
\label{eq:int50}
\end{equation}
Hence $f_{\gamma}^{0}$ for
\muenng~decay is roughly proportional to $(\delta y)^2$.

The other sources of high-energy $\gamma$-rays are annihilation-in-flight of
$e^{+}$s from Michel decays and external bremsstrahlung. The
contribution from positron annihilations in flight depends on the amount of 
material along the positron path.
Its contribution is generally smaller than that coming from 
radiative muon decays, and  becomes important only for extremely good photon 
energy resolutions. Furthermore, it depends on the
materials traversed by the positrons in a specific experimental setup.

From the above, the effective branching ratio of accidental background
is given by
\begin{equation}
B_{acc}=R_{\mu} \cdot (2\delta x) \cdot \left[ \frac{\alpha}{
2\pi}(\delta y)^2({\ln}(\delta y)+7.33) \right]
\times \left( \frac{\delta \theta^2}{ 4} \right) \cdot (2\delta t).
\label{eq:bacc}
\end{equation}
For instance, take some reference numbers such as a $e^{+}$
energy window of 1\%, a photon energy window of 6\%, 
$\delta\theta \sim 15$~mrad (see section~\ref{sec:positrons}), $\Delta
t_{e\gamma}$ = 500 psec, and $R_{\mu} = 10^{8}~\mu^{+}$/sec,
$B_{\rm acc}$ is $\approx 7 \times 10^{-13}$, comparable to the best presently achievable limits.

A few things are worth mentioning at this point based on Eq.~(\ref{eq:bacc}):
\begin{enumerate}
\item the accidental background is dominant at present day resolutions;
\item the rate of accidentals is proportional to the instantaneous muon rate, while the sensitivity is proportional to the total number of stopped muons. Hence a continous beam is preferred over a pulsed one to reject this type of background;
\item the number of accidental coincidences grows with the square of the beam rate, while both the sensitivity and the prompt background increase only linearly. This implies the existence of an optimal rate which maximizes the signal-to-background ratio for given experimental parameters;
\item designing an optimal \meg\ decay search experiment requires the minimization of Eq.~(\ref{eq:bacc}) where it is clear that improving the photon energy and position resolutions has a larger impact due to the quadratic dependence of the background on these windows.
\end{enumerate}

For illustrative purposes we show in Table~\ref{tab:megexps} the characteristics of some of the past experiments that searched for the \meg\ decay together with the limit they set on the branching ratio. This gives us the opportunity of discussing some
\begin{table}[t]
\begin{center}
\begin{tabular}{l c c c c c c c}
\hline
 Experiment  (Year) & Rate [Hz] &Duty f.& $\Delta E_e$&$\Delta E_\gamma $&$ \Delta t_{e\gamma}$&$ \Delta\Theta_{e\gamma}$&
Upper Limit \\
\hline
TRIUMF (1977)     \cite{ale2} &$2\times10^5$     &100\%    & 10.3\% &8.7\% & 6.7 ns      & 80 mrad     &$3.6\times10^{-9}$\\
SIN (1980)     \cite{ale1}    &$5\times10^5$     &100\%    & 8.7\%  &9.3\%    & 1.4 ns  &  --        &$1\times10^{-9}$\\
E328 (1982)    \cite{kinnison}    &$2.4\times10^6$      &6.4\%    & 8.8\%  &8\%    & 1.9 ns  & 37 mrad     &$1.7\times10^{-10}$\\
Crystal Box (1988) \cite{ale4}&    $4\times10^5$     &6.6\%    & 8\%    &8\%     & 1.8 ns  & 87 mrad     &$4.9\times10^{-11}$\\
MEGA (1999) \cite{MEGA2} &$2.5\times10^8$         &6.5\%    & 1.2\%  &4.5\%     & 1.6 ns  & 17 mrad     &$1.2\times10^{-11}$\\
MEG (2013)    \cite{MEGLimit}&    $3\times10^7$             &100\%    & 1.5\%  &4.7\%     & 0.28 ns & 30 mrad     &$4.2\times10^{-13}$\\
$\mathrm{MEG~II}$ (2016)\cite{MEG2}&    $7\times10^7$&100\%    & 0.6\% &2.3\%     & 0.19 ns & 20 mrad     &$5\times10^{-14}$*\\ \hline
\end{tabular}\caption{Historical progress of \meg\,\,searches. Resolutions are expressed as full widths at half maximum (FWHM). 
The duty factor represents the fraction of time in which the muon beam is available for the experiment, and equals $100\%$ for continuous beam, while it is lower for pulsed beams. *expected. \label{tab:megexps}}
\end{center}
\end{table}
%
$\mu \to e \gamma$ search experiments
with the aim of clarifying the previous points and understanding their evolution.

The first experiments used Geiger tubes~\cite{pontecorvo, sard-althaus}, freon bubble chambers~\cite{crittenden} or spark chambers~\cite{ale0,rey} to detect the presence of 
both the positron and the photon\footnote{In reality Sard and Althaus~\cite{sard-althaus} tried to detect just the delayed photon, in a more general search; see the interesting dicsussion in~\cite{pontecorvo}.}
later on the configuration with 
two back-to-back large NaI crystals,
\begin{figure}
\begin{center}
\includegraphics[width=0.95\columnwidth]{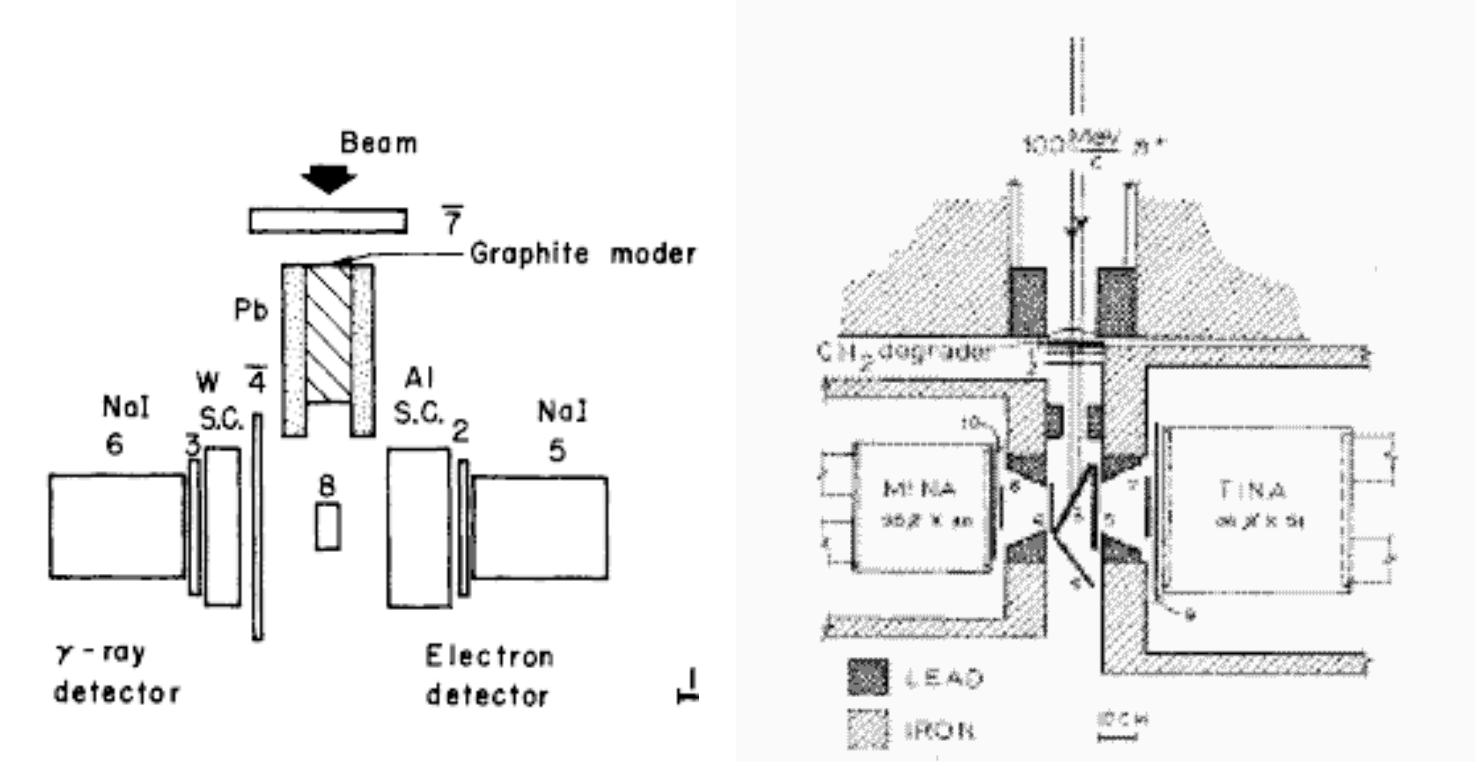}
\end{center}
\caption{Experimental configuration of the early 
\meg\ search experiment~\cite{ale0b}
and~\cite{ale2}.  \label{fig:nainai} In these experiment a pion beam is stopped on a target where it decays into muon and neutrino. The muon does not have enough energy to escape from the target and therefore decays at rest. Its decay products are observed in an almost symmetrical way by the two back-to-back crystals.}
\end{figure}
as in Figure~\ref{fig:nainai}, became common~\cite{ale2,ale1,ale0b,ale1b}
allowing for a better energy and timing determination. 
In these experiments the parent particle is stopped in a target and the detectors are placed such as to be sensitive to a back-to-back decay topology, standing at $90^\circ$ from the impinging beam to minimize beam related background. One or both crystal may be preceded by a thin detector sensitive to positrons.
\begin{figure}
\begin{center}
\includegraphics[width=0.99\textwidth]{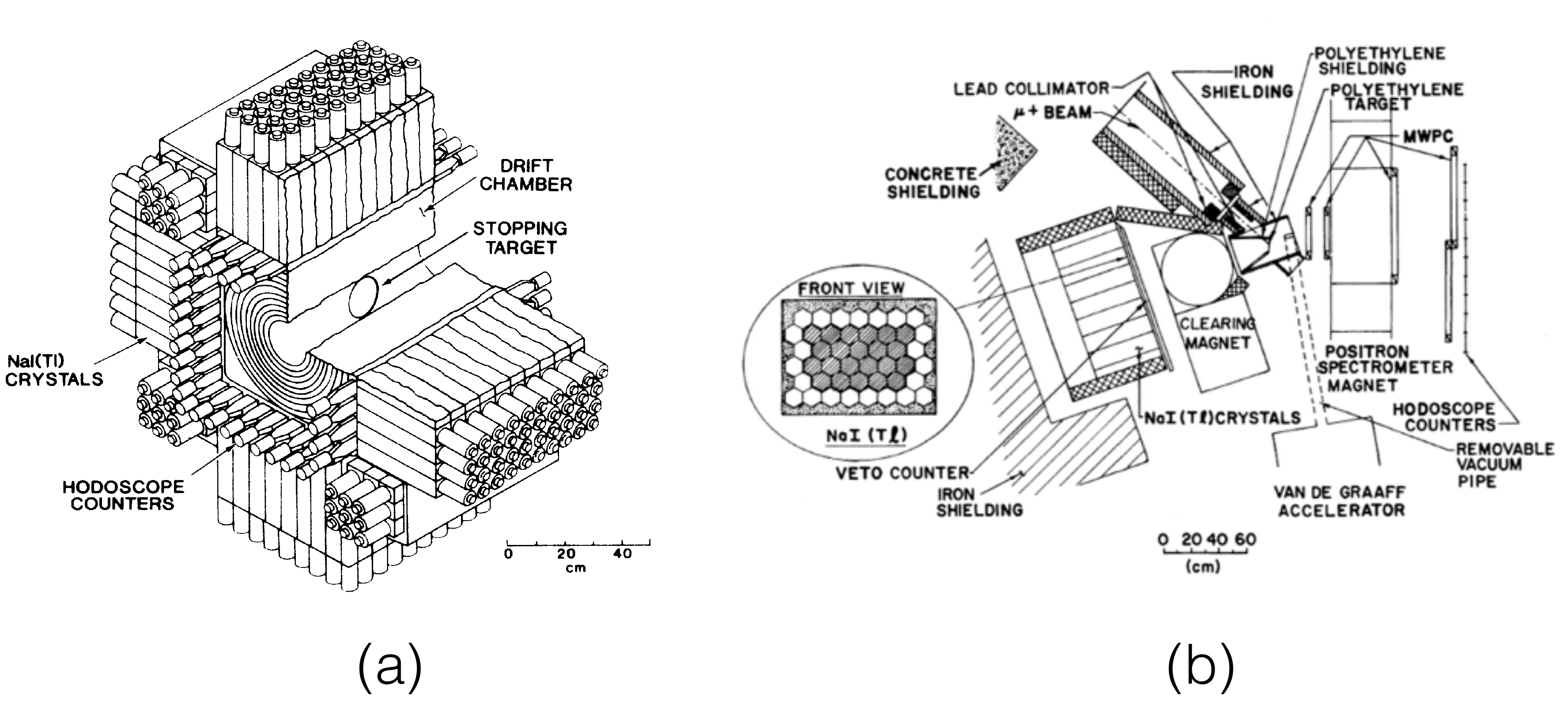}
\end{center}
\caption{(a) The Crystal Box experiment set limits on a host of CLFV decays, including \meg\ and $\mu \to eee$, thanks to its large acceptance and good energy resolution for both neutral and charged particles.\label{fig:crystalbox} (b) The experiment at LAMPF~\cite{ale3b} was the first to couple a dedicated proton accelerator to calibrate the detector.}
\end{figure}

Since the number of produced pions and muons increases with their energy, a larger particle flux requires a higher momentum for the 
particles, which in turn, imply a thick target to bring them to a stop.
The choice of a NaI(Tl) shower detector as a photon detector was dictated by the relatively short duration of the experiments (few weeks). The alternative pair spectrometer with superior energy and angular resolution had an insufficient detection probability. Besides there was also a limited importance in reaching an ultimate resolution on the positron side offered by a magnetic spectrometer:
an extremely good positron energy resolution was not necessary 
since the target thickness contributed heavily to 
the positron energy loss. In the experiment at SIN~\cite{ale1} the energy
loss in the target alone introduced a $4$~MeV uncertainty in the positron 
momentum.
The technique of a quasi symmetrical detection of gamma and positron 
was pursued to an extreme by the Crystal Box experiment where an almost 
$4\pi$ segmented NaI calorimeter surrounded the target region (see Figure~\ref{fig:crystalbox}~\cite{ale4}).

Thinner targets could be used after the introduction of surface muon beams whose lower energy allowed them to be stopped with less material.  

The use of a dipole spectrometer coupled to a hodoscope counter for 
the determination of positron momentum and timing was introduced in the \meg\
search experiment at LAMPF~\cite{ale3b} (see Figure~\ref{fig:crystalbox}b).

The MEGA experiment~\cite{MEGA2} used a surface muon beam at the stopped muon channel at LAMPF, that was brought to rest in a $76~\mu$m Mylar foil centered in the 1.5~T magnetic field of a superconducting solenoid. In comparison to the detectors used to set previous limits, the MEGA detector consisted of a magnetic spectrometer for the positron and three pair spectrometers for the photon, therefore sacrificing larger acceptance and efficiency for better resolution and background rejection. In total 1.2$\times 10^{14}$ muons were stopped during the life of the experiment and the overall efficiency for the observation of the \meg\ event was $\sim 3.9 \times 10^{-3}$. This low figure was partly due to the photon conversion probability ($2.5\%$) and partly to a reduced capability of reconstructing positron tracks in the solenoidal magnetic field compared to the design value. Notwithstanding the good limit posed on the branching ratio ($1.2 \times 10^{-11}$~\cite{MEGA1}) the sensitivity reached by the MEGA experiment was $35$ times worse than written in the proposal, demonstrating how hard is to make progress in this search. The source of many of the experiment shortcomings was analyzed in depth in~\cite{MEGA2}
and proved to be very useful in designing the next experiment: MEG.

\subsubsection{The MEG experiment}
\begin{figure}
\begin{center}
\includegraphics[width=0.99\columnwidth]{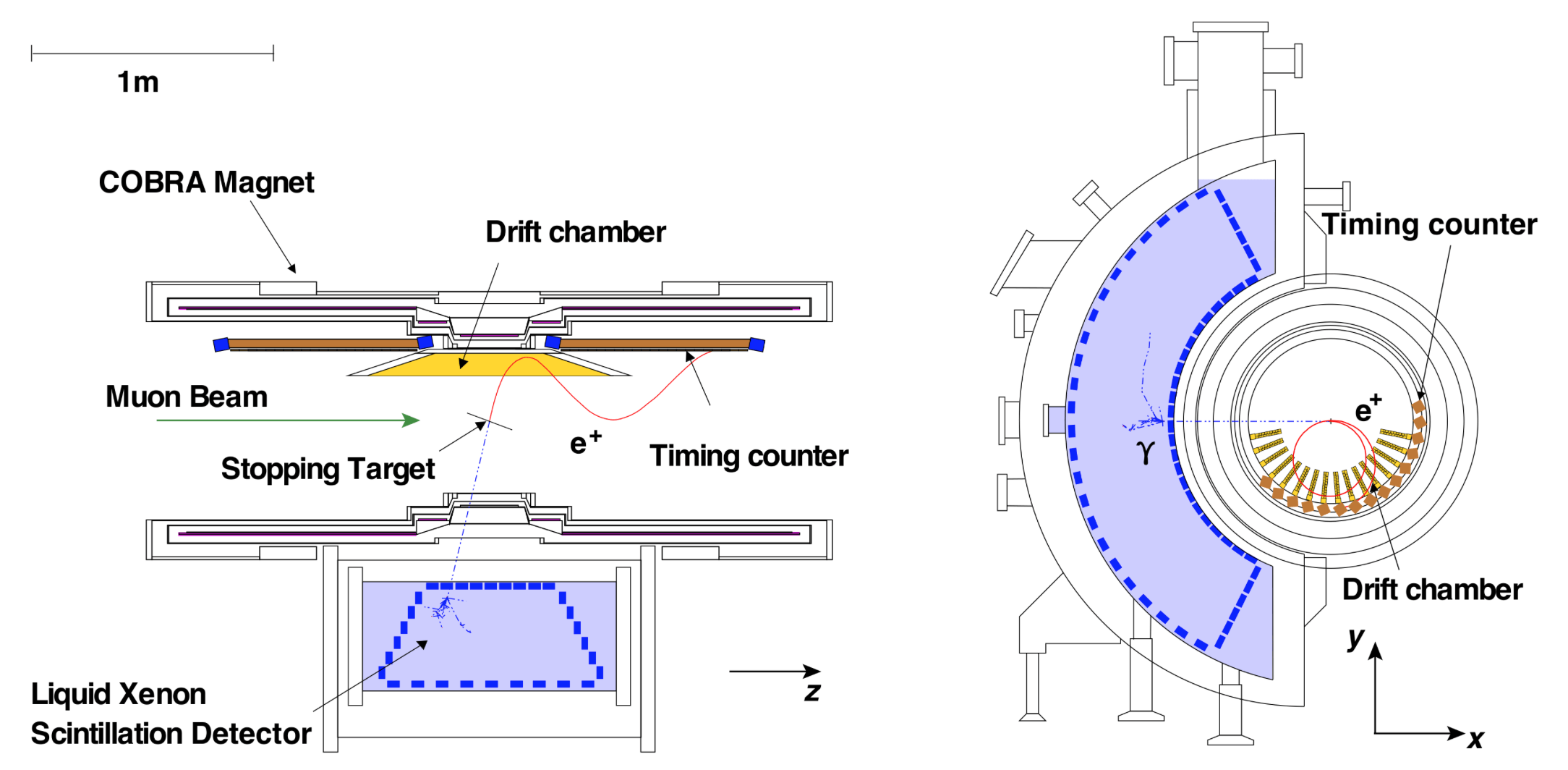}
\end{center}
\caption{\label{fig:meg}The MEG detector at PSI. A beam of surface muons is stopped on a thin target at the center of the COBRA magnet. The positron is measured by a tracking system followed by a timing counter, while the photon is detected in a liquid xenon volume.}
\end{figure}

In order to improve upon the previous experiment, MEG was designed around two concepts: an anti-bottle magnetic field and an innovative liquid xenon detector, for the positron and photon tracking respectively, as is depicted in Figure~\ref{fig:meg}.
A surface muon beam is moderated and then stopped on a thin polyethylene target at the center of the magnet. The positron momentum is measured by a set of drift chambers followed by  plastic scintillator timing counters, while the photon energy and direction are measured in a 800~litre liquid xenon volume viewed by more than 800 photo-multiplier tubes.

In a uniform magnetic field positrons follow helical trajectories whose radius is proportional to the transverse momentum $p_t = p \cos \lambda$ -- the momentum component which is orthogonal to the magnetic field, where $\lambda$ is the helix pitch angle, following the relation
\begin{equation}
p_t ({\rm MeV}) \approx 3 B ({\rm T}) \rho({\rm cm}) 
\end{equation} 
where $B$ is the magnitude of the magnetic field (measured in Tesla) and $\rho$ the radius of the circle in the orthogonal plane (measured in cm). For 52.8~MeV positrons in a uniform 1.4~T magnetic field this gives orbits which are about 25 cm in diameter for completely orthogonal positrons ($\lambda = 0$). The longitudinal speed is given by the parallel momentum $p_\parallel = p \sin \lambda$. Positrons emitted with a low pitch angle make many turns in a uniform magnetic field in few nanoseconds thereby passing many times in the tracker making it blind at a high particle rate. In MEG a non uniform magnetic field, decreasing symmetrically from the center towards the end of the magnet, pushes the positrons away from the center. Furthermore  the field shape is tuned such in a way as to have the helix radius proportional to the {\em absolute} momentum instead of the transverse one to introduce a well defined correlation between an easily measurable geometrical property (the particle's maximum distance from the beam axis) and the particle momentum.
\begin{figure}
\begin{center}
\includegraphics[width=0.99\columnwidth]{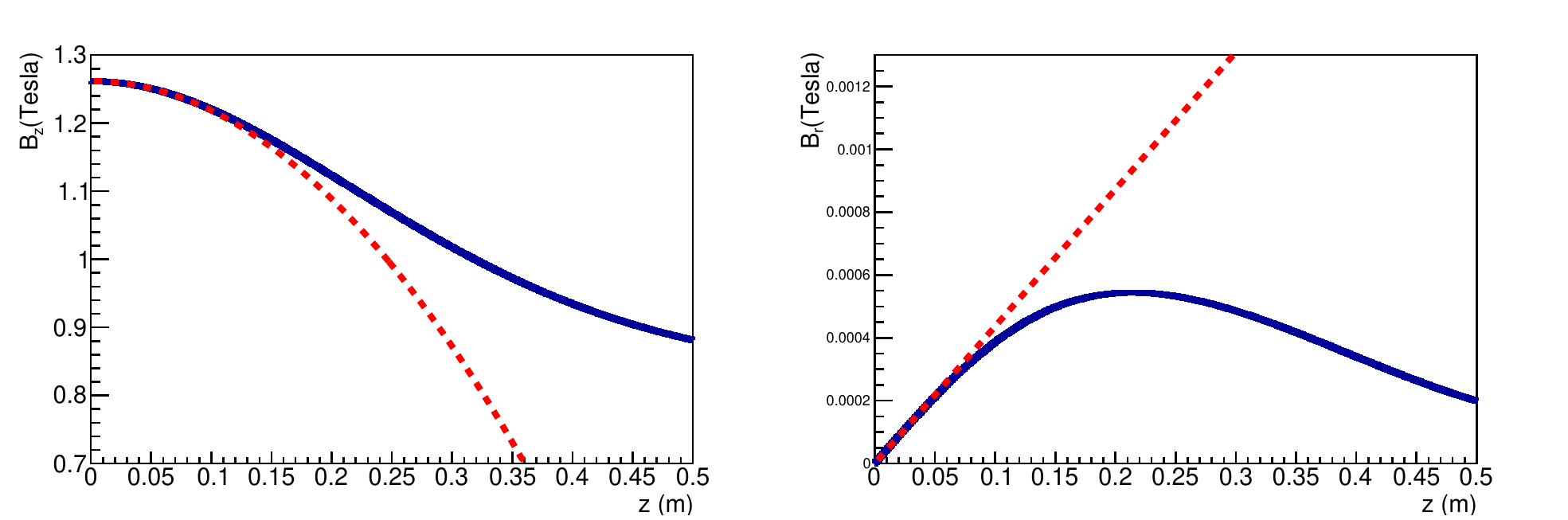}
\caption{\label{fig:cobrafield}Comparison between real (blue continuous line) and approximated according to equations~(\ref{eq:Bz}) and~(\ref{eq:Br}) (red dashed line) field generated by the MEG COBRA magnet to induce an almost constant bending radius. The radial field is computed close to the $z$-axis, since it vanishes for $z=0$. The approximation is good in the central part of the tracker ($|z| < 15 \div 20$~cm).}
\end{center}
\end{figure}
For a quasi-uniform magnetic field and for particles emitted almost at 90$^\circ$ the helix radius is given by 
\begin{equation}
\rho (z) \approx \frac{p \cos \lambda}{3 B(z)}
\end{equation}
where $z$ is the particle displacement along the axis ($z=0$ being the magnet center), which is in turn approximately given by $z \approx \rho(0) \sin \lambda$. It is clear that if the longitudinal component of the magnetic field decreases as 
\begin{equation} 
B_z \sim  B(0) \left( 1 - \frac{\lambda^2}{2} \right) = 
B(0) \left( 1 - \frac{z^2}{2 \rho(0)^2} \right)
\label{eq:Bz}
\end{equation}
the $\lambda$ dependence cancels out to fourth-order and  furthermore,
besides pushing the positrons out of the magnet, the radius of their trajectory depends on the absolute value of the momentum and not on its transverse component, which helps in defining a geometrical acceptance for tracks close to the Michel edge. It is in fact possible to discard low energy positrons by placing the detector in a cylindrical shell at sufficient distance from the magnet axis, as it is visible in Figure~\ref{fig:meg}.
For this reason the magnet of the MEG experiment is called ``COBRA" magnet from ``COnstant Bending RAdius'' magnet. To guarantee the null divergence of the field the radial component is given approximately by 
\begin{equation}
B_r (r, z) = B(0) \frac{r z}{2 \rho(0)^2}.
\label{eq:Br}
\end{equation}
The magnetic field generated by COBRA is somehow more complex than that given by Eq.~(\ref{eq:Bz}) and~(\ref{eq:Br}) but the discrepancy at large $z$ is needed to have a uniform measurement of the tracks in the timing counters.
In Figure~\ref{fig:cobrafield} we show the comparison between the longitudinal and radial component of the COBRA field (blue thick line) and that given by Eq.~(\ref{eq:Bz}) and~(\ref{eq:Br}) (red dashed line).

The photon energy, interaction point and timing are measured by a $\approx 800$~litre liquid xenon scintillating detector (the largest to date) whose single volume is read by more than 800 photo-multiplier tubes. Liquid xenon allows the detector to have a single homogeneous volume with respect to a crystal detector, which is composed of a stack of optically isolated modules, therefore minimizing the passive material that in the latter is unavoidable at the boundary of each crystal. Furthermore it has a high light yield (comparable to that on NaI) with a much shorter decay time of the scintillation light (due to its peculiar emission mechanism, xenon scintillation light is emitted in three distinct processes, the slower having a lifetime of $\tau \approx 45~ns$)~\cite{testfacility, xenon1, xenonieee} leading to a timing resolution on the photon interaction time below 100~ps. Its homogeneous volume allows a peculiar calibration technique: a set of $\alpha$ sources suspended inside the volume by means of thin wires~\cite{alphasources} make it possible to monitor in real time the transparency and the optical properties of the detector.

\begin{figure}
\begin{center}
\includegraphics[width=0.5\columnwidth]{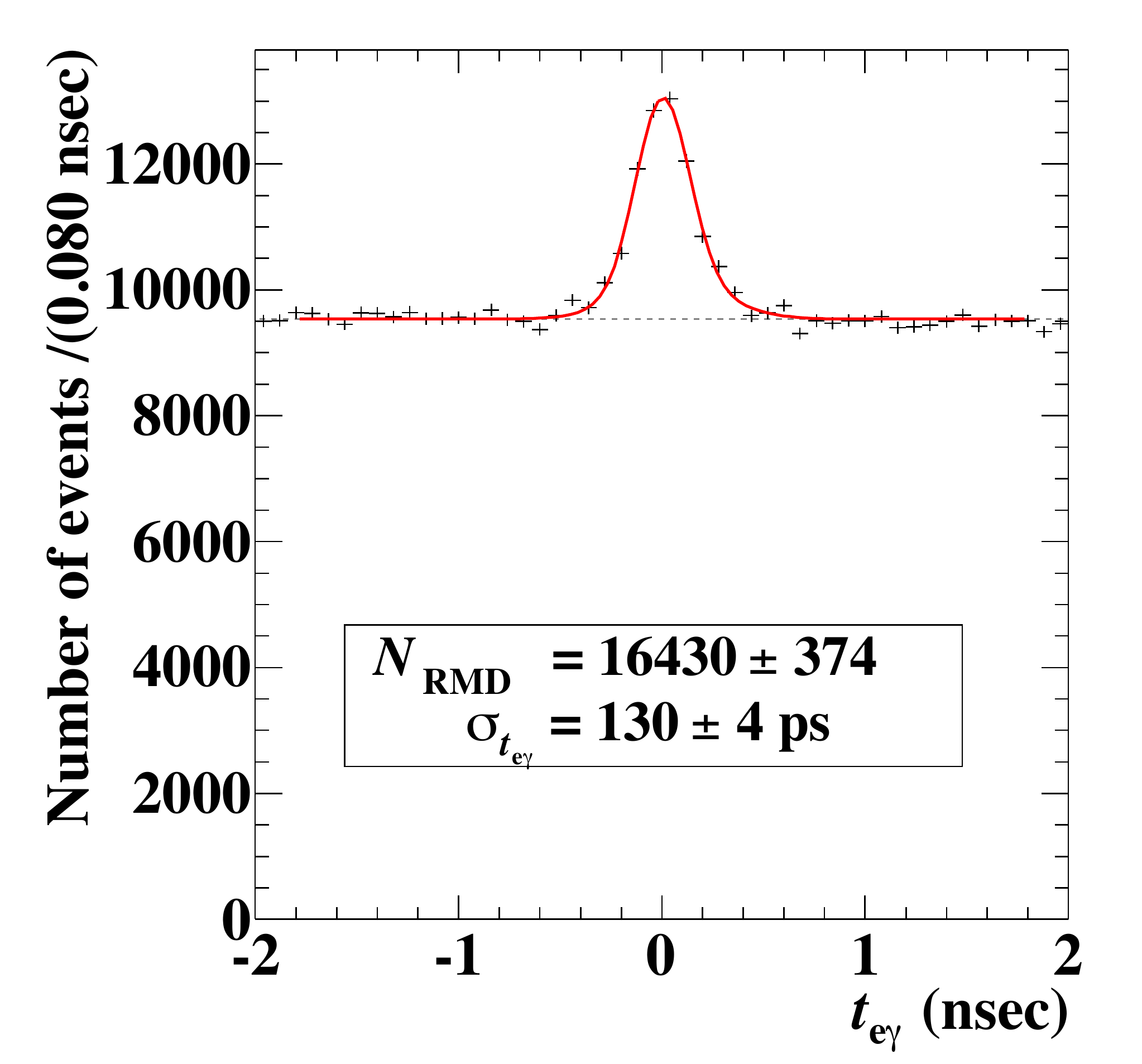}
\end{center}
\caption{\label{fig:megtegamma}Distribution of the relative time between a candidate positron and a candidate gamma ray measured by the MEG experiment. The peak due to radiative decays is visible above the flat background given by random coincidences.}
\end{figure}

MEG took data in the years 2008-2013 totaling $7.5 \times 10^{14}$ muons stopped on target and measuring the decay products with the resolutions reported in table~\ref{tab:megexps}, where we see that besides having optimum energy and position resolutions, the greatest improvement with respect to previous searches was its capability of precisely defining the $e\gamma$ timing coincidence (see also Figure~\ref{fig:megtegamma}). 

A \meg\ candidate event $x_i$ consists in the set of variables describing its kinematics: positron and photon energies, relative time and relative (polar and azimuthal) angles:
\begin{equation}
x_i = \left\{  E_e, E_\gamma, t_{e\gamma}, \theta_{e\gamma}, \phi_{e\gamma} \right\}.
\end{equation}

Since the main background comes from accidental coincidences, a blind maximum likelihood analysis was adopted~\cite{MEGdetector,MEGLimit}, meaning that events that are in a window in the $(E_\gamma, t_{e\gamma})$ plane too close to the signal are put in a separate data stream which is not accessible by the collaboration. In this way it is possible to estimate the number of random coincidences in the signal region by looking at the so-called ``sidebands'' where positron and photon spectra are the same, but there is no time coincidence required. The number of radiative muon decay events in the signal region is on the other hand estimated by extrapolating the number of RMD events at lower energy (used also to define the $t_{e\gamma} = 0$ for the signal) and using the shape of the expected spectrum. In Figure~\ref{fig:megtegamma} the distribution of the relative time between the positron and the photon $t_{e\gamma}$ is shown for events in the analysis region: the constant plateau is given by random coincidences, while RMD pairs show up in the Gaussian peak, which gives a measure of the timing resolution of the experiment. 
After tuning the analysis algorithm the blind box is opened and the number of signal events is extracted by a maximum likelihood fit which takes into account the experimental conditions of different years as well as the existing correlations between variable measurements: the observed spectra in the five variables are fitted to extract the contribution of accidental coincidences, radiative muon decays and \meg\ decays
\begin{figure}
\begin{center}
\includegraphics[width=0.98\columnwidth]{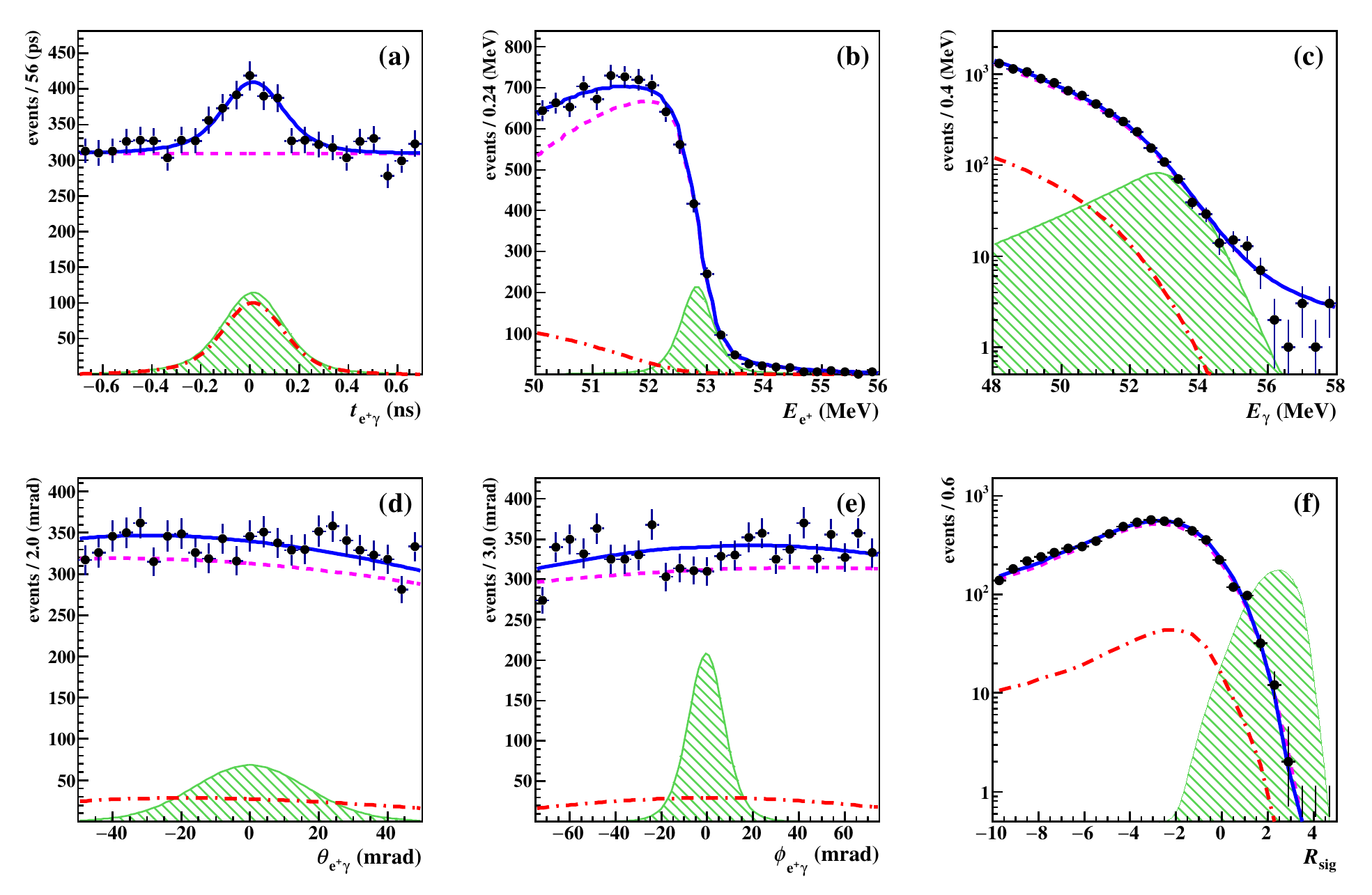}
\end{center}
\caption{The projections of the best fitted likelihood function to the five main 
   observables and $R_{\rm sig}$ together with the data spectra for the full dataset of the MEG experiment.
The magenta dash and red dot-dash lines are individual components of 
the fitted PDFs of accidental and RMD, respectively. The blue solid line is the sum 
of the best fitted PDFs. The green hatched histograms show the signal PDFs 
corresponding to $\sim 500$ events (from~\cite{MEGLimit}).\label{fig:muegamma-fit}}
\end{figure}
as visible from Figure~\ref{fig:muegamma-fit}$a-e$. In the panel~\ref{fig:muegamma-fit}$f$ the kinematic information is condensed in a single variable measuring the likelihood of the event being signal with respect to being background (relative signal likelihood $R_{\rm sig}$) defined as
\begin{equation}
R_{\rm sig} = \log_{10} \left( \frac{S({x}_i)}{f_\mathrm{R}R({x}_i)+f_\mathrm{A}A({x}_i)} \right),
\end{equation}
where $f_\mathrm{R}$ and $f_\mathrm{A}$ are 
the expected fractions of the RMD and accidental background events.

\begin{figure}
\begin{center}
\includegraphics[width=0.99\columnwidth]{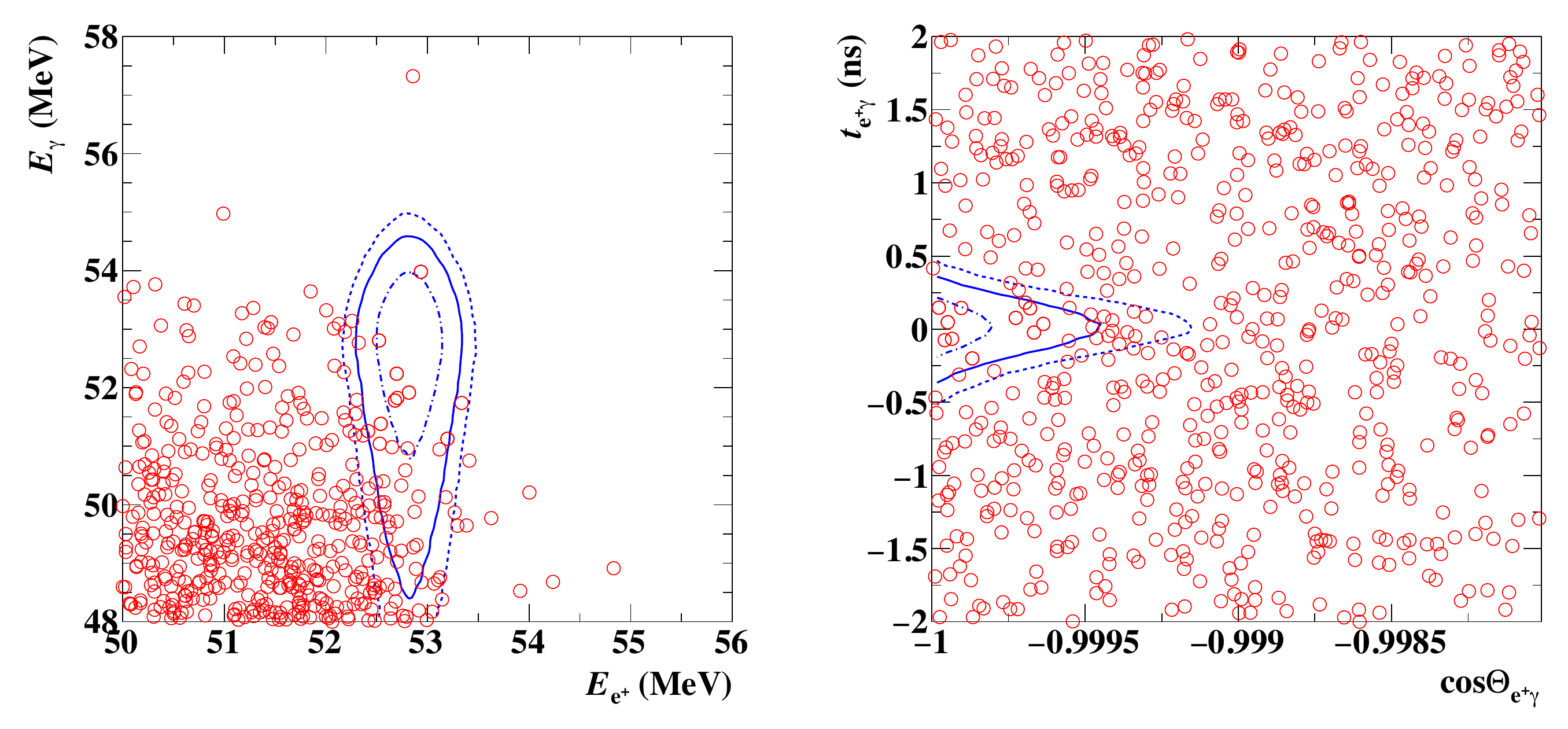}
\caption{\label{fig:meg2Dbox} Distribution of events in the full dataset of the MEG experiment in the ($E_\gamma, E_{e^+}$) and in the  $(\cos \Theta_{e\gamma}, t_{e\gamma})$ plane ($\Theta_{e\gamma}$ is the  angle between the two particles' momenta). The region where the signal should be present are also drawn (at 1, 1.64 and 2$\sigma$). No excess event is seen. Figure from~\cite{MEGLimit}. }
\end{center}
\end{figure}

No excess of events was found and the upper limit on the \meg\ decay was set:
\begin{equation}
{\rm BR}(\meg) < 4.2 \times 10^{-13},
\end{equation}
at 90\% confidence level.  In Figure~\ref{fig:meg2Dbox} the projections of the distribution of the events in the signal box on the $(E_\gamma, E_e)$ and $(t_{e\gamma}, \cos \Theta_{e\gamma})$ planes are shown.

This number represents the most stringent limit on every CLFV decay to date, as visible from Figure~\ref{fig:calando}. The four rightmost red circles there represent the successive limits posed by the MEG experiment, with increasing statistics. It is clear that MEG has reached its final sensitivity and further data taking cannot push the limit much further, as the experiment is becoming dominated by the accidental background. For this reason the collaboration has proposed an upgrade, called MEG~II~\cite{MEG2, MEG2bis}, which reutilizes some of the MEG apparatus and modifies or replaces the sub-detectors to improve the experimental resolutions. In this way it is possible to improve the limit by a further order of magnitude. 

\subsubsection{MEG~II}
\begin{figure}
\begin{center}
\includegraphics[width=0.95\columnwidth]{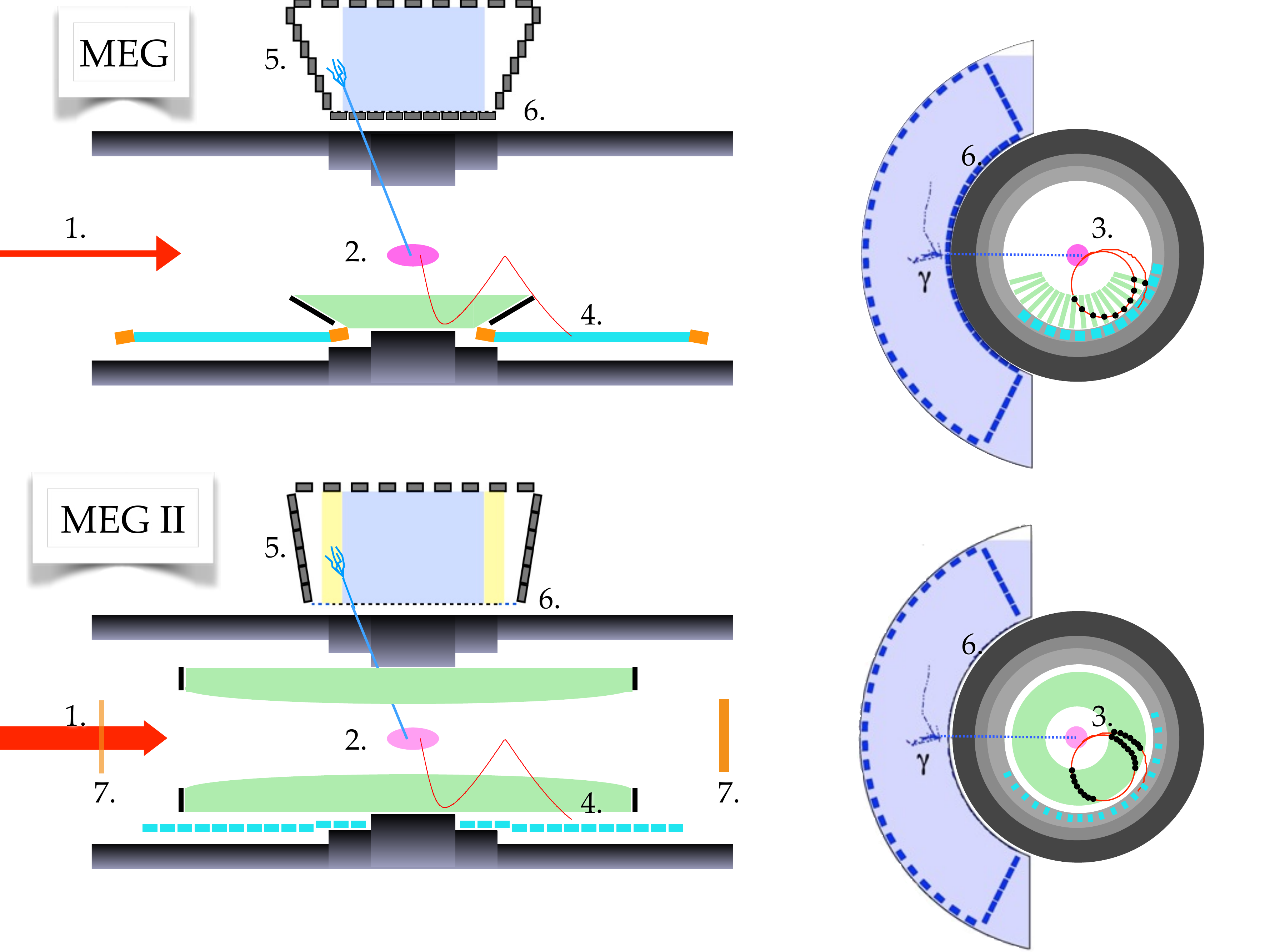}
\end{center}
\caption{\label{fig:MEG2detector}Comparison between the MEG and MEG~II detector. The latter has been improved on several aspects (acceptance, segmentation, \ldots) in order to cope with an increased muon flux (see text for details of the numbers).}
\end{figure}

The major modifications of the apparatus are visible in Figure~\ref{fig:MEG2detector} and can be summarized in:
\begin{enumerate}
\item a larger muon flux ($7 \times 10^7~\mu^+$/s) can be used once the resolutions are improved;
\item a thinner target (140~$\mu$m instead of $205~\mu$m) with a lower contribution to multiple scattering and bremsstrahlung probability on the positron trajectory. The same stopping power will be obtained by slanting the target at a larger angle;
\item a new positron tracker, with higher granularity and transparency, consisting of a cylindrical drift chamber with $2\pi$ coverage in azimuthal angle;
\item a better coupling between the drift chamber and the timing counter, which will be segmented in small tiles in order to withstand the higher rate of decay positrons. Tracking the positron all the way to the timing counter will allow to improve the resolution on the relative timing that was dominated, in MEG, by the uncertainty in the track extrapolation from the drift chambers to the plastic scintillation counters;
\item An increased acceptance of the liquid xenon detector ad the edges with a different distribution of the photo-multiplier tubes on the lateral faces, together with 
\item the complete replacement of the photo-sensors on the gamma entrance face. Smaller silicon photo-detectors (MPPCs) will be used that will help both in having a finer granularity to better reconstruct the photon interaction position energy and timing, and in reducing the traversed material so as to increase the detection efficiency;
\item One or two ``radiative muon detector counters'' (one downstream plus one upstream, the latter still in an evaluation phase due to possible background production) to detect low energy positrons spiralling close to the magnet axis to signal possible coincidences coming from muon radiative decays; 
\item A faster integrated electronics to be able to handle the increased data flux;
\end{enumerate}
The MEG~II detector is in advanced stage of construction and an engineering run is expected at the end of 2017, while the first data taking will be in 2018. The expectation is to reach a sensitivity of $5 \times 10^{-14}$ at 90\% C.L.~with three years of data taking. The expected resolutions for the upgrade are summarized in the last line of Table~\ref{tab:megexps}.

\subsection{$\mu^+ \to e^+e^-e^+$}
In the $\mu^+ \to e^+e^-e^+$ decay one searches for two positrons and one electron coming from a common vertex and with a total energy equal to the muon mass $E_{\rm tot} \approx 105.6$~MeV. Being a three-body decay the daughter particles are emitted in a common plane. Simple relativistic kinematics teaches us that each particle has a maximum momentum of about $m_\mu/2$ and that the decay can be described by two independent variables. 

The energy distribution of each daughter particle depends on the exact dynamics of the underlying unknown physics: in~\cite{okadakuno} it is discussed how the energy is shared among the three ``electrons" assuming different effective operators. 
In general the highest energy particle has a momentum larger than 35~MeV, while the distribution of the lowest energy particle peaks near zero and decreases quickly as its energy tends to its upper limit so that only about one half have an energy larger than 15~MeV~\cite{mu3e1}\footnote{In the simplest case of dipole interaction  the $\mu^+ \to e^+ e^- e^+$ decay is dominated by the diagram in which the materialization of the $e^+ e^-$ pair comes from a bremsstrahlung photon (Dalitz pair), thus exhibiting a sharply decreasing distribution, peaked at low energy.}. 
A $\mu^+ \to e^+e^-e^+$ search experiment must have an excellent tracker as thin as possible since in order to achieve the high acceptance the detector must be able to reconstruct tracks with momenta ranging from a few MeVs up to half of the muon mass.
As a consequence, unlike \meg\ and $\mu \to e$ conversion, to get a limit on the $\mu \to eee$ decay one has to make some assumption on the unknown operator: since the detector is sensitive to particles above a defined momentum, it is necessary to know which is the probability of having all three particles above that momentum threshold, and this depends on the matrix element.

Furthermore unlike the case of the \meg\ search there is no mono-energetic particle in the final state, but the backgrounds are very similar. There is a prompt background due to the allowed muon decay $\mu^+ \to e^+e^-e^+ \bar \nu_\mu \nu_e$ -- whose branching ratio is $(3.4 \pm 0.4) \times 10^{-5}$~\cite{sindrum} -- which becomes serious when the two neutrinos have very little energy. 
\begin{figure}
\begin{center}
\begin{tabular}{cc}
\includegraphics[width=0.48\columnwidth]{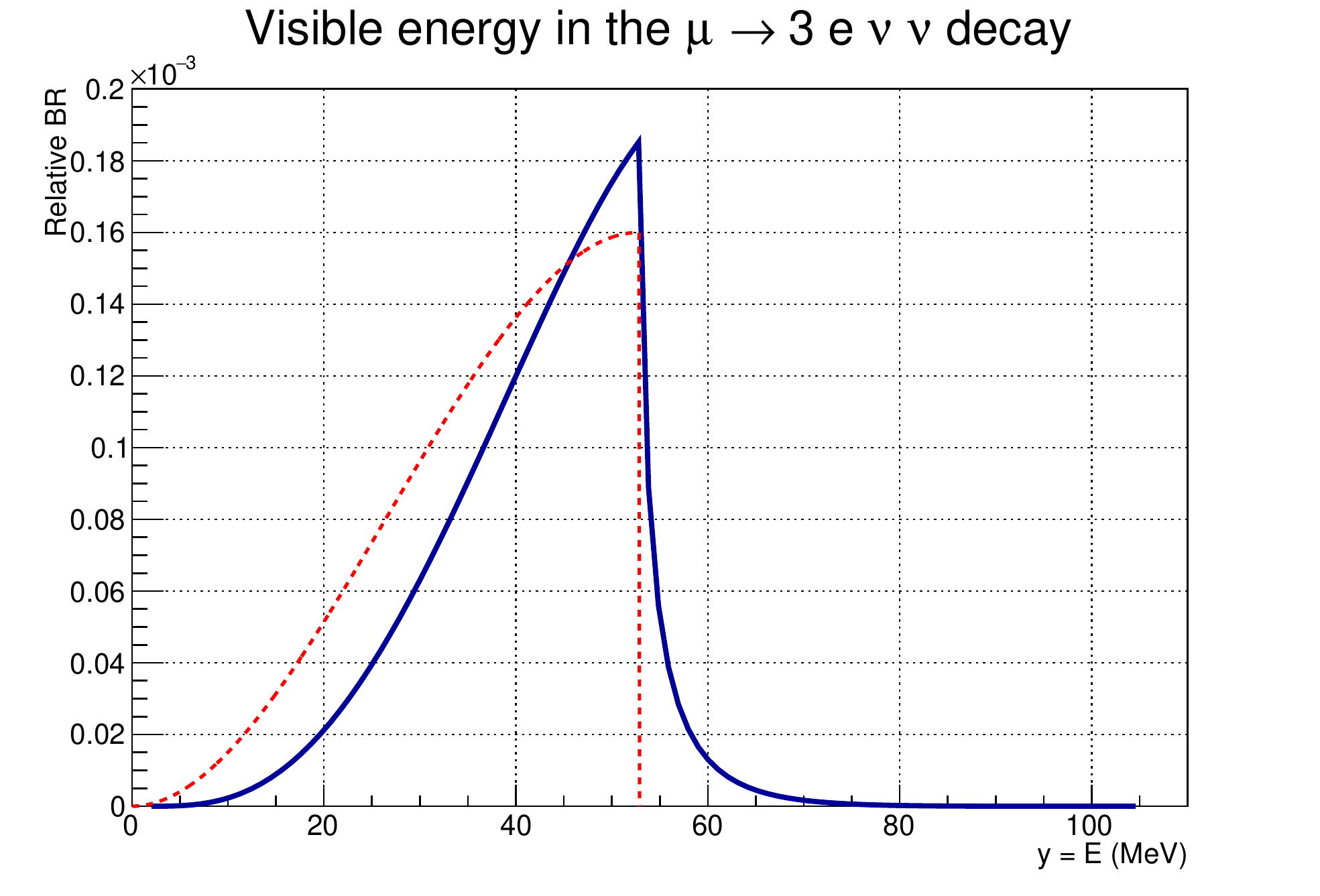} &
\includegraphics[width=0.48\columnwidth]{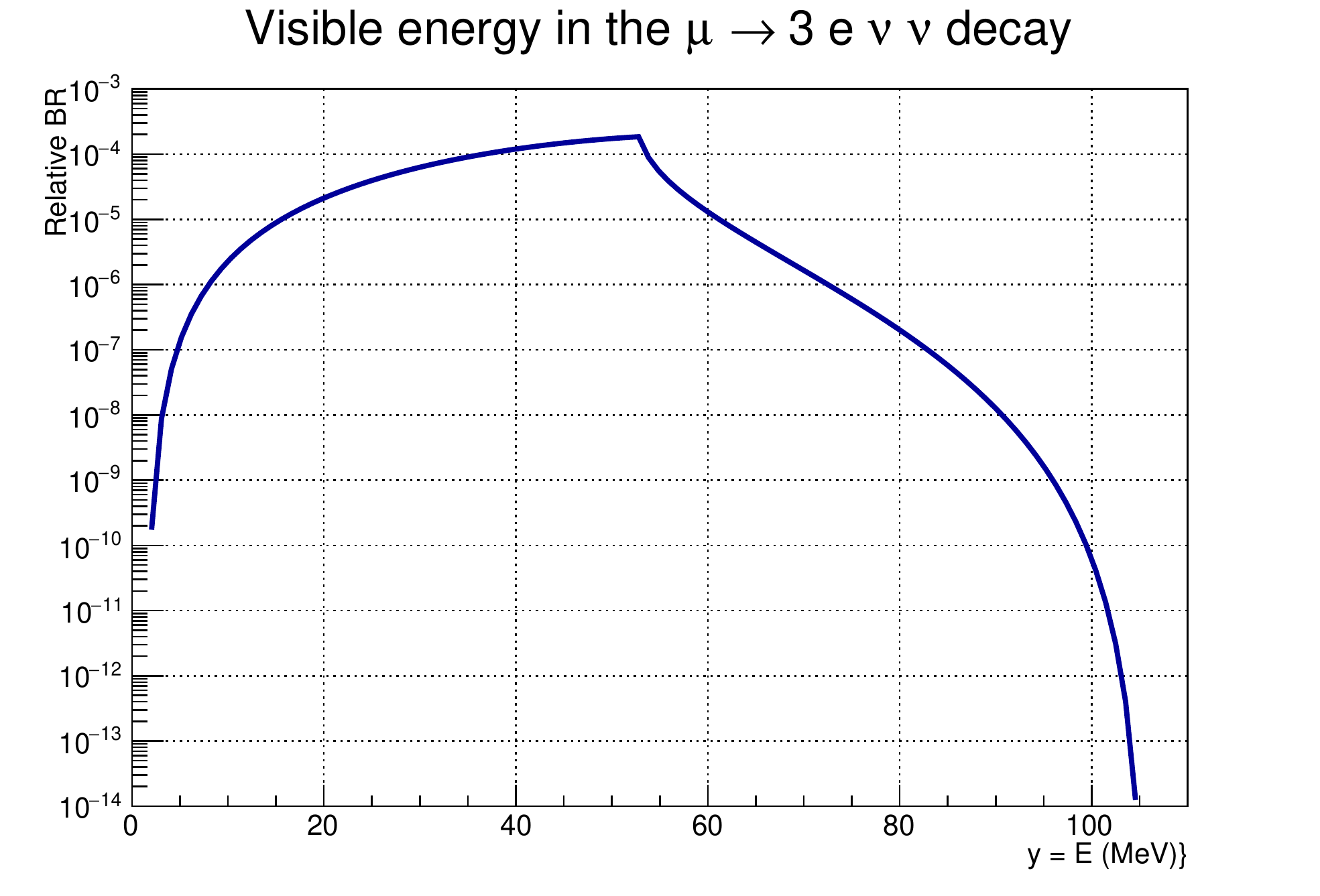} \\
$(a)$ & $(b)$ 
\end{tabular}
\caption{Distribution of the total energy of the charged leptons for the decay $\mu^+ \to e^+e^-e^+ \bar \nu_\mu \nu_e$. In the plot $(a)$ the Michel spectrum is superimposed for comparison. (Thanks to Matteo Fael and Rostislav Konoplich for providing the data used to produce the pictures).\label{fig:matteofael}}.
\end{center}
\end{figure}
The total energy of the three particles closely resembles the Michel spectrum but extends up to $m_\mu$ with a long tail~\cite{djilkibaev}, as visible in Figure~\ref{fig:matteofael}. The similarity to the Michel spectrum is due to the fact that the photon emission from the electron line (photon that materializes in the $e^+ e^-$ pair) dominates over that from the muon line, and the three leptons behave roughly as the single electron of the ordinary muon decay.  In contrast the spectra of individual lepton energies are quite different from the normal spectrum~\cite{fishbane}.

The integral rate of this background as a function of the energy cut has been recently computed by several authors~\cite{mu3e-fael, mu3e-signer} up to the next-to-leading order. In Figure~\ref{fig:djilkibaev} (from~\cite{mu3e-fael}) the integral branching ratio is plotted as a function of the ``invisible energy'' -- that carried away by the two neutrinos. It can be parametrized by~\cite{mu3e-fael}:
\begin{figure}
\begin{center}
\includegraphics[width=0.55\columnwidth]{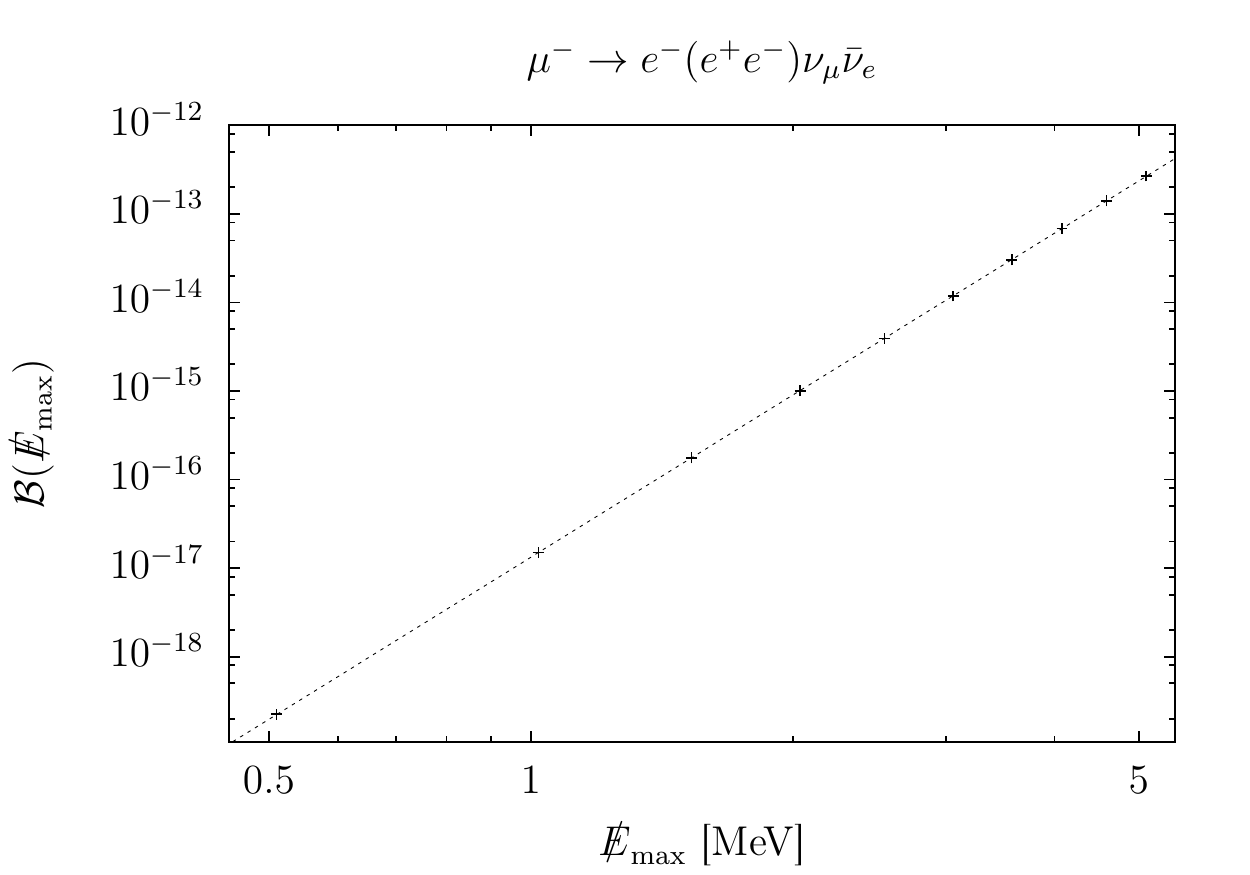} 
\caption{The branching ratios at NLO, as a function of the invisible energy, {\em i.e.} that carried away from neutrinos, denoted $\not \!\!E_{\rm max}= 1,2,\dots,10 \, m_e$, fitted with the ansatz~(\ref{eqn:fit2}).\label{fig:djilkibaev} From~\cite{mu3e-fael}}.
\end{center}
\end{figure}
\begin{equation}
\label{eqn:fit2}
{\rm BR}(E_{\rm cut}) \approx 2.22 \times 10^{-19} \times 
\left( \frac{E_{\rm cut}}{m_e}\right)^{6.08}.
\end{equation}
where the sixth power stems from the fact that near the endpoint the differential branching ratio behaves like $\left( E - m_\mu \right)^5$.
From an experimental point of view this shows that if one could reach an experimental resolution of $\approx 1$~MeV on the sum of the three electron's energies the prompt background would be reduced below the $10^{-16}$ level.

The other background comes from an accidental 
overlay of two or three uncorrelated muon decays. This background strongly depends on the muon rate and can be faked either by one Michel decay superimposed to an accidental $e^+ e^-$ pair (1~MD) or by two Michel decays with a single $e^-$ (2~MD). 
$e^+ e^-$ pairs come from  Bhabha scattering of a positron onto an electron in the detector material or from a photon conversion in the target. Single $e^-$ tracks can be produced by Compton scattering of photons, mis-reconstructed $e^+$ tracks or $e^+ e^-$ pairs in which the positron is not reconstructed. 
The accidental background is therefore quadratic or cubic in the muon rate  while depends to the first or the second power of the acceptance cuts, for the 1~MD and 2~MD cases respectively.
In fact the three particles are neither all simultaneous nor come from the same vertex, hence accidental background can be suppressed with precise measurement of the vertex and timing in addition to the precise momentum reconstruction to ensure that the three particles are on a common plane. As in the case of the \meg\ search the signal-to-background is maximized by reverting to a continuous muon beam.
\begin{figure}
\begin{center}
\includegraphics[width=\columnwidth]{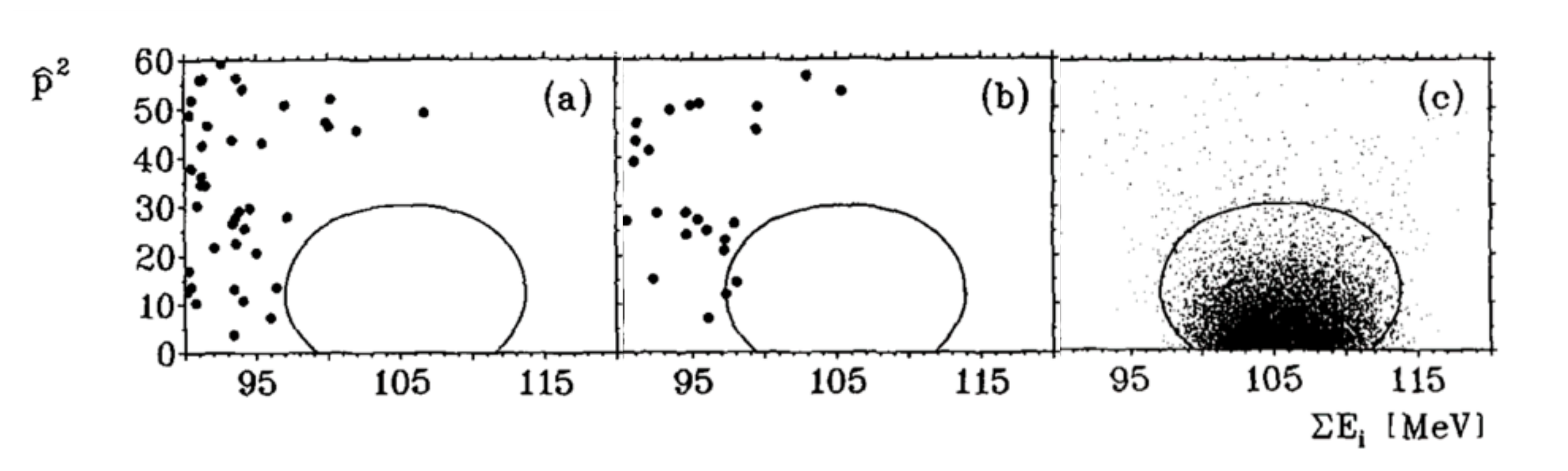}
\caption{\label{fig:mu3e-sindrum}Distribution of $\hat p^2 = (p_\perp / \sigma_{p_\perp})^2 + (p_{||}/\sigma_{p_{||}})^2$  versus  $\sum_{i=1}^{3} E_i$  (planarity of the event vs how much the energy resembles the muon mass) for $\mu \to 3e$ candidate events for the SINDRUM experiment near the signal region for (a)  prompt events, (b) accidental coincidences and (c) Monte Carlo generated $\mu \to 3e$ events. The contour defines a region containin g 95\% of signal events. Adapted from~\cite{sindrum}.}
\end{center}
\end{figure}

The present best limit on the $\mu^+ \to e^+e^-e^+$ process is BR$(\mu^+ \to e^+e^-e^+) < 1.0 \times 10^{-12}$ at 90\% confidence level assuming a constant matrix element\footnote{Recall that the limit depends on the assumption on how the energy is shared among the three electrons -- in particular on the spectrum of the lowest energy particle -- hence on the matrix element.}. It is the most dated limit in Figure~\ref{fig:calando} being set in 1988 by the SINDRUM experiment at PSI~\cite{sindrum} which consisted in five concentric multi-wire proportional chambers surrounded by an array of plastic scintillator counters inside a solenoidal magnetic field. Another experiment was performed later at the Joint Institute for Nuclear Research in Dubna~\cite{baranov} using also concentric multi-wire proportional chambers surrounding a stopped pion beam, but produced a worse limit ($3.6 \times 10^{-11}$) yet comparable to the limit set by the Crystal Box~\cite{ale4} experiment which was, as already mentioned, the almost 4$\pi$ sodium-iodide scintillation detector with no magnetic field capable of setting with a single experiment a lot of limits on rare muon decays.

SINDRUM resolutions were $\sigma_{p_{||}} = 1.4$~MeV/c, $\sigma_{p_\perp} = 0.7$~MeV/c, ($p_{||}$ and $p_\perp$ being the components of $\sum \vec p_i$ in and perpendicular to the decay plane respectively), $\sigma_\theta = 28$~mrad, $\sigma_{\Delta t} < 1$~ns and the vertex resolution was $\sim 1$~mm. They collected about 16\:000 events, about half of which were accidental coincidences. The distribution of events near the signal region is shown in Figure~\ref{fig:mu3e-sindrum} where no candidate event was observed, yielding to the quoted limit of $< 10^{-12}$. 

A new measurement should be sensitive to the $10^{-16}$ level to be competitive with the \meg\ search even if the only source of CLFV is given by the dipole operator (see Eq.~(\ref{eq:m3e}) and Figure~\ref{fig:spiderweb}), therefore an improvement of a factor of $10^4$ in sensitivity is needed. To keep the duration of the experiment within few years (some $3\times 10^7$~sec) a beam intensity of the order of at least $10^9~\mu^+/$sec is mandatory.

\subsubsection{The Mu3e experiment}
The Mu3e experiment to be realized at the Paul Scherrer Institut in Switzerland~\cite{mu3e1, mu3e2} aims at reaching a $10^{-16}$ sensitivity on the $\mu \to e^+ e^- e^+$ decay in three successive phases of the experiment (called phase~Ia, Ib and phase~II).
\begin{sidewaysfigure}
\begin{tabular}{c}
\includegraphics[viewport=0 380 600 480, scale=0.98]{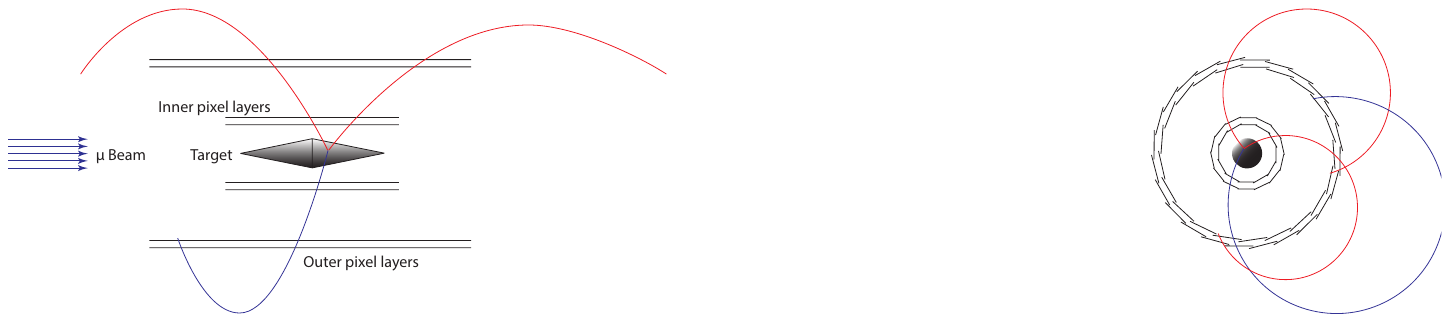} \\ $(a)$ \\
\includegraphics[viewport=0 380 600 480, scale=0.98]{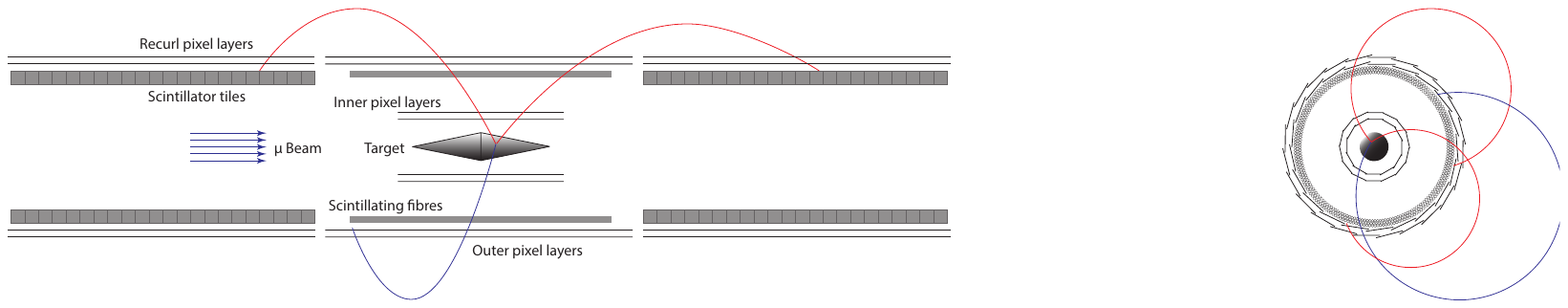} \\ $(b)$ \\
\includegraphics[viewport=0 380 600 480, scale=0.98]{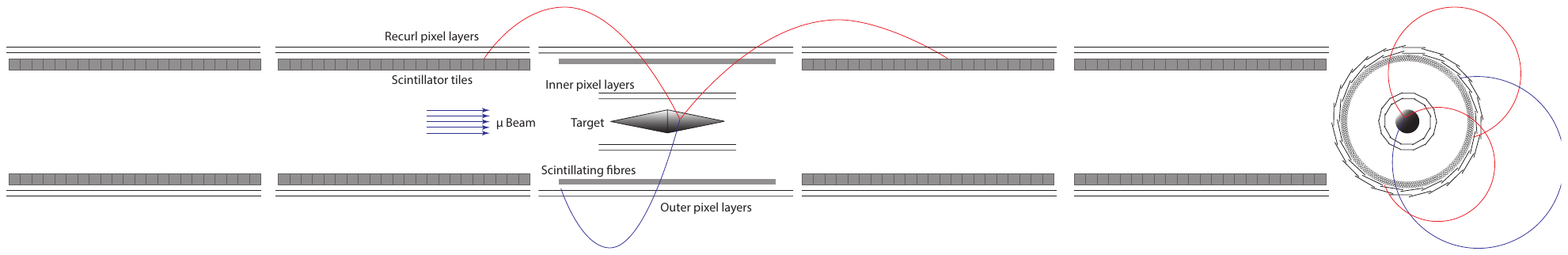} \\ $(c)$ 
\end{tabular}
\caption{Sketch of the Mu3e detector (not to scale) for the different stages of the experiment. From top to bottom the so-called phase~IA (a), Phase~IB (b) and Phase~II (c) detectors (From~\cite{mu3e1})\label{fig:mu3edet}}.
\end{sidewaysfigure}
The same muon beam presently used for the MEG and MEG~II experiments will be transported to a thin (average thickness $85~\mu$m) hollow double-cone Mylar target, with a total length of 10~cm. The target is surrounded by a 2~m long cylindrical detector located inside a 1.5~T solenoidal magnetic field~(see Figure~\ref{fig:mu3edet}) which is segmented in five measuring stations. The central one consists of two double layers of silicon pixel detectors for charged particle tracking complemented by a scintillating fiber tracker for refined particle timing information. The four so-called ``{\em recurl}'' stations located at either side of the central one are made of two layers of pixel sensors surrounding a timing hodoscope made of thicker scintillator tiles.

Since the charged particle momentum will be measured from the particle trajectories in the magnetic field it is clear that in this momentum range the limiting factor to the resolutions is multiple Coulomb scattering (see Figure~\ref{fig:silicon} and~\ref{fig:kapton}). The  tracker consists of an ultra thin silicon pixel tracker, made possible by the High-Voltage Monolithic Active Pixel (HV-MAPS) technology~\cite{hvmaps1, hvmaps2} built up by 50~$\mu m$ silicon sensors glued to an $80~\mu m$ flexible printed circuit board and a $25~\mu m$ kapton support structure, for a total length of less than 0.1\% radiation length. The detector has been designed in such a way as to exploit the fact that the effect of multiple scattering cancels out at first order after half a turn. In fact a charged particle in a magnetic field follows a circular trajectory whose radius is proportional to its momentum. Suppose to estimate the particle's momentum by measuring the diameter $2r$ of the circle: if the particle is scattered at the first measuring position by an angle $\delta \theta$, after half a turn it will cross the tracker again at a ``diameter'' $2r ( 1 \pm \delta \theta^2/2)$%
\footnote{Take a circle of radius $r$ starting from the origin with the center on the $x$-axis $(x-r)^2 + y^2 = r^2$ and rotate it by an angle $\delta \theta$ around the origin: $(x - r \cos \delta \theta)^2 + (y - r\sin \delta \theta)^2 = r^2$. Its new intercepts with the $x-$axis are given by $x=0$ and $x = 2r \cos \delta \theta$ hence the measured "diameter" is approximately $2r ( 1 - \delta\theta^2/2).$   }.

The  resolution of the pixelated $+$ scintillating detector is expected to be of about $\approx 100$~ps in time and $200~\mu$m in space which translates to a momentum resolution between $100$~keV/c and $400$~keV/c in the momentum range  $10 \div 53$~MeV/c, with an expected resolution on the mass and on the (planar and acoplanar) momentum below 1~MeV/c~\cite{andre-bvr}.

The experiment is planned in three stages with increasing sensitivity, each stage corresponding to approximately an order of magnitude improvement. The detector foreseen for the three phases is sketched in Figure~\ref{fig:mu3edet}: in the phase~Ia the experiment will run at a muon rate of a few $\times 10^{7}~\mu^+/$sec with no dedicated timing detector, since the 10~ns resolution of  the tracker detector itself is sufficient to reject the accidental background  to below $10^{-15}$. 

In the phase~Ib the addition of the scintillating fiber tracker in the central detector module will permit  to withstand a rate of $\approx 10^8~\mu^+$/sec, which is the maximum  rate presently deliverable by the PSI beamline,  while the two additional tracking stations will almost double the acceptance for {\em recurling} tracks, improving significanlty the momentum resolution. 

The final phase~II, where the detector will be completed by two more {\em recurling} stations, will allow to reach an ultimate sensitivity of $10^{-16}$ at a muon decay rate of $2 \times 10^9~\mu^+/$sec. To reach this intensity a new beam-line concept is required: there are presently studies ongoing at PSI to reach such an intensity by redesigning the two target stations~\cite{BergMuBeam, klauskirk} and possibly replacing the present graphite target with other materials such as boron carbide.

\begin{figure}
\begin{center}
\includegraphics[width=0.7\columnwidth]{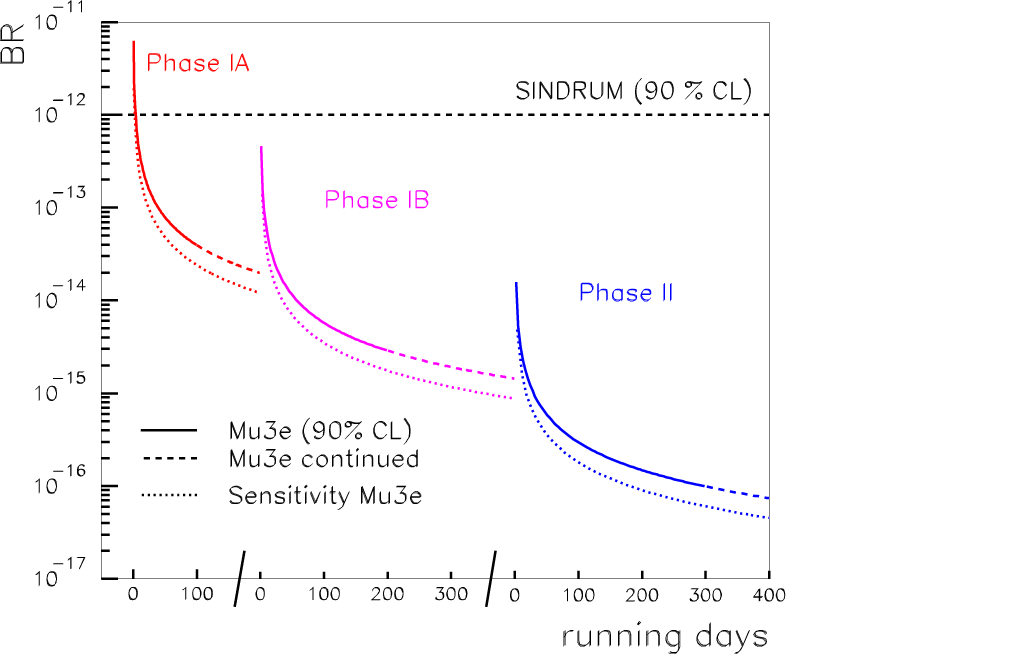}
\caption{\label{fig:mu3elimits}Projected sensitivity of the Mu3e experiment at PSI as a function of the running days for the three subsequent phases of the experiment~\cite{mu3e1}).}
\end{center}
\end{figure}

The experiment has recently entered a phase of detector prototyping and a technical design report is to be published soon~\cite{andre-bvr}. The detector components and the superconducting magnet are not expected before 2018-2019. Promising results have been obtained from a test beam with an 8-layer telescope hosting pixel detector prototypes with an active area of $\approx 10$~cm$^2$~\cite{mu3e-testbeam}. An efficiency $>99\%$ has been  obtained for perpendicular tracks with a timing resolution below 20~ns, while prototype scintillating fiber detectors proved a timing resolution below $500$~ps. The projected sensitivity of the Mu3e experiment is shown in Figure~\ref{fig:mu3elimits}.

\subsection{$\mu^- N \to e^- N$}
Muon to electron conversion is the spontaneous decay of a muon to an 
electron without the emission of neutrinos, within the Coulomb potential of an atomic nucleus: it is therefore only possible for negative muons. For conversions leaving the nucleus in its ground state the nucleons act coherently enhancing its probability relative to the rate of muon capture. The constraint of unchanged nucleus means that all the energy of the muon goes into the kinetic energy of the electron and the recoil of the parent nucleus, hence the signature of such a process is the presence of a monochromatic electron at an energy which is essentially the muon mass, corrected for the binding energy and nuclear recoil, see Eq.~(\ref{eq:muminus}).

\begin{figure}
\begin{center}
\includegraphics[width=0.95\columnwidth]{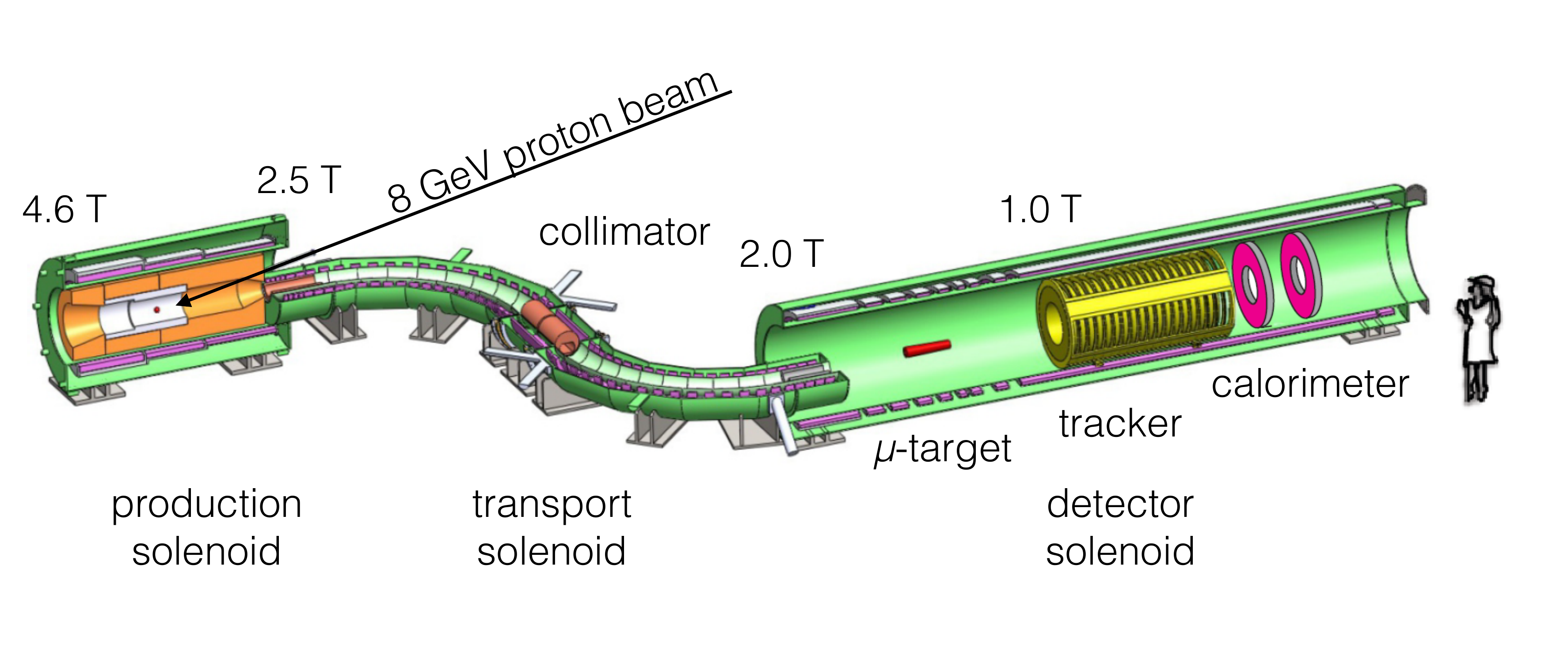}
\caption{\label{fig:mu2e}The Mu2e detector. The backward-going pions produced by the 8~GeV proton beam are captured and decay into muons, which are transported through a bent solenoid to a series of thin aluminum discs where they stop. The electron coming from the muon decay or capture is measured by the straw tube tracker and a pair of crystal calorimeters, all in the shape of hollow cylinders to let low momentum electrons go through undetected. }
\end{center}
\end{figure}

Experimentally, coherent $\mu$ to $e$ conversion offers many advantages over $\mu \to e \gamma$ search: the electron is emitted at the kinematic endpoint of the muon decay in orbit, which constitutes the only intrinsic background. This, as shown in Figure~\ref{fig:czarnecki}, decreases steeply above $m_\mu/2$, a factor of two from the endpoint, allowing detectors with large geometrical acceptance but far from the vast majority of capture and decay events. Doughnut-type detectors  allow all low energy particles to pass through them undetected maintaining a high efficiency on the spectrum endpoint (see Figure~\ref{fig:mu2e}). 

Due to its larger mass, the muon has a large overlap with the nucleus creating some theoretical uncertainty on its wave function. Therefore rather than the full branching ratio, typically $\mu\to e$ conversion experiments discuss the conversion rate 
\begin{equation}
\label{eq:CRdef}
R_{\mu e} ~=~{\rm CR}(\mu N \to e N) ~\equiv~ \frac{\Gamma(\mu-e~{\rm conversion})}{\Gamma({\rm nuclear\  capture})}
\end{equation}
in which theoretical contributions coming from the knowledge of the muon wave function mostly cancel out.

Since no coincidence is required to positively identify an electron as coming from a $\mu$ to $e$ conversion, its energy and trajectory must be measured with the maximum possible resolution.
Primary sources of background are: 
\begin{enumerate}
\item $\mu^-$ decays in orbit (DIO), which produce electrons with an energy up to the kinematic endpoint;
\item radiative muon captures (RMC) in which the high energy photon produces an asymmetric $e^+ e^-$ pair, with the $e^-$ near the kinematic endpoint;
\item radiative captures of negative pions contaminating the muon beam (RPC) which produce photons whose energy is of the order of $m_\pi \approx 140$~MeV . Those photons in turn produce asymmetric $e^+ e^-$ pairs;
\item cosmic rays that interact in the detector material to give a single electron near the kinematic endpoint mimicking the simple event topology;
\item muons decaying in flight that can produce electrons in the right kinematic region, if their momentum is larger than $\approx 76$~MeV.
\end{enumerate}

\begin{figure}
\begin{center}
\includegraphics[width=0.65\columnwidth]{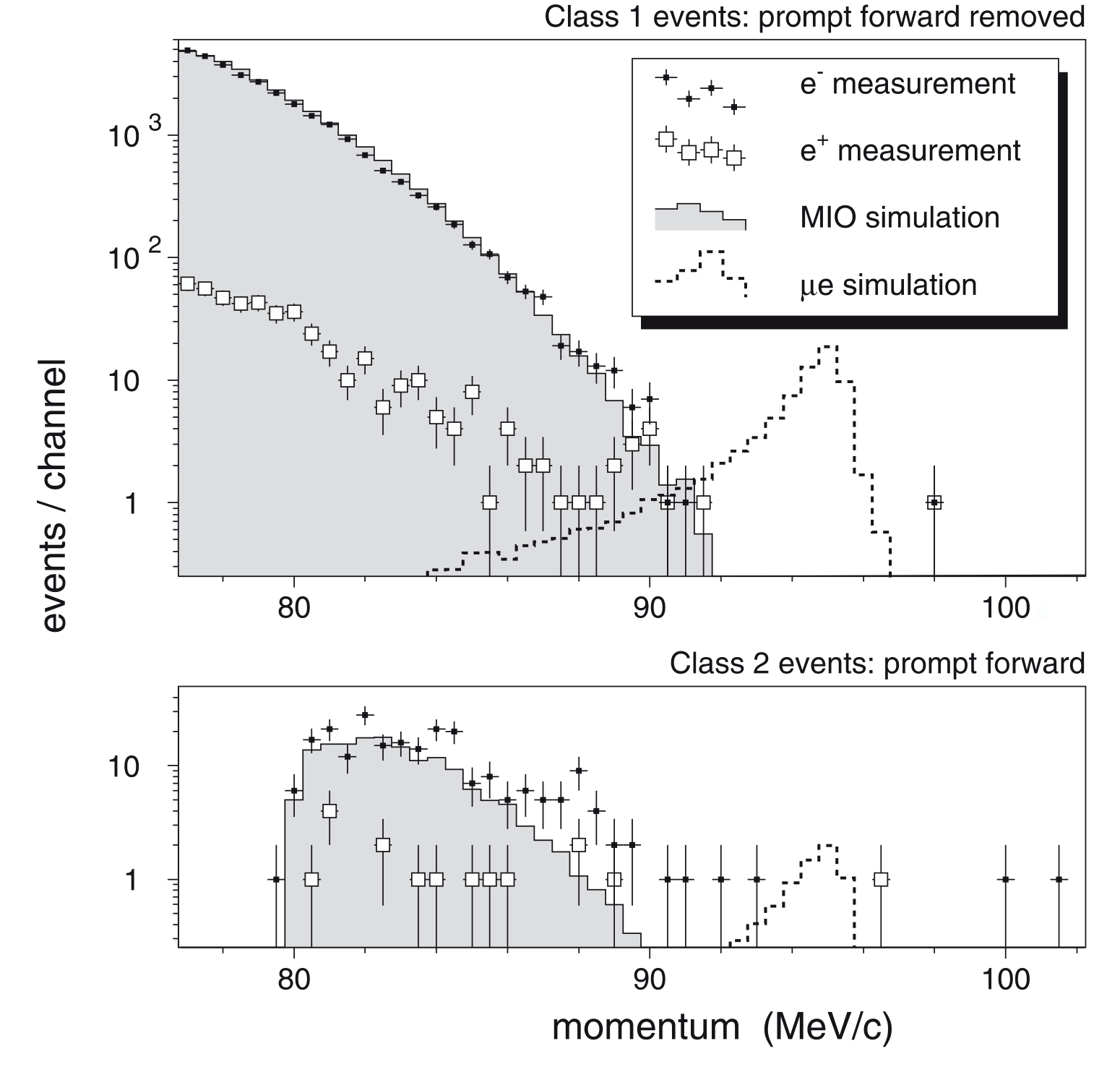}
\caption{\label{fig:sindrum-peak} Momentum distribution of electrons and positrons measured by the SINDRUM~II spectrometer  on a gold target. Events in the forward direction in correlation with the incoming beam are removed from Class~1 events and put in a separate category (Class~2) which includes contamination from prompt beam related background (from~\cite{Bertl}).}
\end{center}
\end{figure}

The muon DIO spectrum decreases near the endpoint as $(E - E_{\rm max})^5$ and within the last MeV its probability is below $10^{-16}$ for sufficiently heavy elements (see Figure~\ref{fig:czarnecki}), hence it is possible to keep it under control with an energy resolution at the percent level.  Radiative pion captures are correlated in time with the incoming beam, and can be reduced by using a pulsed beam and concentrating the measurements between successive spills. Spurious $e^-$ can be generated by cosmic ray muons passing through the target or by antiprotons hitting the experiment material. Cosmic rays can be vetoed with a sufficiently large counter surrounding the target region.
\begin{figure}
\begin{center}
\includegraphics[width=0.9\columnwidth]{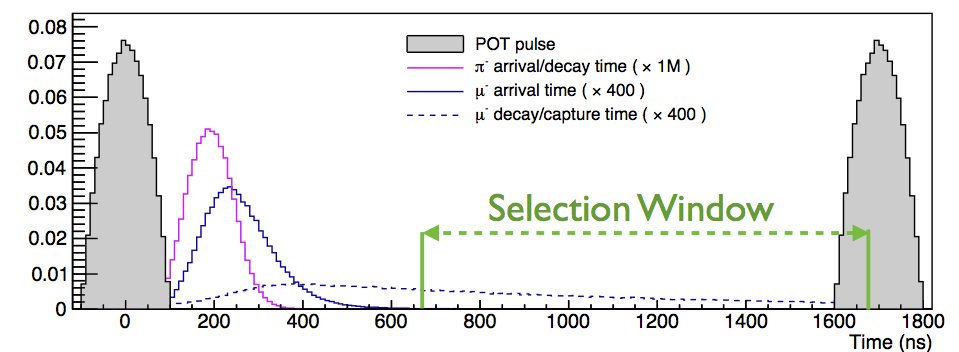}
\caption{\label{fig:obswindow}Time structure of the various particle in the Mu2e experiment. The grey pulses represent the arrival of protons on the pion production target while the purple and the blue pulses represent the arrival time of pions and muons on the capture target, respectively. The muon decay/capture in aluminum is represented by the dashed curve and the selection window is chosen to be far from the particle flash induced by  the primary proton beam (from~\cite{mu2eTDR,mu2e2}).}
\end{center}
\end{figure}
Presently the best limit on the $\mu\to e$ conversion process comes from the SINDRUM-II experiment at PSI~\cite{Bertl} which took the last data in 2000 and published a limit on the conversion rate $R_{\mu e} < 7 \times 10^{-13}$ at $90\%$ confidence level by using a gold target. The positron momentum was measured by two concentrical drift chambers while a scintillator hodoscope situated just inside the inner drift chamber was used for triggering and timing information. SINDRUM-II was performed at the $\pi E5$ beamline, the same of MEG, MEG-II and Mu3e, using a continuous beam of $\approx 10^7 \mu/$sec  with a $\sim 52$~MeV momentum. SINDRUM overall energy resolution was $2.3$~MeV FWHM, and the measured momentum distribution of observed electrons is shown in Figure~\ref{fig:sindrum-peak} together with the magnified version of the expected signal.

Going below SINDRUM-II limit requires an increase of the number of muons, preferably with a pulsed beam structure. In this respect a new beamline idea was put forward by Dzhilkibaev and Lobashev for the MELC experiment that was planned, but later cancelled, at the Moscow Meson Factory~\cite{melc}.
Muons must have a low momentum in order to stop on thin target. They proposed to place a $\pi$ production target in a graded solenoidal field: protons hitting the target produce pions that precess along the magnetic field lines and are partially reflected by the graded field, moving backward with respect to the proton beam. Most of the soft pions decay within few meters and the low energy muons produced are transported, by means of another solenoid, to the detector region through a collimator. The usage of a pulsed proton beam permits the separation of events from RMC and RPC backgrounds. In this way it was estimated that it is possible to transport $10^{11}$ negative muons per second with a 100~$\mu$A proton current, therefore producing $\sim 10^{-4}$~muons per proton instead of the value of $10^{-8}$ typical at surface muon beams (cfr.~10$^8~\mu^+/$sec at PSI with a 2~mA proton current).

Two experiments have been designed following this idea, with some upgrades and modifications: Mu2e at Fermilab~\cite{mu2eTDR} and COMET at J-PARC~\cite{COMET}. We describe their general principle and later discuss them in detail, since they look similar but maintain significant differences. Both experiments can be sliced in three sections: the production target, the pion decay/muon transport system, and the muon target, where the electron detector is located. Pions are produced in bunches which are $\sim 100$~ns long every $1 \div 2~\mu$s, hence there is a time slot which is virtually free from beam-related prompt background, which constitutes the observation window (see Figure~\ref{fig:obswindow}). The curved transport solenoids suppress the prompt background reaching the detector by a factor $10^{10}$ or more, nevertheless one has to be sure that there are no protons hitting the target inside the measurement window. The experiments define as ``extinction" factor the fraction of protons which are out of time with respect to the main bunch: this has to be kept below the $10^{-10}$ level to make this type of background negligible.
Both experiments plan to use, at least in a first stage, an aluminum target ($\tau_{\mu^-} \approx 864~$ns, see equation~\ref{eq:muminustau})\footnote{There are plans to test the conversion on different targets, as this could shed light on the possible non-standard interaction involved~\cite{Cirigliano:2009bz}.
Changing target is less trivial  than it could appear at first sight, since not only the endpoint of the spectrum changes, but also the effective muon lifetime is heavily affected, implying a redesign of the proton beam timing structure.
}. The muon target is followed by a spectrometer in order to precisely measure the momentum of the outgoing electron. 

It is worth mentioning some more details on the curved transport solenoids: charged particles moving through a straight solenoid follow helical trajectories, whose center moves along a magnetic field line with a constant velocity which is given by the component of the particle momentum aligned with the field. In a curved solenoid it is possible to show that at the lowest order, when the curvature radius is sufficiently large, the particle orbits a point which drift slowly in the vertical direction out of the bending plane. The magnitude of the drift velocity is given by~\cite{Jackson}
\begin{equation}
v_D = \frac{m \gamma c}{e B R} \left( v^2_{\parallel} + ^1\!\!/_2 v^2_\perp \right)
\end{equation}
where $m$ and $e$ are the particle mass and charge, $B$ is the magnitude of the magnetic field and $R$ is the bending radius of the solenoid. The parallel and perpendicular component are referred to the magnetic field vector.
If we express the total vertical drift $D$ as a function of the particle path length along the curved solenoid $S$ we obtain:
\begin{equation}
D \propto p S \left( \frac{1}{\cos \theta} + \cos \theta \right) 
\end{equation}
where $p$ is the particle momentum and $\theta$ is the pitch angle of the helix referred to the $B$ field axis ($\theta = 0$ means that the particle travels along the solenoid axis).
Since the dependence on the pitch angle is very weak (the $\theta$ dependence of the term in parentheses is $\approx 2 + \theta^4/4 + {\cal O}(\theta^6)$, so it is below 10\% for angles up to 50 degrees) the dispersion introduced by this drift is mostly governed by particle momentum. By superposing a dipole field one can tune the beam so that a nominal momentum (and pitch angle) is kept on axis, whilst particles with other momenta are removed by the beam pipe and carefully tuned collimators.

\subsubsection{Mu2e}
Mu2e will use an~8~GeV, 25 kW proton beam with 100~ns bunches separated by 1.7~$\mu$s. Its layout is depicted in figure~\ref{fig:mu2e} where the production target, the transport solenoid and the muon target region are highlighted. The magnitude of the magnetic fields is also indicated, ranging from 4.6~T in the production solenoid to the almost constant field of 1~T in the detector region. 
Since it may happen that particles emitted at large angles spend a lot of time in the magnet system, the magnetic field has a negative gradient in the straight sections, analogous to that of the MEG COBRA magnet, that accelerates particles away from the production solenoid and towards the detector solenoid. In these way there are no traps for particles that eventually may scatter towards the detector, in a time frame that can mimic a signal, not being any more correlated with the beam pulse.

In the S-shaped transport solenoid muons from decaying pions follow helicoidal trajectories whose center shifts away from the central plane according to the sign of their charge. A collimator placed midway of the transport solenoid selects negative muons and eliminates other particles. Antiprotons, in particular, are stopped by a thin berillum foil at the center of the collimator, so that only negative muons with a momentum below $100$~MeV reach the stopping target. This is composed by 17 aluminum foils of 200~$\mu$m thickness suspended by thin wires (muons below 50 MeV are stopped) and is followed by a low mass straw tube tracker and a crystal electromagnetic calorimeter. The tracker consists of 22000 straws (a $25~\mu$m sense wire within a 5~mm diameter, 15~$\mu$m thick metalized mylar tube) grouped in 20 measurement stations which allow electrons with a momentum below 52 MeV to escape through the tracker, while being fully efficient above 90~MeV.
A calorimeter consisting in two discs of cesium iodide (CsI) crystals read by silicon photo-multipliers (SiPM) is placed downstream the tracker. Calorimeter information improves track reconstruction and allows particle identification. It will be also used to trigger high energy electrons to reduce the data acquisition rate.
A huge cosmic ray veto surrounds the detector and the passive screens: it consists of scintillator slabs read by SiPM and will veto external muons with an efficiency of $99.99\%$, reducing the fake events in the detector to less than 0.1 event in the full data taking period.
The extinction is provided by a high frequency AC dipole suppressing protons between successive batches, to a level below $10^{-10}$.

Mu2e is expected to start data taking in 2021 and reach a 90\% confidence level sensitivity of $R_{\mu e}<6 \times 10^{-17}$ in three years of running.

\subsubsection{COMET}
\begin{figure}
\begin{center}
\includegraphics[width=0.95\columnwidth]{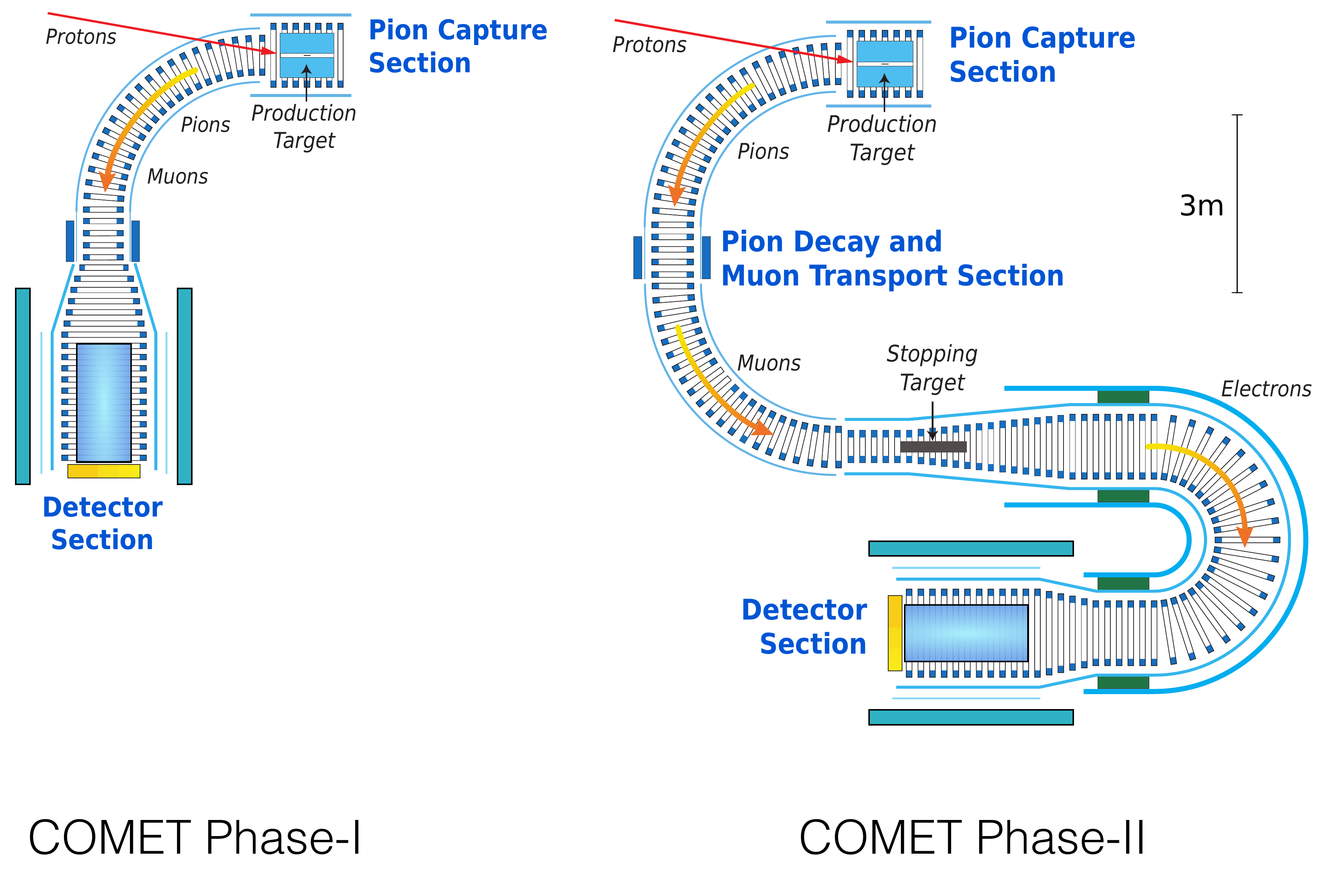}
\caption{\label{fig:comet}The COMET detectors envisaged for the so-called Phase-I and Phase-II. The backward-going pions produced by the impinging proton beam are captured and in turn decay into muons which are transported through a bent solenoid to the stopping target. In Phase-I, envisaged mainly for background evaluation and medium sensitivity search, the stopping target is at the center of the electron detector while in Phase-II the target is followed by a curved electron spectrometer to cope with an increased particle flux and for better momentum measurement.} 
\end{center}
\end{figure}

The COMET (COherent Muon-to-Electron Transition) experiment, being built at the Japanese Proton Accelerator Research Center (J-PARC) in Tokai, Japan, will use an 8~GeV, 56~kW pulsed proton beam with 1.17~$\mu$s bunch separation. With respect to the Mu2e experiment there are two main differences, visible from Figure~\ref{fig:comet}: first the transport solenoid has a C-shape instead of an S-shape, trading a somehow reduced beam intensity ($\sim 30\%$ decrease) with a tighter muon momentum selection: muons stopping in the target will have a momentum between 30 and 50~MeV, compared to Mu2e where there is no lower cut on the momentum.
Second, an extra curved solenoid is used after the stopping target to remove electrons up to the Michel peak and beyond, before the tracker.

COMET will operate in two stages, Phase-I and Phase-II that are schematically reported in Figure~\ref{fig:comet}. Phase-I will help  to understand some of the novel experimental techniques, the beam and the key background rates, as well as make an intermediate measurement of the $\mu-e$ conversion at a 90\% C.L. sensitivity on $R_{\mu e}$ of $7 \times 10^{-15}$. The following Phase-II will allow to achieve the final goal of $7 \times 10^{-17}$.

In Phase-I the proton accelerator power will be limited to 3.2~kW and the muon target will be placed at the end of the first $90^\circ$ bent solenoid.
This will consist of 200~$\mu$m thick aluminum discs, followed by a beam dump.
Given the reduced distance between production target, stopping target and detector a thin cylindrical drift chamber will be used to track the electrons, surrounded by scintillating hodoscopes for triggering and timing.
In Phase-II a straw tube tracker and a crystal electromagnetic calorimeter made of LYSO crystals will allow to cope with the increased particle rate. Furthemore electrons produced in the target will be collected by an additional graded magnetic field and transported around a second section of bent solenoid with larger aperture tuned to transport particles of momentum centered around 100~MeV.
Recent measurements have been able to demonstrate an extinction rate of 10$^{-12}$~\cite{COMET}.

The necessity of proceeding in a staged manner is dictated by the difficulty of simulating all backgrounds at the desired level. As already noted in the general section, when 10$^{19}$ protons are expected to be stopped in the production target the simulation must be highly efficient and highly detailed to insure that no single background event is misinterpreted as signal. Furthermore pion and muon yield produced backward by an 8~GeV proton beam are poorly known, despite some data exists from the HARP experiment~\cite{harp}.

Furthermore data on muon nuclear capture on aluminum is not extensive: in this respect a combined effort between COMET and Mu2e has led to the realization of the AlCap experiment at PSI~\cite{alcap}, whose task is measuring the rate and spectra of particles emitted by muon capture on Al to be fed in the experiment's Monte Carlo codes. Three data taking runs have been completed from 2013 to 2015 and the data analysis is ongoing. 

COMET Phase-I is expected to start its three-years data taking  in 2018, in the meantime the Phase-II detectors are being realized.

Apart from the measurement of the muon conversion to electron, there is also, both for Mu2e and for COMET, the possibility to measure the similar process with $\Delta L = 2$: $\mu^- A(Z,N) \to e^+ A(Z-2,N)$. This decay violates both the conservation of the total lepton number and the lepton flavour numbers, $L_e$ and $L_\mu$ and is closely related to the neutrinoless double-decay (see, {\em e.g.},~\cite{merle}). Theoretical models find $R_{\mu e^+}$ from $10^{-12}$ to $10^{-14}$. The best existing limit is $3.6 \times 10^{-11}$ and has been obtained by the Sindrum II collaboration in the process $\mu^- $Ti $\to e^+$ Ca.

\subsubsection{DeeMe}
DeeMe (Direct emission of electron from Muon to electron conversion) is a $\mu^- N \to e^- N$ search experiment in construction at J-PARC~\cite{DeeMe1,DeeMe2}
\begin{figure}
\begin{center}
\includegraphics[width=0.8\columnwidth]{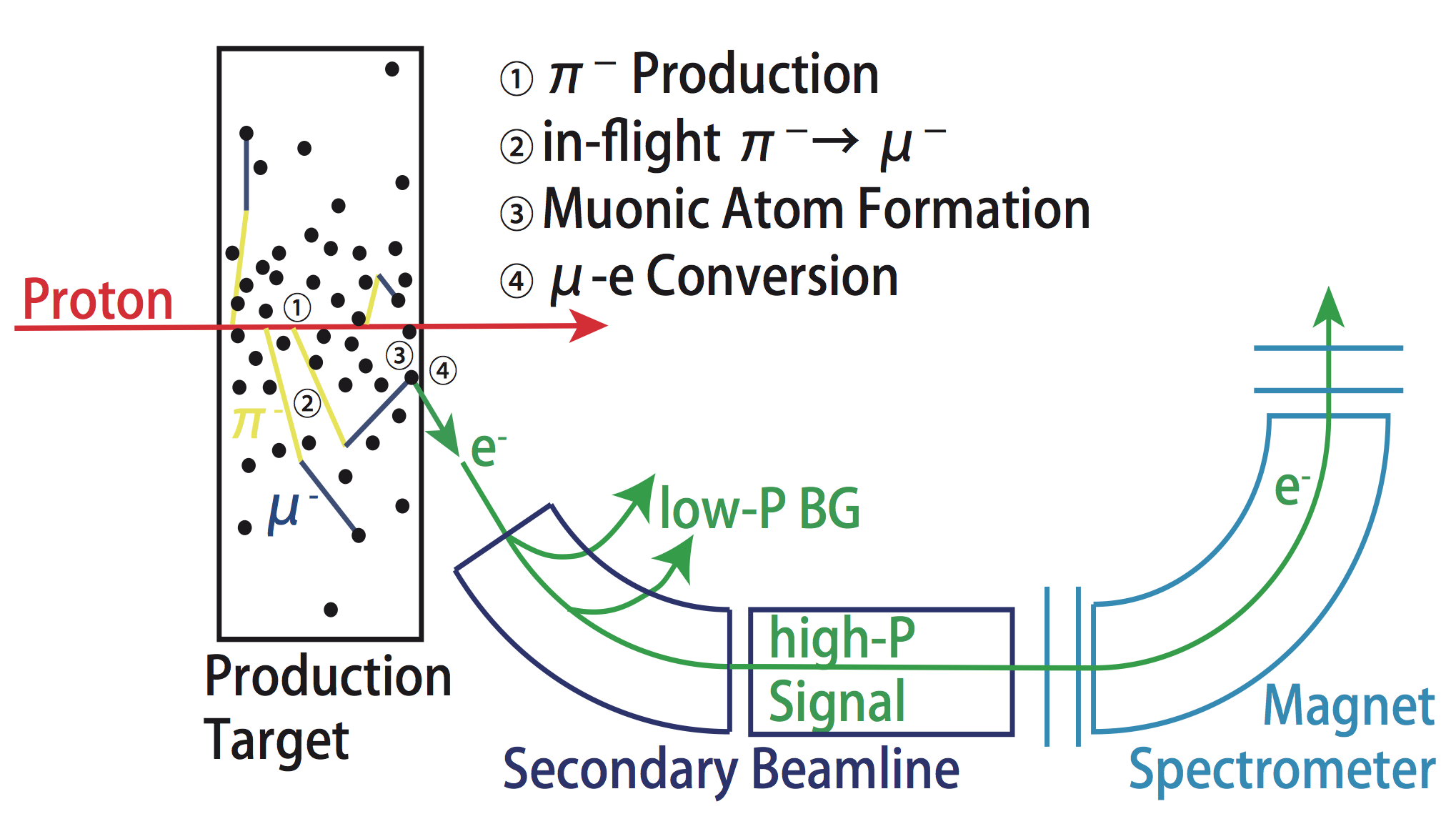}
\caption{\label{fig:deeme}Schematics of the DeeMe experiment. In DeeMe the pion production, decay and muon capture, all happen in the same tagtet (graphite first, then silicon carbide) and emerging electrons are measured by a spectrometer.}
\end{center}
\end{figure}
with a simpler setup allowing, with an early start of the experiment, a moderate sensitivity compared to the other $\mu^- N \to e^- N$
experiments described in the previous sections. The DeeMe experiment will be carried out at a new beamline, H-Line, under construction at the Muon Science Establishment (MUSE) in the Material and Life Science Facility (MLF) at J-PARC.
In DeeMe (see Figure~\ref{fig:deeme}) a 3 GeV proton beam from the Rapid Cycling Synchrotron (RCS) will be stopped in a target creating pions. The produced pions decay to muons in the target itself. Some of the muons are expected to form muonic atoms in the production target, making it possible to search for electrons from $\mu \to e$ conversions generated in the same target. 
The signal electrons will be measured by a magnetic spectrometer based on a dipole magnet borrowed from TRIUMF (the PACMAN magnet) and four multi-wire proportional chambers. Low momentum background particles are removed mainly by dipole magnets in the beam transportation system. The experiment will start with a graphite target to reach a single event sensitivity of $1\times 10^{-13}$ with a running time of 2$\times 10^7$~s at 1 MW operation of RCS. The target will be later on switched to silicon carbide (SiC), whose muonic capture rate is six times higher than that of graphite to improve the sensitivity down to $1\times 10^{-14}$ (SES) for $2\times  10^7$~s running time and further go down to $5 \times 10^{-15}$ (SES) if the running time is extended to $8\times 10^7$~s.
The capture solenoid is already installed and the other parts of the H-line are under construction. The experiment is expected to start data taking in 2017, with a measurement of the high-energy part of the DIO spectrum to be compared with present theoretical predictions.


\section{The tau channels}
\label{sec:tau}
Search for charged lepton flavour violation in decays involving taus is very promising: thanks to its large mass ($m_\tau \approx 1\,777$~MeV) many flavour violating channels are open, in comparison with muon decays. Besides $\tau \to \mu \gamma$,  $\tau \to e \gamma$, $\tau \to 3 \ell$ $\left( \ell \in \{e,\mu\} \right)$ there is a number of channels involving hadrons in the final state such as $\tau \to \ell \pi^0$, $\tau \to \ell \pi^+ \pi^-$ which are all characterized by the presence of an energetic lepton and other charged or neutral particles with a topology which is more or less easy to identify.

All models predicting CLFV in the muon sector imply a violation in the tau sector as well, whose amplitude is often enhanced by several order of magnitude, according to the specific violation pattern, but which is usually a power of the ratio of the muon-to-tau mass $\left( m_\tau / m_\mu \right)^\zeta$~(see section~\ref{sec:susy}).
\begin{figure}[t]
\begin{center}
\includegraphics[width=0.97\columnwidth]{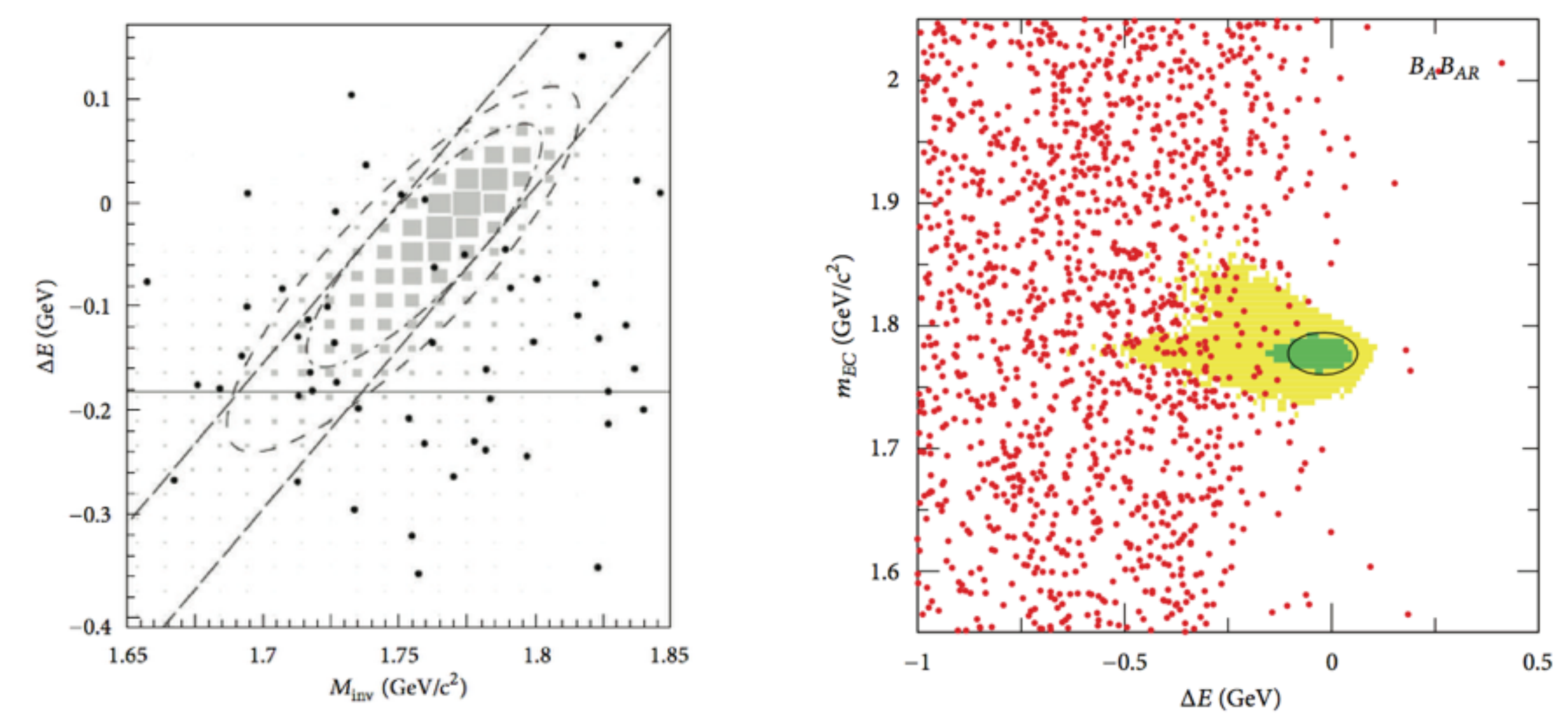}
\caption{\label{fig:belle}Results of the Belle (left) and BaBar (right) experiments search of the $\tau \to e \gamma$ decay.}
\end{center}
\end{figure}

From an experimental point of view some difficulties arise: the $\tau$ lepton has a much shorter lifetime ($2.9 \times 10^{-13}$~s instead of $2.2~\times 10^{-6}$~s for the muon) and is not as copiously produced as muons are, not to mention the fact that it is impossible to produce tau ``beams''.
Tau samples must be obtained at proton or electron accelerators, operating in an energy range where the production cross section is large, with little background, and their decay must be measured with large detectors with good particle identification and tracking capabilities, as well as excellent calorimetry and hermeticity in order to constrain the kinematics.
The enhanced sensitivity due to its larger mass is therefore partly decreased by the reduced number of tau that can be observed and measured since $\tau$ CLFV searches have not been performed with dedicated experiments, but with beams and detectors which are used for a broader physics programme. 

Historycally a great leap forward has been done since the end of the nineties, when the best limits were set by the CLEO experiment at Cornell Electron Storage Ring (CESR)~\cite{CLEO}, thanks to the design and operation of the B-flavour factories at the PEP-II collider at SLAC~\cite{PEPII} and at KEKB in Tsukuba~\cite{KEKB}, whose luminosity and asymmetric beam energy enhanced the capability of reconstructing tau decays. The BaBar and Belle experiments at the two accelerator complexes were built to measure the CP-violating parameters in the B-meson systems. Nevertheless at the center-of-mass energy of the $\Upsilon(4s)$ resonance ($\sqrt{s} =10.58$~GeV) the cross section to produce a $\tau^+ \tau^-$ pair is 90\% of the cross section to produce a $b \bar b$ pair so that the B-factories are $\tau$ factories, too.
Since $\tau$s are produced in pairs one identifies, for candidate events, two hemispheres in the center-of-mass frame where each hemisphere contains the $\tau^+$ or $\tau^-$ decay products. Whereas Standard Model decays of $\tau$ leptons have at least one neutrino, CLFV decay products have a combined energy which equals the total $\tau$ energy  (approximately equal to the beam energy in the center of mass) and a total mass which equals the $\tau$ mass. 
Using a two dimensional signal region in these two variables the signal is separated from the Standard Model $\tau-$decay background with minimal loss of efficiency and the number of events falling in the region of the expected signal are used to compute the limit. The shape of the region and the necessary cuts are optimized at simulation level to give the best ``expected upper limit''. In Figure~\ref{fig:belle} we report, as an example, the results for the Belle and BaBar experiments for the search of the $\tau \to e \gamma$ decay, the plots for other channels being analogous.
\begin{figure}
\begin{center}
\includegraphics[width=0.97\columnwidth]{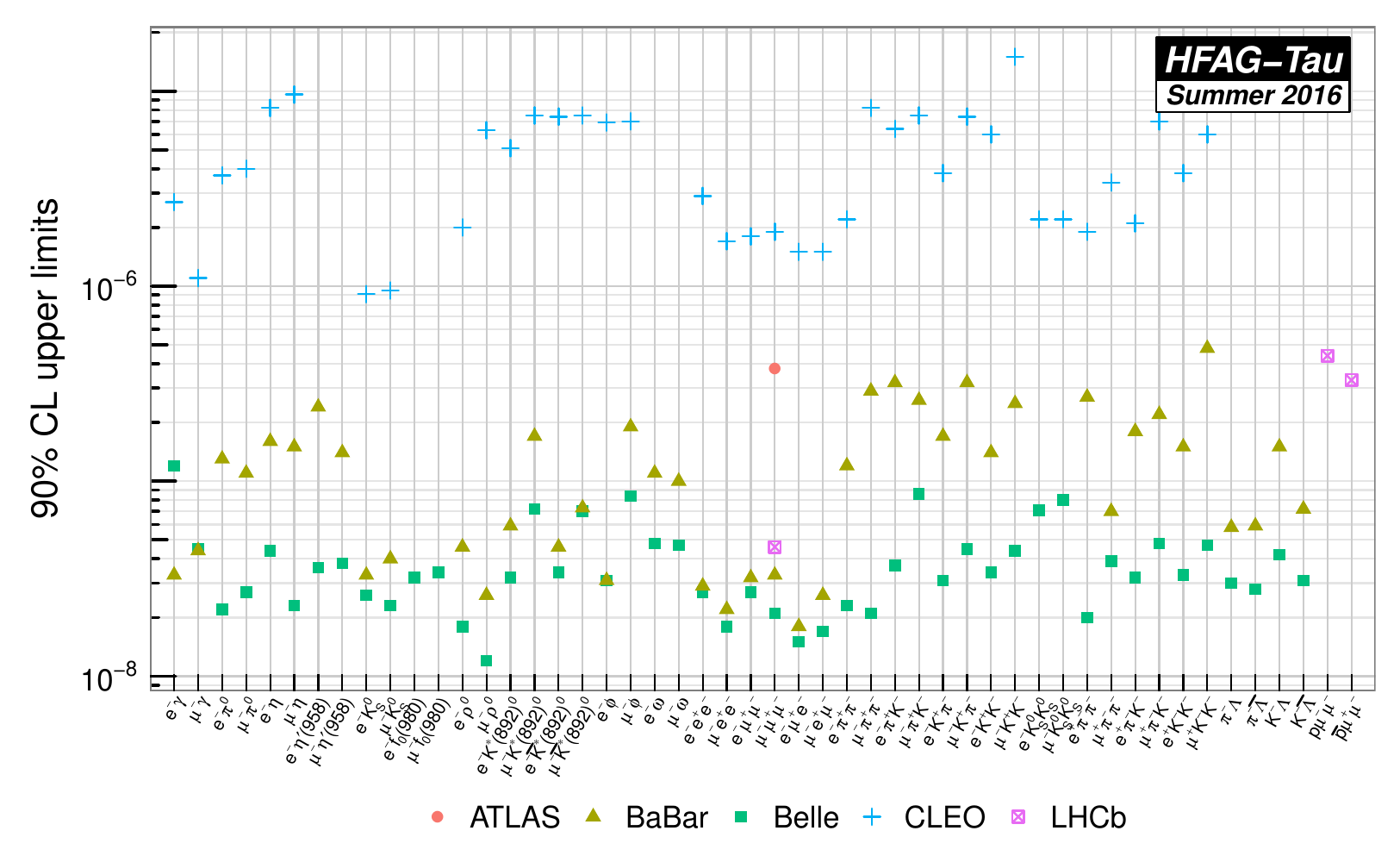}
\caption{\label{fig:taulimits}Limits on CLFV tau decays (taken from the HFAG working group ~\cite{HFLAV}).}
\end{center}
\end{figure}
For instance the present limits on $\tau \to e \gamma$ and $\tau \to \mu \gamma$ are $3.3 \times 10^{-8}$ and $4.4 \times 10^{-8}$ respectively, 
and those regarding decays with three charged particles\footnote{Recall that $\pi^0 \to \gamma \gamma$ and $\rho^0 \to \pi^+ \pi^-$ immediately after production.} in the final state ($\tau \to 3\ell$, $\tau \to \ell \pi^0$, $\tau \to \ell \rho^0$) are a factor two to five better~\cite{bfactories}.

We report in Figure~\ref{fig:taulimits} the limits set by CLEO, BaBar and Belle to a large number of CLFV decays, together with some more recent limits posed by ATLAS and LHCb on selected decay channels. Belle and BaBar samples correspond to $\approx 10^9$ $\tau$ decays which have been thoroughly analyzed to produce those limits.

The absolute normalization is just the inverse of the number of pruduced taus times the various efficiencies, while the main difference between the $\tau \to \ell \gamma$ channels and the three-daughers channels rests in the fact that the latter are background free (they find no candidate event in the signal box) while the two body decay with a photon in the final state is more prone to background events.  In fact the main background arises from $e^+ e^- \to \mu^+ \mu^- \gamma$ or $e^+ e^- \to \tau^+ \tau^- \gamma$ where one of the $\tau$s decays via $\tau \to \ell\nu \bar \nu$. In both cases the photon from initial or final state radiation combines with the muon or the electron to fall accidentally in the signal box, giving rise to an ``irreducible'' background.
\begin{figure}[t]
\begin{center}
\includegraphics[width=0.5\columnwidth]{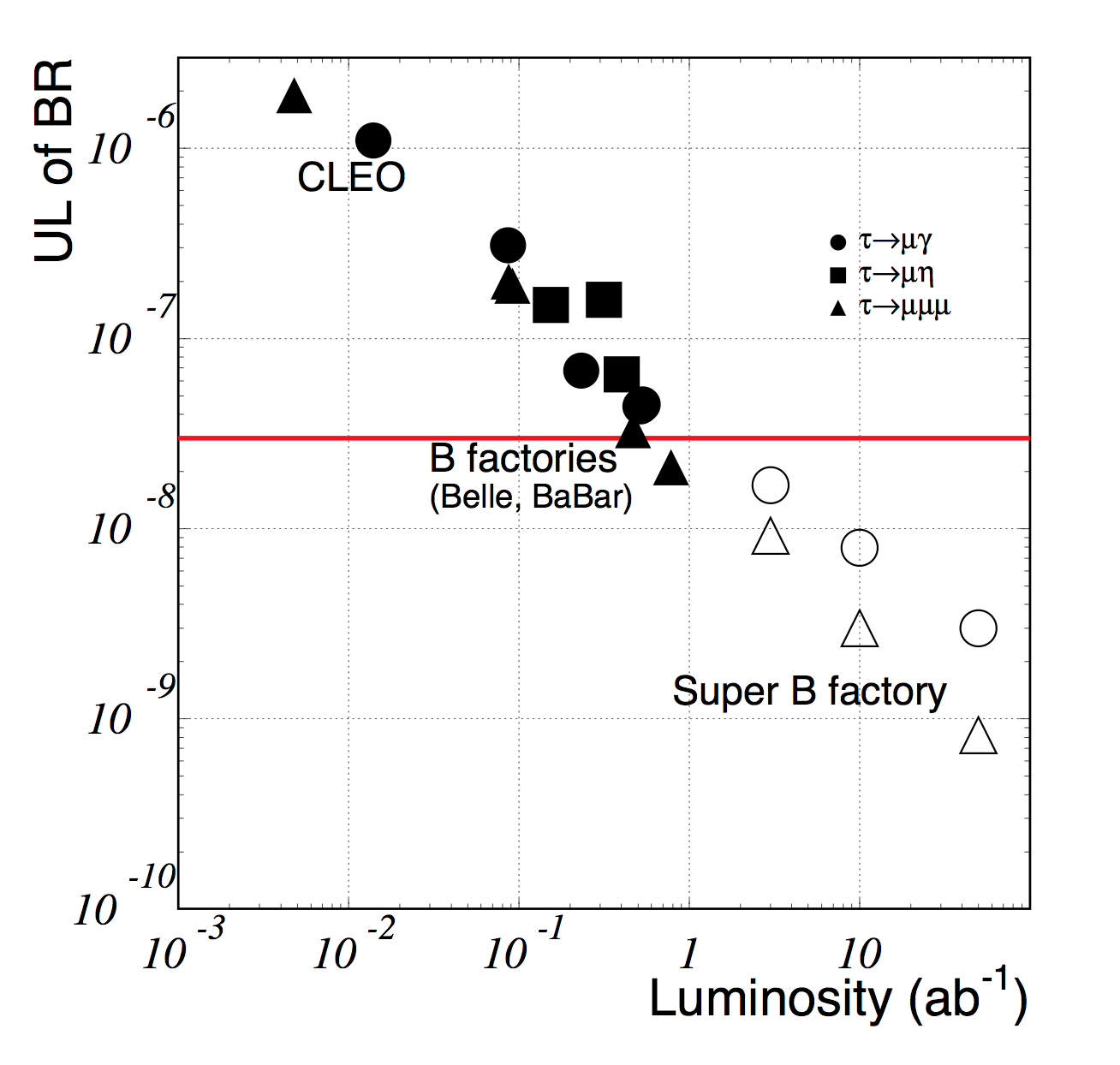}
\caption{\label{fig:taufuture}History and anticipated upper limits on branching fractions for $\tau \to \mu \gamma$, 
$\tau \to \mu \eta$ and $\tau \to \mu \mu \mu$ decays as a function of the integrated luminosity. From~\cite{superB}.}
\end{center}
\end{figure}

In the near future, the Belle~2 experiment at Super KEKB
~\cite{belle2} 
is expected to improve these limits below $5 \times 10^{-9}$ or $10^{-9}$ for the radiative and three body decays respectively~\cite{superB}, for an integrated luminosity of $50$~ab$^{-1}$. At this level  the two body decay  suffers from contamination from the irreducible background, while the three-body decay being on the contrary still background free, as apparent from the different scaling as a function of the luminosity (square root vs linear respectively) shown in Figure~\ref{fig:taufuture} (from~\cite{superB}). However it should be noted that the limit on the $\tau \to \ell \gamma$ final state could be further improved by using the $\tau$ polarization or by accumulating large samples of tau leptons at a lower center-of-mass energy, where initial state radiation (whose energy is proportional to the main beam energy) is negligible in the signal region. 
This could be the case at a $\tau-$charm factory operating just above the $\tau$ production threshold (see, {\em e.g.}~the discussion in~\cite{litau} where, in the framework of a possible experiment at the High Intensity Electron Positron Accelerator (HIEPA) being proposed in China, spectra of photons from signal and initial state radiation are compared at various beam energies or~\cite{charmtau}).

\section{Future Directions}
\label{sec:future}
\begin{figure}[t]
\begin{center}
\includegraphics[width=0.87\columnwidth]{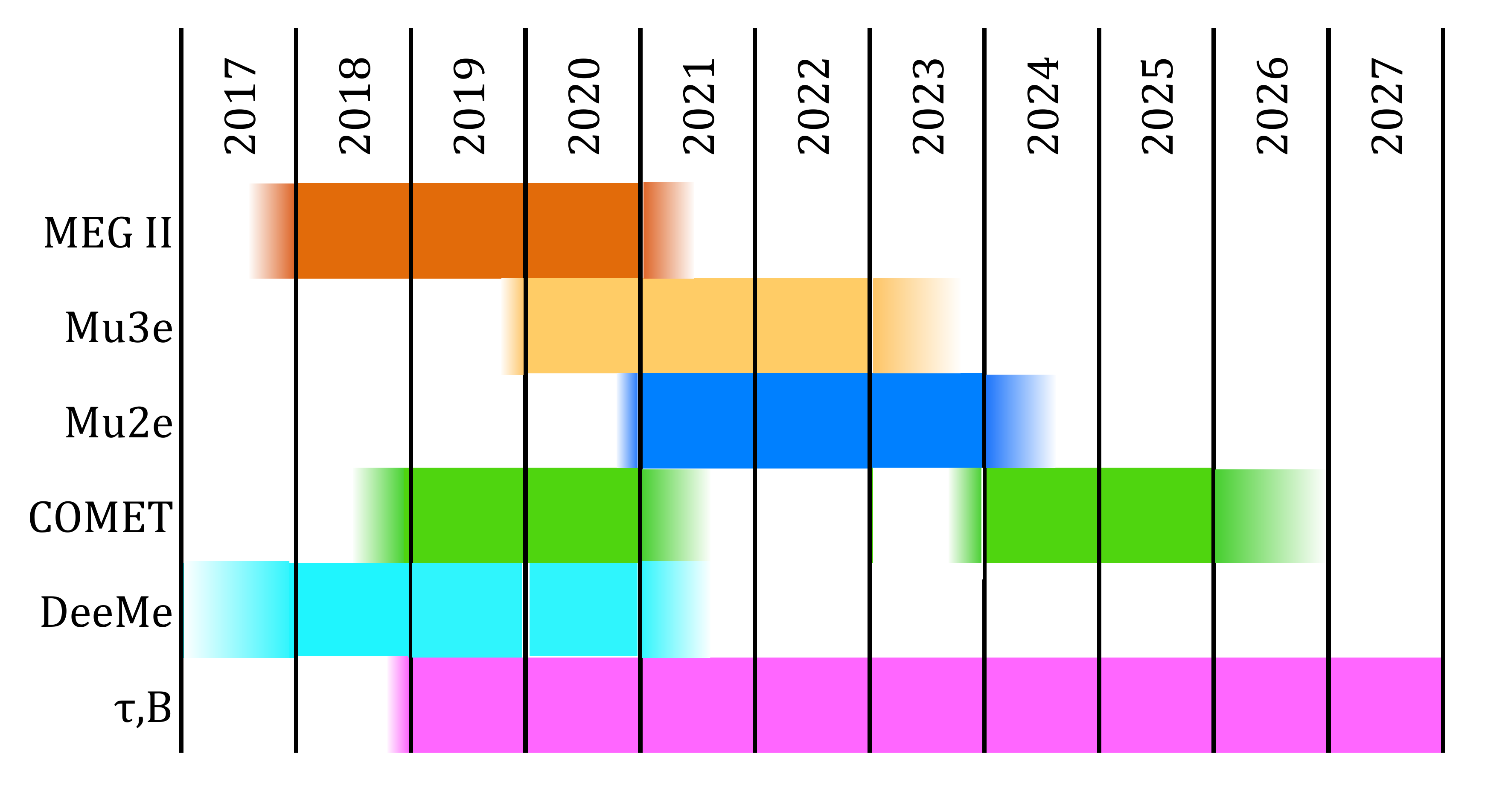}
\caption{\label{fig:timeline}Projected time lines for different projects searching for CLFV decays. MEG~IIis expected to start data taking in 2018 after an engineering run in 2017; Mu3e magnet and detectors are expected at the end of 2019; Mu2e foresees three years of data taking starting in 2021; COMET Phase-I is expected to start commissioning and data taking in 2018 for two-three years, followed by a stop to develop and deploy the beamline and detectors for Phase-II; DeeMe is expected to start soon and take data with graphite and silicon carbide targets in sequence; Belle~II is schedule to start data taking at end 2018.}
\end{center}
\end{figure}

Figure~\ref{fig:timeline} shows a summary time line of CLFV searches in the next decade. It is exciting to see that within the next five to ten years our present knowledge of the fundamental interactions could be disproved or confirmed with a stronger confidence, in particular in the muon sector, where also new results from the g$-$2 experiment at FNAL~\cite{gminus2} are expected.

It is appropriate to examine what could be the next steps towards a sensitivity improvement and to examine if there are objective limitations to these explorations.

If on the one hand the sensitivity scales with the number of muons, on the other the capability to reject the background is related to the experimental acceptances and resolutions on the measurement of daughter particles. Furthermore these two lines should proceed in parallel.

$\mu^+\to e^+\gamma$ and $\mu^+\to e^+e^-e^+$ searches require intense muon beams. While the search of \meg\ is not presently limited by the intensity of the beam the staged approach of Mu3e requires the development of a high intensity beam line which is still under way (see Table~\ref{tab:stefan}). Its estimated flux ($10^{10}$ surface $\mu^+$/s) is two order of magnitude larger than presently available. This does not necessarily maps into a two order of magnitude increase in sensitivity, since with this flux the \meg\ search will be completely overwhelmed by random coincidences at present-day resolutions. With reference to Eq.~(\ref{eq:bacc}) one sees that a $100\times$ increase in muon flux can be compensated only with extreme resolutions: $\delta x \sim \delta y \sim 0.1\%$, $\delta t \sim 50$~ps, $\delta \theta \sim 10$~mrad are necessary to reach an effective ${\rm BR}_{acc} \leq 10^{-16}$, while these conditions may be relaxed, at constant    ${\rm BR}_{acc} $, for a lower beam intensity ($10^9~\mu^+/$s). 
To reach the desired resolutions one could think~\cite{mcooper, whatnext, hitlin,VoenaPapa} of converting the photon and measure the resulting $e^+ e^-$ pair to improve the energy measurement on the photon leg at the percent level. The usage of conversion pairs would allow to track the photon back to the target, permitting a vertex constraint that is 
absent in present and planned experiments. A pointing calorimeter with a degree resolution could play the same role, since the advantage of better resolutions of the pair spectrometer is partly spoiled by the additional multiple scattering on the converting material.
An active target~\cite{atar} pinpointing the parent muon decay position could also help in reducing the number of accidentals, both in \meg\ and in $\mu \to eee$ searches, while spreading the muon beam decay points to several targets, and identifying the starting target of both positron
and photon, would cut the random $e\gamma$ coincidences linearly with the number of targets. 

It must however be taken into account that radiative background is at the level of $10^{-16}$ and $10^{-18}$ for \meg\ and $\mu \to eee$ respectively even at excellent resolutions (see Figure~\ref{fig:phisbg} $a$ and $b$, Figure~\ref{fig:matteofael} ) hence it seems reasonable to assume that $10^{-16} \div 10^{-15}$ represent an ultimate limit for the \meg\ decay search, maybe done with an ``extended'' Mu3e experiment\footnote{In a pre-print appeared while writing this review~\cite{VoenaPapa} an attempt is made to give this simple line of reasoning a more quantitative form, though not considering the possibility of multiple targets.}, while the sensitivity to the $\mu \to eee$ decay itself could be pushed by a further order of magnitude with respect to the projected sensitivity by the usage of an active target~\cite{hitlin} but not much below.  

For $\mu\to e$ conversion experiments there is no contribution from random coincidences and the DIO background could be kept below $10^{-19}$ provided the energy resolution is good enough. Reaching the level of $10^{-18}$ or beyond on $R_{\mu e}$ would require an increase in signal and a reduction in the background induced by beam pions and out-of-time particles. In this respect the PRISM (Phase Rotated Intense Slow Muon-source) project~\cite{prism-prime} is studying the production of a more intense negative muon pulsed beam ($10^{11}~\mu^-$/s) with an energy spread so small (below $0.5$~MeV) to be able to stop the beam in a single thin target, to reduce the target contribution to the electron energy measurement. Instead of selecting monochromatic muons, reducing their number, their phase space is rotated such in a way as to pass from a bunch of particle with a large energy spread but a sharp timing to particles spread in time ($\sim 100~$ns) but with a narrow energy distribution, by means of a few turns in a fixed field, alternating gradient (FFAG) storage ring. Coupled to a COMET-like detector this should allow to reach a SES of $\approx 3 \times 10^{-18}$.

A similar study is being pursued at the FNAL complex to increase by a factor of 10 of the muon beam line intensity using a 1 or 3 GeV proton beam on a production target while delivering the neutrino beam to LBNF. In this environment, a Mu2e-II experiment that reuses a large fraction of the Mu2e apparatus could provide a factor of 10 improved sensitivity.

In any case, as seen in section~\ref{sec:positrons} and~\ref{sec:photons}, the limitation of low energy photon and electron detection pinpoints the  difficulty to go much beyond the present experimental resolutions.
As a consequence  one might ask if innovative detection techniques should be explored. One promising technique is being applied in the field of direct measurement of neutrino mass and/or relic neutrino background detection by means of detecting the cyclotron radiation from electrons (of $\sim 20$~keV energy) spiraling in a homogeneous magnetic field~\cite{project8, ptolemy}. They in fact follow helical orbits with an angular frequency that is independent of the emission angle, during which they loose energy, in form of microwave radiation, due to cyclotron emission. It is a straightforward exercise to scale the numbers given in~\cite{project8} for a possible detection of 50-100 MeV electrons and positrons (see also the respective relativistic counterparts in~\cite{Jackson}). In a 10~T magnetic field the radius of their orbit does not exceed few centimeters, and radiation of 1 to 3~GHz will be emitted with a power of the order of a few pW per turn. 
This is three orders of magnitude more power compared to the $\sim 30$~GHz radiation detected in~\cite{project8}, but emitted in the Ultra-High Frequency radio band. A substantial redesign of the experiments would be needed to explore the possibility of exploiting the superior resolution with which frequency measurements are done in this range, but values such as those presented here are commonly reached magnets used in NMR research.

Finally, among  ideas that are being put forward, we briefly mention the possibility of CLFV production of muons and taus by high-energy lepton beams or electron-ion colliders (see {\em e.~g.}~\cite{original, musolf, chinese}) and we refer our readers to~\cite{abada} and~\cite{bernstein} for a recent theoretical introduction and a discussion on associated experimental issues, respectively. 

\section{Conclusions}
\label{sec:conclusions}
\begin{table}[t]
\renewcommand{\arraystretch}{1.3}
\begin{center}
\begin{tabular}{llccccc}
\hline
Reaction & Present limit & Expected Limit &  Reference  & Experiment \cr
\hline
$\mu^+ \to e^+ \gamma$ & $< 4.2 \times 10^{-13}$  & $ 5 \times 10^{-14}$ & \cite{MEG2} &MEG~II \cr
$\mu^+ \to e^+ e^- e^+$  & $< 1.0 \times 10^{-12}$   &  $ 10^{-16}$ & \cite{mu3e1} & Mu3e \cr
$\mu^{-}{\rm Al}\rightarrow e^{-}{\rm Al}~{}^\dag$ &  $ < 6.1 \times 10^{-13}$  &  $ 10^{-17}$  & \cite{mu2eTDR,COMET} & Mu2e, COMET \cr
$\mu^{-}{\rm Si/C}\rightarrow e^{-}{\rm Si/C}~{}^\dag$ &  \multicolumn{1}{c}{$-$}  &  $ 5 \times 10^{-14}$  & \cite{DeeMe1} & DeeMe \cr\hline
$\tau\rightarrow e\gamma$   & $< 3.3 \times 10^{-8}$ & $5 \times 10^{-9}$  & \cite{superB} & Belle~II\cr
$\tau\rightarrow \mu\gamma$ & $< 4.4 \times 10^{-8}$ &   $10^{-9}$  & \cite{superB} & '' \cr
$\tau\rightarrow eee$ & $< 2.7 \times 10^{-8}$ & $5 \times 10^{-10}$  &  \cite{superB} & '' \cr 
$\tau\rightarrow \mu\mu\mu$ & $< 2.1 \times 10^{-8}$ &  $5 \times 10^{-10}$   & \cite{superB} & '' \cr
$\tau\rightarrow  e\ {\rm had}$ & $< 1.8 \times 10^{-8}~{}^\ddag$ &$3 \times 10^{-10}$ &  \cite{superB} & '' \cr 
$\tau\rightarrow \mu\ {\rm had}$ & $< 1.2 \times 10^{-8}~{}^\ddag$& $3 \times 10^{-10}$  &  \cite{superB} & '' \cr 
\hline
${\rm had}\rightarrow \mu e$ & $< 4.7 \times 10^{-12}~{}^\S$ &  $10^{-12}$   & \cite{moulson} & NA62 \cr
$h\rightarrow e\mu$  & $< 3.5 \times 10^{-4}$ &     $3 \times 10^{-5} ~{}^\P$& \cite{mitra} & HL-LHC \cr
$h\rightarrow \tau\mu$  & $< 2.5 \times 10^{-3}$ &  $3 \times  10^{-4} ~{}^\P$ & \cite{mitra} & "\cr
$h\rightarrow \tau e$  & $< 6.1 \times 10^{-3}$ &   $3 \times 10^{-4} ~{}^\P$  &\cite{mitra}& " \cr
\hline
\end{tabular}
\end{center}
\caption{Present and future limits for selected CLFV processes.
${}^\dag$Rate normalised to the muon capture rate by the nucleus, see Eq.~(\ref{eq:CRdef}). ${}^\ddag$Best limits from $\tau \to e \rho^0$ and $\tau \to \mu \rho^0$ respectively. $^\S$Best limit from $K_L^0$ decay. $^\P$Reference~\cite{mitra} quotes the branching ratio for which one can make a 2$\sigma$ or 5$\sigma$observation; we use the number of expected signal and background events in there to infer 95\% C.L. sensitivities on the three channels, which turn out to be compatible with the scaling for the square root of the relative luminosity - 3000~fb$^{-1}$ assumed in~\cite{mitra} vs $20$~\cite{higgsmue} or 36~\cite{higgsmutau}~fb$^{-1}$. }
\label{tab:futuro}
\end{table}

We hope to have convinced the reader that ours is a subject rich of interesting aspects both theoretically and experimentally, worth to be studied in depth.
On the one side, we are witnessing a wave of experimental efforts involving new ideas and techniques that are and will be pushing the limits to unprecedented levels. On the other side, the violation of the leptonic flavours is a common feature of broad classes of new physics models and it appears to be intimately connected to some of the mysteries of nature that the Standard Model fails to account for: the origin of the tiny masses of neutrinos, the baryon-antibaryon asymmetry of the universe (through leptogenesis), the arising of the hierarchical structure of fermion masses. At least, CLFV searches are very suitable to test several ideas that were proposed to address the aforementioned issues. 

In more general terms, the capability of CLFV searches to explore scales
way beyond the energies of our present (and foreseeably future) colliders -- as shown by Table \ref{tab:bounds} -- could prove crucial in establishing where is the next fundamental scale above the one associated to the electroweak symmetry breaking. In this sense, our searches, as other `high-intensity' experiments, can complement the high-energy machines in the effort to overcome the impasse that particle physics is seemingly finding itself in: no sign of new physics (hopefully yet) after the discovery of the Higgs boson at the LHC.

Another reason why our subject is particularly timely is
the long-standing discrepancy between theoretical and experimental determinations of the anomalous magnetic moment of the muon, $(g-2)_\mu$. In fact, the Muon $g-2$ Experiment currently running at FNAL \cite{Grange:2015fou} is expected to tell within few years whether such a discrepancy is a signal of new physics or is due to some unknown systematics that affected the previous experiment (whose sensitivity will be reached already in 2018). 
As briefly mentioned in our article -- and extensively discussed in the literature, {\em e.g.}~in the recent review \cite{Lindner:2016bgg} -- $(g-2)_\mu$ is an observable closely related to CLFV processes: it is induced by the flavour-conserving counterpart of the dipole operator that would be responsible for decays of the type $\ell_i\to\ell_j\gamma$ (cf.~section \ref{sec:EFT}), and indeed large CLFV rates are typically predicted by those models that can account for the $(g-2)_\mu$ discrepancy, unless they feature a very peculiar flavour structure. Evidence that new physics is responsible for the apparently missing contribution to $(g-2)_\mu$ would then give us high expectations for a similarly striking discovery at CLFV search experiments
(this conclusion is analogous  to that we drew in section \ref{sec:LFU} regarding possible
lepton flavour non-universal effects in $B$ decays).

Finally, as a summary of the presented topics we would like to propose in Table~\ref{tab:futuro} a subset of the processes reported in Table~\ref{tab:limits} with indicated the limit expected in the near future, {\em i.e.}~within $5-10$~years. On the experimental side, the message that we tried to convey throughout our discussion is that the most sensitive searches require the design and realization of dedicated experiments as well as a deep understanding of the processes and of their backgrounds. 
\acknowledgments
LC is pleased to thank his many friends and collaborators, from which he learnt about the topics discussed here, in particular:
A.~Crivellin, J.~Jones-Perez, A.~Masiero, T.~Ota, P.~Paradisi, S.~Pokorski, A.~Romanino, S.~Vempati, O.~Vives, and R.~Ziegler. 
GS would like to thank A.~M.~Baldini, C.~Bemporad, F.~Cei, L.~Galli, M.~Grassi, D.~Nicol\`o and A.~Papa for the useful discussions on the experimental aspects of CLFV searches, some of which are ongoing since more than fifteen years!


\appendix
\section[SUSY amplitudes for $\ell_i \to \ell_{j}\gamma$ in Mass Insertion Approximation]{Supersymmetric amplitudes for $\ell_i \to \ell_{j}\gamma$ in Mass Insertion Approximation}
In this Appendix, we report the MIA expressions of the amplitudes appearing in Eq.~(\ref{eq:BR_LFV}), taken from \cite{Calibbi:2015kja}.
We can write separately the different contributions as follows:
\begin{align}
\mathcal{A}_{L} ~=~ &(\mathcal{A}^{c_1}_L)_{\small SU(2)}  + (\mathcal{A}^{n_1}_L)_{\small SU(2)} + (\mathcal{A}^{n_1}_L)_{\small U(1)} +\nonumber\\
& (\mathcal{A}^{c_2}_L)_{\small SU(2)} +(\mathcal{A}^{n_2}_L)_{\small SU(2)}+(\mathcal{A}^{n_2}_L)_{\small U(1)} +(\mathcal{A}^{n_3}_L)_{ U(1)} ,\\
\mathcal{A}_{R} ~=~& (\mathcal{A}^{n_1}_R)_{\small U(1)} +(\mathcal{A}^{n_2}_R)_{\small U(1)} +(\mathcal{A}^{n_3}_R)_{ U(1)},
\end{align}
where the subscript indicate whether $SU(2)_L$ or $U(1)_Y$ interactions are involved, and the superscript whether charginos
or neutralinos run in the loop.

A first class of contributions is given by diagrams in which the chirality flip is realised 
on the external fermion line: 
\begin{align}
(\mathcal{A}^{c_1}_L)_{\small SU(2)} & = \frac{\alpha_2}{4 \pi} \frac{\Delta_{LL}}{m_L^4}
f_{1c}\left( \frac{|M_2|^2}{m^2_L}\right)\,,
\\
(\mathcal{A}^{n_1}_L)_{\small SU(2)} & =
\frac{\alpha_2}{4 \pi}\frac{\Delta_{LL}}{m_L^4} f_{1n} \left( \frac{|M_2|^2}{m^2_L}\right)\,,
\\
(\mathcal{A}^{n_1}_L)_{\small U(1)} & = \frac{\alpha_Y}{4 \pi} \frac{\Delta_{LL}}{m_L^4}
f_{1n} \left( \frac{|M_1|^2}{m^2_L}\right)\,,
\\
(\mathcal{A}^{n_1}_R)_{\small U(1)} & = \frac{\alpha_Y}{\pi} \frac{\Delta_{RR}}{m_R^4}
f_{1n} \left( \frac{|M_1|^2}{m^2_R}\right)\,,
\end{align} 
where $\alpha_2 =g/4\pi$, $\alpha_Y =g'/4\pi$, $M_1$ ($M_2$) is the Bino (Wino) mass,
$m_L$ ($m_R$) is a common LH (RH) slepton mass.
 
In presence of Higgsinos, the chirality flip can occur on the Yukawa (i.e.~Higgsino-lepton-slepton) vertex. The resulting
contributions are:
\begin{align}
(\mathcal{A}^{c_2}_L)_{\small SU(2)} & =  \frac{\alpha_2}{4 \pi} \frac{\Delta_{LL}}{m_L^4}
\left[ a_2 f_{2c}\left( \frac{|M_2|^2}{m^2_L}\right) - b_2 f_{2c} \left( \frac{|\mu|^2}{m^2_L}\right) \right]\,,
\\
(\mathcal{A}^{n_2}_L)_{\small SU(2)} & =
\frac{\alpha_2}{4 \pi}\frac{\Delta_{LL}}{m_L^4}
\left[a_2 f_{2n} \left( \frac{|M_2|^2}{m^2_L}\right) - b_2 f_{2n} \left( \frac{|\mu|^2}{m^2_L}\right) \right]\,,
\\
(\mathcal{A}^{n_2}_L)_{\small U(1)} & =
- \frac{\alpha_Y}{4 \pi} \frac{\Delta_{LL}}{m_L^4}
\left[a_1 f_{2n} \left( \frac{|M_1|^2}{m^2_L}\right) - b_1 f_{2n} \left( \frac{|\mu|^2}{m^2_L}\right) \right]\,,
\\
(\mathcal{A}^{n_2}_R)_{\small U(1)} & =
\frac{\alpha_Y}{2 \pi} \frac{\Delta_{RR}}{m_R^4}
\left[a_1 f_{2n} \left( \frac{|M_1|^2}{m^2_R}\right) - b_1 f_{2n}\left( \frac{|\mu|^2}{m^2_R}\right) \right]\,.
\end{align}
where
\begin{align}
a_1 &= \frac{(|M_1|^2 + \mu M_1 t_\beta)}{|M_1|^2 - |\mu|^2}\,,\qquad
a_2 = \frac{(|M_2|^2 + \mu M_2 t_\beta)}{|M_2|^2 - |\mu|^2}\,,
\\
b_1 &= \frac{(|\mu|^2 + \mu M_1 t_\beta)}{|M_1|^2 - |\mu|^2}\,,\qquad~~\,
b_2 = \frac{(|\mu|^2 + \mu M_2 t_\beta)}{|M_2|^2 - |\mu|^2}\,.
\end{align}

Finally, the amplitudes corresponding to a chirality flip on the internal sfermion line read
\begin{align}
(\mathcal{A}^{n_3}_L)_{ U(1)} & = \frac{\alpha_Y}{4 \pi} \frac{M_1}{m_{\mu}}
\frac{\Delta_{RL}\Delta_{LL}}{(m_L^2 - m_R^2)}
\left[ \frac{2 f_{2n}\left( \frac{|M_1|^2}{m^2_L}\right)}{m_L^4} + \frac{1}{m_L^2 - m_R^2}
\left( \frac{f_{3n} \left( \frac{|M_1|^2}{m^2_L}\right)}{m_L^2} - \frac{f_{3n} \left( \frac{|M_1|^2}{m^2_R}\right)}{m_R^2} \right) \right]
\nonumber \\
& -  \frac{\alpha_Y}{4 \pi} \frac{M_1}{m_{\mu}}  \frac{\Delta_{RL}}{m_L^2 - m_R^2}
\left[ \frac{f_{3n} \left( \frac{|M_1|^2}{m^2_L}\right)}{m_L^2} - \frac{f_{3n} \left( \frac{|M_1|^2}{m^2_R}\right)}{m_R^2} \right]\,,
\end{align}
\begin{align}
(\mathcal{A}^{n_3}_R)_{ U(1)} & = \frac{\alpha_Y}{4 \pi} \frac{M_1}{m_{\mu}}
\frac{\Delta_{LR}\Delta_{RR}}{(m_R^2 - m_L^2)}
\left[ \frac{2 f_{2n}\left( \frac{|M_1|^2}{m^2_R}\right)}{m_R^4} + \frac{1}{m_R^2 - m_L^2}
\left( \frac{f_{3n} \left( \frac{|M_1|^2}{m^2_R}\right)}{m_R^2} - \frac{f_{3n} \left( \frac{|M_1|^2}{m^2_L}\right)}{m_L^2}\right)
\right]
\nonumber \\
& -  \frac{\alpha_Y}{4 \pi} \frac{M_1}{m_{\mu}}  \frac{\Delta_{LR}}{m_R^2 - m_L^2}
\left[ \frac{f_{3n} \left( \frac{|M_1|^2}{m^2_R}\right)}{m_R^2} - \frac{f_{3n} \left( \frac{|M_1|^2}{m^2_L}\right)}{m_L^2} \right]\,,
\end{align}
where the left-right slepton mixing can be flavour-violating as well, i.e.~given by $(\Delta_{LR})_{i\neq j}$, 
or flavour-conserving, namely $\propto (\Delta_{LR/RL})_{ii} \sim m_{\ell_i} \mu\tan\beta$. In this latter case the 
flavour violation
is provided by $(\Delta_{LL})_{i\neq j}$ or $(\Delta_{LR})_{i\neq j}$.

The loop functions appearing in the above expressions read:
\begin{eqnarray}
f_{1n}(x)&=&\frac{-17x^3+9x^2+9x-1+6x^2(x+3)\ln x}{24(1-x)^5}\,,\\
f_{2n}(x)&=&\frac{-5x^2+4x+1+2x(x+2)\ln x}{4(1-x)^4}\,,\\
f_{3n}(x)&=&\frac{1+2x\ln x-x^2}{2(1-x)^3}\,,\\
f_{1c}(x)&=&\frac{-x^3-9x^2+9x+1+6x(x+1)\ln x}{6(1-x)^5}\,,\\
f_{2c}(x)&=&\frac{-x^2-4x+5+2(2x+1)\ln x}{2(1-x)^4}\,.
\end{eqnarray}
%

%
\section{Expression for the radiative muon decay}
We report here for reference the form of the energy spectrum of the electron from the muon decay including radiative corrections to the first order in $\alpha$~\cite{kinoshita-sirlin2}. This is given by
\begin{equation}
\frac{ {\rm d} \Gamma} {{\rm d} x} = \frac{m_\mu}{24 \pi^3} W_{e \mu}^4 G_F^2 x^2
\left[ \left( 3 - 2x \right) + \frac{\alpha}{2\pi} f(x) \right],
\end{equation}
where $W_{e \gamma} = (m_\mu^2 - m_e^2)/(2 m_\mu)$, $x = E_e/W_{e \mu}$, $E_e$ is the electron/positron energy $m_e \leq E_e \leq W_{e \mu}$ and the small electron mass is neglected.
The radiative correction $f(x)$ is given by 
\begin{align}
f(x) = (6 -4x)R(x) + 6(1-x) \ln x + \frac{1-x}{3x^2} \times \\
\times \left[ (5 + 17x -34x^2) (\omega +\ln x) -22 x +34x^2 \right],
\end{align}
where $\omega = \ln (m_\mu/m_e)$ and $R(x)$ is given by
\begin{align}
R(x) = 2 \sum_{n=1}^{\infty} \frac{x^n}{n^2} - \frac{\pi^2}{3} -2 +
\omega \left[  \frac{3}{2} + 2 \ln \left( \frac{1-x}{x} \right) \right] \\
-\ln x (2 \ln x -1 ) + \left( 3 \ln x - 1 - \frac{1}{x} \right) \ln (1-x).
\end{align}

\end{document}